\documentclass[12pt,usenames]{article}
\pdfoutput=1
\usepackage[utf8]{inputenc}
\usepackage{amssymb}
\usepackage{amsfonts}
\usepackage{amsmath}
\usepackage{svg}
\usepackage{latexsym}
\usepackage{esint}
\usepackage{cite}
\usepackage{subcaption}
\usepackage{framed}
\definecolor{shadecolor}{rgb}{0.9,0.9,0.95}
\usepackage{setspace}
\usepackage{amssymb,graphicx}
\usepackage{float}
\renewcommand{\a}{\alpha}
\newcommand{\ad}{{\dot\alpha}}

\newcommand{\bd}{{\dot\beta}}
\newcommand{\e}{\epsilon}

\newcommand{\ve}{\varepsilon}

\newcommand{\ib}{{\overline i}}
\newcommand{\jb}{{\overline j}}
\newcommand{\xb}{{\overline x}}
\definecolor{darkblue}{cmyk}{0.9,0.9,0,0}
\definecolor{darkred}{rgb}{0.6,0,0.3}
\usepackage{graphicx} 
\usepackage{epsfig}
\usepackage[setpagesize=false,pagebackref=false, linktocpage, bookmarksopen=true, colorlinks=true, linkcolor=darkblue,citecolor=darkblue,urlcolor=darkblue]{hyperref}
\usepackage{hyperref}
\usepackage{comment}

\renewcommand{\thefootnote}{\arabic{footnote}}

\def\eqref#1{(\ref{#1})}

\definecolor{myred}{rgb}{1,0.6,0.6}
\definecolor{myblue}{rgb}{0.6,0.6,1}

\newcommand{\beq}{\begin{equation}}
\newcommand{\eeq}{\end{equation}}

\DeclareMathAlphabet{\mathbbold}{U}{bbold}{m}{n}
\begin{document}
\numberwithin{equation}{section}
\thispagestyle{empty}

\renewcommand{\thefootnote}{\fnsymbol{footnote}}
\setcounter{page}{1}
\setcounter{footnote}{0}
\setcounter{figure}{0}

\newcommand{\tf}[1]{\textcolor{red}{#1}}

\begin{center}
$$$$
{\Large\textbf{\mathversion{bold}
Superstring theory in $AdS_5\times S^5$ background: half-BPS vertex operators
}\par}

\vspace{1.3cm}

\textrm{Thiago Fleury, Lucas N. S. Martins, Dennis Zavaleta}
\\ \vspace{1.2cm}
\footnotesize{\textit{
Instituto Cienovus, \\
Av. Pres. Juscelino Kubitschek, 1327, 4 andar, Conj 41 \\ 
CEP: 04543-011, São Paulo, Brazil.
}}

\par\vspace{1.5cm}

\textbf{Abstract}\vspace{2mm}
\end{center}

In this paper, we construct all vertex operators associated with 
the gravity supermultiplet in the $AdS_5 \times S^5$ background, relevant for worldsheet scattering amplitude computations, via global 
$PSU(2,2|4)$ transformations. 
Although 
one such operator was previously constructed
in the pure spinor formalism using a special superspace parametrization, implementing the full global symmetries proved challenging due to the mixing of bosonic and fermionic coordinates. We resolve this issue by introducing a judicious sequence of reparametrizations that disentangle these coordinates. We then advance the formulation of superstring theory in this background by constructing the corresponding integrated 
vertex operators 
in both the pure spinor and the B-RNS-GSS formalisms.

\noindent

\setcounter{page}{1}
\renewcommand{\thefootnote}{\arabic{footnote}}
\setcounter{footnote}{0}
\setcounter{tocdepth}{2}
\newpage
\tableofcontents

\parskip 5pt plus 1pt   \jot = 1.5ex

\section{Introduction}

Despite progress on the conformal side of the AdS/CFT correspondence, the string side of the correspondence for backgrounds supported by Ramond–Ramond (RR) fluxes remains opaque due to a lack of first-principles computational tools. For example, although 4D $\mathcal{N}=4$ SYM has seen remarkable advances through integrability, bootstrap, and localization, the presence of RR flux has so far obstructed comparable progress in type IIB superstring theory on the $AdS_5\times S^5$ background.
In this paper, we focus on type IIB superstrings in the $AdS_5 \times S^5$ background.

The problem with RR flux in superstring theory is that, in the Ramond–Neveu–Schwarz (RNS) formalism \cite{Friedan:1985ge}, adding integrated RR vertex operators to the sigma‑model action requires half-integer pictures and introduces branch cuts (see, for example, the end of Section 2 in \cite{Witten:1992fb}), which obstructs the straightforward construction of RNS sigma-models for RR backgrounds. In this setting, one is limited to treating the Ramond–Ramond flux perturbatively within the superstring field theory (SFT) framework, as in \cite{Cho:2025coy}. In the Green–Schwarz (GS) formalism \cite{Metsaev:1998it}, a sigma‑model for RR backgrounds does exist, but the action cannot be covariantly quantized due to kappa symmetry \cite{Green:1983wt}. Non-covariant quantizations are possible \cite{Frolov:2002av}, but they involve intricate computations and risk losing states in curved backgrounds \cite{Berkovits:2002zv}.

As a super-Poincaré covariant quantization of the superstring, the pure spinor formalism \cite{BerkovitsPureOriginal} offers a promising alternative. It resolves the covariance problem of the GS formalism and the 
RR-sector idiosyncrasies of the RNS formalism, which are tied to the complicated way in which spacetime supersymmetry appears in the latter. In \cite{BerkovitsPureOriginal, Berkovits:2004xu}, a covariantly quantizable sigma-model action for the $AdS_5\times S^5$ background was constructed using this formalism as the unique worldsheet action that is both 
BRST and $PSU(2,2|4)$ invariant. In flat space, the pure spinor formalism has been related to a non-minimal version of the RNS formalism, known as the B-RNS-GSS formalism \cite{Berkovits:2021xwh,Berkovits:2022qhc}. 
There has been growing interest in the B-RNS-GSS formalism and its extension to curved backgrounds \cite{Berkovits:2022dbm, Chandia:2023eel, Mikhailov:2025kuy}.
The latter also allows for the construction of a worldsheet action that is both BRST and $PSU(2,2|4)$ invariant, which can, in principle, be covariantly quantized; however its quantum consistency has yet to be established. 

As usual in covariant quantization, closed-string vertex operators in the pure spinor lie in the ghost-number-two BRST cohomology, while open-string operators lie in the ghost-number-one cohomology. In flat space, as in the RNS formalism, closed-string vertex operators  in the pure spinor formalism are constructed as left–right products of open ones. This left–right factorization, however, is spoiled in the $AdS_5\times S^5$ background, where no Polyakov–Wiegmann identity exists \cite{Metsaev:1998it}, necessitating a novel construction of vertex operators. Since it has been shown in \cite{IntegratedVertex,LaxIntegrated, Chandia:2025nxx} how to construct both unintegrated and integrated vertex operators in the pure spinor and B-RNS-GSS formalisms from the unintegrated ones in the pure spinor formalism, we will focus on the latter.

In a series of papers \cite{Berkovits:2019ulm,Berkovits:2019rwq,Martins:2019wwi,VertexOld,Mikhailov:2020nlg}, a new method for constructing unintegrated vertex operators in pure spinor formalism for closed‐string states was devised: instead of left–right factorization, one mixes left‐ and right‐moving spinors into analytic and anti‐analytic combinations. This allows the decoupling of 
twenty-four fermionic variables when constructing vertex operators for the bottom states of the supergravity multiplets, drastically simplifying the construction. 
In \cite{VertexOld}, this method 
together with numerical techniques were applied to yield, for the first time, an explicit non-trivial pure spinor vertex operator in this background.

One problem of the vertex operators of \cite{VertexOld} 
is that, due to the $PSU(2,2|4)$ superalgebra, the $SU(2,2)\times SU(4)$ action on them mixes bosonic and fermionic coordinates. This mixing arises in superspace parametrizations that decouple
twenty-four fermionic coordinates from the bottom states of the 
supergravity multiplet. Brute-forcing the fermionic expansion for each $SU(2,2)\times SU(4)$ transformation is computationally intensive because of the curse of signs, which exponentially amplifies the combinatorics of sign tracking and cancellations—an unfortunate drawback, since these transformations are crucial for generating multiple vertex operators from the one constructed in \cite{VertexOld}.

We solve this problem here
through a judicious sequence of reparametrizations that disentangle bosons from fermions under the $SU(2,2)\times SU(4)$ transformations. We then generate multiple unintegrated vertex operators by acting with these transformations on the operator constructed in \cite{VertexOld}. 
The final expression for the vertex operators 
depends on two null polarization vectors 
$(x_0)_{RS}$ and $(y_0)_{IJ}$ where $R,S$ and $I,J$ are 
 $SU(2,2) \times SU(4)$ indices and the transformations only rotate these vectors.
 Knowing the bottom state of the supermultiplet, it is possible to obtain 
 any other state by applying supersymmetry transformations. In the final supercoset parametrization, these transformations are also simple and are at most quadratically 
in the fermionic variables.    
Having obtained multiple unintegrated vertex operators, we construct their integrated versions via descent relations, as in \cite{IntegratedVertex,LaxIntegrated}, which involve worldsheet derivatives and positive conformal weight variables. We also construct the unintegrated and integrated vertex operators for the B-RNS-GSS formalism using \cite{Chandia:2025nxx}. 
Since we focus on the leading supergravity contribution, 
the computation is insensitive to worldsheet normal ordering and $\alpha'$ corrections. Issues of normal ordering and $\alpha'$ corrections will be addressed elsewhere.

One motivation for this work is to compute worldsheet superstring scattering amplitudes in the $AdS_5 \times S^5$ background \cite{workprogress}. To date, pure‑spinor results exist only near the boundary; for example, \cite{TalesNathan} used the near‑boundary approximation \cite{BerkovitsFleury} 
of the closed‑string vertex operators to compute amplitudes between supergravity and super‑Yang–Mills states on D3‑branes at the $AdS_5$ boundary. The results agree with the expectations of the holography dictionary of \cite{Witten:1998qj,Klebanov:1999tb,Maldacena:2003nj}. The results of our paper are a first step toward extending these computations into the interior of $AdS_5\times S^5$, 
aiming to test non-trivial  CFT predictions and producing new results. 
The simplest scattering amplitudes, those involving two or three massless states, are protected by supersymmetry from $\alpha'$ corrections. Conversely, the correlators involving two‑ and three‑point  half‑BPS correlators in $\mathcal{N}=4$ SYM are protected from loop corrections. 
Starting with four massless states, all amplitudes are expected to receive 
non‑trivial
$\alpha'$ corrections and can be compared with the CFT predictions 
\cite{Aprile:2020mus,AldayHansen}. 
Numerous CFT predictions for amplitudes at various multiplicities and orders in $\alpha'$ exist, e.g. \cite{5corrected,Vasco6,Fernandes:2025eqe,Aprile:2026uxe}, for recent results. 

The results in \cite{AldayHansenSilvaI,AldayHansenSilvaII,AldayHansenSilvaIII,
AldayHansen} use a small-curvature (flat-space) expansion. The leading term is the flat-space Virasoro–Shapiro amplitude, and corrections are fixed by demanding consistency of a judicious basis of functions for the worldsheet integrand with the OPE in $\mathcal{N}$=4 SYM. Existing $\mathcal{N}$=4 SYM data obtained from integrability, localization and bootstrap is used for consistency checks and as additional input.
To test these nontrivial predictions in the pure spinor formalism, one requires not only unintegrated vertex operators but at least one integrated vertex operator, and an amplitude prescription that prescribes how to treat the zero modes in the worldsheet correlator; in such a computation, the $\alpha'$ corrections arise from higher‑order poles in the OPEs of world‑sheet variables and from normal ordering. 
This paper focuses on the construction of vertex operators, leaving normal ordering aside, and defers the amplitude prescription and OPEs to future work  \cite{workprogress}.
Note that the amplitude prescription can already be verified by comparing the well-known results for two- and three-point %
amplitudes.  

 The paper is organized as follows.
Section \ref{sectionreviewing} reviews superstrings 
in $AdS_5 \times S^5$ using the pure spinor formalism and the 
worldsheet variables of \cite{VertexOld}, which we call the $g_{\rm{old}}$  parametrization. 
In particular, the BRST transformations and some properties 
of both the unintegrated and integrated vertex operators are reviewed.
The extension of the results to the B-RNS-GSS formalism is made at the end of the section.  
Section \ref{The PSU(2,2|4) transformations} contains 
the $SU(2,2) \times SU(4)$ transformations of the $g_{\rm{old}}$ variables.
They involve several orders in the fermionic variables, and only infinitesimally displaced 
vertex operators are easy to construct in this parametrization.
Starting from Subsection \ref{sucovariance}, we use new sets of variables
that we call $g_{\rm{new}}$ and $g^{\prime}_{\rm{new}}$  parametrizations. 
Their $SU(2,2) \times SU(4)$ transformations are at most quadratic in the fermionic variables. 
The reparametrizations $g_{\rm{old}} \leftrightarrow g_{\rm{new}}  \leftrightarrow g^{\prime}_{\rm{new}}$ are discussed in Subsection 
\ref{reparametrization old to new}. 
The vertex operators obtained by changing the variables of the vertex operators of 
\cite{VertexOld} to the new variables are given in Section
\ref{The ideal coset vertices}. 
 In the final expressions, the vertex operators, 
 are parametrized by two null polarization vectors.   
 Section \ref{The ideal coset vertices} ends with a discussion of several properties 
of the vertex operators.
The paper ends with a discussion of the results and future applications.
Several Appendices provide additional details, explicit expressions, and our conventions.    

\subsection{Overview}

Here we outline the structure of the paper in more detail to guide the reader through our results. As mentioned in the Introduction, our primary goal is to streamline the $SU(2,2)\times SU(4)$ transformations of half-BPS vertex operators in superstring theory, both in the pure spinor formalism and in the B-RNS-GSS formalism. We focus on the pure spinor formalism, since the B-RNS-GSS vertex operators are constructed by acting with covariant derivatives on the bispinor superfield appearing in the unintegrated vertex operators of the pure spinor formalism, in a manner analogous to the pure spinor integrated vertex operators. In other words, we focus on the $SU(2,2)\times SU(4)$ transformations of the bispinor superfield associated with half-BPS states, especially those annihilated by 24 supersymmetries (the bottom component of the half-BPS supermultiplet).

Aligned with our goals, we begin by reviewing the pure spinor formalism in the $AdS_5\times S^5$ background, with minimal discussion of its relation to the B-RNS-GSS formalism. Of course, in \ref{The AdS5S5 superspace}, we start by reviewing the superparametrization of the $AdS_5\times S^5$ superspace chosen for the construction of the half-BPS vertex operator in \cite{VertexOld}, with comments on its bosonic truncation \ref{Bosonic truncation}, stereographic inversions, and its generalization to the full superspace in \ref{Inversion}. We then take the opportunity to clarify common subtleties, including details of the holographic map relating our vertex operators 
to CFT states and local operators in \ref{Boundary-to-bulk propagator}, as well as the role of isotropy gauge transformations in parametrizations of homogeneous superspaces and their covariant superfields in \ref{Covariant superfields}. Subsequently, we introduce the covariant derivatives of these superspaces in \ref{covariant derivatives}. With this toolkit for the $AdS_5\times S^5$ superspace in hand, we begin resolving the distinction between truncation and reduction in \ref{Reductions and truncations}, so that we can define the boundary superspace and its superfields in \ref{Higher componetns transformations}, as well as the two $AdS_5\times S^5$ superfields describing the half-BPS supermultiplet in \ref{superfields for half-BPS supermultiplet}, together with their relation.

After establishing the $AdS_5\times S^5$ superspace, its covariant superfields and covariant derivatives, its boundary superspace, and how these constructions relate to the half-BPS supermultiplet, we begin by reviewing the zero-mode content of the pure spinor formalism in \ref{The pure spinor formalism: zero conformal weight}, which is sufficient for the construction of unintegrated vertex operators and for the computation of both two- and three-point scattering amplitudes. The important point here is that the BRST transformations, when restricted to zero modes, become a linear combination of covariant derivatives dressed with
worldsheet variables that are covariant under the isotropy group 
\ref{BRST transformations of the zero-modes}. In the pure spinor formalism, these worldsheet variables are precisely the pure spinors, which are reviewed in \ref{Pure spinors subsubsection}, together with several equivalent forms of their constraints. Once the superspace parametrization is fixed, the BRST transformations must be accompanied by gauge-restoring transformations that rotate the pure spinors, as reviewed in \ref{Gauge-restoring transformations}. These gauge-restoring transformations are important for maintaining the nilpotency of the BRST transformations after gauge fixing, as shown in \ref{BRST nilpotency on the zero modes}. This allows us to discuss unintegrated vertex operators in \ref{Unintegrated vertex operators} and then focus in particular on the half-BPS bottom components in \ref{previous paper result}. We end our review of the zero-mode content of the pure spinor formalism by commenting on the mass-dimension tower of superfields 
constructed from the covariant derivatives identities 
in \ref{Mass-dimension tower via Jacobi identities}, which prepares the way for moving on to the non-zero-mode sector, where integrated vertex operators live.

We conclude our review in \ref{The integrated vertex operator} by discussing the non-zero-mode sector of the pure spinor formalism and its relation to the B-RNS-GSS formalism. Of course, in \ref{sigma model action} we begin by establishing the sigma-model action for the superstring in the $AdS_5\times S^5$ superspace. We then generalize the BRST transformations to the non-zero modes, while keeping only the leading order in $\alpha'$ in \ref{The BRST transformations of the non-zero modes}. In the non-zero-mode sector, it is important to work modulo the sigma-model equations of motion, so we explicitly present these equations in \ref{equations of motion worldsheet}.%
 With the equations of motion and BRST transformations for the non-zero modes in hand, we review the construction of integrated vertex operators from unintegrated ones via descent relations in \ref{descent relations fundamentals}, obtaining, in \ref{Integrated Vertex Operators} and in \ref{SO to USP}, expressions for the integrated vertex operators in terms of the bispinor superfield appearing in the unintegrated vertex operators. Finally, we briefly explain how the B-RNS-GSS vertex operators can be obtained from this bispinor superfield in \ref{Vertex operators in B-RNS-GSS formalism}. 

We are ready now to tackle our problem, namely, to systematize the $SU(2,2)\times SU(4)$ transformations of half-BPS vertex operators. This is the content of Section \ref{The PSU(2,2|4) transformations}. We begin with the superspace parametrization of \cite{VertexOld}, reviewed in \ref{The AdS5S5 superspace}. To exhibit the problem, we introduce an infinitesimal, non-trivial $SU(2,2)\times SU(4)$ transformation (one that acts non-trivially on our vertex operators), only to find in \ref{Finite transformations} that exponentiating it to a finite transformation produces a complicated mixing with the fermionic coordinates, which we work out up to second order. This boson–fermion mixing is an artifact of 
the chosen parametrization of \cite{VertexOld}.

This motivates the search for a better superspace parametrization, culminating in \ref{sucovariance}. Its finite global transformations are much more transparent, as shown in \ref{TheFiniteTransformations}.
To transfer our vertex operators from the old parametrization to the new ones, and thus take advantage of the more systematic $SU(2,2)\times SU(4)$ action, we must establish the reparametrization map between the two systems of supercoordinates. This is done in \ref{reparametrization old to new}. Without the special matrices $\mathbb{M}$ and $\mathbb{N}$ found in \ref{Rearranging new}, and the rearrangement of the old parametrization placing the bosonic factors on the far left of the supercoset representative in \ref{Rearranging old}, we would not have been able to find the explicit reparametrizations in \ref{Mapping the fermionic variables} and in \ref{Mapping the bosonic variables and compensators}. Furthermore, because of the transparent manner in which the $SU(2,2)\times SU(4)$ transformations act in this new parametrizations, we are free to choose whether the fermionic coordinates appear on the left or on the right of the supercoset representative, as shown in \ref{coordinatesfinal}.

The next step is obvious: apply the reparametrizations found in Section \ref{The PSU(2,2|4) transformations} to the vertex operators constructed in \cite{VertexOld}. This is the most computationally intensive part of the paper and is carried out almost entirely in \emph{Mathematica}. Accordingly, Section \ref{The ideal coset vertices} mainly presents the portions of the result that fit in the main text, with substantial support from the Appendix \ref{vertexgprime}; this is the content of Subsection \ref{vertexoperatorsfromchangingcoordinates}. In Subsection \ref{recovering}, however, we uncover an interesting opportunity: the possibility of expressing the half-BPS vertices 
using two null polarization vectors.  
At this stage, the fermionic coordinates are already $SU(2,2)\times SU(4)$ covariant. The $SU(2,2)\times SU(4)$-covariant bosonic coordinates are introduced in \ref{su22 covariant bosonic variables}, where the null polarization vectors introduced in \ref{Null polarization vectors} can couple. This investment yields, in \ref{su22 covariant vertex operators}, 
vertex operators transforming straightforwardly under the bosonic isometries, which are adorned with the 
covariant 
BRST operator in \ref{purespinordecomposition} and the covariant spinors in \ref{Covariant pure spinors}.

With the half-BPS vertex operators 
parametrized by two null polarization vectors, we are ready to check expected properties of these vertices in \ref{Propertiesofthevertexoperators} as a validation of our results. We comment on $SU(4)$ covariance in \ref{Comments about $SU(4)$ covariance}, the action of supersymmetries 
in \ref{Applying supersymmetries transformations: the supermultiplet}, the flat-space limit in \ref{Flat space limit}, the plane-wave construction in \ref{Plane-wave background}, %
 gauge symmetries in \ref{Further comments about the gauge degrees of freedom} and primary vertex operators in \ref{Primary vertex operators}. In \ref{Discussion} we conclude.

\section{Reviewing the $AdS_5 \times S^5$ superstrings} 
\label{sectionreviewing}

In this first section, we review aspects of the pure spinor formalism in the $AdS_5\times S^5$ background and establish our notation. We do so in a slightly unconventional way, tailored to the present work as  
related to 
\cite{VertexOld}. Up to Subsection \ref{The integrated vertex operator}, we focus on the superstring zero conformal weight variables, 
which suffice for constructing unintegrated supergravity vertex operators; the relevant BRST transformations and operators then appear as supervector fields and superfields on the target superspace, respectively. 

In Subsection \ref{The integrated vertex operator}, we include the non-zero 
conformal weight variables
and present the action functional, the BRST charge, and the integrated vertex operators of the pure spinor formalism. We show how our results extend to this setting—particularly the construction of integrated vertices from unintegrated ones—following \cite{IntegratedVertex}. Finally, as noted in \cite{Chandia:2025nxx}, just as integrated vertices are determined by unintegrated ones, the B-RNS-GSS vertex operators are likewise fixed by the unintegrated vertex operators of the pure spinor formalism.

\subsection{The $AdS_5 \times S^5$ superspace}
\label{The AdS5S5 superspace}

To parametrize the $AdS_5 \times S^5$ superspace, note that it is a homogeneous superspace \cite{Metsaev:1998it}:
\begin{equation}
\frac{PSU(2,2|4)}{USp(2,2) \times USp(4)} \, ,
\label{supercoset}
\end{equation}
where $PSU(2,2|4)$ is the isometry supergroup and $USp(2,2) \times USp(4)$ its isotropy %
group. The bosonic truncation of this homogeneous superspace yields the Cartesian product $AdS_5\times S^5$, where
\begin{equation} 
AdS_5 = \frac{SU(2,2)}{USp(2,2)} \, , \qquad S^5 = \frac{SU(4)}{USp(4)} \, . 
\label{CosetsAdS5S5}
\end{equation}
For the usual real forms, $USp(2,2)\simeq \mathrm{Spin}(4,1)$ and $USp(4)\simeq \mathrm{Spin}(5)$. 
Moreover $SU(2,2) \simeq \mathrm{Spin}(4,2)$ and $SU(4) \simeq \mathrm{Spin}(6)$. 
Since this is a supercoset space, its  parametrization essentially gauge-fixes the isotropy group. The superspace parametrization in which the unintegrated supergravity 
vertex operators of \cite{VertexOld} were constructed is given by
\begin{equation}
g_{\rm{old}} = e^{(\psi^{\alpha}_i q^i_{\alpha} + \theta^{\bar{i}}_{\dot\alpha} q^{\dot\alpha}_{\bar{i}} + x^{\alpha}_{\dot\alpha} K^{\dot\alpha}_{\alpha} + y^{\bar{i}}_{i} K^{i}_{\bar{i}} )} 
e^{(\psi^{\dot\alpha}_i q^i_{\dot\alpha} + \theta^{\bar{i}}_{\alpha} q^{\alpha}_{\bar{i}})}
e^{(\psi^{\alpha}_{\bar{i}} q^{\bar{i}}_{\alpha}+\theta^i_{\dot\alpha} q^{\dot\alpha}_i) }
e^{(\psi^{\dot\alpha}_{\bar{i}} q^{\bar{i}}_{\dot\alpha}+\theta^i_{\alpha} q^{\alpha}_i )}
e^{ z \Delta + w J} \, , 
\label{OldCosetRepresentative}
\end{equation}
where $\alpha = 1,2$ and $\dot\alpha = 3,4$ are the $SL(2,\mathbb{C})$ indices corresponding to the spinorial indices of the $SO(1,3)$ subgroup of  $SO(4,2)$, which acts on the $AdS_5$ factor. Similarly, $i=1,2$ and $\overline i=3,4$ are the $SU(2)\times SU(2)$ indices corresponding to spinorial indices of the $SO(4)$ subgroup of $SO(6)$, which acts on the $S^5$ factor. The supercharges are denoted by the letter $q$. The letter $K$ denotes conformal boosts and certain $SO(6)$ rotations that are not contained in $SO(5)$ (for more details on notation on $\mathfrak{psu}(2,2|4)$ superalgebra, see Appendix \ref{apsupermatrices}). The $\Delta$ denotes the dilatation generator, 
while $J$ denotes $U(1)$ generator in $\mathfrak{so}(6)$ that is not in $\mathfrak{so}(5)$.

An advantage of working with the $g_{\rm{old}}$ superspace parametrization is that, for a specific half-BPS state—defined by a choice of a point on the spacetime boundary and an $R$-charge null orientation— its 
vertex operator \cite{VertexOld} is independent of the fermionic coordinates $\psi^{\alpha}_i$, $\theta^{\bar{i}}_{\dot\alpha}$, $\psi^{\dot\alpha}_i$, $\theta^{\bar{i}}_{\alpha}$, $\psi^{\alpha}_{\bar{i}}$, $\theta^i_{\dot\alpha}$ and the bosonic coordinates $x^{\alpha}_{\dot\alpha}$ and $x^{\bar{i}}_i$. This property was heavily leveraged in \cite{VertexOld}. However, this comes with a trade-off: instead of $SU(2,2)\times SU(4)$ covariant parameters, one has 
\begin{equation}
SO(1,1)_{\Delta}\times SL(2,\mathbb{C})\times S(U(2)\times U(2))    
\end{equation}
covariant ones. Consequently, the $PSU(2,2|4)$ transformations act in a complicated manner in this parametrization, as we will see in Section \ref{The PSU(2,2|4) transformations}.

The $PSU(2,2|4)$ transformations are essential, as they enable the generation of multiple vertex operators from a single one, corresponding to different states. In Section \ref{The ideal coset vertices}, we construct several supergravity vertex operators from the vertex operators constructed in \cite{VertexOld} by first performing the superspace reparametrization of Section \ref{sucovariance} that maps the old parametrization 
$g_{\rm{old}}$ to the new parametrization $g_{\rm{new}}$  of Section \ref{sucovariance}, which is $SU(2,2)\times SU(4)$ covariant.

\subsubsection{Bosonic truncation}
\label{Bosonic truncation}

Before reviewing the $AdS_5\times S^5$ superspace, let us recall its bosonic truncation. Since $AdS_5$ and $S^5$ factorize and have a similar structure, we focus on the $AdS_5$ factor. The bosonic line element obtained by truncating the superspace parametrization of 
\eqref{OldCosetRepresentative} is
\begin{equation}
ds^2_{\rm{AdS}}  = dz^2 + e^{2z} \epsilon_{\alpha \beta}  \epsilon^{\dot\alpha \dot\beta} dx_{\dot\alpha}^{\alpha} dx_{\dot\beta}^{\beta} \, , 
\end{equation} 
where the $AdS_5$ boundary is at $z \rightarrow \infty$. Using the $SO(1,3)$ Pauli matrices, we can eliminate spinorial indices and obtain
\begin{equation}
\label{poincare}
ds^2_{\rm{AdS}}  = \frac{d\rho^2 +dx^{\mu} dx_{\mu}}{\rho^ 2} \, ,\qquad \rho= e^{-z} \, ,     
\end{equation}
where $x_{\dot\alpha}^{\alpha} = x^{\mu}(\sigma_{\mu})_{\dot\alpha}^{\alpha}$ parametrizes the boundary at $\rho \rightarrow 0$. This metric resembles the Poincaré metric, but there is an important difference. The standard Poincaré coordinates are obtained by choosing 
the 
supercoset representative containing 
\begin{equation}
e^{\widehat{x}^{\dot\alpha}_{\alpha} P^{\alpha}_{\dot\alpha}} e^{\widehat z\Delta} \quad\text{instead of} \quad  e^{x_{\dot\alpha}^{\alpha} K_{\alpha}^{\dot\alpha}}e^{z\Delta}\,. 
\label{standardPoincare}
\end{equation}
These choices correspond to distinct coordinate patches on the boundary. Specifically, the standard Poincaré coordinates cover one hemisphere of the $S^4$ conformal compactification of the Euclidean boundary, while the bosonic truncation of $g_{\text{old}}$ covers the complementary hemisphere.
\begin{figure}[t]
    \centering
    \begin{subfigure}{0.48\textwidth}
         \includegraphics[width=\textwidth]{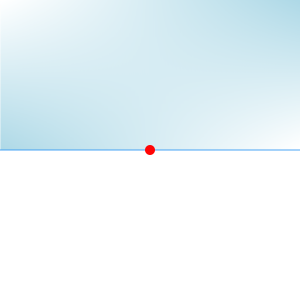}
        \caption{Poincaré patch}
        \label{fig:subfig1}
    \end{subfigure}
    \hfill
    \begin{subfigure}{0.48\textwidth}
                 \includegraphics[width=\textwidth]{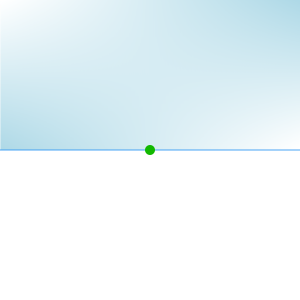}
        \caption{Its stereographic inversion}
        \label{fig:subfig2}
    \end{subfigure}
    \hfill
    \begin{subfigure}{0.48\textwidth}
        \centering
        \includegraphics[width=\textwidth]{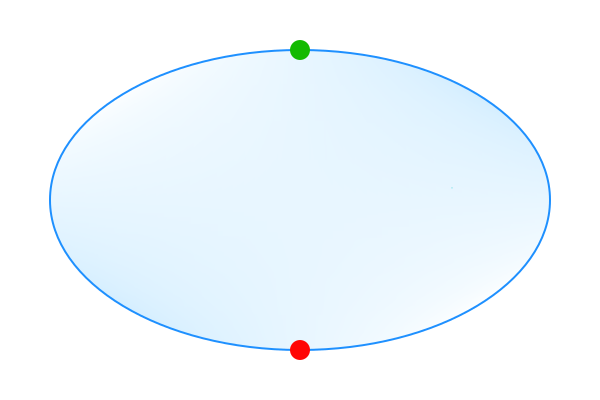}
	\caption{$AdS_2$ slice of $AdS_5$}
        \label{fig:subfig3}
    \end{subfigure}
    \caption{Depiction of the Poincaré patch and its stereographic inversion for an $AdS_2$ slice of $AdS_5$ spacetime.}
    \label{fig:comparison}
\end{figure}
Of course, it is straightforward to relate the two sets of coordinates at zero order in the fermionic components using a stereographic inversion, Figure 1. The challenge lies in keeping track of the fermionic components at all orders while performing super-reparametrizations.

Obtaining the map between \eqref{OldCosetRepresentative} and the standard Poincaré coordinates after bosonic truncation is straightforward. Consider the matrix representation of the coset elements:
\begin{equation}
\label{theequationforchanging}
\begin{bmatrix}
e^{\frac{\widehat z}{2}}\delta_{\alpha}^{\beta}&e^{-\frac{\widehat z}{2}}\widehat x_{\alpha}^{\dot\beta}\\0&e^{-\frac{\widehat z}{2}}\delta_{\dot\alpha}^{\dot\beta}
\end{bmatrix}=
\begin{bmatrix}
e^{\frac{z}{2}}\delta_{\alpha}^{\beta}&0\\e^{\frac{z}{2}} x_{\dot\alpha}^{\beta}&e^{-\frac{z}{2}}\delta_{\dot\alpha}^{\dot\beta}
\end{bmatrix}h_{\text{USp(2,2)}}\,,
\end{equation}
where $h_{\text{USp(2,2)}}$ is a compensator. 
The easiest way of computing the change of variables is to eliminate the compensator by 
contracting both sides of the equation above with the symplectic metric, as follows
\begin{equation}
\label{contracting with simplectic}
e^{z}\varepsilon_{\alpha\beta}= e^{\widehat z}\,\varepsilon_{\alpha\beta}+e^{-\widehat z}\,\widehat x_{\alpha}^{\dot\gamma}\varepsilon _{\dot\gamma\dot\delta}\widehat x^{\dot\delta}_{\beta} \,,  
\qquad 
e^{z} \varepsilon_{\alpha\gamma}x^{\gamma}_{\dot\beta}=e^{-\widehat z}\, \widehat x_{\alpha}^{\dot\gamma}\varepsilon _{\dot\gamma\dot\beta}  \,,
\end{equation}
which yields the stereographic inversion:
\begin{equation}
\label{Standard Poincare reparam}
\widehat x^{\dot\gamma}_{\alpha} = - \frac{\epsilon_{\alpha \beta} \epsilon^{\dot\gamma \dot\beta} x^{\beta}_{\dot\beta}}{(e^{-2z} + ||x||^2)} \, , \quad \quad e^{\widehat z} = \frac{e^{-z}}{(e^{-2z} +  ||x||^2)} \, ,
\end{equation}
where $||x||^2 = -(1/2) \varepsilon_{\alpha \beta} \varepsilon^{\dot\alpha \dot\beta} x^{\alpha}_{\dot\alpha}
x^{\beta}_{\dot\beta}$.  

This is an involution relating the two coordinate systems, mapping the origin of one system to the infinity of the other. The metric using the hatted variables is given by
\begin{equation}
ds_{AdS}^2 = \frac{d\widehat \rho^2 + ||d\widehat x||^2}{\widehat \rho^2} \,,\qquad \widehat \rho = e^{\widehat z}\,,  
\end{equation}
where $\widehat x^{\mu}$ parametrizes the boundary at $\widehat\rho\rightarrow 0$. %
For completeness, we write below the inverse transformations of \eqref{Standard Poincare reparam} and the expression for the compensator in \eqref{theequationforchanging}. They are given by       
\begin{equation}
e^{-z} = \frac{e^{\widehat{z}}}{(e^{2 \widehat{z}} - || \widehat{x} ||^2 )}  \, , \quad \quad
x^{\alpha}_{\dot\alpha} = - \frac{
\epsilon_{\dot\alpha \dot\beta} \epsilon^{\alpha \beta} \widehat{x}_{\beta}^{\dot\beta}}{(e^{2 \widehat{z}} - || \widehat{x} ||^2 )} \, ,  
\label{inversetransformation} 
\end{equation}
and 
\begin{equation}
h_{\text{USp(2,2)}} =
\begin{bmatrix}
e^{\frac{\widehat z}{2}}\delta_{\alpha}^{\beta}&e^{-\frac{\widehat z}{2}}\widehat x_{\alpha}^{\dot\beta}\\0&e^{-\frac{\widehat z}{2}}\delta_{\dot\alpha}^{\dot\beta}
\end{bmatrix}
\begin{bmatrix}
e^{\frac{z}{2}}\delta_{\alpha}^{\beta}&0\\e^{\frac{z}{2}} x_{\dot\alpha}^{\beta}&e^{-\frac{z}{2}}\delta_{\dot\alpha}^{\dot\beta}
\end{bmatrix}^{-1}\,,
\end{equation}
and one can use \eqref{Standard Poincare reparam} to express it only in terms of 
the unhatted variables or use \eqref{inversetransformation} 
to express it in terms of the hatted variables. 

\subsubsection{Stereographic inversion in superspace}
\label{Inversion}

In the previous Subsection we showed how, after bosonic truncation, the coordinates of \eqref{OldCosetRepresentative} can be mapped to the standard Poincaré coordinates through stereographic inversion. In this Section, we extend the stereographic inversion to the full superspace.

After bosonic truncation, the stereographic inversion was obtained by solving \eqref{theequationforchanging}
for $\widehat x$, $\widehat z$ and $h\in USp(2,2)$.
Notice that it effectivelly exchanges the operators according to their scalling dimensions $\Delta$, since $K_{\alpha}^{\dot\alpha}$ have scalling dimension $-1$ and $P_{\dot\alpha}^{\alpha}$ have scalling dimension $+1$.
To generalize this stereographic inversion to the superspace, we do the same for the fermionic generators:
\begin{equation}
q_{\alpha}^{I}\leftrightarrow q_{\dot\alpha}^{I}\,,\qquad q_{I}^{\alpha}\leftrightarrow q_{I}^{\dot\alpha}\,,
\end{equation}
where $I=1,2,3,4$. Acting with this generalized stereographic inversion on \eqref{OldCosetRepresentative}, one obtains a new parametrization of the superspace whose bosonic truncation results in the Poincaré patch. The inverted parametrization is given by
\begin{equation}
g_{\text{old inverted}} = e^{(\widehat 
 \psi^{\hspace{0.04cm} \dot\alpha}_{ i} q^i_{\dot\alpha} + \widehat 
 \theta^{\hspace{0.04cm} \bar{i}}_{ \alpha} q^{\alpha}_{\bar{i}} + \widehat 
 x_{ \dot\alpha}^{ \hspace{0.04cm} \alpha} P_{\alpha}^{\dot\alpha} + \widehat 
 x^{\hspace{0.04cm} \bar{i}}_{ i} K^{i}_{\bar{i}} )} 
e^{(\widehat 
 \psi^{\hspace{0.04cm} \alpha}_i q^i_{\alpha} + \widehat 
 \theta^{\hspace{0.04cm} \bar{i}}_{\dot\alpha} q^{\dot\alpha}_{\bar{i}})}
e^{(\widehat 
 \psi^{\hspace{0.04cm} \dot\alpha}_{\bar{i}} q^{\bar{i}}_{\dot\alpha}+\widehat 
 \theta^{\hspace{0.04cm} i}_{\alpha} q^{\alpha}_i) }
e^{(\widehat 
 \psi^{\hspace{0.04cm} \alpha}_{\bar{i}} q^{\bar{i}}_{\alpha}+\widehat \theta^{\hspace{0.04cm} i}_{\dot\alpha} q^{\dot\alpha}_i )}
e^{ \widehat 
 z \Delta + \widehat 
 w J} \, ,  
\label{stereographic exchange}
\end{equation}
\begin{equation*}
= g_{\text{old}} \,h \,,  
\end{equation*}
where $h$ is a $USp(2,2)\times USp(4)$ compensator, and one solves simultaneously for the hatted coordinates and the compensator. In Euclidean signature, the patch described by \eqref{OldCosetRepresentative}, together with its stereographic inversion, covers the entire superspace. Each patch omits a single point at infinity, which corresponds precisely to the origin of the boundary in the other patch.

\subsubsection{Boundary-to-bulk propagator}
\label{Boundary-to-bulk propagator}

The AdS/CFT conjecture implies that for any half-BPS operator of dimension $\Delta=\ell>0$ in $\mathcal{N}=4$ SYM, there is a 
corresponding physical string state \cite{Banks:1998dd,Polchinski:2010hw,Harlow:2011ke,Berkovits:2019ulm}. 
This one-to-one mapping is expected for other dual pairs;   
see \cite{Eberhardt:2019ywk} for an example in a particular $AdS_3$ background. 
Using the $g_{\rm{old}}$ coordinates of \eqref{OldCosetRepresentative},
the vertex operators of \cite{VertexOld} with $SO(1,1)_{\Delta}$ charge $\Delta$ and annihilated by 24 supersymmetries have the schematic form
\begin{equation}
V_{g_{\rm{old}}} \, \sim \,  \pi^{-2} \frac{\Gamma(\ell)}{\Gamma(\ell-2)} \,e^{- \ell z} f(w, \widetilde\theta^+,\ell) \equiv V^{\ell}_{z} 
f(w, \widetilde\theta^+,\ell)
\,,   
\end{equation}
and the function $f$ was computed in \cite{VertexOld}.
In standard Poincar\'e coordinates defined in \eqref{standardPoincare}, the overall $V^{\ell}_{z}$ factor above takes the form  
\begin{equation}
 V^{\ell}_{z}  \rightarrow \widehat{V}^{\ell}_z = \pi^{-2} \frac{\Gamma(\ell)}{\Gamma(\ell-2)} \left(\frac{\widehat \rho}{\widehat \rho^{2}+||\widehat x||^2}\right)^{\ell} \,,
\end{equation}
where it becomes evident that this factor is essentially the boundary-to-bulk propagator of \cite{Witten:1998qj}. Let us now focus on the near-boundary region, $\widehat \rho = \varepsilon$, with $\varepsilon>0$ a small parameter. Expanding in powers of $\varepsilon$ yields
\begin{equation}
\widehat{V}^{\ell}_z \, \sim \,  \varepsilon^{4-\ell} \delta^{(4)}(\widehat x) + \varepsilon^{\ell} \frac{(\ell -1)(\ell - 2)}{\pi^2} \frac{1}{\widehat x^{2\ell}}\left[1-\frac{\ell\varepsilon^2}{\widehat x^2} + \mathcal{O}(\varepsilon^4)\right]\,.
\end{equation}
The leading term above scales as $\varepsilon^{4-\ell}$ and is accompanied by a Dirac delta supported at the AdS boundary and centered at $\widehat x = 0$. The Dirac delta above follows because the vertex operators of \cite{VertexOld} are 
annihilate by all conformal boosts $K_{\mu}$, which implies that the dual $\mathcal{N}=4$ SYM operator is a conformal primary operator located at 
$\widehat x = 0$. 
Notice that this boundary point $\widehat x=0$ cannot be reached by taking $z\rightarrow \infty$ in the $g_{\rm{old}}$ coordinates of \eqref{OldCosetRepresentative}.

\subsubsection{Covariant superfields}
\label{Covariant superfields}

Covariant superfields arise naturally in vertex operators of worldsheet formalisms that preserve target-space supersymmetry manifestly. In these setups, vertex operators are built from invariant contractions of covariant superfields with covariant worldsheet variables, the latter carrying indices that transform under the gauge (or isotropy) group. %
Just as in the heterotic string with the RNS formalism, where worldsheet fermions serve as placeholders for the $SO(32)$ gauge indices, the bosonic pure spinor variables (and their B-RNS-GSS counterparts) play an analogous role for the Lorentz spin group \cite{DescentII,Berkovits:2022dbm}.

Before introducing gauge–covariant superfields, consider gauge–invariant ones. Since the superspace isotropy group is a gauge group, a gauge–invariant superfield must be right invariant under $H=USp(2,2)\times USp(4)$
\begin{equation}
\Phi(gh) = \Phi(g)\,,\quad \quad h\in H\,.   
\end{equation}
More generally, an $H$–covariant superfield in a representation of $H$ transforms as
\begin{equation}
\Phi_{\mathfrak{a}}(gh) = h_{\mathfrak{a}}^{\mathfrak{b}}\Phi_{\mathfrak{b}}(g)\,,\qquad h\in H\,.
\end{equation}
Upon fixing a gauge, as in \eqref{OldCosetRepresentative}, certain transformations must be accompanied by compensating $H$-gauge transformations to preserve the gauge condition. These compensators rotate the $H$-indices of covariant fields so that, after contraction with $H$-covariant worldsheet variables, the result is $H$-invariant. For concreteness, take the gauge-fixed superspace coordinate $\theta^{i}_{\alpha}$ in \eqref{OldCosetRepresentative} and an $H$-covariant worldsheet variable $\lambda^{\alpha}_{i}$ and form the monomial
\begin{equation}
\theta^{i}_{\alpha}\lambda^{\alpha}_{i}\,.    
\end{equation}
Gauge-restoring transformations act so as to reveal the underlying $H$–covariant superfields represented by this gauge-fixed expression. One can undo the gauge fixing by inserting an arbitrary right transformation $h\in H$ into \eqref{OldCosetRepresentative} and absorbing it into the superfield 
\begin{equation}
\theta^{i}_{\alpha}\lambda^{\alpha}_{i} \rightarrow \theta^{i}_{\alpha} \, h^{\alpha}_{\beta} \, h_{i}^{j} \, \lambda^{\beta}_{j} = \left( \theta^{i}_{\alpha} \, h^{\alpha}_{\beta} \, h_{i}^{j}\right)\lambda^{\beta}_{j} = \Phi^{j}_{\beta}(x,y,z,w,\theta,\psi)\lambda_{j}^{\beta}\,, 
\end{equation}
where $\Phi^{j}_{\beta}(x,y,z,w,\theta,\psi)$ is the underlying covariant superfield that simplifies to $\theta_{\beta}^{j}$ under the gauge choice made in \eqref{OldCosetRepresentative}. 
The gauge transformations are further discussed in the Subsection \ref{Propertiesofthevertexoperators} using our results for the vertex operators.   
    
\subsubsection{Covariant derivatives}
\label{covariant derivatives}

As we will see in Subsection \ref{The integrated vertex operator}, both the B–RNS–GSS vertex operators and the integrated pure spinor ones are obtained by acting with 
covariant derivatives on a bi-spinor superfield and contracting with gauge–covariant worldsheet variables to form gauge–invariant operators. This same superfield appears in the unintegrated pure spinor vertex operator. 
Therefore, once the unintegrated vertex is fixed and the gauge–covariant derivative algebra is known, all (integrated and unintegrated) vertex operators follow by covariant differentiation.

The $\mathfrak{psu}(2,2|4)$ superalgebra admits a $\mathbb{Z}_4$-grading:
\begin{equation}
\mathfrak{g} = \mathfrak{g}_0\oplus\mathfrak{g}_1 \oplus \mathfrak{g}_2\oplus\mathfrak{g}_3\,,    
\end{equation}
where the bosonic truncation consists of picking the even grading 
$\mathfrak{g}_0\oplus
\mathfrak{g}_2$, the isotropy group is in the zero grading $\mathfrak{g}_0$, and the fermionic generators are in the odd graddings $\mathfrak{g}_1 \oplus \mathfrak{g}_3$. For example, the anti-commutator between two fermionic generators—one from $\mathfrak{g}_3$ and the other from $\mathfrak{g}_1$—produces a bosonic generator in $\mathfrak{g}_0$, while the anticommutator of two fermionic generators within the same $\mathbb{Z}_4$ grading yields a bosonic generator in $\mathfrak{g}_2$.  The covariant derivatives are given by the right action of $\mathfrak{g}_1 \oplus\mathfrak{g}_2 \oplus \mathfrak{g}_3$, as follows
\begin{equation}
(1+\varepsilon \cdot\nabla^{(1)} +\varepsilon \cdot \nabla^{(2)} + \varepsilon \cdot\nabla^{(3)} )\Phi (g) = \Phi\left(g[1+\varepsilon \cdot q^{(1)} +\varepsilon \cdot p^{(2)} + \varepsilon \cdot q^{(3)} ]\right)\,,    
\end{equation}
where $q^{(1)}$, $p^{(2)}$ and $q^{(3)}$ correspond to the generators in $\mathfrak{g}_1$ $\mathfrak{g}_2$, and $\mathfrak{g}_3$, respectively, $\nabla^{(1)}$, $\nabla^{(2)}$ and $\nabla^{(3)}$ are the corresponding covariant derivatives, and $\varepsilon$ is a fermionic parameter living in the dual space $(\mathfrak{g}_1 \oplus\mathfrak{g}_2 \oplus \mathfrak{g}_3)^*$.

Using ten-dimensional notation, let the $SO(10)$ vector index split as $\underline a=(a,a')$ with $a$ in $AdS_5$ and $a'$ in $S^5$. 
The index structure of the $\mathbb Z_4$–graded generators is:
\[
\begin{aligned}
\mathfrak g_0 &: \ \widetilde M_{ab},\ \widetilde M_{a'b'} \quad &(10+10=20\,\text{generators}),\\
\mathfrak g_1 &: \ q_{\underline \alpha} \quad &(16\,\text{generators}),\\
\mathfrak g_2 &: \ p_{\underline a}=(p_a, p_{a'}) \quad &(5+5=10\,\text{generators}),\\
\mathfrak g_3 &: \ q_{\underline{\widehat\alpha}} \quad &(16\,\text{generators}),
\end{aligned}
\]
where $\underline \alpha, \underline{\widehat\alpha} =1, \ldots, 16$, correspond to 
Weyl indices. Let us differentiate with respect to the coordinates. Denote the coordinates by $Z^{M}$, 
\begin{equation}
\frac{\partial}{\partial Z^{M}} g = g \left[ E_{M}^{\underline \alpha}q_{\underline \alpha} + E_{M}^{\underline a}p_{\underline a} + E_{M}^{\underline{\widehat\alpha}}q_{\underline{\widehat\alpha}}+ \Omega_{M}\,^{ab}M_{ab} + \Omega_{M}\,^{a'b'}M_{a' b' }\right]\,,
\label{conectionsvielbein}
\end{equation}
where the $E_{M}^{\underline \alpha}$, $E_{M}^{\underline a}$ and $E_{M}^{\underline{\widehat\alpha}}$ are the components of supervielbein, and $\Omega_{A}\,^{ab}$ and $\Omega_{A}\,^{a'b'}$ are the components of the spin connection. The covariant derivatives are obtained by isolating the generators, resulting in
\begin{align}
\nabla_{A} &= E_{A}^{M}\!\left(\frac{\partial}{\partial Z^{M}}
 + \Omega_{M}{}^{ab} M_{ab} + \Omega_{M}{}^{a'b'} M_{a'b'}\right) \, ,
 \label{covariantderivatives} 
\end{align}
where $A = \{\underline \alpha, \underline{a}, \underline{\widehat{\alpha}}\}$ and $E_A^{M}$ are obtained by inverting the supervielbein square matrix.

An important use of covariant derivatives is that they allow the construction of supersymmetric constraints. In general, %
superfields furnish highly reducible representations of the superalgebra. To isolate the irreducible representations, one has to impose supersymmetric constraints. Such constraints reduce the number of independent components of the superfields and may force the decoupling of certain superspace coordinates. 

\subsubsection{%
$AdS_5\times S^5$ boundary}
\label{Reductions and truncations}

By truncation we mean fixing some parameter or field to a specific value—typically zero; for example, bosonic truncation sets the fermionic superspace coordinates to zero. 
A gauge choice in \eqref{OldCosetRepresentative} is a truncation/section of the $h$-coordinates: write $g=g_{\text{old}}h$ and set $h=1$. Reduction, by contrast, keeps only quantities that decouple from—or are invariant under—the truncated directions. 
The $G/H$ superspace is a reduction of $G$ by $H$, while the gauge-choice in \eqref{OldCosetRepresentative} is truncation/section. Focusing on objects independent of this section choice—i.e., gauge-invariant quantities—implements the reduction.

An important example is the description of the $AdS_5\times S^5$ boundary. After bosonic truncation and focusing on the $AdS_5$ factor, the boundary is given by the reduction
\begin{equation}
\frac{SO(4,2)}{(SO(1,1)_{\Delta}\times SO(1,3))\ltimes n}\,,
\end{equation}
where $SO(1,1)_{\Delta}$ denotes dilatations, $SO(1,3)$ is the Lorentz group, $\ltimes$ indicates a semidirect product, and $n$ is the nilpotent radical generated by the special conformal generators. Note that the boundary is not obtained by truncation. In Poincaré coordinates (\ref{poincare}), it corresponds to the limit $\rho\to0$, not the truncation $\rho=0$, since $\rho>0$ in $AdS_5\times S^5$.

Returning to superspaces and superfields, constrained superfields can decouple from certain fermionic parameters. This occurs when the covariant derivatives that annihilate the superfield form an integrable set of vector fields, generating a foliation of superspace. The supersymmetric constraints are then partially absorbed by this reduction. This is what happens with the half-BPS analytic $AdS_5\times S^5$ superfield discussed in the Subsection \ref{superfields for half-BPS supermultiplet}, and the half-BPS boundary superfields discussed in the Subsection \ref{Higher componetns transformations}. Reductions of superspace foliations do not spoil manifest supersymmetry. Practically, to reduce the $AdS_5\times S^5$ superspace, it is convenient to parametrize the supercoset so that the integrable subset of generators appears entirely on the right, and then truncate the parameters on the right to a local section. 

Besides bosonic truncation, one can also consider truncated superspaces such as the Lorentz Harmonic Superspace (LHC) of \cite{Chicherin:2016fac}.

\subsubsection{Boundary superfields}
\label{Higher componetns transformations}

Before discussing constrained superfields in the $AdS_5\times S^5$ superspace, consider the constrained superfields of the gauge theory living on the superspace boundary. The superspace boundary is stabilized by a subgroup larger than $USp(2,2)\times USp(4)$, namely
\begin{equation}
SO(1,1)_{\Delta}\times SL(2,\mathbb{C})\times SU(4)\ltimes N \, ,
\end{equation}
where $SO(1,1)_{\Delta}$ is dilatations, $SL(2,\mathbb{C})\simeq \mathrm{Spin}(1,3)$ 
is the Lorentz group, $\ltimes$ denotes a semidirect product, and $N$ is the nilpotent radical generated by special conformal and superconformal generators. Consequently, the boundary superspace is also homogeneous:
\begin{equation}
\label{boundary superspace}
\frac{PSU(2,2|4)}{SO(1,1)_{\Delta}\times SL(2,\mathbb{C})\times SU(4)\ltimes N}\,.   
\end{equation}
Given a gauge connection on the boundary superspace, its anticommutators define the Sohnius superfields $\overline W_{IJ}$ and $W^{IJ}$ via
\begin{equation}
\{\nabla^{\alpha}_{I},\nabla^{\beta}_{J}\}
= \epsilon^{\alpha\beta}\,\overline W_{IJ},
\qquad 
\{\nabla_{i}^{\alpha},\overline\nabla_{\dot\beta}^{\,j}\}
= \delta_i^{\,j}\,\nabla^{\alpha}_{\dot\beta},
\qquad 
\{\overline\nabla_{\dot\alpha}^{\,J},\overline\nabla_{\dot\beta}^{\,J}\}
= \epsilon_{\dot\alpha\dot\beta}\,W^{IJ},
\end{equation}
whose constraints follow from the graded Jacobi identities and put the gauge multiplet on-shell. Their lowest components are
\begin{equation}
W_{IJ}(x,\theta)=\Phi_{IJ}(x)+(\theta\psi(x))_{IJ}+(\overline\theta\overline\psi(x))_{IJ}+(\theta {F}(x)\theta)_{IJ}+(\overline\theta {F}(x)\overline\theta)_{IJ}+\dots\, ,
\end{equation}
where $\Phi_{IJ}(x)$ are the scalar fields, $\psi(x)$ and $\overline\psi(x)$ are the gluinos, and $F(x)$ denotes the gluon field strengths.

To construct half-BPS supermultiplets, one considers null $SO(6)_R$ vectors and contracts them with the Sohnius superfields. Single-trace operators are then obtained by taking arbitrary powers of the resulting superfield:
\begin{equation}
W_{\ell}(x,\theta,y_0)=\text{Tr}\left[\,(W(x,\theta,y_0))^{\ell}\right]\,,\qquad W(x,\theta,y_0)= y_0^{IJ}W_{IJ}(x,\theta)\,,
\end{equation}
where $y_0^{IJ}$ is a null $SO(6)_R$ vector, projectively stabilized by
\begin{equation}
{\rm{Spin}}(2)\times {\rm{Spin}}(4) = S(U(2)\times U(2))\,,   
\end{equation}
subgroup of the R-symmetry group. Replacing the $SU(4)_R$-factor by its $S(U(2)\times U(2))$ subgroup in the denominator of \eqref{boundary superspace} gives the harmonic superspace
\begin{equation}
\frac{PSU(2,2|4)}{SO(1,1)_{\Delta}\times SL(2,\mathbb{C})\times S(U(2)\times U(2))\ltimes N}\,,
\end{equation}
where the corresponding $R$-symmetry coset is the Grassmannian
\begin{equation}
\frac{SU(4)}{S(U(2)\times U(2))}\cong \mathrm{Gr}(2,4)\,,    
\end{equation}
the space of complex structures on $\mathbb{R}^6$. To construct the whole supermultiplet of half-BPS operators using the harmonic superspace variables $u$, we consider
\begin{equation}
W_{\ell}(x,\theta,u)=\text{Tr}\left[\,(W(x,\theta,u))^{\ell}\right]\,,
\end{equation}
where $W(x,\theta,u)$ is a gauge covariant G-analytic and H-analytic harmonic superfield with one unit of $U(1)$ charge. %
It corresponds to $W(x,\theta, y_0)$ for $y_0^{IJ}=(\overline u\overline u)^{IJ}$. 

The G- and H-analytic superfields $W_{J}$ are on-shell supercurrents in harmonic superspace, and are therefore dual to harmonic prepotentials \cite{Heslop:2000np,Howe:1981qj}. These prepotentials, which are defined off-shell and supported on harmonic superspace, correspond to the boundary limits of on-shell superfields in the bulk superspace. These on-shell covariant bulk-superfields are the ones appearing in the unintegrated vertex operators, as discussed next.

\subsubsection{$AdS_5\times S^5$ superfields for the half-BPS supermultiplet}
\label{superfields for half-BPS supermultiplet}

These half-BPS supermultiplets of the boundary theory are holographically mapped to the Kaluza–Klein excitations of the type IIB supergravity multiplet in the bulk of the $AdS_5 \times S^5$ background. There are two constrained on-shell superfields that isolate the type IIB supergravity multiplet in superspace:
\begin{itemize}
\item the analytic superfield $\Phi$ of \cite{Howe:1983sra};
\item the bi-spinor gauge superfield $A_{\underline{\alpha} \underline{\widehat\beta}}$ of the pure spinor formalism \cite{BerkovitsPureOriginal, BerkovitsChandia}.
\end{itemize}

The analytic superfield is a scalar superfield whose $\theta$-expansion contains gauge-invariant field 
stren\-gths. It satisfies an analytic constraint and a reality condition, schematically written as
\begin{equation}
\label{Analytic constraints}
\nabla = \nabla^{L}+i\nabla^{R}\,,\qquad \overline\nabla\Phi = 0\,,\qquad \nabla^4\Phi \sim \overline \nabla^{4}\overline \Phi\,.    
\end{equation}
In \cite{Heslop:2000np}, it was shown that the boundary values of these superfields correspond to the prepotentials dual to the G- and H-analytic on-shell superfields $W_{J}$ described previously. In the pure spinor formalism, these superfields appear in vertex operators constructed in the $-8$ picture, as described in \cite{Martins:2019wwi}.

The bi-spinor superfield contains gauge potentials—rather than field strengths—in its $\theta$-expansion. In addition to the constraints
\begin{equation}
\gamma_{\underline{abcde}}^{\underline{\alpha} \underline{\beta}}\nabla_{\underline{\alpha}} A_{\underline{\beta} \underline{\widehat\gamma}}  = 0\,,
\qquad \gamma_{\underline{abcde}}^{\underline{\widehat{\alpha}} \underline{\widehat{\beta}}} \nabla_{\underline{\widehat\alpha}}A_{\underline{\gamma} \underline{\widehat{\beta}}}=0 \,,
\end{equation}
it transforms under the gauge invariances
\begin{equation}
\delta A_{\underline{\alpha} \underline{\widehat{\beta}}} = \nabla_{\underline{\alpha}} \chi_{\underline{\widehat{\beta}}} + \nabla_{\underline{\widehat{\beta}}} \chi_{\underline{\alpha}}\,.   
\end{equation}
All zero-picture vertex operators—both integrated and unintegrated, and in both the pure spinor and B-RNS-GSS 
formalisms—are expressed up to gauge transformations as linear combinations of covariant derivatives acting on this bi-spinor superfield, contracted with formalism-dependent worldsheet variables. 

These two superfields are related. Picture-raising the $-8$ picture vertex operator in the pure spinor formalism to picture %
zero
yields a linear combination of covariant derivatives acting on the analytic superfield $\Phi$, thereby producing the bi-spinor superfield $A_{\alpha \widehat\alpha}$. In flat space this was shown in light-cone gauge \cite{Martins:2019wwi},
while in $AdS_5\times S^5$ it has been done explicitly only for the bottom components. 
However,
all other states are obtained via $PSU(2,2|4)$ transformations. 

Using the $g_{\rm{old}}$ parametrization of 
\eqref{OldCosetRepresentative}, both the analytic and bi-spinor superfields for the bottom component were constructed, leveraging the decoupling of 24 fermionic coordinates. However, the $PSU(2,2|4)$ transformations were too cumbersome to generate the remaining states or even change the bottom-component ‘polarization’; we resolve this in Section \ref{The PSU(2,2|4) transformations}.

In $AdS_5\times S^5$, unlike flat space where states are parametrized by polarizations and momenta, states are parametrized by boundary points. For example, the bottom component of the half-BPS supermultiplets is projectively stabilized by
\begin{equation}
SO(1,1)_{\Delta}\times SL(2,\mathbb{C})\times S(U(2)\times U(2))\ltimes N\,,    
\end{equation}
where $\ltimes$ denotes a semidirect product, and $N$ is the nilpotent radical generated by special conformal and superconformal generators. Consequently, the moduli space of half-BPS states is the harmonic superspace
\begin{equation}
g_{\text{Moduli}} \in \frac{PSU(2,2|4)}{SO(1,1)_{\Delta}\times SL(2,\mathbb{C})\times S(U(2)\times U(2))\ltimes N}\,.
\end{equation}
The bosonic trucation of the above space produces the Cartesian product between the boundary of $AdS_5$ and 
the Grassmannian $\mathrm{Gr}(2,4)$. Just as we parametrize the bulk of the superspace by a supercoset element $g_{\text{bulk}}$, one can parametrize these states by a supercoset element $g_{\text{Moduli}}$ and act on the superfields $\Phi^{(\text{bottom})}$ and $A^{(\text{bottom})}_{\alpha\widehat\alpha}$ to obtain the whole superfield with all the states
\begin{equation}
A_{\underline{\alpha} \underline{\widehat\alpha}}= g_{\text{Moduli}}\,A^{(\text{bottom})}_{\alpha\widehat\alpha}\,,\qquad  \Phi = g_{\text{Moduli}}\,\Phi^{(\text{bottom})}\,.   
\end{equation}
In \cite{VertexOld}, this was truncated by setting $g_{\text{Moduli}}=1$. Just as the $SU(2,2)\times SU(4)$ transformations are cumbersome in the parametrization of \eqref{OldCosetRepresentative}, the action of $g_{\text{Moduli}}$ on the truncations $A^{(\text{bottom})}_{\underline{\alpha} \underline{\widehat\alpha}}$ and $\Phi^{(\text{bottom})}$ is likewise cumbersome. In this paper, we solve this by a reparametrization that disentangles bosons from fermions.

\subsection{The pure spinor formalism: zero conformal weight}
\label{The pure spinor formalism: zero conformal weight}

In the pure spinor formalism, physical states are defined as elements of the BRST cohomology. Thus, the data that fully specify a superstring background consist of the 
action $S$ together with the BRST %
operator $Q$, as in \cite{Berkovits:2004xu}. In this paper, we review the worldsheet action in Subsection \ref{The integrated vertex operator}, where we introduce the worldsheet variables of positive conformal weight and the integrated vertex operators. For the moment, we focus on the weight-zero worldsheet variables and their associated vertex operators, since this suffices to construct the unintegrated supergravity vertex operators. In Subsection \ref{The integrated vertex operator}, in addition to introducing the non-zero modes, we show that the integrated vertex operators can be obtained from the unintegrated ones. 

\subsubsection{BRST transformations of the zero-modes}
\label{BRST transformations of the zero-modes}

For %
operators $V$ containing only zero modes, the BRST transformations simplify:
\begin{equation}
Q \cdot V = \left(\lambda_{L}^{\underline{\alpha}} \cdot\nabla_{\underline{\alpha}} + \lambda_{R}^{\underline{\widehat\alpha}}\cdot\nabla_{\underline{\widehat\alpha}} \right) V \,,
\label{BRSTaction}
\end{equation}
where the bosonic variables $\lambda_{L}$ and $\lambda_{R}$ are $SO(1,9)$ pure spinors, as described in the next Subsection, and $\nabla$'s are the %
covariant derivatives along the fermionic directions of the underlying superspace.

In the $AdS_5\times S^5$ background, the vertex operators are given by $V(g,\lambda)$ and are gauge invariant under $H$-right translations where $H=USp(2,2)\times USp(4)$:
\begin{equation}
V[gh,\,h\lambda h^{-1}] = V[g, \lambda] \,,\qquad h\in H\,.
\end{equation}
The BRST transformations of conformal weight-zero operators are generated by constrained infinitesimal right multiplications valued in $\mathfrak{g}_1 \oplus \mathfrak{g}_3$, with compensating gauge transformations in $\mathfrak{g}_0$:
\begin{equation}
\label{BRST on uninitegrated}
Q \cdot V\left[g, \lambda\right] = (\lambda_{L}\cdot\nabla^{(1)} + \lambda_{R}\cdot\nabla^{(3)})V\left[g, \lambda\right]\,.    
\end{equation}
When we gauge-fix, as in \eqref{OldCosetRepresentative}, we must introduce gauge compensators for the BRST transformations to preserve the gauge-fixing conditions, as discussed in Subsection \ref{Covariant superfields}, resulting in the following transformations of the gauge-fixed supercoset representative $g_{\text{old}}$:
\begin{equation}
Q_{\text{fixed}} \cdot g_{\text{old}} = g_{\text{old}} \left(\lambda^{\underline{\alpha}}_{L}\cdot  q^{(L)}_{\underline{\alpha}} + \lambda_{R}^{\underline{\widehat\alpha}} \cdot q^{(R)}_{\underline{\widehat\alpha}} - \Sigma^{(0)}\right) \, , 
\label{BRST}
\end{equation}
where $\Sigma^{(0)}\in\mathfrak{g}_0$ is the gauge compensator fixed to preserve \eqref{OldCosetRepresentative}. 

\subsubsection{Pure spinor variables}
\label{Pure spinors subsubsection}

The pure spinors $\lambda_{L}^{\underline{\alpha}}$ and $\lambda_{R}^{\underline{\widehat\alpha}}$ are bosonic variables with ghost number $+1$. Their index structure is dual to that of their corresponding supercharges. They satisfy the non-linear constraints
\begin{equation}
\{\lambda_{L}\cdot  q^{(L)}, \lambda_{L}\cdot  q^{(L)}\} = 0\,,\qquad \{\lambda_{R}\cdot  q^{(R)}, \lambda_{R}\cdot  q^{(R)}\}=0\,,
\label{purespinor} 
\end{equation}
where the brackets $\{\,,\}$ denote anti-commutation. In ten-dimensional notation, the pure spinor constraints become
\begin{equation}
(\lambda_{L}\gamma^{\underline a}\lambda_{L}) = 0\,,\qquad   (\lambda_{R}\gamma^{\underline a}\lambda_{R}) = 0\,. 
\end{equation}
This presentation of constraints is common in flat space, where $SO(1,9)$ Lorentz symmetry is preserved. In $AdS_5\times S^5$, the Lorentz symmetry breaks to $SO(4,1)\times SO(5)$, whose corresponding spin group is $USp(2,2)\times USp(4)$. The superspace parametrization in \eqref{OldCosetRepresentative} further breaks it to $SO(1,3)\times SO(4)$, with spin group $SL(2,\mathbb{C})\times SU(2)\times SU(2)$. Following \cite{VertexOld}, it it is convenient to embed $SO(1,3)\times SO(4)\subset SO(1,4)\times SO(5)$ into $SO(8)$ (Wick rotated $SO(1,7)$), where the supercharges are grouped as follows:
\begin{equation}
\bar{q}^{\dot{a}}_+ = (q^{\alpha}_{\bar{i}}, q^i_{\dot{\alpha}} ) \, , \quad \bar{q}^{\dot{a}}_- = (q^{\bar{i}}_{\alpha}, q^{\dot\alpha}_i) \, , \quad
q^a_+ = (q^i_{\alpha}, q^{\dot\alpha}_{\bar{i}}) \, , \quad q^a_- =(q^{\bar{i}}_{\dot\alpha}, q^{\alpha}_i) \, .
\label{eq:qinso8} 
\end{equation}
For conventions on dimensional reductions of the $SO(10)$ group, we refer to \cite{VertexOld}. These $SO(8)$ spinors combine into pairs of opposite chirality to form the $SO(1,9)$ spinors:
\begin{equation}
q^{(+)}_{\underline \alpha} = (q_{+}^{a},\overline q_{+}^{\dot a}) \, , \qquad 
q^{(-)}_{\underline \alpha} =  (q_{-}^{a},\overline q_{-}^{\dot a}) \, ,
\end{equation}
which, in turn, are related to the $SO(1,9)$ spinors in \eqref{BRST} and reference \cite{IntegratedVertex} through holomorphic and anti-holomorphic combinations:
\begin{equation*}
q^{(+)}_{\underline \alpha} = q^{(L)}_{\underline \alpha} + i q^{(R)}_{\underline \alpha} \, , \qquad 
q^{(-)}_{\underline \alpha} = q^{(L)}_{\underline \alpha} - i q^{(R)}_{\underline \alpha} \, ,
\end{equation*}
as in \cite{Howe:1983sra}. 
Consequently, the pure spinor variables also organize themselves into the $SO(8)$ notation:
\begin{equation}
\lambda_+^a = (\lambda^i_{\alpha}, \lambda_{\bar{i}}^{\dot{\alpha}})  \, , \quad
\lambda_-^a = (\lambda^{\bar{i}}_{\dot\alpha}, \lambda_{i}^{\alpha}) \, , \quad
\bar\lambda_+^{\dot{a}} = (\lambda^i_{\dot{\alpha}}, \lambda_{\bar{i}}^{\alpha})  \, , \quad
\bar\lambda_-^{\dot{a}} = (\lambda^{\bar{i}}_{\alpha}, \lambda_{i}^{\dot{\alpha}})  \, ,
\label{lambdasso8} 
\end{equation} 
where
\begin{equation*}
\lambda_+^{\underline{\alpha}} = \lambda_L^{\underline{\alpha}} + i \lambda_R^{\underline{\alpha}} \, , \qquad 
\lambda_-^{\underline{\alpha}} = \lambda_L^{\underline{\alpha}} - i \lambda_R^{\underline{\alpha}} \, . 
\end{equation*}
In \cite{VertexOld} the pure spinor constraints in \eqref{purespinor} were writen in the above $SO(8)$ notation, where the components $\lambda_{+}^{a}$ were isolated to be free while the $\lambda_{-}^{a}$ components were completely fixed by the other components. 
We review the constraints in $SO(8)$ notation in Section \ref{The ideal coset vertices}.
In this paper, we will mainly consider the presentation of the pure spinor constraints that preserve the full $USp(2,2)\times USp(4)$ isotropies of $AdS_5\times S^5$. The pure spinor variables arrange as follows
\begin{equation*}
\lambda^{I}_{R} = (\lambda^{i}_{\alpha},-\lambda^{\overline i}_{\alpha},\lambda^{i}_{\dot\alpha},\lambda^{\overline i}_{\alpha}) \,,\qquad \lambda_{I}^{R} = (\lambda_{i}^{\alpha},\lambda_{\overline i}^{\alpha},-\lambda_{i}^{\dot\alpha},\lambda_{\overline i}^{\alpha}) \,.
\end{equation*}
where the pure spinor constraints 
of \eqref{purespinor} can be re-expressed as (see \cite{VertexOld} for details)
\begin{equation}
\label{eq:so6pure}
\begin{aligned} 
& \lambda^I_{R} \lambda^{R}_J + \epsilon^{I K} \epsilon_{J L} {\lambda}^L_{R} {\lambda}^{R}_K = \frac{1}{2} \delta^I_J 
(\lambda^K_{R}  \lambda^{R}_K) \, , \qquad \lambda^{R}_{I}\,\lambda^{I}_{S}
 + \epsilon^{R T}\epsilon_{S U}\,\lambda^{U}_{I}\,\lambda^{I}_{T}
 = \tfrac12\,\delta^{R}_{S}\,(\lambda^{T}_{I}\,\lambda^{I}_{T})
\end{aligned} 
\end{equation}
and
\begin{equation}
\begin{aligned} 
\label{eq:so6pure2}
 \epsilon^{R S} \lambda^I_{R} \lambda^J_{S} + \epsilon^{I K} \epsilon^{J L} \epsilon_{R S} \lambda^{R}_K \lambda^{S}_L 
=  \frac{1}{4} \epsilon^{I J} (\epsilon_{R S} \epsilon^{K L} \lambda^{R}_K \lambda^{S}_L+
\epsilon^{R S} \epsilon_{K L} \lambda_{R}^K \lambda_{S}^L) \, , \\    
\epsilon^{I J}\,\lambda^{R}_{I}\,\lambda^{S}_{J}
 + \epsilon^{R T}\epsilon^{S U}\,\epsilon_{I J}\,\lambda^{I}_{T}\,\lambda^{J}_{U}
 = \frac{1}{4} \,\epsilon^{R S}\!\left(\epsilon_{I J}\epsilon^{T U}\,\lambda^{I}_{T}\,\lambda^{J}_{U}
 + \epsilon^{I J}\epsilon_{T U}\,\lambda_{I}^{T}\,\lambda_{J}^{U}\right).
\end{aligned} 
\end{equation}

\subsubsection{Gauge-restoring transformations}
\label{Gauge-restoring transformations}

Expressing the gauge compensator $\Sigma^{(0)}$ of \eqref{BRST} in terms of the generators listed in Appendix \ref{apsupermatrices}, we have:
\begin{equation}
\Sigma^{(0)} = \widehat{M}^{\alpha}_{\beta} \Sigma^{\beta}_{\alpha} +
\widehat{M}^{\dot{\alpha}}_{\dot{\beta}} \Sigma^{\dot{\beta}}_{\dot{\alpha}} + \widehat{M}^{i}_{j} \Sigma^{j}_{i}+
 \widehat{M}^{\bar{i}}_{\bar{j}} \Sigma^{\bar{j}}_{\bar{i}}   +( P^{\alpha}_{\dot\alpha} +
 \epsilon_{\dot{\alpha} \dot{\beta}} \epsilon^{\alpha \beta} K_{\beta}^{\dot{\beta}}) \Sigma^{\dot{\alpha}}_{\alpha} + 
 ( P^{\bar{i}}_{i} +
 \epsilon_{ij} \epsilon^{\bar{i} \bar{j}} K_{\bar{j}}^{j}) \Sigma^{i}_{\bar{i}} \, . 
 \label{deltalambda}
\end{equation}
The components $\Sigma^{\beta}_{\alpha}$, $\Sigma^{\dot{\beta}}_{\dot{\alpha}}$, $\Sigma^{j}_{i}$, $\Sigma^{\bar{j}}_{\bar{i}}$, $\Sigma^{\dot{\alpha}}_{\alpha}$, and $\Sigma^{i}_{\bar{i}}$ are determined by requiring that the parametrization of the supercoset space is preserved:
\begin{equation}
\Sigma^{(0)} = \lambda_{L}\cdot  q^{(L)} + \lambda_{R}\cdot q^{(R)} -g_{\text{old}}^{-1}\,Q_{\text{fixed}} \cdot g_{\text{old}}\,.     
\end{equation}
Since the pure spinor variables couple to the generators of right actions, they must transform under gauge transformations, so after gauge-fixing, the BRST operator will act as 
\begin{equation}
Q_{\text{fixed}} \cdot \lambda_{L} = [\Sigma^{(0)}, \lambda_{L}]\,,\qquad Q_{\text{fixed}} \cdot \lambda_{R} = [\Sigma^{(0)}, \lambda_{R}]\,.    
\label{purespinortransformation}
\end{equation}
\subsubsection{BRST nilpotency on the zero modes}
\label{BRST nilpotency on the zero modes}

It is straightforward to verify that the BRST transformations defined above are nilpotent, both at gauge-invariant quantities, and at its gauge-fixed version. At gauge-invariant quantities, it follows essentially from the algebra and the pure spinor constraints, where
\begin{equation}
\{\lambda_{L}\cdot \nabla^{(L)},\, \lambda_{L}\cdot \nabla^{(L)}\} =0\,,\qquad   \{\lambda_{R}\cdot\nabla^{(R)},\,\lambda_{R}\cdot\nabla^{(R)}\} = 0 \,,  
\end{equation}
and
\begin{equation}
\{\lambda_{L}\cdot \nabla^{(L)},\,\lambda_{R}\cdot\nabla^{(R)}\}\in H\,,     
\end{equation}
generates a gauge transformation. 

After gauge-fixing, say as in \eqref{OldCosetRepresentative}, nilpotency can be demonstrated by applying the BRST transformations twice to each of the coordinates, as well as in the pure spinor variables. 
The demonstration uses some properties of the supervielbein, their inverses and the connections defined in \eqref{conectionsvielbein}. 
The relevant identities all follow from the algebra of the covariant derivatives.  
The covariant derivatives satisfy the same $\mathfrak{psu}(2,2|4)$ superalgebra as the generators, in particular,   
\begin{equation}
[ \nabla_{A} , \nabla_{B} ] = f^{\mathcal{C}}\,_{A B} \nabla_{\mathcal{C}} \, ,  
\label{comutator}
\end{equation}
where $f^{\mathcal{C}}\,_{A B}$ are the structure constants and the superalgebra indices 
are split as $\mathcal{C} =\{C, \underline{cd} \}$, implying that \eqref{comutator} 
is written in terms of torsions and curvatures. 
Using the expressions of the covariant derivatives in terms of the inverse vielbeins and connections  of \eqref{covariantderivatives}, one has  
\[
E^M_{[A|} \nabla_{M} E^{N}_{| B]} E^{C}_{N} = f^{C}\, _{AB} \, ,  \quad \quad 
\]
Let us call collectively $Z^M$ the worldsheet variables of the gauge-fixed coset representative.
Using the inverse supervielbeins, one can verify that the BRST transformations of the variables and the pure spinors are given by 
\begin{equation}
Q_{\text{fixed}} \cdot Z^M = E^M_A \lambda^A \, ,  \quad  \quad 
Q_{\text{fixed}} \cdot \lambda^A = -(\Omega_M)^A_B E^M_C \lambda^B \lambda^C \equiv -\Sigma^A_B \lambda^B \, ,
\label{purespinortrans}
\end{equation}
Since, 
\begin{equation}
Q_{\text{fixed}} \cdot E^M_A = (\partial_N E^M_A) \, Q_{\text{fixed}} \cdot Z^N \, ,
\label{purespinortrans}
\end{equation} 
one has, because of the pure spinor constraints,
\begin{align}
Q_{\text{fixed}}^2 \cdot Z^M 
        = E_B^N (\nabla_N E_A^P) E_P^C E_C^M  \lambda^A \lambda^B 
        \propto f^{C}\, _{AB} E_C^M \lambda^A \lambda^B = 0 \, . 
\end{align}
The action on the pure spinors are given by,  
\begin{equation}
Q^2_{\text{fixed}} \cdot \lambda^A = 
Q_{\text{fixed}}  \cdot (-(\Omega_M)^A_B  \lambda^B) \, Q_{\text{fixed}} \cdot Z^M   =
E^M_C  \lambda^C Q_{\text{fixed}} \cdot  (-(\Omega_M)^A_B  \lambda^B) = 0 \,  \, , 
\end{equation}
where we have used that the BRST acting on $Z^M$ is nilpotent. 
The BRST acting on the last term containing the connection gives a term proportional to curvature that 
when contracted with the three pure spinors gives zero. 
This can be verified by using the explicit expressions for the structure constants 
of the Appendix \ref{apsupermatrices}.

\subsubsection{The unintegrated vertex operators}
\label{Unintegrated vertex operators}

As in \cite{Witten:1992fb}, closed-string unintegrated physical vertex operators in the pure spinor formalism are defined to have ghost number $+2$. For massless states, the vertices are simpler because they do not involve worldsheet derivatives; consequently, they are functions only of 
the worldsheet variables with zero conformal weight—in this formalism, the superspace coordinates together with the pure spinors. Because worldsheet variables with negative conformal weight are absent, the BRST transformations also simplify and are given by (\ref{BRST on uninitegrated}). 

The only worldsheet variables with non-zero ghost number at conformal weight zero are the pure spinors, which carry ghost number $+1$. Hence the ghost-number assignment completely fixes the powers of the pure spinors:
\begin{equation*}
V = \sum_{n_{\text{ghost}}=n_L+n_R}V_{n_L,n_R}(g,\lambda)\,,    
\end{equation*}
where
\begin{equation}
V_{n_L,n_R}(g,\lambda) = \lambda^{\underline{\alpha}_1}_L\ldots\lambda^{\underline{\alpha}_{n_{L}}}_L \lambda^{\underline{\widehat\alpha}_1}_{R}\ldots\lambda^{\underline{\widehat\alpha}_{n_{R}}}_{R} A_{\underline\alpha_1\ldots \underline\alpha_{n_{L}}; \underline{\widehat\alpha}_1\ldots \underline{\widehat\alpha}_{n_{R}}}(g) \, , 
\end{equation}
and $A_{\underline\alpha_1\ldots \underline\alpha_{n_{L}}; \underline{\widehat\alpha}_1\ldots \underline{\widehat\alpha}_{n_{R}}}(g)$ are gauge-covariant superfields, as explained in Subsection \ref{Covariant superfields}. When we say that vertex operators lie in the $Q$-cohomology, we mean they are BRST-closed with gauge invariances generated by $Q$-exact shifts:
\begin{equation}
Q \cdot V = 0\,,\qquad \delta V = Q \cdot W\,,    
\end{equation}
which impose the linearized equations on the gauge-covariant superfields and generate linearized transformations, respectively.

\subsubsection{Half-BPS bottom component}
\label{previous paper result}

In \cite{VertexOld}, picture-zero, ghost-number-two vertex operators for closed-string states were constructed by acting with the picture-raising operator on the picture $-8$ vertex for the bottom state of a half-BPS supermultiplet with dimension  
$\ell$. Using the superspace parametrization of \eqref{OldCosetRepresentative}, the analytic superfield discussed in Subsection  \ref{superfields for half-BPS supermultiplet}, when truncated to the bottom state of the supermultiplet, becomes
\begin{equation}
\Phi^{(\text{bottom}, \,\ell)}(z+w) = e^{-\ell(z+w)}\,,
\end{equation}
and the associated picture $-8$ vertex is
\begin{equation}
V_{-8\text{ picture}}^{(\text{bottom}, \,\ell)} = \left[\prod_{\dot\alpha, \bar{i} =1,2} \psi^{\dot\alpha}_{\bar{i}} \prod_{ \alpha,i =1,2} 
\theta^i_{\alpha} \prod_{\dot\alpha, \bar{i} =1,2} \delta(\lambda^{\dot{\alpha}}_{\bar{i}}) \prod_{ \alpha,i =1,2} \delta(\lambda^i_{\alpha})\right]\, (\bar{\lambda}_+ \bar{\lambda}_+) \, e^{-\ell(z+w)} \, . 
\label{eq:minus8vertex} 
\end{equation}
The resulted picture-zero vertex obtained was schematically 
\begin{equation}
V^{(\text{bottom}, \,\ell)}(\overline\lambda_{\pm},z+w,\widetilde\theta_+) =  \overline\lambda_{L}^{\dot a}\overline\lambda_{R}^{\dot b}A_{{\dot a\dot b}}^{(\text{bottom}, ,\ell)}(z+w,\widetilde\theta_+) = 
\label{vertextypeII}
\end{equation}
\begin{equation}
 =  \left(\langle\overline\lambda^2_{\pm}\rangle + 
\langle\overline\lambda^2_{\pm} \widetilde\theta_+^2\rangle P_1(\ell)
+\langle\overline\lambda^2_{\pm} \widetilde\theta_+^4\rangle P_2(\ell)
+ \langle\overline\lambda^2_{\pm} \widetilde\theta_+^6\rangle P_3 (\ell)+
\langle\overline\lambda^2_{\pm} \widetilde\theta_+^8\rangle P_4(\ell)\right) e^{-\ell(z+w)}\,   \, .
\label{schematically} 
\end{equation}
As discussed in Subsection \ref{superfields for half-BPS supermultiplet}, from this bottom component one can reconstruct the entire supermultiplet via supersymmetry transformations:
\begin{equation}
V^{(\ell)} = \lambda^{\underline\alpha}_L \lambda^{\underline{\widehat\alpha}}_{R}A_{\underline{\alpha} \underline{\widehat\alpha}}^{(\ell)}(z, w, x, y,\theta,\psi) = g_{\text{Moduli}}V^{(\text{bottom},\ell)}\,.    
\end{equation}

\subsubsection{Mass-dimension tower %
}
\label{Mass-dimension tower via Jacobi identities}

Although we have so far considered vertex operators that distribute ghost number symmetrically between the left- and right-moving pure spinor variables, $\lambda_L$ and $\lambda_R$,
\begin{equation}
V^{(\ell)} = \lambda^{\underline\alpha}_{L} \lambda_{R}^{\underline{\widehat\alpha}}A_{\underline\alpha \underline{\widehat\alpha}}^{(\ell)} \,,   
\label{VonlyAAds} 
\end{equation}
one may more generally start from the vertex:
\begin{equation}
V^{(\ell)} = \lambda^{\underline\alpha}_L \lambda^{\underline{\widehat\alpha}}_R A_{\underline\alpha \underline{\widehat\alpha}}^{(\ell)}+ \lambda_L^{\underline\alpha} \lambda_L^{\underline\delta} B_{\underline\alpha \underline\delta}^{(\ell)} + \lambda_R^{\underline{\widehat\alpha}} \lambda_R^{\underline{\widehat\delta}} B_{\underline{\widehat\alpha} \underline{\widehat\delta}}^{(\ell)} \, .
\label{vertex10d}
\end{equation}

Under the linearized transformations induced by $\delta V = Q \cdot W$, the superfields $B_{\underline{\alpha} \underline{\delta}}$ and $B_{\underline{\widehat\alpha} \underline{\widehat\delta}}$ can be gauged away because of the equations depending  only on the $B$'s of \eqref{conditionsonsuperfields} below, see the discussion on \cite{DescentII} concerning the analysis of the constraints on $H_{\alpha \beta \gamma}$ for details. However, we find it convenient to keep them since the vertex operators computed in this work are not in the gauge \eqref{VonlyAAds} and we can apply the formulas directly without the need to sum $Q$-exact quantities.   
Until the end of this Subsection, we are going to use a more compact notation to avoid cluttering. We are going to omit the $L$ and $R$ from the pure spinors and the superscript $(\ell)$ from the superfields. Moreover, all the indices will be ten-dimensional and we are going to omit the underlines from the Greek indices until the next section. 
The linearized equations implied by $Q \cdot V = 0$ are: 
\begin{equation}
\begin{aligned}
&\lambda^{\alpha} \lambda^{\beta} \lambda^{\gamma} \nabla_{\gamma} B_{\alpha\beta} = 0 \, , \quad  
  \lambda^{\alpha} \lambda^{\widehat\alpha} \lambda^{\beta} \nabla_{\beta} A_{\alpha \widehat\alpha} +  \lambda^{\alpha} \lambda^{\beta} \lambda^{\widehat\gamma} \nabla_{\widehat\gamma}B_{\alpha\beta}=0 \, , \\
& \lambda^{\widehat\alpha} \lambda^{\widehat\beta}  \lambda^{\widehat\gamma} \nabla_{\widehat\gamma} B_{\widehat\alpha \widehat\beta} = 0 \, , \quad \lambda^{\alpha} \lambda^{\widehat\alpha} \lambda^{\widehat\gamma} \nabla_{\widehat\gamma}A_{\alpha \widehat\alpha} +
 \lambda^{\widehat\alpha} \lambda^{\widehat\beta}  \lambda^{\gamma} \nabla_{\gamma} B_{\widehat\alpha \widehat\beta} =0 \, . 
\label{conditionsonsuperfields} 
\end{aligned} 
\end{equation}
Since the pure spinor constraints are equivalent to
\begin{equation}
\lambda^{\alpha} \lambda^{\beta} = \frac{1}{1920}(\lambda \gamma^{\underline{abcde}} \lambda) \gamma_{\underline{abcde}}^{\alpha \beta}  \, , \quad \quad 
\lambda^{\widehat\alpha} \lambda^{\widehat\beta} = \frac{1}{1920}(\lambda \gamma^{\underline{abcde}} \lambda) \gamma_{\underline{abcde}}^{\widehat\alpha \widehat\beta} \, ,
\end{equation}
it follows that the linearized equations are equivalent to
\begin{equation}
\begin{aligned}
& \nabla_{(\gamma} B_{\alpha\beta)} = \gamma^{\underline{a}}_{(\gamma \alpha} B_{|\underline{a}| \beta)} \, ,  \quad 
  \nabla_{(\beta} A_{\alpha) \widehat\alpha} +   \nabla_{\widehat\alpha}B_{(\alpha\beta)} =\gamma^{\underline{a}}_{\alpha \beta} A_{\underline{a} \widehat{\alpha}} \, , \\
&  \nabla_{(\widehat\gamma} B_{\widehat\alpha \widehat\beta)} = \gamma^{\underline{a}}_{(\widehat\gamma \widehat\alpha} B_{|\underline{a}| \widehat\beta)} \, , \quad 
\nabla_{(\widehat\beta}A_{ \widehat\alpha) \alpha} +
  \nabla_{\alpha} B_{(\widehat\alpha \widehat\beta)} =\gamma^{\underline{a}}_{\widehat\alpha \widehat\beta} A_{\underline{a} \alpha} \, ,    
\label{eqAsBs} 
\end{aligned} 
\end{equation}
where symmetrization is unweighted, $B_{(\alpha \beta)} = B_{\alpha \beta}+B_{\beta \alpha}$, and the superfields $B_{\underline a\alpha}$, $B_{\underline a\widehat\alpha}$, $A_{\underline a\alpha}$, and $A_{\underline a\widehat\alpha}$ are defined as above. By 
the defining identities, it follows that these new superfields obey
\begin{equation}
\begin{aligned}
\gamma^{\underline{a}}_{( \alpha \beta} \left(\nabla_{\gamma)} A_{\underline{a} \widehat{\gamma}} - \nabla_{|\underline{a}|} A_{\gamma) \widehat{\gamma}} - i 
 (\gamma_{\underline{a}})_{\widehat\beta \widehat\gamma} \eta^{\sigma \widehat\beta} B_{|\sigma| \gamma)}  
+ 2 \nabla_{\widehat{\gamma}} B_{|\underline{a}| \gamma)} \right)  = 0 \, , \\
\gamma^{\underline{a}}_{( \widehat\alpha \widehat\beta} \left(\nabla_{\hat\gamma)} A_{\underline{a} \gamma}- \nabla_{|\underline{a}|} A_{\widehat\gamma) \gamma} + i 
 (\gamma_{\underline{a}})_{\alpha \gamma} \eta^{\alpha \widehat\delta} B_{|\widehat\delta| \widehat\gamma)}  
+ 2 \nabla_{\gamma} B_{|\underline{a}| \widehat\gamma)} \right)  = 0 \, , 
\end{aligned} 
\end{equation}
and 
\begin{equation}
(\gamma^{\underline{a}})_{(\alpha \beta} (\nabla_{\gamma} B_{\underline{a}| \delta)} - \nabla_{\underline{a}}  B_{\gamma \delta)}) =0 \, , \quad \quad 
(\gamma^{\underline{a}})_{(\widehat\alpha \widehat\beta} (\nabla_{\widehat\gamma} B_{\underline{a}| \widehat\delta)} - \nabla_{\underline{a}}  B_{\widehat\gamma \widehat\delta)}) =0 \, . 
\end{equation}
As in the previous case, one can turn these constraints into definitions of new superfields $W\,_{\widehat\alpha}^{\beta}$, $W\,_{\gamma}^{\widehat\beta}$, $W\,_{\beta}^{\gamma}$, and $W\,_{\widehat\beta}^{\widehat\gamma}$, namely
\begin{equation}
\begin{aligned}
\nabla_{\alpha} A_{\underline{a} \widehat{\alpha}} - \nabla_{\underline{a}} A_{\alpha \widehat{\alpha}} - i (\gamma_{\underline{a}})_{\widehat\alpha \widehat\beta} \eta^{\gamma \widehat\beta} B_{\gamma \alpha}  
+ 2 \nabla_{\widehat{\alpha}} B_{\underline{a}
 \alpha}  = (\gamma_{\underline{a}})_{\alpha \beta} W\,^{\beta}_{\widehat\alpha} \, , \\
 \nabla_{\widehat\gamma} A_{\underline{a} \gamma} - \nabla_{|\underline{a}|} A_{\widehat\gamma \gamma} + i 
 (\gamma_{\underline{a}})_{\alpha \gamma} \eta^{\alpha \widehat\delta} B_{\widehat\delta  \widehat\gamma}
+ 2 \nabla_{\gamma} B_{|\underline{a}| \widehat\gamma}   = (\gamma_{\underline{a}})_{\widehat\gamma \widehat\beta}W\,^{\widehat\beta}_{\gamma}  \, , 
\end{aligned} 
\end{equation}
and
\begin{equation}
\nabla_{(\alpha} B_{\underline{a}| \beta)} - \frac{1}{2}\nabla_{\underline{a}}  B_{(\alpha \beta)} = (\gamma_{\underline{a}})_{(\alpha |\gamma}W \,^{\gamma}_{\beta)} \, , \quad \quad 
\nabla_{(\widehat\alpha} B_{\underline{a}| \widehat\beta)} - \frac{1}{2}\nabla_{\underline{a}}  B_{(\widehat\alpha \widehat\beta)} = (\gamma_{\underline{a}})_{(\widehat\alpha |\widehat\gamma}W\,^{\widehat\gamma}_{\widehat\beta)} \, , 
\end{equation}
where we have used the following gamma-matrix identities:
\begin{equation}
\gamma^{ \underline{a}}_{( \alpha \beta} (\gamma_{\underline{a}})_{\gamma) \delta} =0 \, , \quad \quad  \gamma^{ \underline{a}}_{( \widehat\alpha \widehat\beta} (\gamma_{\underline{a}})_{\widehat\gamma) \widehat\delta} =0 \, . 
\label{gammaidentities}
\end{equation}
The derivatives of the $W$'s results in the following superfields
\begin{equation}
\begin{aligned}
F_{\underline{ab} \gamma} = \nabla_{\underline{a}} A_{\underline{b} \gamma}-
 \nabla_{\underline{b}} A_{\underline{a} \gamma} \, , \quad \quad 
F^{B}_{\underline{ab} \gamma}= \nabla_{\underline{a}} B_{\underline{b} \gamma} - \nabla_{\underline{b}} B_{\underline{a} \gamma} \, , \\
F_{\underline{ab} \widehat\gamma} = \nabla_{\underline{a}} A_{\underline{b} \widehat\gamma}-
 \nabla_{\underline{b}} A_{\underline{a} \widehat\gamma} \, , \quad \quad 
F^{B}_{\underline{ab} \widehat\gamma}= \nabla_{\underline{a}} B_{\underline{b} \widehat\gamma} - \nabla_{\underline{b}} B_{\underline{a} \widehat\gamma} \, .
\end{aligned}
\end{equation}
The equations are  
\begin{equation}
\begin{aligned} 
\nabla_{\alpha}  W\,^{\beta}_{\widehat{\gamma}} - \frac{i}{2} \eta^{\beta \widehat{\delta}} \nabla_{\widehat{\delta}}  A_{\alpha \widehat{\gamma}} + 2 \nabla_{\widehat\gamma} W\,^{\delta}_{\alpha}+ i B_{\underline{a} \alpha} \eta^{\delta \widehat\beta} (\gamma^{\underline{a}})_{\widehat\beta \widehat\gamma}= \frac{1}{4} (\gamma^{\underline{ab}})_{\alpha}^{\phantom{\alpha} \beta}  
F_{\underline{ab} \widehat{\gamma}} - \frac{i}{2} B_{\underline{a} \sigma} \eta^{\sigma \widehat\beta} (\gamma_{\underline{b}})_{\widehat\beta \widehat\gamma} (\gamma^{\underline{ab}})_{\alpha}\,^{\beta} \, ,   \\ 
\nabla_{\widehat{\alpha}}  W\,^{\widehat\beta}_{\alpha} + \frac{i}{2} \eta^{\beta \widehat{\beta}} \nabla_{\beta}  A_{\widehat\alpha \alpha} + 2 \nabla_{\alpha} W\,^{\widehat\beta}_{\widehat\alpha} -  i B_{\underline{a} \widehat\alpha} \eta^{\gamma \widehat\beta} (\gamma^{\underline{a}})_{\gamma \alpha}= \frac{1}{4} (\gamma^{\underline{ab}})_{\widehat\alpha}^{\phantom{\widehat\alpha} \widehat\beta}  
F_{\underline{ab} \alpha} + \frac{i}{2} B_{a \widehat\gamma} \eta^{\gamma \widehat\gamma} (\gamma_{\underline{b}})_{\gamma \alpha} (\gamma^{\underline{ab}})_{\widehat\alpha}\,^{\widehat\beta} \, ,    
\nonumber
\end{aligned} 
\end{equation}
and 
\begin{equation}
\begin{aligned}
\nabla_{(\gamma} W\,^{\alpha}_{\beta)}  - \frac{i}{4} \eta^{\alpha \widehat\alpha} \nabla_{\widehat\alpha} B_{(\beta \gamma)} = \frac{1}{4} (\gamma^{\underline{ab}})_{(\beta}\,^{\alpha} F^{B}_{\underline{ab} |\gamma)} \, ,  \\
\nabla_{(\widehat{\gamma}} W\,^{\widehat{\alpha}}_{\widehat{\beta})} +
\frac{i}{4} \eta^{\alpha \widehat{\alpha}} \nabla_{\alpha} B_{(\widehat\gamma \widehat\beta)}
= \frac{1}{4} (\gamma^{\underline{ab}})_{(\widehat{\gamma}}\,^{\widehat{\alpha}}F_{\underline{ab} | \widehat\beta)} \, . 
\end{aligned} 
\end{equation}
In the next Subsection, we show that this tower of superfields, defined via the Jacobi identities, arises naturally once the non-zero modes are included. Specifically, it appears in the construction of integrated vertex operators via the descent relations.

To get the expansions of all the superfields appearing above, one has 
to act with the covariant derivatives on $A_{\alpha\widehat\alpha}$ and find 
the new superfields as solutions of the given set of equations.
The fermionic covariant derivatives can be read from the BRST transformations  
and the bosonic ones
can be extracted from the commutators 
\begin{equation}
\{\nabla_{\alpha},\nabla_{\beta}\}=\gamma^{\underline{a}}_{\alpha\beta}\nabla_{\underline{a}} \, ,
\end{equation}
by using the decompositions of the gamma matrices of the 
Appendix \ref{apsupermatrices}.

Alternatively, notice that all the equations of this Subsection involve only symmetric combinations of spinor indices. This means that we can contract these indices with bosonic spinors $\rho^{\alpha}$ and $\rho^{\widehat\alpha}$, without losing any information. Also, it is clear that if we impose the pure spinor constraints on $\rho^{\alpha}$ and $\rho^{\widehat\alpha}$, the BRST invariance of the unintegrated vertex operator implies a trivialization of the equations.
Let us consider the second equation of the first line of \eqref{eqAsBs} and contract it with the bosonic spinors  
\begin{equation}
\rho^{\beta} \rho^{\alpha} \rho^{\widehat\gamma} \nabla_{\beta} A_{\alpha\widehat\gamma}(g(Z)) + 2 \rho^{\beta} \rho^{\alpha} \rho^{\widehat\gamma} \nabla_{\widehat\gamma} B_{\beta \alpha}(g(Z)) = \rho^{\beta} \rho^{\alpha} \rho^{\widehat\gamma} \gamma^{\underline{a}}_{\alpha\beta} A_{\underline{a} \widehat\gamma}(g(Z))\,.
\label{contrairho}
\end{equation}
The left-hand-side term containing the superfield $A_{\alpha\widehat\gamma}$ can be obtained by collecting the terms that are linear in $\varepsilon$ in
\begin{equation}
A_{\alpha\widehat\gamma}[g(Z)\left(1+\varepsilon\rho^{\alpha}q_{\alpha}\right)]  = A_{\alpha\widehat\gamma}[g(Z_{\varepsilon})\left(1+\varepsilon\rho^{\alpha}\Omega_{\alpha}\,^{\underline{ab}}\widetilde{M}_{\underline{ab}}\right)]   \,,
\end{equation}
where, for example, 
\begin{equation}
A_{\alpha\widehat\gamma}[g(Z)(1+\varepsilon \widetilde{M}_{\underline{ab}})] \Big|_{\rm{linear \, in \, \varepsilon}}= - \varepsilon\frac{1}{2}(\gamma_{\underline{ab}})_{\alpha}\,^{\beta}A_{\beta\widehat\gamma} (g(Z))  -\varepsilon\frac{1}{2}(\gamma_{\underline{ab}})_{\widehat\gamma}\,^{\widehat\beta}A_{\alpha\widehat\beta}(g(Z))\,,
\end{equation}
and similarly for the covariant derivative acting on the $B$ on 
\eqref{contrairho}.

\subsection{The pure spinor formalism: non-zero conformal weight}
\label{The integrated vertex operator}

Up to this point we have considered only the zero conformal weight worldsheet variables. We now introduce non-zero ones by allowing the worldsheet derivatives $\partial$ and $\overline\partial$ to act on our variables. 
Since including nonzero modes makes vertex operators defined only up to the sigma-model equations of motion (e.o.m), in Subsection \ref{sigma model action} we present the $AdS_5\times S^5$ sigma-model action and in Subsection \ref{equations of motion worldsheet} we derive those equations. In Subsection \ref{descent relations fundamentals} we outline the standard descent relations connecting integrated to unintegrated vertices. 
In Subsection \ref{Integrated Vertex Operators} we present the integrated vertex operator as covariant derivatives 
acting on the bi-spinor superfield of Subsection \ref{previous paper result}, contracted with worldsheet variables to produce a gauge-invariant vertex.
We are going to 
 work at the supergravity level, not considering $\alpha^{\prime}$ corrections, 
and normal ordering is assumed.  
Also, the BRST operator has the simple action of \eqref{BRSTaction}
when $\alpha^{\prime}$ corrections are not considered. 
In Subsection \ref{SO to USP}, we express these integrated vertices in a $USp(2,2)\times USp(4)$ covariant form.
We end this section by constructing the B-RNS-GSS vertex operators.

\subsubsection{
Superstring action for $AdS_5\times S^5$ background}
\label{sigma model action}

Since the integrated vertex operators are defined up to e.o.m's, it is necessary to write
down the worldsheet action of the model. 
The action is completely fixed by imposing BRST invariance. 
It is given by 
\begin{equation}
S = \left\langle \frac{1}{2} J^{(2)} \overline{J}^{(2)} + \frac{3}{4} J^{(3)} \overline{J}^{(1)} + \frac{1}{4} J^{(1)} \overline{J}^{(3)}
+ \omega^{(3)} \overline{\nabla} \lambda^{(1)} + \widehat{\omega}^{(1)} \nabla \widehat{\lambda}^{(3)} - N^{(0)} \widehat{N}^{(0)}\right\rangle \, , 
\label{action}
\end{equation}
where
\begin{equation*}
\langle\,\cdot\,\rangle = \int d^2z \, \text{Str}\left(\,\cdot\,\right)\,,    
\end{equation*}
where $\mathrm{Str}$ denotes the supertrace.
The holomorphic and antiholomorphic worldsheet currents $J$ and $\overline J$ take values in the Lie superalgebra $\mathfrak{psu}(2,2|4)$ and decompose under the $\mathbb{Z}_4$ grading introduced in Subsection \ref{covariant derivatives}. They are
\begin{equation}
J = g^{-1}\partial g\,,\qquad \overline J = g^{-1}\overline\partial g \, , 
\end{equation}    
where the holomorphic current decomposes as
\begin{equation}
J^{(1)} = (g^{-1} \partial g)^{\alpha} q_{\alpha} \, , \qquad 
J^{(3)} = (g^{-1} \partial g)^{\widehat\alpha} q_{\widehat\alpha}  \, ,
\end{equation}
\begin{equation}
J^{(2)} = (g^{-1} \partial g)^{\underline{a}} \widetilde{P}_{\underline{a}} \, ,\qquad
J^{(0)} = (g^{-1} \partial g)^{\underline{ab}} \widetilde{M}_{\underline{ab}} \, ,
\end{equation}
and the antiholomorphic current decomposes as
\begin{equation}
\overline J^{(1)} = (g^{-1} \overline \partial g)^{\alpha} q_{\alpha} \, , \qquad 
\overline J^{(3)} = (g^{-1} \overline \partial g)^{\widehat\alpha} q_{\widehat\alpha}  \, ,
\end{equation}
\begin{equation}
\overline J^{(2)} = (g^{-1} \overline \partial g)^{\underline{a}} \widetilde{P}_{\underline{a}} \, ,\qquad
\overline J^{(0)} = (g^{-1} \overline \partial g)^{\underline{ab}} \widetilde{M}_{\underline{ab}} \, .
\end{equation}
They satisfy the Maurer-Cartan identity
\begin{equation}
\partial \overline J - \overline \partial J  + \{ J, \overline J] =0\,,  
\end{equation}
which decomposes under the $\mathbb{Z}_4$ gradding as
\begin{equation}
\nabla \overline J^{(1)} - \overline\nabla J^{(1)}  + [J^{(2)}, \overline J^{(3)}]  + [ J^{(3)}, \overline J^{(2)}]= 0\,,   
\end{equation}
\begin{equation}
\nabla \overline J^{(3)} - \overline\nabla J^{(3)}  + [J^{(1)}, \overline J^{(2)}]  + [J^{(2)}, \overline J^{(1)}]= 0\,,   
\end{equation}
and
\begin{equation}
\nabla \overline J^{(2)} - \overline\nabla J^{(2)}  + \{ J^{(1)}, \overline J^{(1)}\}  + \{ J^{(3)}, \overline J^{(3)}\}= 0\,,   
\end{equation}
where we have introduced the covariant derivative   
\begin{equation}
\nabla \,\cdot\, = \partial \,\cdot\, + 
[ J^{(0)}, \,\cdot\, ] \, , \quad {\rm{and}} \quad 
\overline{\nabla} \,\cdot\, = \overline{\partial} \,\cdot\, + 
[ \overline{J}^{(0)}, \,\cdot\, ] \, .  
\end{equation}
The worldsheet variables $(\omega,\widehat{\omega})$ are conjugate to the pure spinors $(\lambda,\widehat{\lambda})$ and therefore have conformal weights $(1,0)$ and $(0,1)$, respectively. These variables are valued at elements of $\mathfrak{psu}(2,2|4)$ with odd $\mathbb{Z}_4$ gradding, as follows:
\begin{equation}
\lambda^{(1)} = \lambda^{\alpha} q_{\alpha} \, , \qquad 
\widehat\lambda^{(3)} = \lambda^{\widehat\alpha} q_{\widehat\alpha} \, , \qquad \omega^{(3)}  = \omega_{\alpha} \eta^{\alpha \widehat\alpha}q_{\widehat\alpha}  \, ,\qquad
\widehat\omega^{(1)}  = \widehat\omega_{\widehat\alpha} \eta^{\alpha \widehat\alpha} q_{\alpha} \, , 
\end{equation}
where the Ramond-Ramond background bi-spinor $\eta^{\alpha\widehat\alpha}$ are given by
\begin{equation}
{\rm{Str}} (q_{\alpha} q_{\widehat\alpha}) = \eta_{\alpha \widehat\alpha} \, , \quad \quad \eta^{\beta \widehat\alpha} \eta_{\alpha \widehat\alpha} = - \delta^{\beta}_{\alpha} \, . 
\end{equation}
The ghost currents are obtained from the anti-commutators
\begin{equation}
\begin{aligned}
& N^{(0)} = - \{ \omega^{(3)}, \lambda^{(1)} \} = - \omega_{\alpha} \eta^{\alpha \widehat\alpha} \lambda^{\beta}\{q_{\widehat\alpha} ,   q_{\beta}  \}
=
- \frac{i}{2} (\lambda \gamma^{[ab]} \omega) \widetilde{M}_{[ab]}  \, , \\
& \widehat{N}^{(0)} = - \{ \hat\omega^{(1)}, \widehat\lambda^{(3)} \} 
= - \widehat\omega_{\widehat\alpha}  \eta^{\alpha \widehat\alpha} \widehat\lambda^{\widehat\beta} \{  q_{\alpha} ,  q_{\widehat\beta} \} = \frac{i}{2} (\widehat\lambda \gamma^{[ab]} \widehat\omega) \widetilde{M}_{[ab]} \, . 
\label{TheNgenerators} 
\end{aligned} 
\end{equation}
For comparison with \cite{IntegratedVertex}, we take the $AdS_5$ direction dual to the scaling dimension to be $a=5$ and the $S^5$ direction corresponding to the R-charge to be $a'=10$. Following \cite{IntegratedVertex}, we also contract antisymmetric pairs of vector indices $[ab]$ as
\begin{equation}
A^{[ab]}B_{[ab]} =
A^{ab}B_{ab} - A^{a'b'}B_{a'b'} \, ,
\end{equation}
so that, if 
\begin{equation}
N^{\underline{ab}} =  \frac{1}{2} (\lambda \gamma^{\underline{ab}} \omega) \, , \quad \quad %
\overline{N}^{\underline{ab}} =  \frac{1}{2} (\widehat\lambda \gamma^{\underline{ab}} \hat\omega) \, , 
\end{equation}
then \eqref{TheNgenerators} becomes
\begin{equation}
N^{(0)} = - i N^{ab}\widetilde{M}_{ab}+ i N^{a'b'}\widetilde{M}_{a'b'} \, , \quad \quad 
\widehat{N}^{(0)} = i \widehat{N}^{ab}\widetilde{M}_{ab} - i \widehat{N}^{a'b'}\widetilde{M}_{a'b'} \, . 
\label{TheNequation} 
\end{equation}
For notation details, see Appendix \ref{apsupermatrices}.

\subsubsection{The BRST transformations of the non-zero modes}
\label{The BRST transformations of the non-zero modes}

The BRST transformations of the holomorphic and anti-holomorphic currents follow from the Maurer-Cartan identities
\begin{equation}
Q \cdot J = \partial \lambda  + \{J,  \lambda]\,,  \qquad 
Q \cdot \overline{J} = \overline \partial \lambda  + \{\overline J,  \lambda]\,,
\end{equation}
The $\mathbb{Z}_4$ grading decomposition of the above equations results in
\begin{equation}
\begin{aligned}
& Q \cdot J^{(0)} = [ J^{(3)}, \lambda^{(1)}] + [ J^{(1)}, \widehat\lambda^{(3)}] \, , \qquad 
Q \cdot J^{(1)} = \nabla  \lambda^{(1)} + [ J^{(2)}, \widehat\lambda^{(3)}] \, , \\
& Q \cdot J^{(2)} = [ J^{(1)},\lambda^{(1)}] + [ J^{(3)},  \widehat\lambda^{(3)}] \, , \qquad 
Q \cdot J^{(3)} = \overline{\nabla} \widehat\lambda^{(3)} + [ J^{(2)}, \lambda^{(1)}] \, . 
\end{aligned}
\end{equation}        
The BRST transformations of the variables $(\omega, \widehat\omega)$ conjugate to the pure spinors are
\begin{equation}
Q \cdot \omega^{(3)} = - J^{(3)} \, , \quad \quad  Q \cdot \widehat\omega^{(1)} = - \overline{J}^{(1)} \, , 
\end{equation}
so that the BRST transformations of the currents $N^{(0)}$ and $\widehat N^{(0)}$ are determined to be 
\begin{equation}
\begin{aligned}
Q \cdot N^{(0)} = \{ J^{(3)}, \lambda^{(1)} \}  \, ,  \qquad  
Q \cdot \widehat{N}^{(0)} = \{ J^{(1)}, \widehat\lambda^{(3)} \} \, . 
\end{aligned} 
\end{equation}
Using \eqref{TheNequation} and Appendix \ref{apsupermatrices}, one finds
\begin{equation}
Q  \cdot N^{\underline{ab}} = \frac{1}{2} J^{\widehat{\alpha}} \lambda^{\alpha} (\gamma^{\underline{ab}} \eta)_{\alpha \widehat\alpha} \, , \quad \quad
Q \cdot \widehat{N}^{\underline{ab}} = - \frac{1}{2} J^{\alpha} \lambda^{\widehat\alpha} (\gamma^{\underline{ab}} \eta)_{\alpha \widehat\alpha} \, . 
\end{equation}

\subsubsection{Equations of motion}
\label{equations of motion worldsheet}

The equations of motion are obtained by varying the 
action \eqref{action} with respect to $g\in PSU(2,2|4)$, the pure spinor variables $(\lambda^{(1)},\widehat\lambda^{(3)})$, and their %
conjugates $(\omega^{(3)},\widehat\omega^{(1)})$. 
There is a $USp(2,2|4)$ subgroup of variations that leaves the action invariant. 
It is convenient to parametrize the local variation of $g$ as $g^{-1}\delta g$, 
whose $\mathbb{Z}_4$ decomposition isolates this $USp(2,2|4)$ subgroup. 
The equations of motion are given by
\begin{equation}
\begin{aligned}
& \nabla \bar{J}^{(3)} = - [ J^{(1)}, \bar{J}^{(2)}]
- [ J^{(2)}, \bar{J}^{(1)}]
+ [ N^{(0)} , \bar{J}^{(3)}]
+ [\widehat{N}^{(0)}, J^{(3)} ] \, , \\
& \overline{\nabla} J^{(3)} =  [N^{(0)}, \bar{J}^{(3)}] + [\widehat{N}^{(0)}, J^{(3)}] \, , \\
& \nabla \bar{J}^{(2)} = 
- [J^{(1)}, \bar{J}^{(1)}] + [N^{(0)}, \bar{J}^{(2)}] +
[ \widehat{N}^{(0)}, J^{(2)}] \, , \\
& \overline{\nabla} J^{(2)} =  [J^{(3)}, \bar{J}^{(3)}] 
+ [N^{(0)}, \bar{J}^{(2)}] + [\widehat{N}, J^{(2)}] \, , \\
& \nabla \bar{J}^{(1)} =  [N^{(0)} , \bar{J}^{(1)}] + [ \widehat{N}^{(0)}, J^{(1)}] \, , \\
& \overline{\nabla} J^{(1)} =  [ J^{(2)}, \bar{J}^{(3)}]
+ [J^{(3)}, \bar{J}^{(2)}] + [N^{(0)},\bar{J}^{(1)}]+
[\widehat{N}^{(0)}, J^{(1)}] \, ,
\end{aligned}
\end{equation}
and   
\begin{equation}
\overline{\nabla} \lambda^{(1)} = [ \widehat{N}^{(0)}, \lambda^{(1)}] \, , \qquad \nabla \widehat\lambda^{(3)} = [ N^{(0)}, \widehat\lambda^{(3)}] \, , 
\end{equation}
\begin{equation}
\overline{\nabla} \omega^{(3)} = [ \widehat{N}^{(0)}, \omega^{(3)}] \, , \qquad
\nabla \widehat{\omega}^{(1)} = [N^{(0)}, \widehat{\omega}^{(1)}]\,.    
\end{equation}
Using the equations above, the definitions \eqref{TheNgenerators} and the super-Jacobi identities, it is possible to derive the following equations of motion for $N^{(0)}$ and $\widehat{N}^{(0)}$
\begin{equation}
\overline{\nabla} N^{(0)} = - [ N^{(0)}, \widehat{N}^{(0)}] \, , \qquad
\nabla \widehat{N}^{(0)} = [N^{(0)}, \widehat{N}^{(0)}] \, . 
\end{equation}

\subsubsection{Integrated vertices via descent relations}
\label{descent relations fundamentals}

Having introduced the holomorphic and antiholomorphic derivatives, together with the 
action and its equations of motion, we are ready to construct the integrated vertex operators. After imposing the 
equations of motion, the holomorphic and antiholomorphic derivatives of a $Q$-closed unintegrated vertex $V$ of conformal weight zero are $Q$-exact:
\begin{equation}
\partial V = Q \cdot W\,,\qquad \overline \partial V = Q \cdot \overline{W} \,.   
\end{equation}
Although $W$ and $\overline W$ are not $Q$-closed, one can form $Q$-closed objects by integrating along holomorphic and antiholomorphic cycles %
\begin{equation}
\oint_{C} \frac{dz}{2\pi i}\,Q \cdot W = 0\,,\qquad \oint_{C} \frac{d\overline z}{2\pi i}\,Q \cdot \overline{W} = 0 \,.
\end{equation}
Alternatively, combine them into a curl:
\begin{equation}
Q \cdot \big(\partial \overline W - \overline\partial W\big) = (\partial \overline\partial - \overline\partial\,\partial) V = 0 \,.
\end{equation}
If this curl is $Q$-exact, then there exists $U$ such that
\begin{equation}
\partial \overline W - \overline\partial W = Q \cdot U \,,
\label{firstdescent}
\end{equation}
from which we can form the $Q$-closed integrated object
\begin{equation}
\int_{\Sigma} d^2z \, Q \cdot U = 0 \,,
\end{equation}
with $\Sigma$ the whole worldsheet.
This can be shown either by brute force (as we will do here) or by properly defining the $b$-ghost. The $b$-ghost \cite{Witten:1992fb} is essential in covariant formulations of the superstring; they act as interior-product operators that map $Q$ to the stress–energy tensor (the generator of conformal transformations):
\begin{equation}
\{Q, b\} = T\,,\qquad \{Q,\overline b\}=\overline T\,.    
\end{equation}
For example, $b$-ghost insertions furnish the integration measure on the moduli space of Riemann surfaces. Using these interior products, one shows that
\begin{equation}
\partial V =  L_{-1}\cdot V = \{Q, b_{-1}\} \cdot V = Q \cdot \left(b_{-1}\cdot V\right) \, , 
\end{equation}
and similarly for the antiholomorphic derivative, hence
\begin{equation}
W = b_{-1}\cdot V\,,\qquad \overline W = \overline b_{-1}\cdot V \,.   
\end{equation}
However, as in a flat background, defining the $b$ ghost entails subtleties in the pure spinor formalism, so we prefer to solve for $W$ and $\overline W$ directly, following \cite{DescentI,DescentII}. In the $AdS_5 \times S^5$ background, the direct computation %
of $W$, $\overline W$, and $U$ was carried out in \cite{IntegratedVertex}. 
An alternative method employs the generalized Lax connections \cite{LaxIntegrated}.

One can also obtain the descent equations by starting from the integrated vertex operators viewed as infinitesimal deformations of the action:
\begin{equation}
S = S^{(0)} \rightarrow S^{(0)} +  \varepsilon S^{(1)} \, ,\qquad S^{(1)} = \int dz^2 \, U   
\end{equation}
This term induces a deformation of the BRST operator at the same order in $\varepsilon$, fixed by requiring that the deformed action remains BRST invariant:
\begin{equation}
Q = Q^{(0)} \rightarrow Q^{(0)} + \varepsilon Q^{(1)} \,, \qquad Q^{(0)} \cdot S^{(1)} + Q^{(1)}\cdot S^{(0)} = 0\,.    
\end{equation}
This, in turn, implies that $S^{(1)}$ must be $Q^{(0)}$-closed up to the equations of motion derived from $S^{(0)}$. Because $S^{(1)}$ is an integral of a local term $U$, it follows from $Q^{(0)}$ nilpotency that
\begin{equation}
Q \cdot U = \partial \overline W - \overline\partial W\,,  
\end{equation}
for some $W$ and $\overline W$. Then, by Poincaré lemma and $Q$ nilpotency, it follows that
\begin{equation}
0 = Q^2 \cdot U = Q \cdot (\partial W - \overline\partial W)  
= \partial (Q \cdot \overline{W}) - \overline\partial(Q \cdot W) \implies Q\cdot \overline{W} = \overline\partial V\,,\; Q \cdot W = \partial V\,, 
\end{equation}
for some $V$ in the $Q$ cohomology. 

In what follows, we express $W$, $\overline W$, and $U$ in terms of $V$. Specifically, we construct them from the covariant superfields defined in Subsection \ref{Covariant superfields}, contracted with worldsheet variables to form gauge-invariant vertices. These covariant superfields are realized via the covariant derivatives defined in Subsection \ref{covariant derivatives} acting on the bi-spinor superfield appearing in the unintegrated vertex operator in Subsection \ref{previous paper result}.

\subsubsection{Integrated vertex operators in $AdS_5\times S^5$ background}
\label{Integrated Vertex Operators}

In what follows, %
we construct the integrated vertex operators by solving the last steps of the descent procedure.
We use the notation of the Subsection \ref{Mass-dimension tower via Jacobi identities} for the superfields. 
Some comments on the flat space case are given in the Subsection \ref{Propertiesofthevertexoperators}.   
The first step in establishing the descent equations is finding superfields $W$ and $\overline{W}$ satisfying up to the e.o.m's, 
\begin{equation}
\begin{aligned}
& Q \cdot W =  \partial (\lambda^{\alpha} \lambda^{\widehat\alpha}A_{\alpha\widehat\alpha})+ \partial(\lambda^{\alpha} \lambda^{\beta} B_{\alpha \beta}) + \partial( \lambda^{\widehat\alpha} \lambda^{\widehat\beta} B_{\widehat\alpha \widehat\beta}) \, ,  \\
& Q \cdot \overline{W} = \bar\partial (\lambda^{\alpha} \lambda^{\widehat\alpha}A_{\alpha\widehat\alpha})+ \bar\partial(\lambda^{\alpha} \lambda^{\beta} B_{\alpha \beta}) + \bar\partial( \lambda^{\widehat\alpha} \lambda^{\widehat\beta} B_{\widehat\alpha \widehat\beta}) \, . 
\label{firststep} 
\end{aligned}
\end{equation}
where for example
\begin{equation}
\partial (\lambda^{\alpha} {\lambda}^{\widehat{\alpha}} A_{\alpha \widehat\alpha}) =
\nabla \lambda^{\alpha} {\lambda}^{\widehat{\alpha}} A_{\alpha \widehat\alpha}
+ \lambda^{\widehat\alpha} \nabla{\lambda}^{\widehat{\alpha}} A_{\alpha \widehat\alpha} 
+ \lambda^{\alpha} {\lambda}^{\widehat{\alpha}} (J^{\beta} \nabla_{\beta} +
J^{\widehat\beta} \nabla_{\widehat\beta}+ J^{\underline{a}} \nabla_{\underline{a}}) A_{\alpha \widehat\alpha} \, , 
\end{equation}
and similarly for the other derivatives. We have found the following expressions for 
$W$ by writing it as two contributions  
\begin{equation}
W = W_1 + W_2 \, ,  
\end{equation}
where
\begin{equation}
\begin{aligned}
& W_1 = \lambda^{\widehat\beta} \left(J^{\alpha} A_{\alpha \widehat\beta}+ J^{\underline{a}} A_{\underline{a} \widehat\beta} 
- 2 i \eta_{\alpha \widehat\alpha} J^{\widehat\alpha} W^{\alpha}_{\widehat\beta} - i N^{\underline{a b}} (F_{\underline{a b} \widehat\beta} - 2 i B_{\underline{a} \beta} \eta^{\beta \widehat\alpha} (\gamma_{\underline{b}})_{\widehat{\beta} \widehat{\alpha}} )
\right) \, ,  \\
& W_2 = 2 \lambda^{\alpha}( J^{\beta} B_{\alpha \beta} + J^{\underline{a}} B_{\underline{a} \alpha} - 2 i \eta_{\beta \widehat\alpha} J^{\widehat\alpha} W^{\beta}_{\alpha} - i N^{\underline{ab}} F^B_{\underline{ab}\alpha}) \, . 
\label{W1W2}
\end{aligned}
\end{equation}
Similarly, writing 
\begin{equation}
\overline{W} = \overline{W}_1 + \overline{W}_2 \, ,  
\end{equation}
one has
\begin{equation}
\begin{aligned}
& \overline{W}_1 = \lambda^{\beta} \left(\bar{J}^{\widehat\alpha} A_{\beta \widehat\alpha}+ \bar{J}^{\underline{a}} A_{\beta \underline{a}} 
+ 2 i \eta_{\alpha \widehat\alpha} J^{\alpha} W^{\widehat\alpha}_{\beta} + i \bar{N}^{\underline{a b}} (F_{\underline{a b} \beta} + 2 i B_{\underline{a} \widehat\gamma} \eta^{\gamma \widehat\gamma}(\gamma_{\underline{b}})_{\gamma \beta})
\right) \, , \\
& \overline{W}_2 = 2 \lambda^{\widehat\alpha}( \bar{J}^{\widehat\beta} B_{\widehat\alpha \widehat\beta} + \bar{J}^{\underline{a}} B_{\underline{a} \widehat\alpha} +  2 i \eta_{\alpha \widehat\beta} \bar{J}^{\alpha} W^{\widehat\beta}_{\widehat\alpha} + i N^{\underline{ab}} F^B_{\underline{ab}\widehat\alpha}) \, . 
\label{W1bW2b}
\end{aligned} 
\end{equation}

The procedure to show that \eqref{firststep} are satisfied by both \eqref{W1W2} and \eqref{W1bW2b} follows the same steps as described in detail in \cite{IntegratedVertex}.
The equations only get corrections from the $B$'s and the $W$ and $\overline{W}$ have more terms. For example, it is possible to show the identities   
\begin{equation}
\begin{aligned}
& \lambda^{\widehat\beta} \lambda^{\widehat\gamma} \nabla_{\widehat\beta} A_{\underline{a} \widehat\gamma} =
\frac{i}{2} \lambda^{\widehat\beta} \lambda^{\widehat\gamma} (\gamma_{\underline{a}})_{\widehat\beta \widehat\rho}  \eta^{\rho \widehat\rho} A_{\rho \widehat\gamma} + \lambda^{\widehat\beta} \lambda^{\widehat\gamma}  \nabla_{\underline{a}} B_{\widehat\beta \widehat\gamma} \, ,  \\
& \lambda^{\widehat\beta} \lambda^{\widehat\gamma} \nabla_{\widehat\beta} W^{\alpha}_{\widehat\gamma} = 
- \frac{i}{2} \lambda^{\widehat\beta} \lambda^{\widehat\gamma} (\gamma^{\underline{a}})_{\widehat\gamma \widehat\rho} \eta^{\alpha \widehat\rho} A_{\underline{a} \widehat\beta} + \frac{i}{2} \lambda^{\widehat\beta} \lambda^{\widehat\gamma} 
\eta^{\alpha \widehat\rho} \nabla_{\widehat\rho} B_{\widehat\beta \widehat\gamma} \, ,
\end{aligned} 
\end{equation}
where the terms with the $B$'s are absent from the expressions of \cite{IntegratedVertex}.

Having determined $W$ and $\overline{W}$, we can solve up to e.o.m's, the equation $Q \cdot U = \partial \overline{W} - \bar{\partial} W$ which fixes the integrated vertex operator $U$. The expression for $U$ is 
\begin{equation}
\begin{aligned}
\label{TheUsuperfield}
U & =2 i\eta_{\beta\widehat\gamma}J^{\alpha}\overline{J}^{\beta}W_{\alpha}^{\widehat\gamma}-2 i\eta_{\gamma\widehat\alpha}J^{\widehat\alpha}\overline{J}^{\widehat\beta}W^{\gamma}_{\widehat\beta}+J^{\alpha}\overline{J}^{\widehat\beta}A_{\alpha\widehat\beta}+J^{\alpha}\overline{J}^{\underline a}A_{\alpha\underline a}-J^{\underline a}\overline{J}^{\widehat\alpha}A_{\underline a\widehat\alpha}    \\
& +i J^{\alpha}\overline N^{\underline{ab}}(F_{\alpha\underline{ab}} + 2 i B_{\underline{a} \widehat{\gamma}} \eta^{\gamma \widehat{\gamma}}(\gamma_{\underline{b}})_{\gamma \alpha}) 
+ i N^{\underline{ab}}\overline{J}^{\widehat\alpha} (F_{\underline{ab}\widehat\alpha} - 2 i B_{\underline{a} \beta} \eta^{\beta \widehat{\alpha}} (\gamma_{\underline{b}})_{\widehat{\alpha} \widehat{\beta}}) 
 \\ & +J^{\widehat\beta}\overline{J}^{\alpha}\overline{\mathcal{V}}_{\alpha\widehat\beta}+J^{\underline{a}}\overline{J}^{\alpha}\overline{\mathcal{V}}_{\underline a\alpha}+J^{\widehat\alpha}\overline{J}^{\underline{a}}\mathcal{V}_{\underline{a}\widehat\alpha}+J^{\underline{a}}\overline{J}^{\underline{b}}\mathcal{V}_{\underline{ab}}    \\
& +\frac{1}{4}N^{\underline{ab}}\overline N^{\underline{cd}}\mathcal{V}_{\underline{ab}\underline{cd}}+\frac{1}{2}N^{\underline{ab}}\overline{J}^{\alpha}\overline{\mathcal{V}}_{\underline{ab}\alpha}+\frac{1}{2}J^{\widehat\alpha}\overline N^{\underline{ab}}\mathcal{V}_{\underline{ab}\widehat\alpha}+\frac{1}{2}J^{\underline{a}}\overline N^{\underline{bc}}\mathcal{V}_{\underline{a}\underline{bc}}+\frac{1}{2}N^{\underline{bc}}\overline{J}^{\underline{a}}\overline{\mathcal{V}}_{\underline{a}\underline{bc}} \,.   
\end{aligned}
\end{equation}

All the superfields $\mathcal{V}$ and  $\overline{\mathcal{V}}$  are determined by acting with the covariant derivatives on the superfields 
of the first and second lines. 
The expressions defining them, which follows from the descent equations, can be found in the Appendix  
\ref{Integrated vertex operators superfields}.

\subsubsection{Integrated vertex in $USp(2,2)\times USp(4)$ notation %
}
\label{SO to USP}

The goal of this Subsection is the decomposition of the vertices $V$ of \eqref{vertex10d}  given in $SO(10)$ notation, into both $USp(2,2)\times USp(4)$ and 
the $\mathfrak{su}_2^{\oplus 4}$-component notation used in this work. 
The decompositions of the supercharges, gamma matrices, and 
$\eta_{\alpha \widehat\alpha}$ are given in the Appendix \ref{apsupermatrices}.
The decompositions of the pure spinors are discussed in the Subsection \ref{Covariant pure spinors}.
The vertex operators decompose as 
\begin{equation}
V_{USp(2,2)\times USp(4)} =  
(\lambda_L)^R_I (\lambda_R)^S_J 
A_{R S}^{I J} + (\lambda_L)^R_I (\lambda_L)^S_J 
B_{R S}^{I J} +(\lambda_R)^R_I (\lambda_R)^S_J 
\widehat{B}_{R S}^{I J}  \, . 
\end{equation}

As an example of decomposition of the superfields, let us consider the second equation on the first line of \eqref{eqAsBs}, which we reproduce here for the reader's convenience 
\begin{equation}
\nabla_{(\beta} A_{\alpha) \widehat\alpha} +   \nabla_{\widehat\alpha}B_{(\alpha\beta)} =\gamma^{\underline{a}}_{\alpha \beta} A_{\underline{a} \widehat{\alpha}} \, . 
\end{equation}
This equation determines the superfield $A_{\underline{a}\widehat\alpha}$, i.e., it can be inverted by multiplying both sides by $\gamma_{\underline{b}}^{\alpha \beta}$.
Decomposing the equation, the components $A_{5}\,_{I}^R$ and $A_{5'}\,_{I}^R$ can be extracted from
\begin{equation}
\begin{aligned}
& \nabla_{\alpha}^{i}A_{\beta}^{j}\,_{I}^R +\nabla_{\beta}^{j}A_{\alpha}^{i}\,_{I}^R + 2 \nabla_{I}^R 
B_{\alpha}^{i}\, _{\beta}^{j}=\left(A_{5}\,_{I}^R +A_{5'}\,_{I}^R \right)\varepsilon^{ij}\varepsilon_{\alpha\beta}\,, \\ & \nabla_{\dot\alpha}^{\overline i}A_{\dot\beta}^{\overline j}\,\,_{I}^R+\nabla_{\dot\beta}^{\overline j}A_{\dot\alpha}^{\overline i}\,\,_{I}^R + 2 \nabla_{I}^R 
B_{\dot\alpha}^{\bar{i}}\, _{\dot\beta}^{\bar{j}}=-\left(A_{5}\,_{I}^R+A_{5'}\,_{I}^R \right)\varepsilon^{\overline i\overline j}\varepsilon_{\dot\alpha\dot\beta}   \,,
\end{aligned} 
\end{equation}
and
\begin{equation}
\begin{aligned}
\nabla_{\dot\alpha}^{i}A_{\dot\beta}^{j}\,_{I}^R+\nabla_{\dot\beta}^{j}A_{\dot\alpha}^{i}\,_{I}^R+ 2 \nabla_{I}^R 
B_{\dot\alpha}^{i}\, _{\dot\beta}^{j}=\left(A_{5}\,_{I}^R-A_{5'}\,_{I}^R \right)\varepsilon^{ij}\varepsilon_{\dot\alpha\dot\beta}\,, \\\qquad  \nabla_{\alpha}^{\overline i}A_{\beta}^{\overline j}\,_{I}^R+\nabla_{\beta}^{\overline j}A_{\alpha}^{\overline i}\,_{I}^R+ 2 \nabla\,_{I}^R
B_{\alpha}^{\bar{i}}\, _{\beta}^{\bar{j}}=\left(A_{5'}\,_{I}^R-A_{5}\,_{I}^R\right)\varepsilon^{\overline i\overline j}\varepsilon_{\alpha\beta}   \,.    
\end{aligned} 
\end{equation}
Moreover, the components $A_{\mu}\,_{I}^R$ can be extracted from
\begin{equation}
\begin{aligned}
\nabla_{\alpha}^{i}A_{\dot\beta}^{j}\,_{I}^R+\nabla_{\dot\beta}^{j} A_{\alpha}^{i}\,_{I}^R + 2 \nabla_{I}^R
B_{\alpha}^{i}\,_{\dot\beta}^{j} = \varepsilon^{ij}A_{\alpha\dot\beta}\,_{I}^R\,, \\
 \nabla_{\dot \alpha}^{\overline i}A_{\beta}^{\overline j}\,_{I}^R+\nabla_{\beta}^{\overline j}A_{\dot \alpha}^{\overline i}\,_{I}^R + 2 \nabla_{I}^R
B_{\dot\alpha}^{\bar{i}}\,_{\beta}^{\bar{j}} = \varepsilon^{\overline i\overline j}A_{\beta\dot\alpha}\,_{I}^R\,,  
\end{aligned} 
\end{equation}
while the $A_{\mu'}\,_{I}^R$
component can be extracted from
\begin{equation}
\begin{aligned}
\nabla_{\alpha}^{i}A_{\beta}^{\overline j}\,_{I}^R+\nabla_{\beta}^{\overline j}A_{\alpha}^{i}\,\,_{I}^R    + 2 \nabla_{I}^R
B_{\alpha}^{i}\,_{\beta}^{\bar{j}} =\varepsilon_{\alpha\beta}A^{i\overline j}\,_{I}^R\,, \\
\nabla_{\dot\alpha}^{\overline i}A_{\dot \beta}^{ j}\,_{I}^R+\nabla_{\dot \beta}^{ j}A_{\dot\alpha}^{\overline i}\,_{I}^R+ 2 \nabla_{I}^R
B_{\alpha}^{i}\,_{\beta}^{\bar{j}} =\varepsilon_{\dot\alpha\dot\beta}A^{j\overline i}\,_{I}^R \, .
\end{aligned} 
\end{equation}

Similar analyses can be performed to all equations of the previous Subsections. Finally, all the expansions of the superfields are replaced in \eqref{TheUsuperfield} and one gets the integrated vertex operators. 

\subsubsection{Vertex operators in the B-RNS-GSS formalism}
\label{Vertex operators in B-RNS-GSS formalism}

Vertex operators for type IIB superstrings both in a flat spacetime background and in an $AdS_5 \times S^5$ background were constructed in the B-RNS-GSS formalism in \cite{Chandia:2025nxx} in terms of 
superfields. The strategy used in that paper considers as starting point the most general expression
with conformal weight (1,1), without ghost number and carrying non-positive charge with respect to a set of $U(1)$ generators, see \cite{Chandia:2025nxx} for details. This last condition is necessary because  
the B-RNS-GSS has additional worldsheet fields, and without the projection to the positive charge terms the formalism would have more states than the other formalisms. From this general expression one inverts the descent procedure to get 
the unintegrated vertex operators and the equations that the superfields of the integrated vertex operators  need to satisfy.      

The integrated vertex operators for $AdS_5 \times S^5$ in the B-RNS-GSS formalism have more superfields than the one reviewed above for the pure spinor formalism, see \eqref{TheUsuperfield}. Moreover, in this new formalism the unintegrated vertex operators are functions of the superfields appearing in the integrated ones. Notice that 
the superfield $A_{\alpha \widehat\alpha}$ appearing in both formalisms obeys the same superspace equations of motions. 
In addition, in both formalisms the integrated vertex operators are linear differential operators acting on the same superfields of the Subsection \ref{Mass-dimension tower via Jacobi identities}. In conclusion, 
using the unintegrated vertex operators computed in this paper, both types of vertex operators in the B-RNS-GSS formalism are obtained.      

\pagebreak

\section{The $PSU(2,2|4)$ transformations} 
\label{The PSU(2,2|4) transformations} 

In this Section, we study finite global $PSU(2,2|4)$ transformations and different coset parame- trizations from $g_{\text{old}}$. While the infinitesimal ones in the parametrization $g_{\text{old}}$ of \eqref{OldCosetRepresentative} are straightforward (see Subsection~\ref{slcovariance}), the finite transformations are cumbersome due to boson–fermion mixing. 
We exemplify the difficulties by computing translations up to the second order in the fermions for $g_{\text{old}}$.  
In Subsection~\ref{sucovariance}, we introduce %
$SU(2,2)\times SU(4)$ fermionic covariant coordinates: a reparametrization mapping \eqref{OldCosetRepresentative} to \eqref{sucovariance} that disentangles bosons from fermions and renders the finite $PSU(2,2|4)$ action tractable. The $PSU(2,2|4)$ transformations are crucial: starting from the vertex operator for the bottom half-BPS state, 
acting with $PSU(2,2|4)$ generates 
all polarizations 
and
the entire supermultiplet. We also compute the BRST transformations of 
the new coset parametrizations in the Appendix \ref{BRSTideal}. 
 
\subsection{The old parametrization}
\label{slcovariance}

The $PSU(2,2|4)$ global transformations $M$ that 
leave the sigma-model action of Subsection~\ref{sigma model action} invariant are, in the parametrization \eqref{OldCosetRepresentative},
\begin{equation}
M\,g_{\text{old}}(z,w,\theta, \psi, x,y)=g_{\text{old}}(z',w',\theta', \psi', x',y')\,h\,,\qquad h\in H\,,
\label{GlobalTransformations}
\end{equation}
with $H=USp(2,2)\times USp(4)$. Generically $h\neq1$, so the global action moves us off the gauge slice of 
$g_{\text{old}}$;
to restore it, act with a compensating $H$ transformation setting $h\to1$. Determining the transformation amounts to solving for the primed coordinates and $h$, yielding both the transformed coordinates and the compensator. Tracking this compensator is essential: although the vertex operator is gauge invariant, we constructed it in a fixed gauge rather than in terms of covariant superfields; see Subsection~\ref{Covariant superfields}.

Upon bosonic truncation, see Subsection \ref{Bosonic truncation}, the $SU(2,2)\times SU(4)$ transformations are straightforward. The difficulty arises when fermions are retained. For certain transformations, however, solving for the primed coordinates and 
$h$ remains straightforward even with fermions; these are essentially shifts of the leftmost coordinates in \eqref{OldCosetRepresentative} and precisely the transformations that leave
the vertex operators of \cite{VertexOld} equivariant. This is no accident: 
the parametrization $g_{\text{old}}$ was chosen 
for that reason. By contrast, transformations generated by elements on the right in the exponential of 
$g_{\text{old}}$
make solving for the primed coordinates and the compensator 
much more involved. These are exactly the transformations of interest, since they move 
the vertex operator polarization and 
the vertex traverses the supermultiplet.
 
\subsubsection{Infinitesimal global transformations} 
\label{Infinitesimal transformations}

To obtain the infinitesimal global transformations $\delta Z^M$, we introduce the infinitesimal 
parameters $\eta^A_0$ and $\eta_0^{[\underline{ab}]}$, parametrizing 
the left-translations:
\begin{equation}
\left(1 + \eta_0^A t_A + \eta_0^{[\underline{ab}]}t_{[\underline{ab}]}\right) g(Z) = \left(1+ \delta Z^M \frac{\partial }{\partial Z^M}\right)g(Z)\left(1+  \Sigma(Z)\right) \, ,
\end{equation} 
where $A=(\underline\alpha,\underline{a},\underline{\widehat\alpha})$ labels the generators with non-zero $\mathbb Z_4$ grading, while $[\underline{ab}]$ labels the generators of zero grading. The above equation is to be understood up to first order in the infinitesimal parameters. 
Multiplying both sides by $g(Z)^{-1}$, 
we arrive at
\begin{equation}
\delta Z^M E_M^A t_A + \Sigma = g(Z)^{-1} \, (\eta^A_0 t_A + \eta_0^{[\underline{ab}]}t_{[\underline{ab}]}) \, g(Z) = \eta^{A}(Z)t_{A} + \eta^{[\underline{ab}]}(Z)t_{[\underline{ab}]}\, ,
\label{transformationsinfinitesimal}
\end{equation}
where the $E_{M}^{A}$ are the vielbeins components obtained in \eqref{conectionsvielbein}, and the $\eta^{A}(Z)$ are the Killing spinor supervectors. We have
\begin{equation}
\delta Z^{M} = \eta^{A}(Z)E_{A}^{M}\,,\qquad \Sigma(Z) =  \eta^{[\underline{ab}]}(Z)t_{[\underline{ab}]}\,.  
\end{equation}
In particular, one can obtain the Killing spinors by restricting the index of $\eta^A_0$ to take values at odd gradings. %
The above equations are easily worked out through {\it{Mathematica}}. It is also not complicated to obtain a set of infinitesimally translated vertices depending on $\eta_0^{\underline{a}}$ starting from the ones
computed in \cite{VertexOld}. In Appendix \ref{InfinitesimalVertices}, we wrote explicitly one such family of vertex operators. 
The expression for them is more compact than the general finite translated vertices. 
However, the computation of non-trivial three-point amplitudes requires knowledge of vertex operators at least at the second order 
in $\eta_0^{\underline{a}}$. Nevertheless, the infinitesimally translated vertex operators of Appendix \ref{InfinitesimalVertices} can be used for the computation of both two-point amplitudes and OPE's.     
  
\subsubsection{Finite global transformations} 
\label{Finite transformations}

Having all the 32+30 Killing supervectors, 
one can take linear combinations of the infinitesimal transformations parametrized by $\eta_0^{I}$,
\begin{equation}
\delta_{\eta_0} Z^M = \eta_0^{I}\eta_{I}^{A}(Z)E_{A}^{M} = \eta_0^{I} \eta_I^M(Z) \, , \qquad I=(A,[\underline{ab}]) \, . 
\label{inftrans}
\end{equation}
One way of obtaining the finite transformations is by integrating the infinitesimal ones.
We found that this method is computationally more efficient than directly solving \eqref{GlobalTransformations} in {\it{Mathematica}}.
The method for integrating the infinitesimal transformations is as follows. 
Consider the notation $|M|=0$ for bosonic and $|M|=1$ for fermionic indices; then
 \begin{equation}
 |\eta_0^I|  = |I| \, , \quad \quad 
 |\eta^M_I|  = |M| + |I| \, . 
 \end{equation}
  
  In general, any bosonic vector field $\mathcal V_i^M(Z)$ admits the following $\theta$-expansions   
  \begin{equation}\label{bosonic}
  \left\{
 \begin{aligned}
  \mathcal V_1^a(x,\theta) & = \mathcal V^a(x) + \theta^{\alpha}\theta^{\beta} \mathcal V^a_{\beta \alpha}(x) + \ldots \, , \\
 \mathcal V_2^a(x,\theta) & = \theta^\alpha \mathcal V^a_\alpha(x) + \theta^{\alpha}\theta^{\beta}\theta^{\gamma} \mathcal V_{\gamma \beta \alpha}^a(x) + \ldots \, ,
\end{aligned}
  \right.
  \end{equation}
and we are collectively denoting by $\theta$ all the fermionic variables of the coset $g_{\text{old}}$ to keep the notation compact.  
For fermionic vector fields, one has the possibilities

    \begin{equation}\label{fermionic}
  \left\{
  \begin{aligned}
  \mathcal V_1^\alpha(x,\theta) & = \mathcal V^\alpha(x) + \theta^{\gamma}\theta^{\beta} \mathcal V^\alpha_{\beta \gamma}(x) + \ldots \, ,   \\
  \mathcal V_2^\alpha(x,\theta) & = \theta^\beta \mathcal V^\alpha_\beta(x) + \theta^{\delta}\theta^{\gamma}\theta^{\beta} \mathcal V_{\beta \gamma \delta}^\alpha(x) + \ldots \, . \\
  \end{aligned}
  \right.
  \end{equation}

In what follows, we are going to consider only bosonic transformations such as the ones generated by $P^\a_\ad$ or $P^i_{\overline j}$. In this case all the parameters 
$\eta_0^I$ in \eqref{inftrans}
are also bosonic, i.e., $|I| = 0$, 
and the relevant vector fields of \eqref{inftrans} have $\theta$-expansions of only the type $\mathcal{V}_1$. 
The case of transformations with fermionic parameters %
can be treated analogously. 
For any infinitesimal transformations of the form \eqref{inftrans}, 
the finite transformations
are given by the \emph{unique} solution $Z^M(t)=(z^a(t), \rho^{\alpha}(t))$ to the differential equations
  \begin{equation}
 \dot{Z}^M(t) = \mathcal V^M(Z(t)) \, , 
 \label{difequations}
  \end{equation}
with initial conditions
  \begin{equation}
  z^a(0) = x^a \, ,  \qquad \rho^\a(0) = \theta^\alpha \, ,
  \end{equation}
where $\mathcal V^M(Z)$ is the fixed rescaled linear combination $\mathcal{V}^M(Z) =  \widehat\eta_0^{I} \eta_I^M(Z)$.
In the particular case where $\eta_0^{I}$ is non-zero for only one $I_0$, we can take $t=\eta_0^{I}$.     
The finite transformations $Z^M(t)$ depend on the initial conditions $(x^a,\theta^{\alpha})$ and also admit a 
$\theta$-expansion. Thus the differential equations \eqref{difequations} can be organized in
powers of the fermionic variables. We can write 
  \begin{equation}
  \begin{aligned}
  z^a(x,\theta | t) & = (\theta^0\cdot z^a)(x,t) + (\theta^2 \cdot z^a) (x,t) + (\theta^4 \cdot z^a)(x,t) + \ldots \, ,    \\
  \rho^\a(x,\theta |t) & = (\theta^1 \cdot \rho^\alpha)(x,t) + (\theta^3\cdot \rho^\alpha)(x,t) + (\theta^5\cdot \rho^\alpha)(x,t) + \ldots \, ,  
  \end{aligned}
  \end{equation}
where the superscript on the $\theta$'s on the right-hand side of the equation above indicates its order.  
Similarly, there is a $\theta$-expansion for the vector fields $(\mathcal V^a(z, \rho), \mathcal V^\alpha(z,\rho))$, where we omit the dependence on $t$. From (\ref{bosonic}) and (\ref{fermionic}) we get the expansion of the vector fields in terms of $\rho$. The expansions are 
\begin{equation}
  \begin{aligned}
  \mathcal V^a(z,\rho) & = (\rho^0\cdot \mathcal V^a)(z) + (\rho^2\cdot \mathcal V^a)(z) + (\rho^4\cdot\mathcal V^a)(z) + \ldots \, , \\
  \mathcal V^\a(z,\rho) & = (\rho^1\cdot \mathcal V^\a)(z) + (\rho^3\cdot \mathcal V^\a)(z) + (\rho^5\cdot\mathcal V^\a)(z) + \ldots \, . 
  \end{aligned}
  \end{equation}
 Each term in the previous equations has their own $\theta$-expansions. Using the notation
\begin{equation}
z^a_k(t) := (\theta^k\cdot z^a)(x,t) \, ,\qquad \rho^\alpha_k(t) := (\theta^k\cdot\rho^\alpha)(x,t) \, , 
  \end{equation}
one has 
\begin{align}
  (\rho^k\cdot \mathcal V^M)(Z) & = \mathcal V_{(k, k)}^M(z_0; \rho_1) + \mathcal V_{(k,k+2)}^M(z_0, z_2;\rho_1,\rho_3) \\
&  + \ldots + \mathcal V_{(k,k+\ell)}^M(z_0,\ldots, z_\ell; \rho_1,\ldots, \rho_{\ell+1}) + \ldots \, . \nonumber
  \end{align}
The two labels $(k,k +\ell)$ in the generic term $\mathcal V^M_{(k,k+\ell)}$ are read as follows. The first entry indicates the order in the $\rho$-expansion, and the second one the total order in $\theta$. Therefore, the components of the vector fields are $\theta$-expanded as
  \begin{equation}
  \begin{aligned}
  \mathcal V^a(Z) & =  \mathcal V^a_{(0,0)}(z_0)  + \mathcal V^a_{(0,2)}(z_0,z_2)   + \mathcal V^a_{(2,2)}(z_0;\rho_1)  \\
  & + \mathcal V^a_{(0,4)}(z_0,z_2,z_4)  + \mathcal V^a_{(2,4)}(z_0,z_2;\rho_1,\rho_3)  + \mathcal V^a_{(4,4)}(z_0;\rho_1) + \ldots \, ,  \\
 \end{aligned}
  \end{equation}
  and
\begin{equation}
  \begin{aligned}
  \mathcal V^\a(Z) & =  \mathcal V^{\alpha}_{(1,1)}(z_0;\rho_1)  + \mathcal V^{\alpha}_{(1,3)}(z_0,z_2;\rho_1,\rho_3)   + \mathcal V^{\alpha}_{(3,3)}(z_0,\rho_1)   \\
  & + \mathcal V^{\alpha}_{(1,5)}(z_0,z_2,z_4; \rho_1,\rho_3,\rho_5)  + \mathcal V^{\alpha}_{(3,5)}(z_0,z_2;\rho_1,\rho_3)  + \mathcal V^{\alpha}_{(5,5)}(z_0,\rho_1) + \ldots 
  \end{aligned}
  \end{equation}
where the terms with the same first entry in $\mathcal{V}$ come from the same term in the $\rho$-expansion of (\ref{bosonic}) or (\ref{fermionic}). 
After replacing the $\theta$-expansions for both the coordinate functions $(z^a(t),\rho^\a(t))$ and the vector fields $(\mathcal V^a, \mathcal V^\alpha)$ above in the differential equation \eqref{inftrans}, we arrive at the following system of differential equations to be integrate in order to compute the finite transformations from the infinitesimal ones 
 \begin{itemize}
\item 
$\theta^0$: 
 $\quad \quad \quad \dot{z}^a_0 
= \mathcal{V}^a_{(0,0)}( z_0 )  \, , $ 
\item 
  $\theta^1$
$ \, \quad \quad \quad \, \dot{\rho}^\alpha_1 = \mathcal V^\a_{(1,1)}( z_0 ; \rho_1 ) \, , $
\item 
  $\theta^2$:
$ \quad \quad \quad  \dot{z}^m_2 = \mathcal V^m_{(0,2)}( z_0, z_2) +  \mathcal V^m_{(2,2)}( z_0; \rho_1 ) \, , 
$
\item 
  $\theta^3$: 
$ \quad \quad \quad   \dot{\rho}^\alpha_3 = \mathcal V^{\alpha}_{(1,3)}(z_0, z_2;\rho_1,\rho_3) + \mathcal V^{\alpha}_{(3,3)}(z_0;\rho_1) \, , $
\item \ldots
\end{itemize}
The system of equations continue until the entire $\theta$-expansion is exhausted or cancellations due to symmetry occur. 

Recall that vertex operators with different polarizations suitable for computing amplitudes can be obtained by applying both space-time and $R$-charge translations $P$
to the vertices 
of \cite{VertexOld}.
Notice that the global transformations of the variables under the conformal boosts $K$ are very simple and they are just a shift of the variables $x$'s and $y$'s for $g_{\rm{old}}$. 
However, the vertices of \cite{VertexOld} are all annihilated by the $K$'s and acting with them do not change their polarizations.  

In order to get the finite translations 
by $P$, we need to solve \eqref{GlobalTransformations} with $M = \exp( \eta^\ad_\a P^\a_{\dot{\a}} + \eta^i_{\overline i} P^{\overline i}_{i} )$ for finite $\eta$'s.     
Using \emph{Mathematica}, this equation can be solved independently for any of the eight $P$'s  generators by computing the infinitesimal transformations and then integrating the equations 
as described previously. Treating all the eight $P$'s generators at once is a heavy symbolic computation.  
Considering that all eight generators $P^\a_{\dot{\a}}$ and $P^{\overline i}_{i}$ commute, we separate the exponential as 
  \begin{equation}
  \begin{aligned}
  \exp( \eta^\ad_\a P^\a_{\dot{\a}} + \eta^i_{\overline i} P^{\overline i}_{i} ) = \exp(\eta^2_{\overline 2} P^{\overline 2}_{2}) & \exp(\eta^1_{\overline 2} P^{\overline 2}_{1})\exp(\eta^2_{\overline 1} P^{\overline 1}_{2})\exp(\eta^1_{\overline 1} P^{\overline 1}_{1}) \\
  & \times \exp( \eta^{\dot 2}_2 P^2_{\dot{2}})\exp( \eta^{\dot 1}_2 P^2_{\dot{1}})\exp( \eta^{\dot 2}_1 P^1_{\dot{2}})\exp( \eta^{\dot 1}_1 P^1_{\dot{1}}) \, , 
  \end{aligned}
  \end{equation}
and solve equation \eqref{GlobalTransformations}  for each of the eight transformations one by one. The final result  
for the transformations of the variables is obtained by the composition of the transformations and 
the final compensator is given by the product of the eight compensators.
Next, we give the full solution for the transformations of the variables at order zero in $\theta$.
Since the expressions for higher orders in $\theta$'s are quite complicated, a few more 
results up to quadratic order in the $\theta$'s are given in the Appendix \ref{finiteOldcoset}.
Notice that the transformations do not truncate at quadratic order.
The infinitesimal transformations terminate only at order $\theta^{16}$ and it is computationally demanding to apply 
the full transformations to the vertices of \cite{VertexOld}.  
For some applications where not all orders in $\theta$ are necessary, the expressions 
for the vertex operators in $g_{\rm{old}}$ are more compact than the expressions appearing in new parametrizations below.  
In any case, in the next Subsection, we are going to introduce $SU(2,2)\times SU(4)$–covariant coordinates.
For them, all the global transformations truncate at second order in the fermionic variables.   

Under $\eta^\ad_\a P^\a_{\dot{\a}}$ translations at order $\theta^0$, 
\begin{equation}
 e^{z^{\prime}}  = e^{z}\ \Omega \, , \quad  
 e^{w^{\prime}}  = e^w \, , \quad 
 (x^{\prime})^\alpha_\ad  = \Omega^{-1} \left(  x^\alpha_\ad + (\ve\ve\eta)^\alpha_\ad\ X^2 \right) \, , \quad
 (x^{\prime})^{\overline i}_{i}  = x^{\overline i}_{i} \, ,
 \label{adstrans}
  \end{equation}
  and under $\eta^i_{\overline i} P^{\overline i}_{i}$  translations,
  \begin{equation}
 e^{z^{\prime}}  = e^{z} \, , \quad  
  e^{w^{\prime}}  = e^w\ \overline\Omega^{-1}  \, , \quad  
  (x^{\prime})^\alpha_\ad  = x^\alpha_\ad  \, , \quad  
  (x^{\prime})^{\overline i}_{i}  = \overline\Omega^{-1} \left( x^{\overline i}_{i} + (\ve\ve\eta)^{\overline i}_i\ \overline X^2 \right) \, ,
 \label{strans}
 \end{equation}
where the conformal factors are 
  \begin{align}
  \label{OmegaFactor}
  \Omega(\eta, x, z) & = 1 + (x \eta) + \frac12 (\eta \cdot \eta) \left( \frac12 (x\cdot x) - e^{-2z} \right) \, ,  \\
  \overline\Omega(\overline\eta, \xb, w) & = 1 + (\xb \overline\eta) + \frac12 (\overline\eta \cdot \overline\eta) \left( \frac12 (\xb\cdot \xb) - e^{2w} \right) \, , \nonumber
  \end{align}
and $X^2 = \frac12 (x\cdot x) - e^{-2z}$ and $\overline X^2 = \frac12 (\overline x\cdot \overline x) - e^{2w}$. 
We have used the following definitions  
\begin{equation}
  \begin{aligned}
  (\eta \cdot\eta) = \ve_{\ad\bd} \ve^{\alpha\beta} \eta^\ad_\alpha \eta^\bd_\beta & \quad, \qquad(\overline\eta \cdot \overline\eta) = \ve_{i j} \ve^{\ib\jb} \eta_\ib^i \eta^j_\jb \, , \\
  (x\cdot x) = \ve^{\ad\bd} \ve_{\alpha\beta} x_\ad^\alpha x_\bd^\beta & \quad, \qquad(\xb\cdot\xb) = \ve^{i j} \ve_{\ib\jb} x^\ib_i x_j^\jb \, , \\
  (x \eta) = x^\a_\ad \eta^\ad_\a &\quad ,\qquad (\xb \overline\eta) = x^\ib_i \eta^i_\ib \, . 
  \label{fewdefinitions}
  \end{aligned}
  \end{equation}
Notice that if we had defined a variable $\overline z = - w$, then the transformations \eqref{adstrans} and \eqref{strans} would be identical. 

The above transformations parametrized by $\eta^\ad_\a$  
reduce for $z \rightarrow \infty$  to the expected 4d conformal transformations at the boundary.
Recall that in our parametrization the $x$'s are contracted with the $K$'s in the exponentials.  
It is well known that the transformations can be written in a more compact form by using the 
inversion operator $I$. Its action on our coordinates is    
\begin{equation}
I \cdot x^\alpha_\ad = \frac{x^\alpha_\ad}{X^2} \, , \qquad  I \cdot z = z + {\rm{log}} \left(X^2\right) \, , \qquad I \cdot x^\ib_i =
\frac{x^\ib_i}{\overline X^2} \, , \qquad I \cdot w= w -  {\rm{log}} \left(\overline X^2\right) \, . 
\nonumber
\end{equation}

The transformations by $\eta^\ad_\a P^\a_{\dot{\a}} + \eta^i_{\overline i} P^{\overline i}_{i}$
produce a non-zero compensator $h$ in \eqref{GlobalTransformations}. It is necessary to perform a compensating gauge transformation that rotates the pure spinors. At order $\theta^0$ for $\eta^i_{\overline i} =0$, the rotations are  
\begin{equation}
\begin{aligned}
 (\lambda^{\prime})^\a_I & = \sqrt\Omega \left( (\delta^\alpha_\beta + (x^{\prime} \eta)^\alpha_\beta) \lambda^\beta_I + e^z ( (x^{\prime})^\alpha_{\dot{\beta}} - x^\alpha_{\dot{\beta}} - (x^{\prime} \eta x)^\alpha_{\dot{\beta}} ) \lambda^{\dot{\beta}}_I \right) \, , \nonumber \\
(\lambda^{\prime})^{\dot{\alpha}}_I & = \frac{1}{\sqrt\Omega} \left( e^{-z} \eta^{\dot{\alpha}}_\beta \lambda^\beta_I + (\delta^{\dot{\alpha}}_{\dot\beta} - (\eta x)^{\dot\alpha}_{\dot\beta}) \lambda^{\dot\beta}_I \right) \, ,
\end{aligned}
\end{equation}
and 
\begin{equation}
\begin{aligned}
 (\lambda^{\prime})^I_\alpha & = \frac{1}{\sqrt\Omega} \left( \lambda^I_\beta(\delta^\beta_\alpha + (x \eta)^\beta_\alpha) + e^{-z} \lambda^I_\bd \eta^\bd_\alpha \right) \, ,   \nonumber \\
  (\lambda^{\prime})^I_\ad & = \sqrt{\Omega} \left( \lambda^I_\bd (\delta^\bd_\ad - (\eta x^{\prime})^\bd_\alpha) + e^z \lambda^I_\beta (x^\beta_\ad - (x^{\prime})^\beta_\ad - (x \eta x^{\prime})^\beta_\ad) \right) \, .  \nonumber \\
\end{aligned}
\end{equation}

\subsection{The new parametrizations
}
\label{sucovariance}

As discussed in the previous Subsection, the action of some global $PSU(2,2|4)$ transformations on $g_{\rm{old}}$ produces transformations of the coordinates  whose series expansions 
in the fermionic variables have several orders. 
Therefore, the expressions for the vertex operators for finite $\eta^\ad_\a$ and $\eta^i_{\overline i}$ 
are complicated using $g_{\rm{old}}$. 
Fortunately, there are coordinates in which all the $PSU(2,2|4)$ transformations are at most
quadratic in the fermionic variables. 
Thus it is convenient to apply a change of variables to the vertex operators of \cite{VertexOld} before applying  $PSU(2,2|4)$ transformations.  
In what follows, these new coset representatives will be called $g_{\text{new}}$ and $g'_{\text{new}}$, and 
they are given by  
 \begin{equation}
g_{\text{new}} = e^{\Psi^S_I q^I_S} e^{\Theta_S^I q_I^S} U V 
= e^{\Psi} e^{\Theta} U V \, , \quad {\rm{where}} \quad  U= e^{X^{\alpha}_{\dot\alpha} K^{\dot\alpha}_{\alpha}} e^{Z \Delta} \, , \quad  V = 
e^{Y^{\bar{i}}_i K^i_{\bar{i}}} e^{W J} \, , 
\label{CosetIdeal}
\end{equation} 
and 
\begin{equation}
g^{\prime}_{\text{new}} = 
e^{Z \Delta} e^{W J}
e^{\widetilde{X}^{\alpha}_{\dot\alpha} K^{\dot\alpha}_{\alpha}} 
e^{\widetilde{Y}^{\bar{i}}_i K^i_{\bar{i}}}
e^{\widehat\Psi 
} 
e^{\widehat\Theta } \, ,  \quad {\rm{with}} \quad
\widetilde{X}^{\alpha}_{\dot\alpha} = e^Z X^{\alpha}_{\dot\alpha} \, , \quad 
\widetilde{Y}^{\bar{i}}_{i} = e^{-W} Y^{\bar{i}}_{i} \, .  
\label{gprime}
\end{equation} 
In the expressions above, the following compact notations for the indices were used: $S =\{\alpha, \dot\alpha\}$ and $I = \{i, \bar{i}\}$.
In this Subsection, we are going to 
compute the finite global transformations. The BRST transformations 
of the variables of $g_{\text{new}}$ 
are given in the Appendix \ref{BRSTideal}. 
In the next Subsection, we consider the %
reparametrization 
$g_{\text{old}} \leftrightarrow g_{\text{new}}$ necessary to transform the vertex operators to the new
variables.

\subsubsection{Finite transformations} 
\label{TheFiniteTransformations} 

All the infinitesimal transformations of the variables of $g_{\rm{new}}$ are computed in the Appendix \ref{psu224ideal}. 
As explained in the previous Subsection, the infinitesimal transformations can be integrated to produce the 
finite ones. 
The final form of the vertex operators as given in Section \ref{The ideal coset vertices} depends only 
on $SU(2,2)\times SU(4)$ covariant variables such as $\Psi^S_I$ and $\Theta^I_S$. 
Consider global transformations of the form $N={\rm{exp}}(\eta^R_S M^S_R)$ and $T={\rm{exp}}(\eta^I_J M^J_I)$ with finite parameters $\eta$'s and both $M^S_R$ and $M^J_I$ bosonic generators of
$\mathfrak{psu}(2,2|4)$. The finite transformations of the fermionic variables of 
$g_{\rm{new}}$ are 
\begin{equation}
\Psi^{\prime S}_I = (N^{-1})^S_{R} \Psi^{R}_J T^J_I \, , \quad  \quad
\Theta^{\prime I}_S= N_S^{R} \Theta^{J}_R (T^{-1})_J^I \, . 
\end{equation}

The transformations of the variables parametrizing the $U$ and $V$ of 
\eqref{CosetIdeal} can be deduced from the formulas given below for $g^{\prime}_{\text{new}}$.      
Here we also focus only on translations with parameters $\eta^{\dot\alpha}_{\alpha}$ and $\eta^{i}_{\bar{i}}$ which correspond to the $C^{\dot\alpha}_{\alpha}$ and $D^{i}_{\bar{i}}$ in the notation of the Appendix \ref{psu224ideal}. 
In order to achieve the final form of the vertex operators, as motivated in the next Subsection, it was convenient at an intermediary stage to write the vertex operators using the variables of the coset representative $g^{\prime}_{\text{new}}$. 
Notice that the super-reparametrization $g^{\prime}_{\text{new}} \leftrightarrow g_{\text{new}}$ is simple, being the identity 
transformation for the bosonic variables, while for the fermions one has       
\begin{equation}\label{change-coord}
  \begin{aligned}
  \Psi^{R}_{I} = (U^{-1})^{ R}_{\widetilde{S}}\ \widehat\Psi^{\widetilde{S}}_{\widetilde{J}}\ V^{\widetilde{J}}_{ I} \, , \quad \quad 
  \Theta_{R}^{I} = (V^{-1})_{ \widetilde J}^{ I}\ \widehat\Theta_{ \widetilde S}^{\widetilde J} \ U_{R}^{\widetilde S} \, .
  \end{aligned}
  \end{equation}

The finite transformations for the variables of $g^{\prime}_{\text{new}}$ are going to be used in the next Section and we are going to write them explicitly below. 
The transformations of the bosonic variables are 
\begin{equation}
  \label{Translations2}
  \begin{aligned}
  Z' &  
  = Z + \log(\Omega_n) \, ,  \\
  \widetilde{X}'^{\alpha}_{\dot\beta} & 
  =\widetilde{X}^{\alpha}_{\dot\beta} + (\varepsilon\varepsilon\eta)^{\alpha}_{\dot\beta} e^{-Z}\left(\frac12 \widetilde{X}\cdot \widetilde{X} - 1\right) \, , \\
 W' & 
 =W - \log(\overline\Omega_n) \, , \\
  \widetilde{Y}'^{\overline i}_j & 
  =\widetilde{Y}^{\overline i}_j  + (\varepsilon\varepsilon\eta)^{\overline i}_j e^W \left(\frac12 \widetilde{Y}\cdot \widetilde{Y} - 1  \right) \, , 
 \end{aligned}
  \end{equation}
 with 
 \begin{equation}
 \widetilde{X} \cdot \widetilde{X} =  
 \widetilde{X}^{\alpha}_{\dot\alpha} \varepsilon^{\dot\alpha \dot\beta}  \varepsilon_{\alpha \beta}  \widetilde{X}^{\beta}_{\dot\beta} \, , \quad  
 \widetilde{Y} \cdot \widetilde{Y} =  
 \widetilde{Y}^{\bar{i}}_{i} \varepsilon^{i j}  \varepsilon_{\bar{i} \bar{j}}  \widetilde{Y}^{\bar{j}}_{j} \, , \quad  
(\varepsilon\varepsilon\eta)^{\alpha}_{\dot\alpha} = \varepsilon^{\alpha \beta} 
 \varepsilon_{\dot\alpha \dot\beta} \eta^{\dot\beta}_{\beta} \, , \quad
(\varepsilon\varepsilon\eta)^{\bar{i}}_{i} = 
\varepsilon^{\bar{i} \bar{j}} 
 \varepsilon_{i j} \eta^{j}_{\bar{j}} \, . 
 \nonumber
\end{equation}
 In the formulas above, 
we see the appearance of the conformal factors $\Omega_n$ and $\overline\Omega_n$, they are given by   
  \begin{equation}
  \begin{aligned}
  \Omega_n & = 1 + e^{-Z} \widetilde{X}^\alpha_{\dot\beta} \eta^{\dot\beta}_{\alpha} + e^{-2Z} \frac12 (\eta\cdot\eta)_x \left(\frac12 \widetilde{X}\cdot \widetilde{X} - 1  \right) \, , \\
  \overline\Omega_n & = 1 + e^W \widetilde{Y}^{\overline i}_j \eta^{j}_{\overline i} + e^{2W} \frac12 (\eta\cdot\eta)_ y \left( \frac12 \widetilde{Y}\cdot \widetilde{Y} - 1 \right) \, , 
  \end{aligned}
  \end{equation}
where
\begin{equation}
(\eta \cdot \eta)_x  =   \eta_{\alpha}^{\dot\alpha} \varepsilon_{\dot\alpha \dot\beta}  \varepsilon^{\alpha \beta}  \eta_{\beta}^{\dot\beta} \, , \quad \quad  
(\eta \cdot \eta)_y  = \eta_{\bar{i}}^{i} \varepsilon_{i j}  \varepsilon^{\bar{i} \bar{j}}  \eta_{\bar{j}}^{j} \, . 
\end{equation}

The finite transformations generate non-zero compensators. This implies that the pure spinors also transform. 
The trasformations of the fermionic variables and pure spinors can be written  
as $USp(2,2) \times USp(4)$ rotations,
  \begin{equation}
  \label{TransformPsiTheta}
 (\widehat{\Psi}^{\prime})^{\widetilde{R}}_{\widetilde{I}} = (h_1^{-1})^{\widetilde{R}}_{\widetilde{S}} \, \widehat{\Psi}^{\widetilde{S}}_{\widetilde{J}}\ (h_2)^{\widetilde{J}}_{\widetilde{I}} \, , \quad \quad  
  \widehat\Theta_{\widetilde{R}}^{\widetilde{I}} = (h_2^{-1})^{\widetilde{I}}_{\widetilde{J}}\ \widehat\Theta_{\widetilde{S}}^{\widetilde{J}}\ (h_1)^{\widetilde{S}}_{\widetilde{R}} \, , 
  \end{equation}
and
  \begin{equation}
   \label{TransformLambda}
  (\lambda^{\prime})^{\widetilde{R}}_{\widetilde{I}} = (h_1^{-1})^{\widetilde{R}}_{\widetilde{S}}\ \lambda^{\widetilde{S}}_{\widetilde{J}}\ (h_2)^{\widetilde{J}}_{\widetilde{I}}  \, , \quad \quad 
  (\lambda^{\prime})_{\widetilde{R}}^{\widetilde{I}}= (h_2^{-1})^{\widetilde{I}}_{\widetilde{J}}\ \lambda_{\widetilde{S}}^{\widetilde{J}}\ (h_1)^{\widetilde{S}}_{\widetilde{R}} \, , 
  \end{equation}
where the 
rotations are
  \begin{equation}
  \begin{aligned}
  h_1 & = \frac{1}{\Omega_n^{1/2}}\left(
    \begin{array}{cc}
    \delta^\alpha_\beta + (\widetilde{X}^{\alpha}_{\dot\alpha} \eta^{\dot\alpha}_{\beta}) e^{-Z} &  \eta^{\dot\alpha}_\beta e^{-Z} \\
(\varepsilon\varepsilon\eta)^\alpha_{\dot\beta} e^{-Z} & \delta^{\dot\alpha}_{\dot\beta} + \varepsilon^{\dot\alpha\dot\sigma}\varepsilon_{\dot\beta\dot\gamma} (\eta^{\dot\gamma}_{\gamma} \widetilde{X}^{\gamma}_{\dot\sigma} ) e^{-Z}
    \end{array}
  \right) \, , \\
  h_2 & = \frac{1}{\overline\Omega_n^{1/2}} \left(
    \begin{array}{cc}
    \delta^i_j + \varepsilon^{i k}\varepsilon_{j \ell} (\eta^{\ell}_{\bar{i}} \widetilde{Y}^{\bar{i}}_k) e^W &  (\varepsilon\varepsilon \eta)^{\overline i}_{j} e^W \\
  \eta^i_{\overline j} e^W  & \delta^{\overline i}_{\overline j} + (\widetilde{Y}^{\overline i}_i \eta^i_{\overline j}) e^W
    \end{array}
  \right) \, ,
   \end{aligned}
\end{equation}
with the respective inverses 
 \begin{equation}
  \begin{aligned}
  (h_1^{-1}) & = \frac{1}{\Omega_n^{1/2}}\left(
    \begin{array}{cc}
    \delta^\alpha_\beta + \varepsilon^{\alpha\sigma}\varepsilon_{\beta\gamma}(\widetilde{X}^\gamma_{\dot\alpha} \eta^{\dot\alpha}_{\sigma}) e^{-Z} & -\eta^{\dot\alpha}_\beta e^{-Z}  \\
 -(\varepsilon\varepsilon\eta)^\alpha_{\dot\beta} e^{-Z}     & \delta^{\dot\alpha}_{\dot\beta} + (\eta^{\dot\alpha}_{\alpha} \widetilde{X}^{\alpha}_{\dot\beta}) e^{-Z}
    \end{array}
  \right) \, ,\\
  (h_2^{-1}) & = \frac{1}{\overline\Omega_n^{1/2}} \left(
    \begin{array}{cc}
    \delta^i_j + (\eta^i_{\bar{i}} \widetilde{Y}^{\bar{i}}_j) e^W &  -(\varepsilon\varepsilon \eta)^{\overline i}_{j} e^W\\
  -\eta^i_{\overline j} e^W  & \delta^{\overline i}_{\overline j} + \varepsilon^{\overline i \overline k}\varepsilon_{\overline j \overline\ell}(\widetilde{Y}^{\overline \ell}_i \eta^i_{\overline k}) e^W 
    \end{array}
  \right) \, . 
  \end{aligned}
  \end{equation}

The finite translations of the bosonic variables  \eqref{Translations2}
are independent of the fermionic variables and 
they are of quadratic order in the $\widetilde{X}$'s and $\widetilde{Y}$'s.
We should mention for completeness that there is a further transformation of variables 
that linearizes the finite translation transformations of the bosonic variables. In fact, they become a simple shift of the variables. 
These variables $\widehat{x}$ and $\widehat{z}$ were considered in the Subsection \ref{Bosonic truncation}. 
In that Subsection, only reparametrizations of the $AdS$ variables were considered. 
However, it is possible to define similar $S^5$ variables that we call $\widehat{y}$ and $\widehat{w}$. 
Their finite translations are    
\begin{equation}
\delta \widehat{x}_{\alpha}^{\dot\alpha} = \eta_{\alpha}^{\dot\alpha} \, , \qquad
\delta \widehat{y}\,^{\bar{i}}_i = {\eta}^{\bar{i}}_i \, , \qquad
\delta \widehat{z} = 0 \, , \qquad 
\delta \widehat{w} = 0 \, . \quad    
\end{equation}

\subsection{Reparametrizations: disentangling bosons from fermions}
\label{reparametrization old to new}

In this Subsection, we are going to derive the super-reparametrizations from the old coordinates $g_{\text{old}}$ to the new ones $g_{\text{new}}$. 
The derivation uses supermatrices and a good introduction to them with applications 
is the textbook \cite{Cornwell}. 
The map can be used to find the vertex operators for $g_{\text{new}}$ by transforming the ones of \cite{VertexOld}. 
The derivation of the super-reparametrizations involves several steps, as described in detail below.   

\subsubsection{Rearranging $g_{\text{old}}$}
\label{Rearranging old}

First, we are going to rearrange $g_{\text{old}}$ by moving all the bosonic generators to the left. Notice that, trivially,  
\begin{equation}
e^{(\psi^{\alpha}_i q^i_{\alpha} + \theta^{\bar{i}}_{\dot\alpha} q^{\dot\alpha}_{\bar{i}} + x^{\alpha}_{\dot\alpha} K^{\dot\alpha}_{\alpha} + y^{\bar{i}}_{i} K^{i}_{\bar{i}} )} = e^{x^{\alpha}_{\dot\alpha} K^{\dot\alpha}_{\alpha}} e^{y^{\bar{i}}_{i} K^{i}_{\bar{i}}}  e^{(\psi^{\alpha}_i q^i_{\alpha} + \theta^{\bar{i}}_{\dot\alpha} q^{\dot\alpha}_{\bar{i}} )} \, ,  
\end{equation} 
because all the generators commute. So the non-trivial part is to move $e^{ z \Delta + w J}$ to the left. The result is 
\begin{equation}
\begin{aligned}
g_{\text{old}} \, \rightarrow \, g_{\text{old}}^{\prime} & = 
e^{x^{\alpha}_{\dot\alpha} K^{\dot\alpha}_{\alpha}} e^{y^{\bar{i}}_{i} K^{i}_{\bar{i}}} e^{ z \Delta + w J} 
e^{(\widetilde\psi^{\alpha}_i q^i_{\alpha} + \widetilde\theta^{\bar{i}}_{\dot\alpha} q^{\dot\alpha}_{\bar{i}})} 
e^{(\widetilde\psi^{\dot\alpha}_i q^i_{\dot\alpha} + \widetilde\theta^{\bar{i}}_{\alpha} q^{\alpha}_{\bar{i}})}
e^{( \widetilde\psi^{\alpha}_{\bar{i}} q^{\bar{i}}_{\alpha} + \widetilde\theta^i_{\dot\alpha} q^{\dot\alpha}_i )}
e^{( \widetilde\psi^{\dot\alpha}_{\bar{i}} q^{\bar{i}}_{\dot\alpha} + \widetilde\theta^i_{\alpha} q^{\alpha}_i )} \, \\
& = u \, v \, e^{(\widetilde\psi^{\alpha}_i q^i_{\alpha} + \widetilde\theta^{\bar{i}}_{\dot\alpha} q^{\dot\alpha}_{\bar{i}})} 
e^{(\widetilde\psi^{\dot\alpha}_i q^i_{\dot\alpha} + \widetilde\theta^{\bar{i}}_{\alpha} q^{\alpha}_{\bar{i}})}
e^{( \widetilde\psi^{\alpha}_{\bar{i}} q^{\bar{i}}_{\alpha} + \widetilde\theta^i_{\dot\alpha} q^{\dot\alpha}_i )}
e^{( \widetilde\psi^{\dot\alpha}_{\bar{i}} q^{\bar{i}}_{\dot\alpha} + \widetilde\theta^i_{\alpha} q^{\alpha}_i )} \, ,
\label{cosetreo}
\end{aligned} 
\end{equation}  
where 
\begin{equation}
u = e^{x^{\alpha}_{\dot\alpha} K^{\dot\alpha}_{\alpha}} e^{z \Delta} \, , \quad \quad  v = 
e^{y^{\bar{i}}_i K^i_{\bar{i}}} e^{w J} \, , 
\label{defininguv}
\end{equation}
and the fermions with tilde are  
\begin{equation}
\begin{aligned}
& (\widetilde\theta^i_{\alpha}, \widetilde\psi^{\dot\alpha}_{\bar{i}}) = e^{\frac{w-z}{2}}  (\theta^i_{\alpha}, \psi^{\dot\alpha}_{\bar{i}}) \, , \quad (\widetilde\theta^{\bar{i}}_{\dot\alpha}, \widetilde\psi^{\alpha}_i) = e^{- \frac{w-z}{2}}  (\theta^{\bar{i}}_{\dot\alpha}, \psi^{\alpha}_i) \, , \\
& (\widetilde\theta^i_{\dot{\alpha}} , \widetilde\psi^{\alpha}_{\bar{i}}) =  
e^{\frac{w+z}{2}} (\theta^i_{\dot{\alpha}} , \psi^{\alpha}_{\bar{i}}) \, ,  \quad 
(\widetilde\theta^{\bar{i}}_{\alpha}, \widetilde\psi^{\dot{\alpha}}_i) = e^{- \frac{w+z}{2}} (\theta^{\bar{i}}_{\alpha}, \psi^{\dot{\alpha}}_i)  \, . 
\label{tildefermions}
\end{aligned} 
\end{equation}
In order to derive the expression above, one uses that when $[X,Y] = s Y$, then $e^Y e^{X} = e^{X} e^{{\rm{exp}}(-s)Y}$.
Notice that the tilde variables do not carry any charge under $\Delta$ and $J$.
The vertex operators of \cite{VertexOld} were constructed in terms of the tilde variables just defined, and 
this is the reason for the overall factor of $e^{n(z+w)}$ appearing on the vertex operators of that paper.    

To achieve any change of variables using exponential representations, it is possible in principle to apply BCH identities several times. However, the intermediate expressions can be very lengthy, and it could be hard to implement the manipulations even on a computer. Here we are going to use a supermatrix representation, which simplifies the manipulations in our case.   
Recall that the superalgebra $\mathfrak{psu}(2,2|4)$ does not admit a supermatrix representation. 
This follows because some of the commutators would produce a term proportional to the identity matrix, which is not a generator of the superalgebra. 
It is possible to enlarge the algebra by adding the extra generator $\mathbb{C}$,
the central charge generator, and the superalgebra becomes  $\mathfrak{su}(2,2|4)$, which admits 
a matrix representation. In what follows, we are going to add this additional generator 
to the cosets and work with supermatrix representations. After obtaining the reparametrizations between the
variables of these extended cosets, we project the result back to $\mathfrak{psu}(2,2|4)$.  
Thus, we are going to consider the representative   
\begin{equation}
g_{\text{old}}^{\mathbb{C}}  = e^{\mathbb{C}_o} g_{\text{old}}^{\prime} \, ,  
\label{extendedOld}
\end{equation}
with $g_{\text{old}}^{\prime}$ defined in \eqref{cosetreo}. Notice that
the position of the additional exponential factor is arbitrary, since $\mathbb{C}$, being a
central charge element, commutes with all the other generators.  
The conventions for the superalgebras are given in the Appendix 
\ref{apsupermatrices}. 
Consider the following definitions  
\begin{table}[H]
\begin{equation}
\centering
\left[
\begin{tabular}{cc}
$\widetilde\theta_{--} $ & $\widetilde\theta_{-+}$ \\
$\widetilde\theta_{+-} $ & $\widetilde\theta_{++}$
\end{tabular}
\right]  =
\left[
\begin{tabular}{cc}
$\widetilde\theta_{\alpha}^j q_j^{\alpha} $ & $\widetilde\theta_{\alpha}^{\bar{j}} q_{\bar{j}}^{\alpha}$ \\
$\widetilde\theta_{\dot\alpha}^j q_j^{\dot\alpha} $ & $\widetilde\theta_{\dot\alpha}^{\bar{j}} q_{\bar{j}}^{\dot\alpha}$
\end{tabular}
\right]  \, , \quad 
\left[
\begin{tabular}{cc}
$\widetilde\psi_{++} $ & $\widetilde\psi_{-+}$ \\
$\widetilde\psi_{+-} $ & $\widetilde\psi_{--}$
\end{tabular}
\right] =
\left[
\begin{tabular}{cc}
$\widetilde\psi^{\alpha}_i q^i_{\alpha} $ & $\widetilde\psi^{\dot\alpha}_{i} q^{i}_{\dot\alpha}$ \\
$\widetilde\psi^{\alpha}_{\bar{i}} q^{\bar{i}}_{\alpha} $ & $\widetilde\psi^{\dot\alpha}_{\bar{i}} q^{\bar{i}}_{\dot\alpha}$
\end{tabular}
\right] \, , 
\end{equation}
\end{table}
\noindent where each $\widetilde{\theta}_{\pm \pm}$ and $\widetilde{\psi}_{\pm \pm}$ are two by two matrices, and for example, 
\begin{table}[H]
\begin{equation}
\centering
\widetilde\theta_{\alpha}^j q_j^{\alpha}  =
\left[ 
\begin{tabular}{cc}
$\widetilde\theta_1^{1} q_1^1$ & $\widetilde\theta^{2}_{1} q^1_2$ \\
$\widetilde\theta^1_{2} q^2_1$ & $\widetilde\theta^2_{2} q^2_2$  \\
\end{tabular}
\right] \, . 
\end{equation}
\end{table}
\noindent In addition, both the $u^R_S$ and $v^I_J$ can be computed from \eqref{defininguv} as follows 
\begin{table}[H]
\begin{equation}
\left[
\begin{tabular}{cc}
$ u^{\alpha}_{\beta} $ & $ u^{\dot\alpha}_{\beta}  $ \\
$ u^{\alpha}_{\dot\alpha}  $ & $ u^{\dot\alpha}_{\dot\beta} $
\end{tabular}
\right] = \left[
\begin{tabular}{cc}
$\delta^\alpha_\beta$ & $0 $ \\
$x^{\alpha}_{\dot\alpha} $ & $ \delta^{\dot{\alpha}}_{\dot{\beta}}$
\end{tabular}
\right]\left[
\begin{tabular}{cc}
$ e^{\frac{z}{2}} \delta^\beta_\gamma  $ & $ 0 $ \\
$0 $ & $ e^{- \frac{z}{2}} \delta^{\dot{\beta}}_{\dot{\gamma}}$
\end{tabular}
\right] =
\left[
\begin{tabular}{cc}
$ e^{\frac{z}{2}} \delta^\alpha_\beta  $ & $ 0 $ \\
$ e^{\frac{z}{2}} x^{\alpha}_{\dot\alpha}  $ & $ e^{-\frac{z}{2}} \delta^{\dot{\alpha}}_{\dot{\beta}}$
\end{tabular}
\right] \, , 
\end{equation}
\end{table}
\noindent and 
\begin{table}[H]
\begin{equation}
\left[
\begin{tabular}{cc}
$ u^i_k  $ & $ u^{\bar{i}}_i  $ \\
$u^{i}_{\bar{k}} $ & $u^{\bar{i}}_{\bar{k}}$
\end{tabular}
\right]  = \left[
\begin{tabular}{cc}
$\delta^i_k$ & $x^{\bar{i}}_i  $ \\
$0 $ & $ \delta^{\bar{i}}_{\bar{k}}$
\end{tabular}
\right]\left[
\begin{tabular}{cc}
$ e^{\frac{w}{2}} \delta^k_j  $ & $ 0 $ \\
$0 $ & $ e^{-\frac{w}{2}} \delta^{\bar{k}}_{\bar{j}}$
\end{tabular}
\right]  =  \left[
\begin{tabular}{cc}
$ e^{\frac{w}{2}} \delta^i_k  $ & $ e^{-\frac{w}{2}} x^{\bar{i}}_i  $ \\
$0 $ & $ e^{-\frac{w}{2}} \delta^{\bar{i}}_{\bar{k}}$
\end{tabular}
\right] \, . 
\end{equation}
\end{table}
\noindent Using the matrix representations above the $g_{\text{old}}^{\mathbb{C}}$ of 
\eqref{extendedOld} takes the following form 
\begin{table}[H]
\begin{equation}
\begin{aligned}
\label{supermatrices}
g_{\text{old}}^{\mathbb{C}} &= e^{\mathbb{C}_o}  \begin{bmatrix}
 u &0\\0& v  
\end{bmatrix} 
\begin{bmatrix}
     1& \widetilde\theta_{++} 
     \\ \widetilde\psi_{++} &1
 \end{bmatrix}  
 \begin{bmatrix}
     1& \widetilde\theta_{-+}  \\
     \widetilde\psi_{-+} &1
 \end{bmatrix} 
\begin{bmatrix}
     1& \widetilde\theta_{+ -} \\
     \widetilde\psi_{+-}  &1
 \end{bmatrix}   
 \begin{bmatrix}
    1&\widetilde\theta_{--}\\ \widetilde\psi_{--}&1
     \end{bmatrix}    \\
 & = e^{\mathbb{C}_o} \begin{bmatrix}u&0\\0&v
 \end{bmatrix}  
 \begin{bmatrix}
     1&\widetilde\theta_{+}\\\widetilde\psi_{+}&1- \widetilde\psi_{++}\widetilde\theta _{-+}
\end{bmatrix}   \begin{bmatrix}
     1&\widetilde\theta_{-}\\\widetilde\psi_{-}&1- \widetilde\psi_{+-}\widetilde\theta_{--}
  \end{bmatrix} \\
 & = e^{\mathbb{C}_o} \begin{bmatrix}u&0\\0&v
\end{bmatrix}
\begin{bmatrix}
     1- \widetilde\theta_{+}\widetilde\psi_{-}&\widetilde\theta \left(1-\widetilde\psi_{+-}\widetilde\theta_{--}\right)\\\left(1-\widetilde\psi_{++}\widetilde\theta_{-+}\right)\widetilde\psi& - \widetilde\psi_{+}\widetilde\theta_{-}+\left(1-\widetilde\psi_{++}\widetilde\theta_{-+}\right)\left(1-\widetilde\psi_{+-}\widetilde\theta_{--}\right)
 \end{bmatrix} \, . 
\end{aligned}
\end{equation} 
\end{table}
\noindent In the equation above, the following notation was used
\begin{equation}
\widetilde\theta_{\pm} = \widetilde\theta_{\pm \pm} + \widetilde\theta_{\mp \pm} \, , \quad \quad  \widetilde\psi_{\pm} = \widetilde\psi_{\pm \pm} + \widetilde\psi_{\mp \pm} \, , \quad \quad 
\widetilde\theta = \widetilde\theta_- + \widetilde\theta_+ \, , \quad \quad  \widetilde\psi = \widetilde\psi_- + \widetilde\psi_+ \, . 
\end{equation}

\subsubsection{Rearranging $g_{\text{new}}$}
\label{Rearranging new}

Let us enlarge the coset representative $g_{\text{new}}$ of \eqref{CosetIdeal} by also adding the generator $\mathbb{C}$ and multiplying the result by 
compensators. One has   
\begin{equation}
g_{\text{new}}^{\mathbb{C},h} = e^{\mathbb{C}_n} \, g_{\text{new}} \, h_{\mathfrak{usp}(2|2)} \, h_{\mathfrak{usp}(4)} \, . 
\label{CosetIdealenlarge}
\end{equation}
The compensators are necessary, as we will see, because their results are nonzero when performing the super-reparametrizations.    
As matrices, the $\Psi$ and the $\Theta$ inside $g_{\text{new}}$ are   
\begin{table}[H]
\begin{equation}
\centering
\Psi= 
\left[ 
\begin{tabular}{cccc}
$\Psi^1_{1} q^1_1$ & $\Psi^2_{1} q^1_2$ & $\Psi^{\dot{1}}_{1} q^1_{\dot{1}}$ & $\Psi^{\dot{2}}_{1} q^1_{\dot{2}}$ \\
$\Psi^1_{2} q^2_1$ & $\Psi^2_{2} q^2_2$ & $\Psi^{\dot{1}}_{2} q^2_{\dot{1}}$ & $\Psi^{\dot{2}}_{2} q^2_{\dot{2}}$ \\
$\Psi^1_{\bar{1}} q^{\bar{1}}_1$ & $\Psi^2_{\bar{1}} q^{\bar{1}}_2$ & $\Psi^{\dot{1}}_{\bar{1}} q^{\bar{1}}_{\dot{1}}$ & $\Psi^{\dot{2}}_{\bar{1}} q^{\bar{1}}_{\dot{2}}$ \\
$\Psi^1_{\bar{2}} q^{\bar{2}}_1$ & $\Psi^2_{\bar{2}} q^{\bar{2}}_2$ & $\Psi^{\dot{1}}_{\bar{2}} q^{\bar{2}}_{\dot{1}}$ & $\Psi^{\dot{2}}_{\bar{2}} q^{\bar{2}}_{\dot{2}}$ 
\end{tabular}
\right] \, , \quad 
\Theta = 
\left[ 
\begin{tabular}{cccc}
$\Theta_1^{1} q_1^1$ & $\Theta^2_{1} q^1_2$ & $\Theta^{\bar{1}}_{1} q^1_{\bar{1}}$ & $\Theta^{\bar{2}}_{1} q^1_{\bar{2}}$ \\
$\Theta^1_{2} q^2_1$ & $\Theta^2_{2} q^2_2$ & $\Theta^{\bar{1}}_{2} q^2_{\bar{1}}$ & $\Theta^{\bar{2}}_{2} q^2_{\bar{2}}$ \\
$\Theta^1_{\dot{1}} q^{\dot{1}}_1$ & $\Theta^2_{\dot{1}} q^{\dot{1}}_2$ & $\Theta^{\bar{1}}_{\dot{1}} q^{\dot{1}}_{\bar{1}}$ & $\Theta^{\bar{2}}_{\dot{1}} q^{\dot{1}}_{\bar{2}}$ \\
$\Theta^1_{\dot{2}} q^{\dot{2}}_1$ & $\Theta^2_{\dot{2}} q^{\dot{2}}_2$ & $\Theta^{\bar{1}}_{\dot{2}} q^{\dot{2}}_{\bar{1}}$ & $\Theta^{\bar{2}}_{\dot{2}} q^{\dot{2}}_{\bar{2}}$ 
\end{tabular}
\right] \, . 
\end{equation}
\end{table}

\noindent Thus  
\begin{table}[H]
\begin{equation}
e^{\Psi} = \left[
\begin{tabular}{cc}
$1$ & $0 $ \\
$\Psi $ & $ 1$
\end{tabular}
\right] \, , \quad \quad  e^{\Theta} = \left[
\begin{tabular}{cc}
$1$ & $\Theta $ \\
$ 0 $ & $ 1$
\end{tabular}
\right] \, , 
\end{equation}
\end{table}

\noindent and
\begin{table}[h!]
\begin{equation}
e^{\Psi} \, e^{\Theta} = \left[
\begin{tabular}{cc}
$1$ & $\Theta $ \\
$ \Psi $ & $ 1 - \Psi \Theta$
\end{tabular}
\right] \, . 
\label{FermionicPart}
\end{equation}
\end{table}

In approximating toward the old variables of \eqref{extendedOld}, two matrices are of particular importance
\begin{equation*}
\label{MN}
 \mathbb{M}\in SU(2,2)\,,\qquad  \mathbb{N}\in SU(4)\,,    
\end{equation*}
\noindent satisfying    
\begin{equation}
\begin{bmatrix}U&0\\0&V
\end{bmatrix} \begin{bmatrix}
    e^{\mathbb{C}_n}h_{AdS_5}&0\\0&e^{\mathbb{C}_n}h_{S^5}
\end{bmatrix}=\begin{bmatrix}u&0\\0&v
\end{bmatrix} \begin{bmatrix}
    e^{\mathbb{C}_n} \mathbb{M}&0\\0& e^{\mathbb{C}_n}\mathbb{N} 
\end{bmatrix}  \equiv 
\begin{bmatrix}u&0\\0&v
\end{bmatrix} \begin{bmatrix}
     \mathbb{\widetilde{M}} &0\\0&  \mathbb{\widetilde{N}}
\end{bmatrix} \, .
\label{DefinitionsMN}
\end{equation}
 
Substituing \eqref{FermionicPart}  and \eqref{DefinitionsMN} into \eqref{CosetIdealenlarge}, one has 

\begin{equation}
\begin{aligned}
g_{\text{new}}^{\mathbb{C},h}  & =  \left[
\begin{tabular}{cc}
$1$ & $\Theta $ \\
$ \Psi $ & $ 1 - \Psi \Theta$
\end{tabular}
\right]
\begin{bmatrix}u&0\\0&v
\end{bmatrix} \begin{bmatrix}
 \mathbb{\widetilde{M}}&0\\0& \mathbb{\widetilde{N}}
\end{bmatrix}   \,  \\
& =  \begin{bmatrix}u&0\\0&v
\end{bmatrix} \begin{bmatrix}
    1&\widetilde \Theta \\\widetilde \Psi &1 - \widetilde \Psi \widetilde \Theta
\end{bmatrix} 
 \begin{bmatrix}
    \mathbb{\widetilde{M}}&0\\0& \mathbb{\widetilde{N}}
\end{bmatrix} =  \begin{bmatrix}u&0\\0&v
\end{bmatrix} 
\begin{bmatrix}
    \mathbb{\widetilde{M}}&\tilde \Theta \mathbb{\widetilde{N}}\\\tilde \Psi \mathbb{\widetilde{M}}&(1-\tilde \Psi \tilde \Theta)\mathbb{\widetilde{N}}
\end{bmatrix} \, , 
\label{matrixIdealCoset}
\end{aligned} 
\end{equation}
where the capital tilde variables are related to the original fermionic variables by 
\begin{equation}
\label{TildedFermionsMatrices}
\begin{bmatrix}
    0& \widetilde\Theta\\  \widetilde\Psi&0
\end{bmatrix}  =\begin{bmatrix}u^{-1}&0\\0&v^{-1} 
\end{bmatrix} \begin{bmatrix}
    0& \Theta \\ \Psi &0
\end{bmatrix}\begin{bmatrix}u&0\\0&v
\end{bmatrix}   \,.  
\end{equation}

Notice that the above definition of the capital tilde variables $\widetilde\Psi$ and  $\widetilde\Theta$ mix the new fermionic variables with  
the old bosonic variables. 
Later we are going to disentangled the two set of coordinates and define fermionic coordinates 
which are combinations of only the variables of the new coset.

\subsubsection{Mapping the fermionic variables}
\label{Mapping the fermionic variables}

After the manipulations of the two previous Subsections, we are now in position to find the change of coordinates  
between the capital tilde variables $\widetilde\Psi$ and  $\widetilde\Theta$ of \eqref{TildedFermionsMatrices}
and the old variables $\widetilde\psi$ and 
$\widetilde\theta$. This follows from equating \eqref{supermatrices} with \eqref{matrixIdealCoset}, which gives       
\begin{equation}
\begin{bmatrix}
     1 -\widetilde\theta_{+}\widetilde\psi_{-}&\widetilde\theta \left(1-\widetilde\psi_{+-}\widetilde\theta_{--}\right)\\\left(1-\widetilde\psi_{++}\widetilde\theta_{-+}\right)\widetilde\psi&- \widetilde\psi_{+}\widetilde\theta_{-}+\left(1-\widetilde\psi_{++}\widetilde\theta_{-+}\right)\left(1-\widetilde\psi_{+-}\widetilde\theta_{--}\right)
 \end{bmatrix} =
\begin{bmatrix}
    \mathbb{\widetilde{M}}&\tilde \Theta \mathbb{\widetilde{N}}\\\widetilde \Psi \mathbb{\widetilde{M}}&(1-\widetilde \Psi \widetilde \Theta)\mathbb{\widetilde{N}}
\end{bmatrix} \, .
 \end{equation}
Notice that we have set $\mathbb{C}_o =0$ above without losing any generality, since
we can absorb any contribution into $\mathbb{C}_n$, which is part of the definition 
of $\mathbb{\widetilde{M}}$ and $\mathbb{\widetilde{N}}$.    
 It is easy to see that the equation above implies 
\begin{equation}
\label{infinitesimals}
\mathbb{\widetilde{M}}=1- \widetilde{\theta}_+ \widetilde\psi_-\,, 
\end{equation}
\begin{equation*}
\mathbb{\widetilde{N}}= - \widetilde\psi_{+}\widetilde\theta_{-} +\left(1-\widetilde\psi_{++}\widetilde\theta_{-+}\right)\left(1+ \widetilde\psi \mathbb{\widetilde M}^{-1}\widetilde\theta \right)\left(1-\widetilde\psi_{+-}\widetilde\theta_{--}\right)\,,
\end{equation*}
\begin{equation*}
\widetilde \Psi =\left(1-\widetilde\psi_{++}\widetilde\theta_{-+}\right)\widetilde\psi \, \mathbb{\widetilde{M}}^{-1}  \,, 
\end{equation*}
\begin{equation*}
\widetilde \Theta =\widetilde\theta \left(1-\widetilde\psi_{+-}\widetilde\theta_{--}\right)\mathbb{\widetilde{N}}^{-1}\,,   
\end{equation*}
where the above constructions has a natural time order, i.e., we first need to construct $\mathbb{\widetilde{M}}$ and $\mathbb{\widetilde{M}}^{-1}$ to construct $\widetilde\Psi$ and $\mathbb{\widetilde{N}}$, and then construct $\mathbb{\widetilde{N}}^{-1}$ to construct $\widetilde\Theta$. 
Notice that it is possible to invert the map  above  and write $\widetilde\psi$ and $\widetilde\theta$ in terms of $\widetilde\Psi$ and $\widetilde\Theta$ by solving the following equations recursively 
\begin{equation}
\label{iteration}
\widetilde\psi =\left(1-\widetilde\psi_{++}\widetilde\theta_{-+}\right)^{-1}\widetilde \Psi \mathbb{\widetilde{M}}\,,   
\end{equation}
\begin{equation*}
\widetilde\theta  =\widetilde \Theta \mathbb{\widetilde{N}}\left(1-\widetilde\psi_{+-}\widetilde\theta_{--}\right)^{-1} \,. 
\end{equation*}
For example, at order one in $\widetilde\Psi$ and $\widetilde\Theta$, one has
\begin{equation}
\widetilde\psi_1 = \widetilde\Psi \, , \quad  \widetilde\theta_1 = \widetilde\Theta \, ,  
\label{order1}
\end{equation}  
where the subscript denotes the fermionic order. 
Notice also that the matrices  $\mathbb{\widetilde{M}}$ and  $\mathbb{\widetilde{N}}$ are equal to the identity
matrix at order zero.  Then one uses both equations of \eqref{iteration} to generate $\widetilde\psi_3(\widetilde\Psi,\widetilde\Theta)$ 
and $\widetilde\theta_3(\widetilde\Psi,\widetilde\Theta)$ by substituting 
\eqref{order1} into \eqref{iteration}. The iteration procedure is guaranteed to stop after some finite number of steps because the variables are fermionic.

\subsubsection{Mapping the bosonic variables and compensators} 
\label{Mapping the bosonic variables and compensators}

Let us consider the coordinate transformations of the bosonic variables.
We want to isolate the gauge compensators $h_{\mathfrak{usp}(2|2)}$ and $h_{\mathfrak{usp}(4)}$ of \eqref{CosetIdealenlarge} from $\mathbb{M}$ and $\mathbb{N}$.  
We also want to find the variables 
$X, Y, Z$ and $W$ as a function of $x, y, z$ and $w$. The definitions of $\mathbb{M}$ and $\mathbb{N}$ 
 are given in \eqref{DefinitionsMN}, and we recall them here for the reader's convenience 
\begin{equation}
\label{MNrelatedtoall}
\begin{bmatrix}
\mathbb{M}&0\\0& \mathbb{N} 
\end{bmatrix} =
\begin{bmatrix}u^{-1}&0\\0&v^{-1}
\end{bmatrix} 
\begin{bmatrix}U&0\\0&V
\end{bmatrix} \begin{bmatrix}
    h_{\mathfrak{usp}(2|2)}&0\\0& h_{\mathfrak{usp}(4)}
\end{bmatrix}  \equiv
\begin{bmatrix} \mathbb{U} &0\\0& \mathbb{V}
\end{bmatrix} \begin{bmatrix}
    h_{\mathfrak{usp}(2|2)}&0\\0& h_{\mathfrak{usp}(4)}
\end{bmatrix}  \, . 
\end{equation}
Notice that the factors of $e^{\mathbb{C}_n}$ appearing in \eqref{DefinitionsMN} are proportional to the identity matrix and it does not contribute to the equation above. 
It was also convenient to define the auxiliary matrices $\mathbb{U}$ and $\mathbb{V}$ depending on the variables $\mathbb{X}, \mathbb{Y}, \mathbb{Z}$ and $\mathbb{W}$.  
At the origin of the old coordinates, the auxiliary variables are direct related to the new bosonic
coordinates.  
One has, by breaking the equation above into four by four blocks   
\begin{equation}
\begin{bmatrix}
\mathbb{M}_{\alpha}^{\beta}&\mathbb{M}_{\alpha}^{\dot\beta}\\\mathbb{M}_{\dot\alpha}^{\beta}&\mathbb{M}_{\dot\alpha}^{\dot\beta}
\end{bmatrix}=
\begin{bmatrix}
e^{\frac{\mathbb{Z}}{2}}\delta_{\alpha}^{\beta}&0\\e^{\frac{\mathbb{Z}}{2}} \mathbb{X}_{\dot\alpha}^{\beta}&e^{-\frac{\mathbb{Z}}{2}}\delta_{\dot\alpha}\,^{\dot\beta}
\end{bmatrix}h_{\mathfrak{usp}(2|2)}\,, \quad \quad 
\begin{bmatrix}
\mathbb{N}_{i}^{j}&\mathbb{N}_{i}^{\overline j}\\\mathbb{N}_{\overline i}^{j}&\mathbb{N}_{\overline i}^{\overline j}
\end{bmatrix}=
\begin{bmatrix}
e^{\frac{\mathbb{W}}{2}}\delta_{i}^{j}&e^{-\frac{\mathbb{W}}{2}}\mathbb{Y}\,_{i}^{\overline j}\\0&e^{-\frac{\mathbb{W}}{2}}\delta_{\overline i}^{\overline j}
\end{bmatrix}h_{\mathfrak{usp}(4)}\,.
\label{TheBosonic1}
\end{equation}

Recall that the compensators are $USp(2,2)$ and $USp(4)$ group elements, so they satisfy the identity
\begin{equation}
h_{\mathfrak{usp}(2|2)} \cdot \sigma_6 \cdot h_{\mathfrak{usp}(2|2)}^T = \sigma_6 \, , \quad \quad  {\rm{where}} \quad \quad  \sigma_6 =\left[ 
\begin{tabular}{cc}
$  \varepsilon_{\alpha \beta}$ & 0 \\
0 & -   $  \varepsilon_{\dot{\alpha} \dot{\beta}}$ 
\end{tabular}
\right] \, ,
\label{sigma6}
\end{equation}
with $\varepsilon_{1 2}=\varepsilon_{\dot{1} \dot{2}}=1$\,. The matrix $\sigma_6$ 
is one of the 6d Pauli matrices, see the Subsection \ref{6dPauliMatrices} for 
all matrices.   
Thus  in order to eliminate $h_{\mathfrak{usp}(2|2)}$ and $h_{\mathfrak{usp}(4)}$ from \eqref{TheBosonic1}, we multiply 
both sides of the equation with $\sigma_6$ and the transposed equation. One has the following result for the variables
\begin{equation}
e^{\mathbb{Z}}\varepsilon_{\alpha\beta}=  \mathbb{M}_{\alpha}^{\gamma} \, \varepsilon _{\gamma\delta} \, \mathbb{M}^{\delta}_{\beta} -  \mathbb{M}_{\alpha}^{\dot\gamma} \, \varepsilon _{\dot\gamma\dot\delta} \, \mathbb{M}^{\dot\delta}_{\beta} \,,  \quad \quad 
e^{\mathbb{Z}} \varepsilon_{\alpha\gamma}\mathbb{X}^{\gamma}_{\dot\beta}=   \mathbb{M}_{\alpha}^{\gamma}\, \varepsilon _{\gamma\delta} \, \mathbb{M}^{\delta}_{\dot\beta} -  \mathbb{M}_{\alpha}^{\dot\gamma} \, \varepsilon _{\dot\gamma\dot\delta} \, \mathbb{M}^{\dot\delta}_{\dot\beta}  \,,
\label{BosonicM} 
\end{equation}
and 
\begin{equation}
e^{-\mathbb{W}}\varepsilon_{\bar i\bar j} 
=  - \mathbb{N}_{\bar i}^{k} \, \varepsilon_{kl} \, \mathbb{N}^{l}_{\bar j} +  \mathbb{N}_{\bar i}^{\bar k} \, \varepsilon_{\bar k\bar l} \, \mathbb{N}^{\bar l}_{\bar j}\,,
\quad \quad 
e^{-\mathbb{W}}\varepsilon_{\bar i\bar k}\mathbb{Y}^{\bar k}_{j}
=  - \mathbb{N}_{\bar i}^{k} \, \varepsilon_{kl} \, \mathbb{N}^{l}\,_{j} + \mathbb{N}_{\bar i}^{\bar k} \, \varepsilon_{\bar k\bar l} \, \mathbb{N}^{\bar l}_{j}\,.
\label{BosonicN} 
\end{equation}

Departing from the origin of \eqref{cosetreo}, by means of the left translations parametrized by $x,y, z$ and $w$, the matrix equation 
\begin{equation}
\begin{bmatrix} U &0\\0& V
\end{bmatrix} =
\begin{bmatrix}u &0\\0&v 
\end{bmatrix} 
\begin{bmatrix} \mathbb{U}&0\\0& \mathbb{V}
\end{bmatrix} \, , 
\end{equation}
implies 
\begin{equation}
\label{moving bosons}
e^{Z}=e^{z}e^{\mathbb{Z}}\,, \quad e^{W}=e^{w}e^{\mathbb{W}}\, ,  \quad  X_{\dot\alpha}^{\beta}=x_{\dot\alpha}^{\beta}+e^{-z}\mathbb{X}_{\dot\alpha}^{\beta}\, ,
\quad Y_{i}^{\bar j}=y_{i}^{\bar j}+e^{w}\mathbb{Y}_{i}^{\bar j}\,.     
\end{equation} 

Having determined the variables $\mathbb{X}$, $\mathbb{Z}$ and $\mathbb{W}$, we can recover the 
compensators. The equation \eqref{TheBosonic1} implies 
\begin{equation}
\begin{bmatrix}
e^{-\frac{\mathbb{Z}}{2}}\delta_{\alpha}^{\beta}&0\\-e^{\frac{\mathbb{Z}}{2}} \mathbb{X}_{\dot\alpha}^{\beta}&e^{\frac{\mathbb{Z}}{2}}\delta_{\dot\alpha}^{\dot\beta}
\end{bmatrix}\begin{bmatrix}
\mathbb{M}_{\beta}^{\gamma}&\mathbb{M}_{\beta}^{\dot\gamma}\\\mathbb{M}_{\dot\beta}^{\gamma}&\mathbb{M}_{\dot\beta}^{\dot\gamma}
\end{bmatrix}=\begin{bmatrix}
h_{\mathfrak{usp}(2|2)}\, _{\alpha}^{\gamma}&h_{\mathfrak{usp}(2|2)}\,_{\alpha}^{\dot\gamma}\\h_{\mathfrak{usp}(2|2)}\,_{\dot\alpha}^{\gamma}&h_{\mathfrak{usp}(2|2)}\,_{\dot\alpha}^{\dot\gamma} 
\end{bmatrix} \, , 
\end{equation}
and 
\begin{equation}
\begin{bmatrix}
e^{-\frac{\mathbb{W}}{2}}\delta_{i}^{j}&-e^{-\frac{\mathbb{W}}{2}}\mathbb{Y}_{i}^{\bar j}\\0&e^{\frac{\mathbb{W}}{2}}\delta_{\bar i}^{\bar j}
\end{bmatrix}\begin{bmatrix}
\mathbb{N}_{i}^{j}&\mathbb{N}_{i}^{\bar j}\\\mathbb{N}_{\bar i}^{j}&\mathbb{N}_{\bar i}^{\bar j}
\end{bmatrix}=\begin{bmatrix}
h_{\mathfrak{usp}(4)}\,_{i}^{j}&h_{\mathfrak{usp}(4)}\,_{i}^{\bar j}\\h_{\mathfrak{usp}(4)}\,_{\bar i}^{j}&h_{\mathfrak{usp}(4)}\,_{\bar i}^{\bar j}
\end{bmatrix} \, . 
\end{equation}

Equivallently, 
\begin{equation}
\begin{aligned}
h_{\mathfrak{usp}(2|2)}\,_{\alpha}^{\gamma}=e^{-\frac{\mathbb{Z}}{2}}\mathbb{M}_{\alpha}^{\gamma} \, , \quad 
& h_{\mathfrak{usp}(2|2)}\,_{\alpha}^{\dot\gamma}=e^{-\frac{\mathbb{Z}}{2}}\mathbb{M}_{\alpha}^{\dot\gamma}  \, , \quad 
h_{\mathfrak{usp}(2|2)}\,_{\dot\alpha}^{\gamma}=e^{\frac{\mathbb{Z}}{2}}\mathbb{M}_{\dot\alpha}^{\gamma}-e^{\frac{\mathbb{Z}}{2}} \, \mathbb{X}_{\dot\alpha}^{\beta} \, \mathbb{M}_{\beta}^{\gamma} \, , \quad \\
& h_{\mathfrak{usp}(2|2)}\,_{\dot\alpha}^{\dot\gamma}=e^{\frac{\mathbb{Z}}{2}}\mathbb{M}_{\dot\alpha}^{\dot\gamma}-e^{\frac{\mathbb{Z}}{2}} \mathbb{X}\,_{\dot\alpha}^{\beta}\mathbb{M}_{\beta}^{\dot\gamma} \, , 
\end{aligned}
\end{equation}
finally, 
\begin{equation}
\begin{aligned}
h_{\mathfrak{usp}(4)}\,_{\bar i}^{j}=e^{\frac{\mathbb{W}}{2}}\mathbb{N}_{\overline i}^{j} \,, \quad 
& h_{\mathfrak{usp}(4)}\,_{\bar i}^{\bar j}=e^{\frac{\mathbb{W}}{2}}\mathbb{N}_{\bar i}^{\bar j}\,, \quad 
h_{\mathfrak{usp}(4)}\,_{i}^{j}=e^{-\frac{\mathbb{W}}{2}}\mathbb{N}_{i}^{j}-e^{-\frac{\mathbb{W}}{2}}\mathbb{Y}_{i}^{\overline k}\mathbb{N}_{\overline k}^{j}\,, \\
& h_{\mathfrak{usp}(4)}\,_{i}^{\bar j}=e^{-\frac{\mathbb{W}}{2}}\mathbb{N}_{i}^{\bar j}-e^{-\frac{\mathbb{W}}{2}}\mathbb{Y}_{i}^{\bar k}\mathbb{N}_{\bar k}^{\bar j}\,.
\end{aligned}
\end{equation}

The final step to compute the transformations is the determination of the matrices $\mathbb{M}$ and $\mathbb{N}$. Recall that 
$ e^{-\mathbb{C}_n} \, \mathbb{\widetilde{M}} = \mathbb{M}$ and $e^{-\mathbb{C}_n} \, \mathbb{\widetilde{N}} = \mathbb{N}$
and the matrices $\mathbb{\widetilde{M}}$ and $\mathbb{\widetilde{N}}$ were computed in \eqref{infinitesimals}.
Utilizing the fact that products of supermatrices preserve superdeterminants, one has
\begin{equation}
e^{4 \mathbb{C}_n}=\text{det} \, \mathbb{\widetilde{M}}=\text{det} \, \mathbb{\widetilde{N}}\,.   
\end{equation}
The equation above enables one to extract $e^{- \mathbb{C}_n}$ and consequently the matrices 
$\mathbb{M}$ and $\mathbb{N}$. In fact, it is possible to write it in closed form by using the Cayley-Hamilton theorem.
Denoting 
\begin{equation}
p_1 =  \text{tr}(\widetilde\theta_+ \widetilde\psi_-) \, , \quad 
p_2 =  \text{tr}(\widetilde\theta_+ \widetilde\psi_- \widetilde\theta_+ \tilde\psi_-)  \, , \quad \ldots \quad  \, , \quad
p_n =  \text{tr}\left((\widetilde\theta_+ \widetilde\psi_-)^n \right) \, , 
\end{equation}
one has
\begin{equation}
e^{4C} = 1+ p_1 +  \frac{1}{2}(p_1^2 - p_2)+ \frac{1}{6}(p_1^3  - 3 p_1 p_2  + 2 p_3)+\frac{1}{24}(p_1^4  - 6 p_1^2 p_2 + 3 p_2^2 + 8 p_1 p_3 - 6 p_4) \, . 
\end{equation}

Notice that since $\mathbb{X}, \mathbb{Y}, \mathbb{Z}$ and $\mathbb{W}$ only depends 
on the fermions $\widetilde\theta$ and $\widetilde\psi$, it is possible to express 
$x,y, z$ and $w$ in terms of the bosonic coordinates $X,Y, Z,W$ and the fermionic 
coordinates $\widetilde \Psi$ and $\widetilde \Theta$ by using \eqref{iteration}
and 
\begin{equation}
\label{movingbosonsinvert}
\begin{aligned}  
e^{z}=e^{Z}e^{-\mathbb{Z}(\Psi,\Theta)}\,, \quad \quad   e^{w}=e^{W}e^{-\mathbb{W}(\Psi,\Theta)}\, ,  \quad  \quad \quad \quad \quad \\ 
 x_{\dot\alpha}^{\beta}=X_{\dot\alpha}^{\beta}-e^{-Z}e^{\mathbb{Z}(\Psi,\Theta)} \mathbb{X}_{\dot\alpha}^{\beta}\, ,
\quad \quad  y_{i}^{\bar j}=Y_{i}^{\bar j} - e^{W}e^{-\mathbb{W}(\Psi,\Theta)}\mathbb{Y}_{i}^{\bar j}\,,     
\end{aligned} 
\end{equation} 
and the equations above can be solved order by order in $\{ \Psi, \Theta \}$ recursively. 

\subsubsection{The fermionic variables $\widehat{\Psi}$ and $\widehat{\Theta}$}
\label{coordinatesfinal}

Recall that the variables $\widetilde{\Psi}$ and $\widetilde\Theta$ of \eqref{TildedFermionsMatrices} mix 
the new fermionic variables with the old bosonic variables parametrizing $u$ and $v$.  
In particular the equations \eqref{iteration} can be used to find  
$\widetilde{\psi}$ and $\widetilde\theta$ iteratively in terms of the latter.
To get the vertices expressed in terms of only the new coordinates, it 
is necessary to map $\widetilde{\psi}$ and $\widetilde\theta$ to a different set of fermionic variables, precisely the ones appearing in $g^{\prime}_{\rm{new}}$ of \eqref{gprime}.
As explained below, there is a map 
\begin{equation}
\widetilde{\psi} = g(\widehat{\Psi},\widehat{\Theta}) \, , \quad \quad 
\widetilde{\theta} = f(\widehat{\Psi},\widehat{\Theta}) \, ,  
\label{transformfg}
\end{equation}
where 
\begin{equation}
\begin{aligned} 
\label{PsitildedN}
& \quad \quad  \quad \quad \quad \widehat{\Psi}_{i}^{\alpha}=e^{\frac{Z-W}{2}}\Psi_{i}^{\alpha}-e^{\frac{Z-W}{2}}X_{i}^{\overline j}\Psi_{\overline j}^{\alpha}+e^{\frac{Z-W}{2}}\Psi_{i}^{\dot\beta}X_{\dot\beta}^{\alpha} 
-e^{\frac{Z-W}{2}}X\,_{i}^{\overline j} \Psi_{\overline j}^{\dot\beta}X\,_{\dot\beta}^{\alpha}\,, \\
& \widehat{\Psi}_{\overline i}^{\alpha}=  e^{\frac{Z+W}{2}} \Psi_{\overline i}^{\alpha} +e^{\frac{Z+W}{2}}\Psi_{\overline i}^{\dot\beta}X\,_{\dot\beta}^{\alpha}\,,  \quad 
\widehat{\Psi}_{i}^{\dot\alpha}= e^{-\frac{Z+W}{2}}\Psi_{i}^{\dot\alpha}-e^{-\frac{Z+W}{2}}X\,_{i}^{\overline j}\Psi_{\overline j}^{\dot\alpha} \,,  \quad 
\widehat{\Psi}_{\overline i}^{\dot\alpha}= e^{\frac{W-Z}{2}}\Psi_{\overline i}^{\dot\alpha}   \,.
\end{aligned} 
\end{equation}
and 
\begin{equation}
\begin{aligned} 
\label{ThetatildedN}
& \quad \quad  \quad \quad \quad \widehat\Theta_{\dot\alpha}^{\overline i}=e^{\frac{Z-W}{2}} \Theta_{\dot\alpha}^{\overline i}-e^{\frac{Z-W}{2}}X_{\dot\alpha}^{\beta} \Theta_{\beta}^{\overline i}+e^{\frac{Z-W}{2}} \Theta_{\dot\alpha}^{j}X_{j}^{\overline i}-e^{\frac{Z-W}{2}}X_{\dot\alpha}^{\beta} \Theta_{\beta}^{j}X_{j}^{\overline i}\,,   \\
& \widehat\Theta_{\dot\alpha}^{i}=e^{\frac{Z+W}{2}}\Theta_{\dot\alpha}^{i} -e^{\frac{Z+
W}{2}}X_{\dot\alpha}^{\beta}\Theta_{\beta}^{i}\,,   \quad 
\widehat\Theta_{\alpha}^{\overline i}=e^{-\frac{Z+W}{2}}\Theta_{\alpha}^{\overline i}-e^{-\frac{Z+W}{2}} \Theta_{\alpha}^{j}x_{j}^{\overline i}\,, \quad 
\widehat\Theta_{\alpha}^{i}=e^{\frac{W-Z}{2}} \Theta_{\alpha}^{i} \,.
\end{aligned} 
\end{equation}
 
The functions $f$ and $g$ of \eqref{transformfg} correspond to the reparametrization $g_{\text{old}} \rightarrow g^{\prime}_{\text{new}}$. 
One way of finding these functions is to use the results of the previous Subsections 
to determine $f^{-1}$ and $g^{-1}$ and then invert the functions.
The inverse functions are obtained by replacing the new variables by the old 
ones in the right-hand side of  $\eqref{PsitildedN}$ and $\eqref{ThetatildedN}$ and inverting the resulting equations iteratively. 
In {\it{Mathematica}}, one efficient way of obtaining the inverse functions is by solving 
the equations resulting from the chain of transformations $g^{\prime}_{\text{new}} \rightarrow g_{\text{old}} 
\rightarrow g^{\prime}_{\text{new}}$. The final results can be written in terms of $\mathfrak{su}_2^{\oplus 4}$ invariants.  
We give some of the results for these transformations at lower order in the fermionic variables below.
One has 
\begin{equation}
\label{transformatheta}
\begin{aligned}
& \widetilde{\theta}^i_{\alpha}= \widehat{\Theta }_{\alpha}^{i}
- \widehat{\Theta }_{\alpha }^i \widehat{\Theta }_{\beta }^{\bar{j}} \widehat{\Psi
   }_{\bar{j}}^{\beta } - \epsilon ^{\gamma  \delta } \epsilon _{\alpha  \beta }
  \widehat{\Theta }_{\gamma }^{\bar{j}} \widehat{\Theta }_{\delta }^i \widehat{\Psi
   }_{\bar{j}}^{\beta } + \widehat{\Theta }_{\alpha }^{\bar{j}} \widehat{\Theta
   }_{\dot{\beta }}^i \widehat{\Psi }_{\bar{j}}^{\dot{\beta }} + \ldots \, , \quad \\ 
& \widetilde{\theta}_{\bar{i}}^{\dot\alpha}=\widehat{\Theta }_{\bar{i}}^{\dot\alpha}
 - \widehat{\Theta }_{\beta }^{\bar{j}} \widehat{\Psi }_{\bar{i}}^{\beta } \widehat{\Psi
   }_{\bar{j}}^{\dot{\alpha }} + \frac{1}{2} \epsilon ^{\dot{\alpha } \dot{\delta }}
   \epsilon _{\dot{\beta } \dot{\gamma }} \widehat{\Theta }_{\dot{\delta }}^{\bar{j}}
   \widehat{\Psi }_{\bar{i}}^{\dot{\beta }} \widehat{\Psi }_{\bar{j}}^{\dot{\gamma
   }} + \frac{1}{2} \epsilon ^{\bar{k} \bar{l}} \epsilon _{\bar{i} \bar{j}}
   \widehat{\Theta }_{\dot{\beta }}^{\bar{j}} \widehat{\Psi }_{\bar{k}}^{\dot{\alpha }}
   \widehat{\Psi }_{\bar{l}}^{\dot{\beta }} + \ldots , 
\end{aligned}
\end{equation}
and 
\begin{equation}
\label{zwtransforme} 
\begin{aligned}
& e^z = e^Z + \frac{1}{2} e^Z \widehat{\Theta}^{\bar{i}}_{\alpha} \widehat{\Psi}^{\alpha}_{\bar{i}} 
- \frac{1}{2} e^Z \widehat{\Theta}^{\bar{i}}_{\dot\alpha} \widehat{\Psi}^{\dot\alpha}_{\bar{i}} + \ldots \, , \\
& e^w =e^W  - \frac{1}{2} e^W \widehat{\Theta}^{\bar{i}}_{\alpha} \widehat{\Psi}^{\alpha}_{\bar{i}} 
- \frac{1}{2} e^W \widehat{\Theta}^{\bar{i}}_{\dot\alpha} \widehat{\Psi}^{\dot\alpha}_{\bar{i}} + \ldots \, . 
\end{aligned}
\end{equation}

Since the compensators are nonzero, the pure spinors also rotates  

\begin{equation}
\begin{aligned}
 \lambda'^{i}_{\alpha} \, = \,  &  \lambda^{i}_{\alpha} - \frac{1}{2} \widehat{\Theta}^{\bar{i}}_{\beta} 
 \widehat{\Psi}^{\beta}_{\bar{i}} \lambda^{i}_{\alpha} +
\widehat{\Theta}^{\bar{i}}_{\alpha} \widehat{\Psi}^{\beta}_{\bar{i}}  \lambda^i_{\beta} 
+ \widehat{\Theta}^{\bar{i}}_{\alpha} \widehat{\Psi}^{\dot\beta}_{\bar{i}}  \lambda^i_{\dot\beta} 
- \widehat{\Theta}^{i}_{\beta} \widehat{\Psi}^{\beta}_{\bar{i}}  \lambda^{\bar{i}}_{\alpha} + \ldots \, , \\
 \lambda'^{\bar{i}}_{\alpha} \, = \, &  \lambda^{\bar{i}}_{\alpha} + \frac{1}{2} 
 \widehat{\Theta}^{\bar{j}}_{\dot\beta} \widehat{\Psi}^{\dot\beta}_{\bar{j}} \lambda^{\bar{i}}_{\alpha}
 + \widehat{\Theta}^{\bar{j}}_{\alpha} \widehat{\Psi}^{\beta}_{\bar{j}} \lambda^{\bar{i}}_{\beta}
+ \widehat{\Theta}^{\bar{j}}_{\alpha} \widehat{\Psi}^{\dot\beta}_{\bar{j}} \lambda^{\bar{i}}_{\dot\beta}
- \widehat{\Theta}^{\bar{i}}_{\dot\alpha} \widehat{\Psi}^{\dot\beta}_{\bar{j}} \lambda^{\bar{j}}_{\alpha} \\
&  + \varepsilon_{j k} \varepsilon^{\bar{i} \bar{j}} \widehat{\Theta}^{j}_{\dot\alpha} \widehat{\Psi}^{\dot\beta}_{\bar{j}} \lambda^{k}_{\alpha} - \widehat{\Theta}^{\bar{i}}_{\beta} \widehat{\Psi}^{\beta}_{\bar{j}} \lambda^{\bar{j}}_{\alpha}  + \ldots  \, , 
\end{aligned} 
\end{equation}
and 
\begin{equation}
\begin{aligned}
\lambda'^{i}_{\dot\alpha} \, = \,  & \lambda^{i}_{\dot\alpha} -\frac{1}{2} \widehat{\Theta}^{\bar{i}}_{\dot\beta} \widehat{\Psi}^{\dot\beta}_{\bar{i}} \lambda^{\bar{i}}_{\dot\alpha} 
+ \widehat{\Theta}^{\bar{i}}_{\dot\alpha} \widehat{\Psi}^{\dot\beta}_{\bar{i}} \lambda^{i}_{\dot\beta} 
- \varepsilon^{\alpha \beta} \varepsilon_{\dot\alpha \dot\beta} \widehat{\Theta}^{\bar{i}}_{\alpha} \widehat{\Psi}^{\dot\beta}_{\bar{i}} \lambda^{i}_{\beta} 
- \widehat{\Theta}^{i}_{\bar\alpha} \widehat{\Psi}^{\bar\alpha}_{\bar{i}} \lambda^{\bar{i}}_{\dot\alpha} + \ldots,  \\ 
\lambda'^{\bar{i}}_{\dot\alpha} \, = \, & \lambda^{\bar{i}}_{\dot\alpha} + \frac{1}{2}
\widehat{\Theta}^{\bar{j}}_{\beta} \widehat{\Psi}^{\beta}_{\bar{j}} \lambda^{\bar{i}}_{\dot\alpha} 
+ \varepsilon_{i j} \varepsilon^{\bar{i} \bar{j}} \widehat{\Theta}^{i}_{\dot\beta} \widehat{\Psi}^{\dot\beta}_{\bar{j}} \lambda^{j}_{\dot\alpha} 
-  \varepsilon^{\alpha \beta} \varepsilon_{\dot\alpha \dot\beta} \widehat{\Theta}^{\bar{i}}_{\alpha} \widehat{\Psi}^{\dot\beta}_{\bar{j}} \lambda^{\bar{j}}_{\beta}  \\
& - \widehat{\Theta}^{\bar{i}}_{\dot\beta} \widehat{\Psi}^{\dot\beta}_{\bar{j}} \lambda^{\bar{j}}_{\dot\alpha}
- \widehat{\Theta}^{\bar{i}}_{\beta} \widehat{\Psi}^{\beta}_{\bar{j}} \lambda^{\bar{j}}_{\dot\alpha}
+ \widehat{\Theta}^{\bar{j}}_{\dot\alpha} \widehat{\Psi}^{\dot\beta}_{\bar{j}} \lambda^{\bar{i}}_{\dot\beta} + \ldots . 
\end{aligned} 
\end{equation}
Similar expressions can be computed for the other sixteen $\lambda^R_I$ pure spinors. 

The BRST transformations of the variables $\widehat{\Psi}$ and $\widehat{\Theta}$ can be
obtained by acting with $Q_n$ of \eqref{BRSTnewvariables} on all the terms of \eqref{PsitildedN} and \eqref{ThetatildedN}.
Alternatively, it is possible to compute it by using the general formulas of the Subsection  
\ref{BRSTreparametrizations}, and 
both procedures are also valid for the BRST transformations of the tilde variables of 
\eqref{gprime}.

\pagebreak

\section{ 
The vertex operators %
} 
\label{The ideal coset vertices}

In this section, we start by presenting the vertices using $g^{\prime}_{\text{new}}$ of \eqref{gprime}, directly obtained 
from the change of variables from $g_{\rm{old}}$ of  \eqref{OldCosetRepresentative}, without any simplification.  
As a consistency check, the resulting vertices were verified to be closed under the BRST operator 
of $g^{\prime}_{\text{new}}$.
Then we apply finite translations on them to get a set of vertices parametrized 
by $\eta^{\dot\alpha}_\beta$, and $\eta^i_{\bar{j}}$. 
The final expressions can be recast using only $SU(2,2)\times SU(4)$ bosonic 
covariant variables 
and the fermionic variables of $g_{\text{new}}$ of  \eqref{CosetIdeal} after a few manipulations. 
We end the Section by discussing several properties of the vertex operators. 

\subsection{The super-reparametrization of the vertex operators} 
\label{vertexoperatorsfromchangingcoordinates}

The vertex operators of \cite{VertexOld} only depend on $\widetilde{\theta}_+ = e^{\frac{w-z}{2}} \{\theta^i_{\alpha}, \theta^{\dot\alpha}_{\bar{i}} \}$, see  \eqref{OldCosetRepresentative}. 
This follows only because $g_{\text{old}}$ was
carefully chosen and it is the bottom of the supermultiplet preserving 24 supersymmetries. As a consequence, its fermionic expansion stops at order $\widetilde{\theta}^8_+$. 
After the change of variables to $g^{\prime}_{\rm{new}}$, the half-BPS vertex operators have more orders in the fermionic expansion, ending at order twelve. 
Thus, the vertices using $g^{\prime}_{\text{new}}$ are given by 
\begin{equation}
V_{\text{new}^{\prime}} = V^0_{\text{new}^{\prime}} + V^2_{\text{new}^{\prime}} +V^4_{\text{new}^{\prime}} 
+V^6_{\text{new}^{\prime}} + V^8_{\text{new}^{\prime}} +V^{10}_{\text{new}^{\prime}} +V^{12}_{\text{new}^{\prime}} \, . 
\label{vertexnew}
\end{equation}
Here, we are going to write explicitly only a few parts of the vertex operators, and the 
complete results are in the ancillary {\it{Mathematica}} notebook. Namely, we are going
to write down the full expression up to order two in the fermions and then write 
the terms with maximum power of $n$ for the higher orders in the fermionic expansion in the Appendix \ref{vertexgprime}.  Notice that these are the only ones surviving the flat space limit. The starting point for the transformation of the variables is the vertices given in the Appendix \ref{DilatonOldCoset}, which are obtained from the ones in 
\cite{VertexOld} by adding $Q$-exact terms. A distinguishing property
of these vertices is that they reduce to the usual dilaton vertex when $n=0$, i.e., they have no dependence on the fermionic variables \cite{BerkovitsSimplifying}. 
At order zero, the transformations of the bosonic variables are trivial. Thus       
\begin{equation}
\label{thedilatonvertex}
V^0_{\text{new}^{\prime}}  = e^{n(W+Z)} V_{\rm{dilaton}} \, , 
\end{equation}
where 
\begin{equation}
V_{\rm{dilaton}} = (\bar{\lambda}_+ \bar{\lambda}_+) + (\lambda_+ \lambda_+) \, ,  
\end{equation}     
and 
\begin{equation}
(\bar\lambda_+ \bar\lambda_+) = \lambda^{\alpha}_{\bar{i}} \lambda^{\beta}_{\bar{j}} \epsilon_{\alpha \beta} \epsilon^{\bar{i} \bar{j}} - \lambda_{\dot{\alpha}}^{i} \lambda_{\dot{\beta}}^{j} \epsilon^{\dot{\alpha} \dot{\beta}} \epsilon_{i j} \, , \quad \quad    
(\lambda_+ \lambda_+) = \lambda^i_{\alpha} \lambda^j_{\beta} \epsilon^{\alpha \beta} \epsilon_{ij} -
\lambda^{\dot\alpha}_{\bar{i}} \lambda^{\dot\beta}_{\bar{j}} \epsilon_{\dot\alpha \dot\beta} \epsilon^{\bar{i} \bar{j}} \, . 
\end{equation}

At order two, the vertices are obtained from the terms $V_D^0$ and $V_D^2$ of the Appendix 
\ref{DilatonOldCoset} by using \eqref{transformatheta} and \eqref{zwtransforme}. At any order, we only need to consider the transformations of the exponential of $V_D^0$ because the pure spinors of $V_{\rm{dilaton}}$ are 
invariant under $SU(2,2) \times SU(4)$ rotations. The result is
\begin{equation}
\begin{aligned}
V^2_{\text{new}^{\prime}}  & = e^{n (W+Z)} n \widehat{\Theta }_{\rho }^j \widehat{\Psi }_{\bar{i}}^{\dot{\alpha }}  (2 \epsilon ^{\alpha  \rho } \epsilon _{i j} \lambda _{\alpha }^{\bar{i}} \lambda _{\dot{\alpha }}^i 
- \epsilon ^{\alpha  \rho } \epsilon _{i j} \lambda _{\alpha }^i \lambda _{\dot{\alpha }}^{\bar{i}}) \\
& + e^{n (W+Z)} n \widehat{\Theta }_{\alpha }^i \widehat{\Psi }_{\bar{j}}^{\dot{\gamma }} (2 \epsilon ^{\bar{i} \bar{j}} \epsilon _{\dot{\alpha } \dot{\gamma }} \lambda _i^{\dot{\alpha }} \lambda _{\bar{i}}^{\alpha }  +  \epsilon ^{\bar{i} \bar{j}} \epsilon _{\dot{\alpha } \dot{\gamma }} \lambda _i^{\alpha } \lambda _{\bar{i}}^{\dot{\alpha }}) \\
& + \frac{1}{2} e^{n (W+Z)} n \widehat{\Theta }_{\beta }^k \widehat{\Theta }_{\gamma }^j (\epsilon ^{\alpha  \gamma } \epsilon _{j k} \lambda _i^{\beta } \lambda _{\alpha }^i - \epsilon ^{\beta  \gamma } \epsilon _{i k} \lambda _j^{\alpha } \lambda _{\alpha }^i) \\
& + e^{n (W+Z)} n \widehat{\Theta }_{\beta }^i \widehat{\Theta }_{\gamma }^j (\epsilon ^{\alpha  \gamma } \epsilon _{i j} \lambda _{\alpha }^{\bar{i}} \lambda _{\bar{i}}^{\beta } 
- \epsilon ^{\beta  \gamma } \epsilon _{k i} \lambda _j^{\dot{\alpha }} \lambda _{\dot{\alpha }}^k) \\
& + \frac{1}{2} e^{n (W+Z)} n \widehat{\Psi }_{\bar{i}}^{\dot{\gamma }} \widehat{\Psi }_{\bar{j}}^{\dot{\beta }} (\epsilon ^{\bar{j} \bar{k}} \epsilon _{\dot{\beta } \dot{\gamma }} \lambda _{\bar{k}}^{\dot{\alpha }} \lambda _{\dot{\alpha }}^{\bar{i}} 
- \epsilon ^{\bar{i} \bar{j}} \epsilon _{\dot{\gamma } \dot{\alpha }} \lambda _{\bar{k}}^{\dot{\alpha }} \lambda _{\dot{\beta }}^{\bar{k}} ) \\
& + e^{n (W+Z)} n \widehat{\Psi }_{\bar{i}}^{\dot{\gamma }} \widehat{\Psi }_{\bar{j}}^{\dot{\beta }}(\epsilon ^{\bar{j} \bar{s}} \epsilon _{\dot{\beta } \dot{\gamma }} \lambda _{\alpha }^{\bar{i}} \lambda _{\bar{s}}^{\alpha } 
- \epsilon ^{\bar{i} \bar{j}} \epsilon _{\dot{\beta } \dot{\alpha }} \lambda _i^{\dot{\alpha }} \lambda _{\dot{\gamma }}^i ) \, . \\
& + e^{n(Z+W)} n \widehat{\Theta }_{\dot\alpha}^{\bar{i}} \widehat{\Psi }_{\bar{i}}^{\dot{\alpha}} 
(\epsilon_{\bar{i} \bar{j}} \epsilon^{\alpha \beta} \lambda^{\bar{i}}_{\alpha} \lambda^{\bar{j}}_{\beta}
- \epsilon_{\dot{\alpha} \dot{\beta}} \epsilon^{i j} \lambda_{i}^{\dot\alpha} \lambda_{j}^{\dot\beta}-\epsilon_{i j} \epsilon^{\alpha \beta} \lambda^{i}_{\alpha} \lambda^{j}_{\beta}+\epsilon_{\bar{i} \bar{j}} \epsilon^{\dot\alpha \dot\beta} \lambda^{\bar{i}}_{\dot\alpha} \lambda^{\bar{j}}_{\dot\beta}) \, ,
\end{aligned} 
\end{equation}
and the remaining orders are in the Appendix \ref{vertexgprime}.

\subsubsection{Finite translated vertices}

In terms of the hatted variables,  
it is straightforward to apply finite translations parametrized by $\eta^{\dot\alpha}_\beta$ and $\eta^k_{\bar{j}}$ 
on the vertex operators by using \eqref{Translations2}, 
\eqref{TransformPsiTheta}
and 
\eqref{TransformLambda}. So, for example, the term with no fermionic variables of any 
of the vertex operators $V^0_{\text{new}^{\prime}}$ transforms as 
\begin{equation}
V^0_{\text{new}^{\prime}}  \,  \rightarrow \,  V^0_{\eta^{\dot\alpha}_\beta, \eta^k_{\bar{j}}}
=  \left(\frac{\Omega_n}{\overline\Omega_n} \right)^n  V^0_{\text{new}^{\prime}} \, , 
\end{equation}
where we have used 
\begin{equation}
\label{TheExp} 
e^{n(W+Z)} \,  \rightarrow \, 
e^{n(W - \log(\overline\Omega_n) + Z+ \log(\Omega_n))} = 
\left(\frac{\Omega_n}{\overline\Omega_n} \right)^n e^{n(W+Z)} \, . 
\end{equation}    

The same procedure can be straightforwardly applied to all
terms of the vertex operators by additionally transforming
both the fermionic variables and the pure spinors.

\subsection{%
The covariant coordinates and the null polarizations vectors} 
\label{recovering}

The vertex operators of \eqref{vertexnew} are written in terms of the variables of 
$g^{\prime}_{\rm{new}}$. It is possible to write them using only $SU(2,2)\times SU(4)$ covariant variables 
by performing further transformations and by using null polarization vectors. 
We start this Subsection by defining both the $SU(2,2)\times SU(4)$ bosonic 
covariant variables and the null polarization vectors that we are going to use.     
The fermionic variables $ \Psi^{R}_{I}$ and $\Theta_{R}^{I}$ of $g_{\rm{new}}$ are already  
both $SU(2,2)\times SU(4)$ covariant and gauge invariant. 
Moreover, 
we can define covariant pure spinors by contracting the pure spinors with the bosonic variables as  
\begin{equation}
\Lambda_I^R = V_I^{\widetilde{I}} \lambda_{\widetilde{I}}^{\widetilde{R}} (U^{-1})_{\widetilde{R}}^R \, , \quad \quad \quad 
\bar\Lambda^I_R = U_R^{\widetilde{R}} \lambda^{\widetilde{I}}_{\widetilde{R}} (V^{-1})_{\widetilde{I}}^I \, .
\label{defLambda}
\end{equation}
The final expression for our vertices will depend on those variables as well.  

\subsubsection{$SU(2,2)\times SU(4)$ covariant bosonic variables}
\label{su22 covariant bosonic variables}

In this Subsection, we are going to define the so-called 
embedding coordinates 
$X_{RS}$ and $Y_{IJ}$. 
Given any element $g\in SU(2,2)\times SU(4)$,
we define a matrix representation of its action on the generators of $\mathfrak{g}_1\oplus\mathfrak{g}_3$ as follows. 
Notice that the group element can be decomposed as $g = u v$, where $u\in SU(2,2)$ and $v\in SU(4)$. In particular for 
\begin{equation}
u = e^{\Sigma^R_S M^S_R} :  \quad \quad 
u\ q_I^R\ u^{-1} =: u^R_S\ q^S_I \, , \quad u\ q_R^I\ u^{-1} =: (u^{-1})^S_R\ q^I_S \, , 
\end{equation}
where
  \begin{equation}
  \begin{aligned}
 (u^{-1})^R_S & = \delta^R_S - \Sigma^R_S + \frac12 \Sigma^R_T \Sigma^T_S + \ldots \, , \quad \quad 
 u^R_S & = \delta^R_S + \Sigma^R_S + \frac12 \Sigma^R_T \Sigma^T_S 
  + \ldots \, . 
   \end{aligned}
  \end{equation}
 with similar expressions for the $v = {\rm{exp}} \, \Sigma^I_J M^J_I$. 
Under right multiplication $g \rightarrow g h$ by another element $h\in SU(2,2)\times SU(4)$, we have the following transformation properties of the coefficients (for simplicity, we assume below 
that $h\in SU(2,2)$ only):
  \begin{equation}
  u \,h \, q^I_R \, h^{-1} \,u^{-1} 
  =  (h^{-1})^S_R \, (u^{-1})^T_S \, q^I_T \, ,  \quad \quad  
 u \, h \, q^R_I \, h^{-1} \, u^{-1} 
 = h^R_S \, u^S_T \, q^T_I \, . 
 \label{transformuv}
  \end{equation}

Given any representation $u^R_S$ and $v^I_J$, define the coefficients
  \begin{equation}
  u^{\widetilde{R}}_{S} := u^R_S\, , \qquad (u^{-1})^{R}_{\widetilde S} := (u^{-1})^R_S\, ,\qquad v^{\widetilde I}_{ J} := v^I_J\, , \qquad (v^{-1})^{I}_{\widetilde J} := (v^{-1})^I_J \, , 
  \end{equation}
where the tilde over the indices denotes the indices that are going to transform under the gauge transformations
of $USp(2,2) \times USp(4)$.
Finally, we use these coefficients to define the gauge-invariant embedding coordinates
  \begin{equation}
  \label{embeddingcoordinates}
  \begin{aligned}
  X_{RS} & = u^{\widetilde R}_R \,  (\sigma_6)_{\widetilde R \widetilde S} \,  u^{\widetilde S}_{S} \,  , \quad \quad \quad \quad  Y_{IJ} = v^{\widetilde I}_I \, (\sigma_6)_{\widetilde I\widetilde J} \, v^{\widetilde J}_J \, ,  \\
  X^{RS} & = (u^{-1})^R_{\widetilde R} (\sigma_6)^{\widetilde R \widetilde S} (u^{-1})_{\widetilde S}^{S} \, ,\quad  Y^{IJ} = (v^{-1})_{\widetilde I}^I (\sigma_{6})^{\widetilde I\widetilde J} (v^{-1})_{\widetilde J}^J \, . 
  \end{aligned}
  \end{equation}

In particular, for a coset representative of the form 
\begin{equation}
g =e^{x^{\alpha}_{\dot\alpha} K^{\dot\alpha}_{\alpha}} e^{z \Delta} \, , 
\nonumber
\end{equation}
one has  
\begin{equation}
[X_{RS}] =
\left(
\begin{tabular}{cc}
$e^z \varepsilon_{\alpha\beta}$ & $e^z \varepsilon_{\alpha\beta} x^\beta_{\dot{\beta}}$ \\
$e^z x^{\gamma}_{\dot{\alpha}} \varepsilon_{\gamma\beta}$ & $-\left(e^z x^2 + e^{-z}\right)\varepsilon_{\dot{\alpha}\dot{\beta}}$
\end{tabular}
\right) \, , 
\label{XRS}
\end{equation}
and for the standard Poincaré patch,   
\begin{equation}
\widehat{g} =e^{\widehat{x}^{\dot\alpha}_{\alpha} P^{\alpha}_{\dot\alpha}} e^{\widehat{z} \Delta} \, , 
\end{equation}
the embedding coordinates are 
\begin{equation}
[\widehat{X}_{RS}] =
\left(
\begin{tabular}{cc}
$\left(e^{\widehat{z}} - e^{- \widehat{z}} \widehat{x}^{2} \right) \varepsilon_{\alpha \beta}$ & $-e^{- \widehat{z}} \widehat{x}^{\dot{\gamma}}_{\alpha} \varepsilon_{\dot{\gamma}\beta}$ \\
$-e^{- \widehat{z}} \widehat{x}^{\gamma}_{\beta} \varepsilon_{\dot{\alpha}\dot{\gamma}}$ & $-e^{-\widehat{z}} \varepsilon_{\dot{\alpha}\dot{\beta}}$
\end{tabular}
\right) \, . 
\label{XprimeRS}
\end{equation}
Given a coset representative for $v$, similar expressions can be obtained for $Y_{IJ}$, see below. Notice that, using the definitions  \eqref{embeddingcoordinates}, the following identities hold 
\begin{equation}
X^{RS} = \frac{1}{2} \epsilon^{RSTP} X_{TP} \, , \quad \quad  Y^{IJ} = \frac{1}{2} \epsilon^{IJKS} Y_{KS} \, . 
\end{equation}

\subsubsection{Null polarization vectors} 
\label{Null polarization vectors}

The vertex operators constructed in \cite{VertexOld} and reviewed in the Subsection \ref{previous paper result} describe the bottom of a half-BPS supermultiplet.
These physical states
are specified by
\begin{itemize}
    \item a point on the $AdS_5$ boundary (equivalently, a projective null vector transforming under $SO(n,m)$);
    \item an embedding $S(U(2)\times U(2))\subset SU(4)$ (equivalently, a projective null vector of $SO(6)$).
\end{itemize}

This is the bosonic truncation of the harmonic boundary superspace of \eqref{boundary superspace}. The full supermultiplet is obtained by retaining the fermionic coordinates of the boundary superspace, as explained in Subsection \ref{superfields for half-BPS supermultiplet}. The bottom-state polarizations form projective null vectors, 
and thus transform under $SU(2,2)\times SU(4)$. 
Let $x_0$ and $y_0$ denote these null two-forms; explicitly,
\begin{equation}
\varepsilon^{RSTU}(x_0)_{RS}(x_0)_{TU}=0\,,\qquad
\varepsilon^{IJKL}(y_{0})_{IJ}(y_0)_{KL}=0\,.
\end{equation}
\begin{figure}[t]
    \centering
     \includegraphics[width=0.3\linewidth]{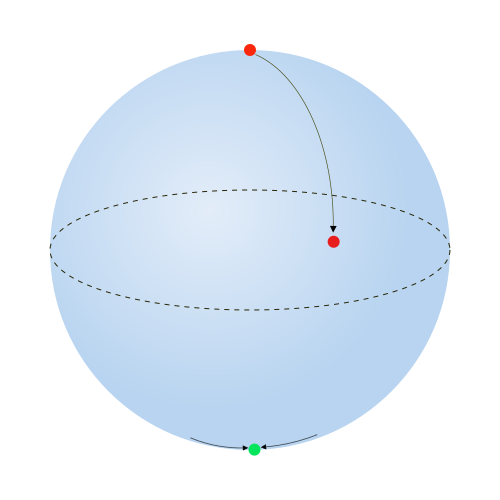} 
    \caption{Depiction of the transformations generated by the special conformal generator $K^{\dot\alpha}_{\alpha}$ on the boundary of an $AdS_3$ slice of $AdS_5$ spacetime. The green point represents the origin of the Poincaré patch, while the red point represents the origin of its stereographic inversion. The green point is a fixed point under transformations generated by $K^{\dot\alpha}_{\alpha}$.} 
    \label{fig:comparison}
\end{figure}
The point at infinity of the Poincaré patch, which correspond to the origin of the parametrization of  \eqref{OldCosetRepresentative}, corresponds to the following projective null vectors
``at infinity",
\begin{equation}
(x_{\infty})_{RS} \propto \begin{bmatrix}
   0&0\\0& \varepsilon_{\dot\alpha\dot\beta}
\end{bmatrix} \,,\qquad    (y_{\infty})_{IJ} \propto \begin{bmatrix}
   - \varepsilon_{i j} &0\\0&0
\end{bmatrix}\,.
\label{polarizationsatinfinity}
\end{equation}
Here, we choose an embedding of $S(U(2)\times U(2))$ into $SU(4)$, determined by the stabilizer of the Levi–Civita symbol $\varepsilon_{\overline i\overline j}$. These are the projetive null vectors associated with the polarization of the half-BPS vertex operators constructed in \cite{VertexOld}. By acting with $SU(2,2)\times SU(4)$ transformations on, $(v_{R}^{\alpha},\overline v_{R}^{\dot\alpha})$ and $(u_{I}^{i},\overline u_{I}^{\overline i})$, we can obtain all polarizations of the half-BPS bottom states, whose corresponding projective null vectors are
\begin{equation}
(x_0)_{RS} \propto v_{R}^{\alpha}\varepsilon_{\alpha\beta}v_{S}^{\beta}\,,\qquad (y_0)_{IJ} \propto  \overline u_{I}^{\overline i}\varepsilon_{\overline i\overline j}\overline u_{J}^{\overline j}\, .
\label{xoyo}
\end{equation}
Let us consider an example of \eqref{xoyo}, namely, the case of translations 
generated by both $\eta^{\dot\alpha}_\beta$ and $\eta^k_{\bar{j}}$. In this case, 
\begin{equation}
u_P = \left(
      \begin{array}{cc}
1 &  \eta^{\dot\alpha}_{\alpha} \\
  0  & 1
      \end{array}
    \right) \, , \quad \quad 
v_P =  \left(
      \begin{array}{cc}
1 & 0\\
   \eta^{i}_{\bar{i}}   & 1
      \end{array}
    \right) \, .      
\end{equation}
The resulting null vectors are 
 \begin{equation}
    (x_0)_{RS} = (u_P)_R^T (u_P)_S^U   (x_0^{\infty})_{TU} = \left(
      \begin{array}{cc}
      - \frac12 (\eta\cdot\eta) \varepsilon_{\alpha\beta} &  \eta^{\dot\gamma}_{\beta} \varepsilon_{\dot\gamma\dot\alpha} \\
     \varepsilon_{\dot\beta\dot\gamma}  \eta^{\dot\gamma}_\alpha & \varepsilon_{\dot\alpha\dot\beta}
      \end{array}
    \right) \, ,
\end{equation} 
and    
\begin{equation}     
    \qquad (y_0)_{IJ} = (v_P)_I^K (v_P)_J^L (y_0^{\infty})_{KL} =  \left(
      \begin{array}{cc}
      -\varepsilon_{ij} & -  \varepsilon_{jk} \eta^k_{\overline i}\\
    - \eta^k_{\overline j}  \varepsilon_{ki}   & \frac12 (\eta\cdot\eta) \varepsilon_{\overline i \overline j}
      \end{array}
    \right) \, .  
    \end{equation} 

In the dual gauge theory, the projective null vector $(x^D_0)_{RS}$ gives rise to the spacetime coordinates $x_{\dot\alpha}^{\alpha} = x^{\mu}\sigma_{\mu\alpha}^{\dot\alpha}$ as follows
\begin{equation}
(x^D_0)_{RS} =  \left(
\begin{tabular}{cc}
$\varepsilon_{\alpha\beta}$ & $\varepsilon_{\alpha\beta} x^\beta_{\dot{\beta}}$ \\
$x^{\gamma}_{\dot{\alpha}} \varepsilon_{\gamma\beta}$ & $- x^2\varepsilon_{\dot{\alpha}\dot{\beta}}$
\end{tabular}
\right) \,,
\end{equation}
the projective null vector $(y^D_0)_{IJ}$ selects an analytic scalar $Z$, which enters the construction of the single-trace operator:
\begin{equation}
O_{J}(x)=\text{Tr}Z^{J}(x)\,,\qquad Z(x)  = (y^D_0)^{IJ} \Phi_{IJ}(x) \equiv \varepsilon^{\overline i\overline j}\overline u^{I}_{\overline i}\overline u^{J}_{\overline i}\Phi_{IJ}(x)  \,,  
\end{equation}
where $\Phi_{IJ}(x)=-\Phi_{JI}(x)$ are the six scalars fields of $\mathcal{N}=4$ SYM, evaluated at the point $x$ on the $AdS_5$ boundary, and $(u^{I}_{i},\overline {u}^{J}_{\overline j})$ are the $\frac{SU(4)}{S(U(2)\times U(2))}$ harmonic variables parametrizing null $SO(6)$ vectors. The $SO(6)$ null vectors are constructed by contracting these harmonic variables with the Levi-Civita symbol $\varepsilon_{\overline i\overline j}$. Their complex conjugates are obtained instead by contracting with $\varepsilon_{i j}$. 

\subsection{Vertex operators in terms of null polarization vectors}
\label{su22 covariant vertex operators}

The half-BPS vertex operators have so far been written using $\mathfrak{su}_2^{\oplus 4}$-component notation. 
They describe vertex operators located at infinity,
characterized by the two null polarization vectors $(x_\infty, y_\infty)$ introduced in \eqref{polarizationsatinfinity}. 
To move these polarization vectors to arbitrary ones,
one can consider 
$SU(2,2)\times SU(4)$ transformations. 
This can be achieved by following the procedure described below 

\paragraph{$\bullet$} 
Consider any term of the vertex operators given in the previous Subsection.
It is written in $\mathfrak{su}_2^{\oplus 4}$-component  notation and depends on the 
$\lambda$'s, $\widehat{\Psi}$ and $\widehat{\Theta}$, as well as the tensors $\varepsilon^{\alpha\beta},\ \varepsilon^{\dot\alpha\dot\beta},\ \varepsilon^{ij},\ \varepsilon^{\overline i\overline j}$. 
It is possible to rewrite it, still using $\mathfrak{su}_4^{\oplus 2}$-component notation, in a convenient way in terms of the fixed null polarizations vectors $(x_\infty, y_\infty)$ and the symplectic forms $\Omega_{R S} = (\sigma_6)_{R S}$, $\Omega_{I J} = (\sigma_6)_{I J }$, where the matrix $\sigma_6$ is given in \eqref{sigma6}.
 For example, we can project out any of the components of $\lambda^{R}_{I}$ as 
  \begin{equation}\label{projected}
  \begin{aligned}
&  \left[ (x_\infty)^{R R_1} \Omega_{R_1 S}\ \lambda^{S}_{J}\  \Omega^{J J_1} (y_\infty)_{J_1 I} \right] = \left(
    \begin{array}{cc}
    \lambda^{\alpha}_i & 0 \\
    0 & 0 \\
    \end{array}
  \right)  \, , \\
& \left[ \Omega^{R R_1} (x_\infty)_{R_1 S}\ \lambda^{S}_{J}\ \Omega^{J J_1} (y_\infty)_{J_1  I} \right] = \left(
    \begin{array}{cc}
    0 & 0 \\
    \lambda^{\dot\alpha}_i & 0 \\
    \end{array}
  \right)\ , \\
&  \left[ (x_\infty)^{R R_1} \Omega_{R_1 S}\ \lambda^{S}_{ J}\ (y_\infty)^{J J_1} \Omega_{ J_1 I} \right] = \left(
    \begin{array}{cc}
    0 & \lambda^{\alpha}_{\overline i} \\
    0 & 0 \\
    \end{array}
  \right) \ ,\quad \\
&  \left[ \Omega^{R R_1} (x_\infty)_{R_1 S}\ \lambda^{S}_{ J}\ (y_\infty)^{J J_1} \Omega_{J_1 I} \right] = \left(
    \begin{array}{cc}
    0 & 0 \\
    0 & \lambda^{\dot\alpha}_{\overline i} \\
    \end{array}
  \right) \, .
  \end{aligned}
  \end{equation}
To better illustrate the rewritten of the terms, consider the following example
  \begin{equation}
  \lambda^{\alpha}_i\ \varepsilon^{ij}\ \widehat\Psi_j^\beta\ \varepsilon_{\beta\alpha} = (x_\infty\Omega\lambda\Omega y_\infty)^{R}_{I}\ (-\Omega^{I  J})\ (x_\infty\Omega\widehat\Psi\Omega y_\infty)^{S}_{J}\ \Omega_{S R} \, , 
 \label{exampleinfinity}
  \end{equation}
where we have used the convention that the variables are contracted while preserving the order inside the parentheses, so that
\begin{equation}
(x_\infty\Omega\widehat\Psi\Omega y_\infty)^{S}_{J} = 
(x_\infty)^{S S_1} \Omega_{S_1 S_2} \widehat\Psi^{S_2}_{I_1} \Omega^{I_1 I_2} (y_\infty)_{I_2 J} \, . 
\end{equation}

\paragraph{$\bullet$} Consider any term written as on the right-hand side of 
\eqref{exampleinfinity}. It is possible to write it in terms 
of the fermionic variables $\Psi$ and $\Theta$  of $g_{\rm{new}}$ and the gauge-invariant bosonic variables $\Lambda$, $\bar{\Lambda}$, $X_{RS}$, and 
$Y_{IJ}$ defined in \eqref{defLambda} and \eqref{embeddingcoordinates}, respectively.  
Recall that these variables are written in terms of the variables of $g^{\prime}_{\rm{new}}$ by contracting with appropriate 
factors of $U$ and $V$ and their inverses. Using the following properties of the polarizations      
\begin{equation}
    \begin{aligned}
    (x_\infty)_{R S} = \frac{1}{e^{Z}} U_{R}^{\widetilde{T}}\ (x_\infty)_{\widetilde{T}\widetilde{U}} U_{S}^{\widetilde{U}}\quad & ,\qquad  (y_\infty)_{I J} = \frac{1}{e^{-W}} V_{I}^{\widetilde{K}}\ (y_\infty)_{\widetilde{K} \widetilde{L}} V_{J}^{\widetilde{L}}  \, , \\
    (x_\infty)^{R S} = \frac{1}{e^{Z}} (U^{-1})^{R}_{\widetilde{T}}\ (x_\infty)^{\widetilde{T} \widetilde{U}}\ (U^{-1})^{S}_{\widetilde{U}}\quad & ,\qquad  (y_\infty)^{I J} = \frac{1}{e^{-W}} (V^{-1})^{I}_{\widetilde{K}}\ (y_\infty)^{\widetilde{K} \widetilde{L}}\ (V^{-1})^{ J}_{\widetilde{L}}  \, , \\
    e^{Z} = -\frac{1}{2} (x_\infty)_{R S} X^{S R} \quad & , \qquad e^{-W} = - \frac{1}{2} (y_\infty)_{I J} Y^{J I} \, , 
 \label{propertiespolarization}   
    \end{aligned}
    \end{equation}
all dependence  on the $U$'s and $V$'s disappears because the original terms have 
all indices contracted. Thus, for \eqref{exampleinfinity} one has the  
expression,
 \begin{equation}
  \lambda^{\alpha}_i\ \varepsilon^{ij}\ \widehat\Psi_j^\beta\ \varepsilon_{\beta\alpha} = - \frac{(x_\infty X\Lambda Y y_\infty)^{R}_{I}}{4(x_\infty\cdot X)(y_\infty\cdot Y)}\ Y^{IJ}\ \frac{(x_\infty X\Psi Y y_\infty)^{S}_{J}}{4(x_\infty\cdot X)(y_\infty\cdot Y)}\ X_{SR} \, . 
  \end{equation}
 Using the two steps described above, we have succeeded in changing the coordinates 
from $g'_{\text{new}}$ of  \eqref{gprime} to $g_{\text{new}}$ of \eqref{CosetIdeal}. 
Notice that all the bosonic and fermionic variables above 
have simple transformations under
$SU(2,2) \times SU(4)$. 
To see the 
transformation 
of the bosonic variables, notice that under a global $SU(2,2) \times SU(4)$ given by $\mathcal R = \exp(\eta^R_S t^S_R + \eta^I_J t^J_I)$, the $U$- and $V$-coefficients transform in a way that satisfies  \eqref{transformuv}
  $$
  \mathcal R^R_S\ U^S_{\widetilde T} = U'{}^R_{\widetilde S}\ h^{\widetilde S}_{\widetilde T} \quad ,\qquad \mathcal R^I_J\ V^J_{\widetilde K} = V'{}^I_{\widetilde J}\ h^{\widetilde J}_{\widetilde K}  \quad , 
  $$
which guarantees the cancellation of all compensating transformations $h$, leaving only $\mathcal R^R_S$ and $\mathcal R^I_J$ acting on the null vectors 
  \begin{equation}
  (x_0)_{RS} := \mathcal R_R^T (x_\infty)_{TU} \mathcal R^U_S \quad , \qquad (y_0)_{IJ} := \mathcal R_I^K (y_\infty)_{KL} \mathcal R^L_J \, . 
  \end{equation}
In conclusion, using the method described above, we can apply a finite translation to move the vertex operators away from infinity, i.e., from the polarization $(x_{\infty}, y_{\infty})$  to any null polarization  $(x_0, y_0)$, schematically as
 \begin{equation}
  (\mathcal R \cdot V)(\Psi, \Theta, X, Y, \Lambda,\overline\Lambda \,|\, x_{\infty}, y_{\infty}) = V(\Psi,\Theta, X,Y, \Lambda,\overline\Lambda\,|\,\ \mathcal R\cdot x_{\infty} = x_0,\ \mathcal R \cdot y_{\infty} = y_0) \, . 
  \end{equation}

\paragraph{Dilaton vertex operator}
The dilaton vertex operator 
is obtained, up to pure spinor constraints, as the $n \rightarrow 0$  limit of our vertices. 
It is given in $\mathfrak{su}_2^{\oplus 4}$ notation by 
  \begin{equation}
  V_{\rm dil} = \varepsilon_{ij}\varepsilon^{\alpha\beta} \lambda{}^i_\alpha \lambda{}^j_\beta - \varepsilon_{\overline i\overline j}\varepsilon^{\alpha\beta} \lambda{}^{\overline i}_\alpha \lambda{}^{\overline j}_\beta - \varepsilon_{\dot\alpha\dot\beta} \varepsilon^{\overline i\overline j} \lambda^{\dot\alpha}_{\overline i}\lambda^{\dot\beta}_{\overline j} + \varepsilon_{\dot\alpha\dot\beta} \varepsilon^{ij} \lambda^{\dot\alpha}_i \lambda^{\dot\beta}_j \, .
  \end{equation}
Following the method described above to recover the null polarization vectors, 
this vertex operator can be written as
  \begin{equation}
  \begin{aligned}
  V_{\rm dil} & = \frac{1}{4(x_0\cdot X)(y_0\cdot Y)}\Big[(\overline\Lambda x_0 \overline\Lambda y_0) + (\overline\Lambda x_0 \overline\Lambda Y y_0 Y) - (\Lambda y_0 \Lambda x_0) - (\Lambda Y y_0 Y \Lambda x_0) \Big] \\
  & = \frac{1}{2(x_0\cdot X)}\Big[(\overline\Lambda x_0 \overline\Lambda Y) - (\Lambda Y \Lambda x_0) \Big] \, .
  \label{Vdilcase1} 
  \end{aligned}
  \end{equation}
Furthermore, if we were to impose the pure spinor constraints on $\Lambda$ and $\overline\Lambda$, we would arrive at
  \begin{equation}
  V_{\rm dil} = \frac12\left[(\overline\Lambda X \overline \Lambda\ Y) - (\Lambda Y \Lambda X)\right] \,. 
 \label{dilatonvertex}
  \end{equation}
The expression above is independent of the null polarizations vectors $(x_0, y_0)$ and it is manifestly $SU(2,2) \times SU(4)$-invariant.

\paragraph{Half-BPS vertex operator %
parametrized by null polarization vectors}
Here we write only the  $V_{2,n}$ part of the vertex operators. The full expression is provided in the accompanying  {\it{Mathematica}} file, 

\begin{equation}
V_{0,n}(x_0,y_0) = \left(\frac{x_0\cdot X}{y_0\cdot Y}\right)^n V_{\rm dil}  \, , 
\label{dilatoncovariant}
\end{equation}
where $V_{\rm dil}$ is given in  \eqref{dilatonvertex} and 

\begin{equation}
  \begin{aligned}
  V_{2,n}(x_0,y_0) = n \left(\frac{x_0\cdot X}{y_0\cdot Y}\right)^n & \Big[ -V_{\rm dil}  (\Theta X \dot{x}_0 \Psi \dot{y}_0 Y) \\
  & \quad - 2 (\overline\Lambda \dot{x}_0\Theta \dot{y}_0 \overline\Lambda X \dot{x}_0 \Psi \dot{y}_0 Y) - (\overline\Lambda X \dot{x}_0 \Psi \dot{y}_0 Y) (\overline\Lambda \dot{x}_0 \Theta \dot{y}_0) \\
  & \quad + 2 (\Lambda Y \dot{y}_0 \Theta \dot{x}_0 X \Lambda \dot{y}_0 \Psi \dot{x}_0) + (\Lambda \dot{y}_0 \Psi \dot{x}_0) (\Lambda Y \dot{y}_0 \Theta \dot{x}_0 X) \\
  & \quad - \frac12 (\Lambda Y \dot{y}_0 \overline\Lambda \dot{x}_0 \Theta \dot{y}_0 \Theta \dot{x}_0 X) + \frac12 (\Lambda Y \dot{y}_0 \Theta \dot{x}_0 \Theta \dot{y}_0 \overline\Lambda \dot{x}_0 X) \\
  & \quad - (\Lambda \dot{y}_0 Y \overline\Lambda \dot{x}_0 \Theta \dot{y}_0 \Theta \dot{x}_0 X) + (\Lambda Y \dot{y}_0 \Theta \dot{x}_0 \Theta \dot{y}_0 \overline\Lambda X \dot{x}_0) \\
  & \quad -\frac12 (\Lambda \dot{y}_0 \Psi \dot{x}_0 \Psi \dot{y}_0 Y \overline\Lambda X \dot{x}_0) + \frac12 (\Lambda \dot{y}_0 Y \overline\Lambda X \dot{x}_0 \Psi \dot{y}_0  \Psi \dot{x}_0) \\
  & \quad - (\Lambda \dot{y}_0 \Psi \dot{x}_0 \Psi \dot{y}_0 Y \overline\Lambda \dot{x}_0 X) + (\Lambda Y \dot{y}_0 \overline\Lambda X \dot{x}_0 \Psi \dot{y}_0 \Psi \dot{x}_0)\Big] \, ,
  \end{aligned}
  \label{V2old} 
  \end{equation}
where $\dot{x}_0$ and $\dot{y}_0$ are normalized polarizations, defined as 
\begin{equation}
(\dot{x}_0)_{RS} = \frac{(x_0)_{RS}}{2(x_0 \cdot X)} \, , \quad \quad  
(\dot{y}_0)_{IJ} = \frac{(y_0)_{IJ}}{2(y_0 \cdot Y)} \, . 
\end{equation}

\paragraph{Another representative for the vertex operators}
The $V_{2,n}(x_0,y_0)$ was obtained by following the procedure described above for 
writing the vertex operators in terms of the null polarization vectors.   
Depending on the vertex representative, the final expression can contain spurious poles
in $y_0$. For example, $V_{2,n}(x_0,y_0)$ of \eqref{V2old} has a pole at $n=-1$.   
Recall that the vertex operator 
can be modified by adding $Q$-exact terms. 
We will show that $V_{2,n}(x_0,y_0)$ can be made polynomial in $y_0$ for %
any integer value of $L = -n > 0$ by explicitly constructing the relevant 
$Q$-exact terms.
We divide the calculation into two steps. 
To facilitate the computation, consider a reordering of $V_{2,n}(x_0,y_0)$ in which only the first line has the desired property
  \begin{equation}\label{v2}
  \begin{aligned}
 V_{2,n}(x_0,y_0) & = n  \left(\frac{x_0\cdot X}{y_0\cdot Y}\right)^n \Big[ -V_{\rm dil} 
 (\Theta X \dot{x}_0 \Psi \dot{y}_0 Y) \\
   & - 2 (\overline\Lambda \dot{x}_0\Theta \dot{y}_0 \overline\Lambda X \dot{x}_0 \Psi \dot{y}_0 Y) - (\overline\Lambda X \dot{x}_0 \Psi \dot{y}_0 Y) (\overline\Lambda \dot{x}_0 \Theta \dot{y}_0) + (\Lambda Y \dot{y}_0 \Theta \dot{x}_0 \Theta \dot{y}_0 \overline\Lambda X \dot{x}_0)\\
  & + 2 (\Lambda Y \dot{y}_0 \Theta \dot{x}_0 X \Lambda \dot{y}_0 \Psi \dot{x}_0) + (\Lambda \dot{y}_0 \Psi \dot{x}_0) (\Lambda Y \dot{y}_0 \Theta \dot{x}_0 X) - (\Lambda \dot{y}_0 \Psi \dot{x}_0 \Psi \dot{y}_0 Y \overline\Lambda \dot{x}_0 X) \\
  & - \frac12 (\Lambda Y \dot{y}_0 \overline\Lambda \dot{x}_0 \Theta \dot{y}_0 \Theta \dot{x}_0 X) + \frac12 (\Lambda Y \dot{y}_0 \Theta \dot{x}_0 \Theta \dot{y}_0 \overline\Lambda \dot{x}_0 X) - (\Lambda \dot{y}_0 Y \overline\Lambda \dot{x}_0 \Theta \dot{y}_0 \Theta \dot{x}_0 X) \\
  &  -\frac12 (\Lambda \dot{y}_0 \Psi \dot{x}_0 \Psi \dot{y}_0 Y \overline\Lambda X \dot{x}_0) + \frac12 (\Lambda \dot{y}_0 Y \overline\Lambda X \dot{x}_0 \Psi \dot{y}_0  \Psi \dot{x}_0) + (\Lambda Y \dot{y}_0 \overline\Lambda X \dot{x}_0 \Psi \dot{y}_0 \Psi \dot{x}_0)\Big] \, . 
  \end{aligned}
  \end{equation}
It can be checked that the second-order poles of $(y_0\cdot Y)$ in the last four lines of (\ref{v2}) can be removed, respectively, by each line of the following $Q$-exact expression
  \begin{equation}
  \begin{aligned}
\Omega_1=  Q \Bigg\{\ n \left(\frac{x_0\cdot X}{y_0\cdot Y}\right)^n & \Big[ (\Theta \dot{x}_0 \Theta \dot{y}_0 \overline\Lambda X \dot{x}_0\Psi \dot{y}_0 Y) \\
  &  + (\Lambda \dot{y}_0 \Psi \dot{x}_0 \Psi    \dot{y}_0 Y \Theta \dot{x}_0 X) \\
  &  -\frac12 (\Lambda Y \dot{y}_0 \Theta \dot{x}_0\Theta \dot{y}_0\Theta \dot{x}_0 X) -\frac12 ( \Psi Y \dot{y}_0 \Theta \dot{x}_0 \overline\Lambda \dot{y}_0 \Theta \dot{x}_0 X) \\
  & + \frac12 (\Psi \dot{y}_0 \Psi \dot{x}_0 \Psi \dot{y}_0 Y \overline\Lambda X \dot{x}_0) + \frac12 ( \Psi \dot{y}_0 \Lambda \dot{x}_0 \Psi \dot{y}_0 Y \Theta X \dot{x}_0) \Big] \Bigg\} \, . 
  \end{aligned}
  \end{equation}
The resulting expression without poles is 
  \begin{equation}
  \begin{aligned}
  & V^{\Omega_1}_{2,n}(x_0,y_0) =  V_{2,n}(x_0,y_0) + \Omega_1 \\
 &  = 
  n \left(\frac{x_0\cdot X}{y_0\cdot Y}\right)^n  \Big[ -V_{\rm dil} 
  (\Theta X \dot{x}_0 \Psi \dot{y}_0 Y) \\
&   + (\Theta \dot{x}_0\Theta \dot{y}_0\overline\Lambda X \dot{x}_0\Lambda) - (\overline\Lambda X \dot{x}_0\Psi)(\overline\Lambda \dot{x}_0\Theta \dot{y}_0) - (\overline\Lambda \dot{x}_0\Theta \dot{y}_0\overline\Lambda X \dot{x}_0\Psi) - (\Theta \dot{x}_0\overline\Lambda \dot{y}_0\overline\Lambda X \dot{x}_0\Psi) \\
 & + 2 (\Lambda\Theta x_0 X\Lambda y_0\Psi x_0) + (\Lambda y_0 \Psi x_0)(\Lambda\Theta x_0 X)
 - (\Lambda \overline\Lambda \dot{x}_0 \Theta \dot{y}_0 \Theta \dot{x}_0 X) 
+ (\Lambda \overline\Lambda X \dot{x}_0 \Psi \dot{y}_0 \Psi \dot{x}_0)\Big] \, . 
  \end{aligned}
  \end{equation}
It is possible to make the expression for $ V_{2,n}(x_0,y_0)$ even more compact by adding further additional $Q$-exact terms 
\begin{equation}
\begin{aligned}
\Omega_2 = Q \Bigg\{ n \left(\frac{x_0\cdot X}{y_0\cdot Y}\right)^n \Big[(\Psi \bar{\Lambda} x_0 \Theta y_0 \Theta x_0 X) - (\Theta x_0 \Theta y_0 \bar{\Lambda} X x_0 \Psi) - (\Lambda \Theta X x_0 \Psi y_0 \Psi x_0) 
\Big] \Bigg\} \, , 
\end{aligned}
\end{equation}
and, the final expression is 
 \begin{equation}
\begin{aligned}
&V^{\Omega_{1,2}}_{2,n}(x_0,y_0) =  V^{\Omega_1}_{2,n}(x_0,y_0)  + \Omega_2 \\
& =  n \left(\frac{x_0\cdot X}{y_0\cdot Y}\right)^n   \Big[ 2(\Lambda \Theta \Lambda \dot{y}_0 \Psi \dot{x}_0) -2 (\bar{\Lambda} \dot{x}_0 \Theta \dot{y}_0 \bar{\Lambda} \Psi) - (\bar{\Lambda} \Psi) (\Theta \dot{x}_0 \bar{\Lambda} \dot{y}_0) - (\Theta \Lambda)(\Lambda \dot{y}_0 \Psi \dot{x}_0)\\
&  
 + \frac{1}{2} (\Lambda Y \Lambda X) (\Theta X \dot{x}_0 \Psi \dot{y}_0 Y) - \frac{1}{2} (\bar{\Lambda} X \bar{\Lambda} Y) 
(\Theta X \dot{x}_0 \Psi \dot{y}_0 Y) \Big] \, . 
\label{TheV2covariant} 
\end{aligned} 
\end{equation}  

We end this part by rewriting the normalized null polarization vectors appearing on the vertex operators above in way that it is easier 
to compare to the $AdS$ part 
in \cite{Adamo:2016rtr, Bhat:2021dez} after an appropriated gauge-fixing.  
The formulas below uses $Y_{IJ}$ defined in \eqref{embeddingcoordinates}, whose matrix form is 
\begin{equation}
Y_{IJ} = \left(
\begin{tabular}{cc}
$(e^w + e^{-w} y^2) \varepsilon_{i j} $ & $ - e^{-w} y^{\bar{i}}_{i} \varepsilon_{\bar{i} \bar{j}}$ \\
$ - e^{-w} \varepsilon_{\bar{i} \bar{j}} y^{\bar{j}}_{i} $ & $  - e^{-w} \varepsilon_{\bar{i}\bar{j}}$
\end{tabular}
\right) \, .
\end{equation}
Using \eqref{propertiespolarization} and \eqref{XRS}, one has
\begin{equation}
\begin{aligned}
[\dot{x}_\infty]_{RS} &= \begin{bmatrix}
0&0\\0& e^{-z} \varepsilon_{\dot\alpha\dot\beta}
\end{bmatrix} \\
& = -\left(
\begin{tabular}{cc}
$e^z \varepsilon_{\alpha\beta}$ & $e^z \varepsilon_{\alpha\beta} x^\beta_{\dot{\beta}}$ \\
$e^z x^{\gamma}_{\dot{\alpha}} \varepsilon_{\gamma\beta}$ & $-\left(e^z x^2 + e^{-z}\right)\varepsilon_{\dot{\alpha}\dot{\beta}}$
\end{tabular}
\right)+ \frac{1}{2} (x_\infty \cdot X) \left(
\begin{tabular}{cc}
$\varepsilon_{\alpha\beta}$ & $\varepsilon_{\alpha\beta} x^\beta_{\dot{\beta}}$ \\
$x^{\gamma}_{\dot{\alpha}} \varepsilon_{\gamma\beta}$ & $- x^2\varepsilon_{\dot{\alpha}\dot{\beta}}$
\end{tabular}
\right) \, , 
\end{aligned} 
\label{xinfinity}
\end{equation}
and 
\begin{equation}
\begin{aligned}
[\dot{y}_\infty]_{IJ} &= \begin{bmatrix}
- e^{w} \varepsilon_{i j} &0\\0& 0
\end{bmatrix} \\
& = -\left(
\begin{tabular}{cc}
$(e^w + e^{-w} y^2) \varepsilon_{i j} $ & $ - e^{-w} y^{\bar{i}}_{i} \varepsilon_{\bar{i} \bar{j}}$ \\
$ - e^{-w} \varepsilon_{\bar{i} \bar{j}} y^{\bar{j}}_{i} $ & $  - e^{-w} \varepsilon_{\bar{i}\bar{j}}$
\end{tabular}
\right)+ \frac{1}{2} (y_\infty \cdot Y) \left(
\begin{tabular}{cc}
$ y^2 \varepsilon_{i j} $ & $ - y^{\bar{i}}_{i} \varepsilon_{\bar{i} \bar{j}}$ \\
$ - \varepsilon_{\bar{i} \bar{j}} y^{\bar{j}}_{i} $ & $  - \varepsilon_{\bar{i}\bar{j}}$
\end{tabular}
\right) \, , 
\end{aligned} 
\end{equation}
where the first matrix in the right-hand side above is $[Y_{IJ}]$ and 
$y^2 = - \frac{1}{2} y^{\bar{i}}_i y^{\bar{j}}_j \varepsilon_{\bar{i} \bar{j}} \varepsilon^{i j}$. 
It follows from \eqref{xinfinity} that the normalized null polarization vector is playing the role of the 
{\it{infinity twistor}} of  \cite{Adamo:2016rtr, Bhat:2021dez}.  

\subsubsection{Covariant BRST operator}   
\label{purespinordecomposition}

We have been working with a supercoset that is locally parametrized by ten bosonic coordinates and thirty-two fermionic ones. Alternatively, the action of the BRST operator on the vertex operators parametrized by the null polarization vectors   
can be derived using a purely 
fermionic coset given by
\begin{equation}
\frac{PSU(2,2|4)}{SU(2,2) \times SU(4)} \, , 
\label{covariantsupercoset}
\end{equation}
together with embedding coordinates parametrizing both $AdS_5$ and $S^5$ as,  
\begin{equation}
\epsilon^{RSTP} X_{RS} X_{TP} = 
\epsilon^{IJKL} Y_{IJ} Y_{KL} = -8 \, , 
\end{equation} 
and the gauge-invariant pure spinors $\Lambda$ and $\bar{\Lambda}$. 
This fermionic coset already appeared in 
\cite{Berkovits:2016xnb} in the context of twistor strings in $AdS_5 \times S^5$.  
Another example where a coset with a bigger gauge symmetry group was used together with harmonic variables is \cite{BerkovitsFleury}. 
The coset \eqref{covariantsupercoset} can be parametrized as   
\begin{equation}
g_{\text{fermionic}} = e^{\Psi} e^{\Theta}  \, .  
\label{fermioniccoset}
\end{equation} 

The BRST transformations of the fermionic variables of the coset representative \eqref{fermioniccoset} 
are as in $g_{\rm{new}}$ and given in \eqref{BRSTPsi} and 
\eqref{BRSTTheta}. The compensator $\Sigma$ 
produced 
by the action of the BRST operator on the coset representative    
follows from $m_Q$ of  \eqref{mQ}, reproduced below,
\begin{equation}
m_Q = [ \Theta, \Lambda] =[ \Theta^I_R q^R_I, \Lambda^S_J q^J_S ] =  \Theta^I_R \, \Lambda^S_J \, (\delta^R_S M^J_I + \delta^J_I M^R_S) \, , 
\end{equation}
and this implies 
\begin{equation}
\Sigma_{\rm{su}(2,2)} = \Theta^I_R \Lambda^S_I (M_{\rm{su}(2,2)})^R_S \, , \quad \quad  \Sigma_{\rm{su}(4)} = \Theta^I_R \, \Lambda^R_J (M_{\rm{su}(4)})^J_I \, . 
\end{equation}

Thus the BRST transformations of the remaining variables follow from the expressions for $\Sigma$ above and are given, for example, by 
\begin{equation}
Q_{\text{fermionic}}  \cdot X_{R S} = \Theta^I_R \Lambda^T_I X_{T S} + 
\Theta^I_S \Lambda^T_I X_{R T} - \frac{1}{2} (\Theta^I_T \Lambda^T_I) X_{R S} \, .  
\end{equation}

The BRST operator can be written compactly as  
\begin{equation}
\begin{aligned}
Q_{\text{fermionic}}  ={}& \Lambda^{R}_{I}\frac{\partial}{\partial \Psi^{R}_{I}} + \left(\overline{\Lambda}^{I}_{R} - (\Theta\Lambda\Theta)^{I}_{R}\right)\frac{\partial}{\partial \Theta^{I}_{R}} + \left((\overline{\Lambda}\Lambda\Theta)^{I}_{R} - (\Theta\Lambda\overline{\Lambda})^{I}_{R}\right)\frac{\partial}{\partial \overline{\Lambda}^{I}_{R}} \\
&+ F^{T}_{R}X_{TS}\frac{\partial}{\partial X_{RS}} + F^{K}_{I}Y_{KJ}\frac{\partial}{\partial Y_{IJ}} \, , 
\end{aligned}
\end{equation} 
where 
\begin{equation}
\quad F := [\Theta,\Lambda] \, . 
\end{equation}
 
\subsubsection{Covariant pure spinors} 
\label{Covariant pure spinors}
 
The vertex operators depend on the spinor variables $\Lambda$ and $\bar{\Lambda}$ defined in \eqref{defLambda}. 
Here, we are going only to initiate the study of the geometry of the space generated by these two spinors and their  invertible combinations. A complete analysis including the worldsheet action will appear elsewhere.   
The spinors $\Lambda$ and $\bar{\Lambda}$    
are both $SU(2,2) \times SU(4)$-covariant and gauge-invariant. 
We will argue below that these variables are constrained.  
Their properties were first analysed in Appendix B 
of \cite{VertexOld}. In that Appendix, it was demostrated that it is possible to write some components of
$\Lambda$ and $\bar{\Lambda}$ in terms of the remaining components and the bosonic variables $U$ and $V$,   
using an $SO(8)$ notation. 
The derivation is as follows. 
First, let us consider the usual non-gauge-invariant pure spinors with the notation of \eqref{lambdasso8} and solve the pure spinor constraints using an $SO(8)$ notation. 
The constraints imply \cite{Martins:2019wwi},  
\begin{equation}
  \lambda_{-}^{a}
  = \frac{(\bar{\lambda}_{+}\,\sigma^{mn}\,\bar{\lambda}_{-})}
         {4(\bar{\lambda}_{+} \bar{\lambda}_{+})}
    \,(\sigma^{mn}\,\lambda_{+})^{a} \, . 
    \label{lambdaminus}
\end{equation}
In other words, the $\lambda_{-}^{a}$ variables are written in terms of the remaining $\lambda$ variables. 
There remain further constraints on $(\bar{\lambda}^a_+,\bar{\lambda}^{\dot{a}}_-)$ after solving 
for $\lambda_{-}^{a}$ as above, but the $\lambda_+^a$ can be taken as free variables.
The remaining constraints are 
\begin{equation}
\bar{\lambda}^{\dot{a}}_+ \bar{\lambda}^{\dot{a}}_+ +  
\bar{\lambda}^{\dot{a}}_- \bar{\lambda}^{\dot{a}}_-=0 \, , \quad \quad
\bar{\lambda}^{\dot{a}}_+ \bar{\lambda}^{\dot{a}}_- =0 \, . 
\end{equation}
These equations are solved by the variables    
\begin{equation} 
\bar{\lambda}^{\dot{a}}_+ = \rho \bar{e}^{\dot{a}}_+ \, , \quad \quad
\bar{\lambda}^{\dot{a}}_- = i \rho \bar{e}^{\dot{a}}_- \, , \quad \quad 
\rho^2 = \bar{\lambda}^{\dot{a}}_+ \bar{\lambda}^{\dot{a}}_+ \, ,
\label{rhoangles}
\end{equation}
with
\begin{equation}
\bar{e}^{\dot{a}}_{\pm} \bar{e}^{\dot{a}}_{\pm} = 1 \, , \quad \quad 
\bar{e}^{\dot{a}}_{\pm} \bar{e}^{\dot{a}}_{\mp} = 0 \, . 
\end{equation}
The unit vectors $\bar{e}^{\dot{a}}_{\pm}$ are parametrized by fourteen angles, but the constraints eliminate one of them, leaving thirteen angles plus $\rho$. In conclusion, a full solution of the pure spinor constraints in $SO(8)$ notation is given by \eqref{lambdaminus} and \eqref{rhoangles}.
Consider the combination
\begin{equation}
\Lambda q = \Lambda^R_I q_R^I + \bar{\Lambda}^I_R q^R_I 
= (U V) \, \lambda q \, 
(U V)^{-1} \, . 
\label{changeofvariables} 
\end{equation}
By evaluating the last equation above using the $U$ and $V$ of \eqref{CosetIdeal}, one verifies 
that $\Lambda_{\alpha}^i$ and $\bar\Lambda_{\bar{i}}^{\dot\alpha}$ are proportional to $\lambda_{\alpha}^i$ and $\lambda_{\bar{i}}^{\dot\alpha}$. Thus they can be taken as free variables.
Then equation \eqref{lambdaminus} implies that 
$\Lambda_{\dot\alpha}^{\bar{i}}$ and $\bar\Lambda_{i}^{\alpha}$ are functions of the remaining components of $\Lambda$ and $\bar{\Lambda}$ for fixed values of the bosonic coordinates, see Appendix B 
of \cite{VertexOld} for precise expressions. 
One way of verifying that $\Lambda$ and $\bar{\Lambda}$ are constrained variables is by writing the pure spinor constraints of \eqref{eq:so6pure} and \eqref{eq:so6pure2} 
in terms of them, contracting the equations with appropriate factors of $U$ and $V$ and their inverses.      
Defining the dual spinors as 
  \begin{equation}
  (\Lambda^\vee)^I_R = X_{RS}\ \Lambda^S_J\ Y^{JI} \,  ,\qquad (\overline\Lambda^\vee)^R_I = X^{RS}\ \overline\Lambda{}^J_S\ Y_{JI} \, , 
  \end{equation}
the pure spinor constraints take the form 
 \begin{equation}
  \begin{aligned}
  X_{TR}\ \langle\Lambda \overline\Lambda\rangle^R_S - X_{SR} \langle\Lambda\overline\Lambda\rangle^R_T = 0 \, , \\
  Y_{KI}\ \langle\overline\Lambda\Lambda\rangle^I_J - Y_{JI}\ \langle\overline\Lambda\Lambda\rangle^I_K = 0  \, ,  \\
  \langle\Lambda \Lambda^\vee\rangle^R_S + \langle\overline\Lambda{}^\vee\overline\Lambda\rangle^R_S = 0 \, ,  \\
  \langle\Lambda{}^\vee \Lambda\rangle^I_J + \langle\overline\Lambda \overline\Lambda{}^\vee\rangle^I_J = 0 \, , 
\label{covariantpure}
  \end{aligned}
\end{equation}
where the angle brackets denote the traceless part of the product inside e.g.
\begin{equation}
\langle\Lambda \overline\Lambda\rangle^R_S = \Lambda^R_I \overline\Lambda{}^I_S - \frac14 \delta^R_S\ (\Lambda\overline\Lambda) \, . 
\end{equation}
Note that the first two relations of \eqref{covariantpure} are equivalent to 
\begin{equation}
   X^B\ (\gamma_{AB})^S{}_R\ \Lambda^R_I\overline\Lambda{}^I_S = 0 \, , \qquad Y^B\ (\gamma_{AB})^J{}_I\ \overline\Lambda{}^I_R \Lambda^R_J = 0 \, .
  \end{equation}
This means that the six by six matrices $(\gamma_{AB})^S{}_R\ \Lambda^R_I\overline\Lambda{}^I_S$ and $(\gamma_{AB})^J{}_I\ \overline\Lambda{}^I_R \Lambda^R_J$ have zero as an eigenvalue and are therefore singular. This implies that they have a vanishing determinant or Pfaffian, which in turn gives the polynomial constraints
  \begin{equation}
  \begin{aligned}
  P^{S_1 S_2 S_3}_{R_1 R_2 R_3}\ \Lambda^{R_1}_I\overline\Lambda{}^I_{S_1} \Lambda^{R_2}_J\overline\Lambda{}^J_{S_2} \Lambda^{R_3}_K\overline\Lambda{}^K_{S_3} = 0 \, ,\qquad Q^{J_1 J_2 J_3}_{I_1 I_2 I_3}\ \overline\Lambda{}^{I_1}_R \Lambda^R_{J_1} \overline\Lambda{}^{I_2}_S \Lambda^S_{J_2} \overline\Lambda{}^{I_3}_T \Lambda^T_{J_3} = 0 \, , 
  \end{aligned}
  \end{equation}
where
  \begin{equation}
  \begin{aligned}
  P^{R_1 R_2 R_3}_{S_1 S_2 S_3} & = \varepsilon^{A_1 B_1 A_2 B_2 A_3 B_3} (\gamma_{A_1 B_1})^{R_1}{}_{S_1} (\gamma_{A_2 B_2})^{R_2}
  {}_{S_2} (\gamma_{A_3 B_3})^{R_3}{}_{S_3} \, ,  \\
  Q^{I_1 I_2 I_3}_{J_1 J_2 J_3} & = \varepsilon^{A_1 B_1 A_2 B_2 A_3 B_3} (\gamma_{A_1 B_1})^{I_1}{}_{J_1} (\gamma_{A_2 B_2})^{I_2}{}_{J_2} (\gamma_{A_3 B_3})^{I_3}{}_{J_3} \, . 
  \end{aligned}
  \end{equation}
Equivalently,
  \begin{equation}
  \langle\Lambda \overline\Lambda\rangle^R_S \langle\Lambda \overline\Lambda\rangle^S_U \langle\Lambda \overline\Lambda\rangle^U_R = 0 \, . 
  \end{equation}
 Alternativelly, we can obtain the same constraint and others (not necessarily independent) by manipulating the expression
  \begin{equation}
  X_{TR}\ \langle\Lambda \overline\Lambda\rangle^R_S - X_{SR} \langle\Lambda\overline\Lambda\rangle^R_T = 0 \quad \Longrightarrow \quad \langle\Lambda \overline\Lambda\rangle^R_S = - X_{SU}\ \langle\Lambda\overline\Lambda\rangle^U_T \ X^{TR} \, , 
  \end{equation}
which is only true when $X_{RS}$ represents a point in the bulk of $AdS_5$. Finally, the last equation implies that the trace of any odd powers of $\langle \Lambda\overline\Lambda\rangle^R_S$ vanishes
  \begin{equation}
  \langle \Lambda\overline\Lambda\rangle^{R_1}_{R_2} \langle \Lambda\overline\Lambda\rangle^{R_2}_{R_3} \ldots \langle \Lambda\overline\Lambda\rangle^{R_{2k+1}}_{R_1} = 0 \, . 
  \end{equation}
Similarly, for the other analogous pure spinors constraints with $Y$, we obtain
  \begin{equation}
  \langle \overline\Lambda\Lambda\rangle^{I_1}_{I_2} \langle \overline\Lambda\Lambda\rangle^{I_2}_{I_3} \ldots \langle \overline\Lambda\Lambda\rangle^{I_{2k+1}}_{I_1} = 0 \, . 
  \end{equation}

Even using an explicit parametrization for the pure spinors as in \eqref{lambdaminus} and \eqref{rhoangles}, it is involved to   
compute the dimension of the space spanned by $\Lambda$ and $\bar{\Lambda}$.
Notice that a naive counting of degrees of freedom suggests that they are unconstrained. The two pure spinors $\lambda_L$ and $\lambda_R$ have in total twenty-two free components, and both $U$ and $V$ are parametrized by five parameters. Combined, the total number of parameters is thirty-two. 
However, the constraints derived above imply that this counting is not correct and that $\Lambda$ and $\bar{\Lambda}$ span a lower dimensional space. 
Using {\emph{Mathematica}}, we have computed the Jacobian of the transformation of variables \eqref{changeofvariables} and, after evaluating it at different points, we have found different ranks, with thirty-one being the maximum value.
The appearance of different ranks at different points is expected, since the pure spinor constraints are nonlinear. 
Additionally, we can extend the analysis to supercosets with different isotropy groups.  

We can also consider the counting of the degrees of freedom for $\Lambda$ and $\bar{\Lambda}$ at the boundary. The boundary is the intersection of the embedding coordinates $X_{RS}$ with $\mathbb{R}^{4,2}$ at infinity \cite{Witten:1998qj}. The way of approaching the boundary $z \rightarrow \infty$ is to rescale $U$ and, consequently, $X_{RS}$ so as to obtain a finite result in this limit. 
In case of a $D_3$-brane at the boundary of the space, there are also boundary conditions for the pure spinor variables.   
We have performed numerical analysis by rescaling $U$ only and, upon trying several boundary points, we have found a maximum rank of twenty for the Jacobian of the change of variables. However, we have not performed an exhaustive analysis at this stage, and it is possible that analytical results could be obtained in the boundary limit.    
  
\noindent {\bf{The $\Lambda_L$ and $\Lambda_R$ spinors and the $\lambda_L$ and $\lambda_R$ pure spinors}}\footnote{We would like to thank Ulisses M. Portugal for comments about the $\Lambda_L$, $\Lambda_R$ space.
Part of the results follows from discussions with him.}    
\vspace{1mm}

The spinors $\Lambda$ and $\bar{\Lambda}$ analysed above can be related to the gauge-invariant spinors $\Lambda_L$ and $\Lambda_R$ defined below and in \cite{BerkovitsVafa,BerkovitsNew}. 
Relating the two sets of spinors is also important for defining integration on the pure spinor space \cite{workprogress}. 
The spinors $\Lambda_L$ and $\Lambda_R$ are constructed by contracting the pure spinors $\lambda_L$ and $\lambda_R$ with factors of $U$ and $V$, 
\begin{equation}
(\Lambda_L)^R_I = V^{\widetilde{I}}_I (\lambda_L)_{\widetilde{I}}^{\widetilde{R}} (U^{-1})^R_{\widetilde{R}} \, , \quad \quad
(\Lambda_R)^R_I = V^{\widetilde{I}}_I (\lambda_R)_{\widetilde{I}}^{\widetilde{R}} (U^{-1})^R_{\widetilde{R}} \, . 
\label{capitaLR}
\end{equation}
To relate the two sets of spinors, consider the following combination of pure spinors and supercharges appearing, for example, in the computation of the BRST transformations,  
\begin{equation}
\frac{1}{2} \lambda_L q_L + \frac{1}{2}\lambda_R q_R =
\widetilde\lambda \widetilde{q} + \widetilde{\bar{\lambda}} 
\widetilde{\bar{q}} \, ,  
\end{equation}
thus in our conventions 
\begin{equation}
i q_R = \widetilde{q} + \widetilde{\bar{q}} \, , \quad 
q_L = \widetilde{q} - \widetilde{\bar{q}} \, , \quad
\lambda_R/i = \widetilde\lambda+ \widetilde{\bar{\lambda}} \, , \quad 
\lambda_L = \widetilde\lambda - \widetilde{\bar{\lambda}} \, , 
\label{lambdaRlambdaL}
\end{equation}
and
\begin{equation}
\begin{aligned}
\widetilde{\lambda} = \lambda^R_I \, , \quad
\widetilde{\bar{\lambda}} = \sigma^{R S}_6 (\sigma_6)_{IJ} \lambda^J_S \, , \quad  
\widetilde{q} = q^R_I \, , \quad
\widetilde{\bar{q}} = \sigma^{R S}_6 (\sigma_6)_{IJ} q^J_S \, .
\label{lambdaRlambdaLtilde}
\end{aligned}
\end{equation}

It follows from \eqref{lambdaRlambdaL} and \eqref{lambdaRlambdaLtilde}, by contracting with appropriate factors of $U$ and $V$, that 
\begin{equation}
    (\Lambda_R)^R_I = i [ \Lambda^R_I + X^{RS} Y_{IJ} \bar\Lambda_S^J] \, , \quad \quad 
    (\Lambda_L)^R_I =  [ \Lambda^R_I -X^{RS} Y_{IJ} \bar\Lambda^J_S]  \, .
\label{defLR}
\end{equation}

Let us derive the consequences of the pure spinor constraints in terms of these new variables, starting from the 
pure spinor constraints for $\lambda_L$ and $\lambda_R$. In our conventions, they are given by
\begin{equation}
\begin{aligned}
&(\lambda_L)^R_I (\sigma_6)_{R S} (\lambda_L)^S_J = -\frac{1}{4} (\lambda_L \lambda_L)  
(\sigma_6)_{I J} \, , \quad  \\
&(\lambda_L)^R_I (\sigma_6)^{I J} (\lambda_L)^S_J = -\frac{1}{4} (\lambda_L \lambda_L)  
(\sigma_6)^{R S} \, , 
\label{RLvinculos}
\end{aligned} 
\end{equation} 
and similarly for $\lambda_R$. 
Let us analyse the equations above; their properties will be important for the discussions below. 
The space of solutions $\overline{\mathcal C}$ of these equations can be separated into orbits of $USp(2,2) \times USp(4)$,
  \begin{equation}
  \overline{\mathcal C} = \mathcal C_{\rm reg}\ \cup\ \mathcal C_{\rm sing}^2 \cup \mathcal C_{\rm sing}^1 \cup \{0\} \, , 
  \end{equation}
where each component is defined as follows:
  \begin{itemize}
  \item If $(\lambda\lambda)\neq 0$, then both equations in \eqref{RLvinculos} are equivalent and admit the generic solution  
  \begin{equation}
  \mathcal C_{\rm reg} := \mathbb C \times Sp(4,\mathbb C) \, . 
 \label{purespinorA}
  \end{equation}
  \item If $(\lambda\lambda) = 0$, then those two equations are independent and both need to be enforced
    \begin{equation}
    \lambda^R_I (\sigma_6)^{IJ} \lambda^S_J = 0\,, \qquad \lambda^R_I (\sigma_6)_{RS} \lambda^S_J = 0 \, . 
    \end{equation}
    These two equations imply that the matrix $[\lambda^R_I]$ is not invertible and hence can only have rank $3$, $2$, $1$ or $0$.
    \begin{itemize}
    \item For matrices $[\lambda^R_I]$ of rank $3$, there are no solutions, since this would imply that we could find totally isotropic subspaces of dimension $3$ in a space of dimension $4$.
    \item If $[\lambda^R_I]$ is a solution of rank $2$, then it can be written as
      \begin{equation}
      \lambda^R_I = \mu^R \mu_I + \sigma^R\sigma_I
      \end{equation}
    where $(\mu^R)$ and $(\sigma^R)$ are linearly independent, as are $(\mu_I)$ and $(\sigma_I)$. In this case, the pure spinor constraint, together with linear independence, implies that the solutions are
      \begin{equation}
      \mathcal C_{\rm sing}^2 := \Big\{\ \lambda^R_I = \mu^R \mu_I + \sigma^R\sigma_I\ \Big| \quad {\rm span}\langle \mu^R, \sigma^R\rangle \text{ and } {\rm span}\langle \mu_I, \sigma_I\rangle \text{ are tot. isotropic} \Big\}
      \end{equation}
    \item If $[\lambda^R_I]$ is a solution of rank $1$, then it can be written as
      \begin{equation}
      \lambda^R_I = \mu^R \mu_I
      \end{equation}
    where $(\mu^R)$ and $(\mu_I)$ are nonzero vectors. In this case, the pure spinor constraints are satisfied trivially
      \begin{equation}
      \mathcal C_{\rm sing}^1 := \Big\{\ \lambda^R_I = \mu^R \mu_I \ \Big| \quad (\mu^R) \neq 0\ ,\ (\mu_I) \neq 0\ \Big\} \, . 
      \end{equation}
    \end{itemize}
  \end{itemize}

By contracting \eqref{RLvinculos} with appropriate factors of $U$ and $V$, the constraints in terms of $\Lambda_L$ and $\Lambda_R$ are 
\begin{equation}
\begin{aligned}
&(\Lambda_L)^R_I X_{R S} (\Lambda_L)^S_J = -\frac{1}{4} (\Lambda_L \Lambda_L)  
Y_{I J} \, , \quad \quad \\
& (\Lambda_L)^R_I Y^{I J} (\Lambda_L)^S_J = -\frac{1}{4} (\Lambda_L \Lambda_L)  
X^{R S} \, , 
\label{Lspinorcoinstraint}
\end{aligned} 
\end{equation} 
and we can also consider the inverses for nonsingular cases, 
\begin{equation}
\begin{aligned}
& \frac{1}{4} (\Lambda_L^{-1})^I_R Y_{I J} (\Lambda_L^{-1})^J_S = -(\Lambda_L \Lambda_L)^{-1} X_{RS} \, , \quad \quad \\ 
&  \frac{1}{4}  (\Lambda_L^{-1})^I_R X^{RS} (\Lambda_L^{-1})^J_S = -(\Lambda_L \Lambda_L)^{-1} Y^{IJ} \, . 
\label{Linversespinor}
\end{aligned} 
\end{equation}

Analogously to the $\Lambda$ and $\bar{\Lambda}$ case, one important question is the dimension of the space generated by $(\Lambda_L)^R_I$ and $(\Lambda_R)^R_I$. The vertex operators can also be written in terms of them because 
\eqref{defLR} is invertible. 
Here we only argue that they are constrained spinors. 
First, let us analyse
$\Lambda_L$. 
The sympletic form will be denoted by $\Omega$ below;  
it can always be related to $\sigma_6$ in \eqref{sigma6} by a change of basis. 
Given a general complex matrix $M$ of $\operatorname{GL}(4,\mathbb{C})$, it can always  be decomposed as 
\begin{equation}
M = U^{\prime} S  (V^{\prime})^{-1} \,   
\end{equation}
where $S$ is a symplectic matrix, $S^T \Omega S = \Omega$, and both $U^{\prime}$ and $V^{\prime}$ satisfy
$(U^{\prime})^T \Omega U^{\prime} = - \Omega$. 
To see this, note that a polar decomposition of a general matrix $M^{\prime} = S^{\prime} W$ always exists 
by using an exponential parametrization.   
Now suppose 
that some reality conditions on the parameters are imposed.
In the definition of $\Lambda_L$ in \eqref{capitaLR}, 
both $U$ are $V$ can be taken as the cosets \eqref{CosetsAdS5S5}. 
Recall that the groups $SU(n,p)$ preserve an indefinite inner product, i.e., consider the diagonal
matrix $\eta_{n,p} = \{I_n, -I_p\}$, with $I_i$ the $i \times i$ identity matrix; thus, an element $A \in SU(n,p)$  satisfies $A^{\dag} \eta_{n,p} A =\eta_{n,p}$. 
This implies that $\Lambda_L$ obeys the following quadratic constraints       
\begin{equation}
\Lambda_L^{\dag} \eta_V \Lambda_L  = (U^{-1})^{\dag} \lambda_L^{\dag} \eta_V  \lambda_L U^{-1} \, , \quad 
\Lambda_L \eta_U \Lambda_L^{\dag}  = V \lambda_L^{\dag} \eta_U  \lambda_L  V^{\dag} \, , 
\end{equation}
which in the particular case where both $U$ and $V$ are in $SU(4)$, implies that $\Lambda_L$ has the same singular values 
as $\lambda_L$.  

Let us consider the space spanned by $\Lambda_L$ and $\Lambda_R$. 
Consider the following matrix
\begin{equation}
S^R_S=(\Lambda_L)^R_I (\Lambda_R^{-1})^I_S \, , 
\end{equation}   
which is well-defined when $\Lambda_R$ is nonsingular. We will show that $S$ cannot be an arbitrary complex four by four matrix. 
Using \eqref{Lspinorcoinstraint} and \eqref{Linversespinor}, one derives 
\begin{equation}
\begin{aligned}
S^R_S X_{RT} S^T_U = (\Lambda_L \Lambda_L)(\Lambda_R \Lambda_R)^{-1} X_{SU} \equiv c X_{SU} \, .\label{nontrivialS}
\end{aligned} 
\end{equation}
The analysis of the equation above can be divided into two cases: either $\rm{Pf}(X) \neq 0$ or $\rm{Pf}(X)=0$.
First, let us consider an equation for $S^{\prime}$ without the scaling on the right-hand side and with $\rm{Pf}(X) \neq 0$, i.e.,
\begin{equation}
S^{\prime R}_S X_{RT} S^{\prime T}_U = X_{SU} \, . 
\label{ThePrimeEquation}
\end{equation}
Using the properties of the Pfaffian, one concludes 
\begin{equation}
{\rm{Pf}}(S^{\prime} X S^{\prime T})
= {\rm{det}}(S^{\prime}) \, {\rm{Pf}} X 
\; \rightarrow \; {\rm{det}}(S^{\prime}) =1 \, . 
\label{detS}
\end{equation}
Also given that $X$ is invertible, it is easy to verify that $S^{\prime -1} = X S^{\prime T} X^{-1}$. Thus, the matrices $S^{\prime}$ form the symplectic group $Sp(4,X)$ which is related to the usual sympletic group by a change of basis. We can also ask the reverse question:
given an arbitrary $S^{\prime}$, what restrictions does it impose on $X$ for \eqref{ThePrimeEquation} to be satisfied? 
To address this question, it is convenient to rewrite \eqref{ThePrimeEquation} as the vanishing of a characteristic polynomial.
This reformulation is also very convenient for performing numerical checks. 
Since $X$ is an anti-symmetric matrix, we can use 
the upper-triangular entries of $X$ to vectorize it as 
\begin{equation}
x=(X_{12},X_{13},X_{14},X_{23},X_{24},X_{34})^T \, . 
\end{equation}
Suppose $e_i$ are four dimensional unit vectors with only the $i$-th component nonzero. 
Alternatively, $x = X_{ij} \, e_{i} \wedge e_{j}$. In this basis,    
\begin{equation}
 A = (S^{\prime} e_i) \wedge (S^{\prime} e_j) = \sum_{k<l} (S^{\prime}_{ki} S^{\prime}_{lj}-S^{\prime}_{li} S^{\prime}_{kj}) \, e_{k} \wedge e_l \, , 
\label{matrixAsix}
\end{equation}  
and the equation \eqref{ThePrimeEquation} becomes 
\begin{equation}
{\rm{det}}(A - I_6) = 0 \, .
\label{det} 
\end{equation}

Consider the case where the matrix $S^{\prime}$ is diagonalizable, the non-diagonalizable case will be treat  below.  
Then, if the eigenvalues of $S^{\prime}$ are denoted by $\lambda_i$, the eigenvalues of the matrix $A$ above are 
of the form $\lambda_i \lambda_j$ with $i < j$. Notice that the dimension of the space $X$ is the number of unit eigenvalues
of the matrix $A$. 
If there is only one unit eigenvalue, then $X$ is of the form $v \wedge w$ for some $v$ and $w$, which would  imply ${\rm{Pf}}(X)=0$, contradicting the initial hypothesis; thus, the space generated by $X$ necessarily has dimension greater than or equal to two. For the case where $S^{\prime}$ is non-diagonalizable, 
the dimension of the space of solutions of \eqref{det} is also greater than or equal to two, again because of the 
supplementary condition ${\rm{Pf}}(X) = X_{12} X_{34}- X_{13} X_{24} + X_{14} X_{23} \neq 0$. Before 
analyzing the case ${\rm{Pf}}(X) = 0$, let us consider \eqref{nontrivialS} for $S$ with nontrivial scaling. The class of matrices $S$ is the same as the regular solution given in     
\eqref{purespinorA}, but now with respect to $X$. 
The determinant of $S$ is ${\rm{det}}(S) =c^2$, which follows from an equation similar to \eqref{detS}.
The dimension of the space $X$ is computed similarly, 
and in the case of diagonalizable $S$, the eigenvalues have to satisfy $\lambda_i \lambda_j = c$ for some $i$ and $j$ for the equation to have a solution. The conclusion here is similar to before:
the space of allowed $X$ with ${\rm{Pf}}(X) \neq 0$ has dimension greater than or equal to two.   

Having analysed the case ${\rm{Pf}}(X) \neq 0$, let us consider the case ${\rm{Pf}}(X) = 0$. We first consider \eqref{ThePrimeEquation} for $S^{\prime}$. 
The zero-Pfaffian condition implies $X=0$ or ${\rm{rank}}(X)=2$. The case $X=0$ is trivial, and any $S^{\prime}$ solves the equation. 
In the ${\rm{rank}}(X)=2$ case, there is a basis where $X$ is written as two by two blocks with the upper-diagonal block nonzero and given by $J$, where $J^2=I_2$. Geometrically, the matrices $S^{\prime}$ solving the equation preserve a two-plane. The general solution is easy to parametrize. Consider 
\begin{equation}
S^{\prime} = \left( \begin{tabular}{cc}
 $A$ & $B$ \\
$C$ & $D$ 
\label{blockmatrices}
\end{tabular} 
\right) \, , 
\end{equation}
then, in that particular basis, the solution for invertible $S^{\prime}$ is,  
\begin{equation}
A \in Sp(2) \sim Sl(2) \, , \quad \quad B = 0 \, , \quad \quad {\rm{det}}(D)=0 \, ,  
\end{equation}         
and $C$ uncoinstrained. The condition on $D$ can be dropped when $S^{\prime}$ is not invertible. 
We can ask whether a given $S^{\prime}$ fully determines $X$ in the case where ${\rm{Pf}}(X) = 0$. 
Again, we can analyze the characteristic polynomial of the matrix $A$ in \eqref{det}.
It is possible to have a minimum of zero and a maximum of six unit eigenvalues. 
Having no unit eigenvalues implies that the equation has no solution.
For the other cases, the number of unit eigenvalues gives the dimension of $X$. In particular, it is possible to have a one-dimensional solution.
The case with nontrivial scaling in \eqref{nontrivialS} can be treated similarly.
The nontrivial scaling changes the values of the eigenvalues of the matrix $A$ in \eqref{det} and the conditions on the block matrices when $S$ is decomposed similarly to 
\eqref{blockmatrices}.   

In conclusion, the discussion above implies that both sets of spinors $\{\Lambda, \bar{\Lambda}\}$ and 
$\{\Lambda_L,\Lambda_R\}$ have constraints. Notice that there is an invertible map relating them, and the constraints appear from different perspectives. Also, the boundary limit differs from the bulk analysis because it is necessary to rescale the coordinates to obtain a nontrivial limit.  
We have performed some preliminary numerical analysis of the geometry of both spaces using particular representations for the pure spinors and the $U$ and $V$.
We have found that the rank of the Jacobian of the transformation depends on the points of the parametrization, as expected, and it is smaller than thirty-two, which implies the existence of constraints. It is possible to analyze the Jacobian analytically at special points and for infinitesimal variations around them by using several representations. 

\subsection{Properties of the vertex operators} 
\label{Propertiesofthevertexoperators}

In this Subsection, we describe several properties of the vertex operators.
We start by reviewing the argument of 
\cite{VertexOld} that the vertex operators are not $Q$-exact. Then, we comment on $SU(4)$ representations, the supersymmetry transformations of the vertex operators, the flat space limit, 
the plane-wave limit, the gauge transformations, and finally the primary condition. 

\subsubsection{The vertex operators are not $Q$-exact} 

Setting $n =0$ in the vertex operators, one reproduces the fully $PSU(2,2|4)$-invariant dilaton vertex operator $V_{\rm{dil}}$ in  \eqref{dilatonvertex}. As demostrated in \cite{BerkovitsSimplifying}, the integrated form of $V_{\rm{dil}}$ corresponds to the worldsheet action; thus, it is a physical state in the BRST cohomology. 
Alternatively, it is possible to test whether a state is $Q$-exact by generating a basis of $Q$-exact terms with {\it{Mathematica}} and verifying
whether  the system has a numerical solution. 
This was done for the dilaton vertex operator and the other half-BPS states of \cite{VertexOld} using the BRST transformations in Appendix B of that paper.
Another approach is to use the global isometries. 
Consider the dilaton vertex operator for $g_{\rm{old}}$ in the two equivalent forms below, which are related by the pure spinor constraints,
\begin{equation} 
V_{\rm{old, dil}} = (\bar\lambda_+ \bar\lambda_+) + (\lambda_+ \lambda_+) = 
(\bar\lambda_- \bar\lambda_-)+(\lambda_- \lambda_-) \, ,   
\label{olddilaton}
\end{equation}     

with 
\begin{equation}
(\bar\lambda_- \bar\lambda_-) = \lambda^{\dot\alpha}_{i} \lambda^{\dot\beta}_{j} \epsilon_{\dot\alpha \dot\beta} \epsilon^{i j} - \lambda_{\alpha}^{\bar{i}} \lambda_{\beta}^{\bar{j}} \epsilon^{\alpha \beta} \epsilon_{\bar{i} \bar{j}} \, , \quad \quad    
(\lambda_- \lambda_-) = \lambda^{\bar{i}}_{\dot\alpha} \lambda^{\bar{j}}_{\dot\beta} \epsilon^{\dot\alpha \dot\beta} \epsilon_{\bar{i} \bar{j}} -
\lambda^{\alpha}_{i} \lambda^{\beta}_{j} \epsilon_{\alpha \beta} \epsilon^{i j} \, .  
\end{equation}
The only possible way to write $V_{\rm{old, dil}}$ in terms of $\{\bar\lambda_-,\lambda_-\} $ as $Q$-exact would be, at lowest order in the fermions, as the exact term  
\begin{equation}
 Q \cdot \Omega \equiv Q \cdot ((\bar\lambda_- \widetilde{\bar\theta}_-) + (\lambda_- \widetilde{\theta}_-)) \, ,
 \label{TheOmegaQ}
\end{equation}
where $\widetilde{\bar\theta}_- = e^{- \frac{w+z}{2}}(\theta^{\bar{i}}_{\alpha}, \psi^{\dot{\alpha}}_i)$ 
and $\widetilde\theta_- = e^{- \frac{w-z}{2}}(\theta^{\bar{i}}_{\dot\alpha}, \psi^{\alpha}_i)$.

 Recall that the BRST transformations commute with all the global $PSU(2,2|4)$ transformations, and $V_{\rm{old, dil}}$ is invariant under all of them. Moreover, the supersymmetry transformations with parameters 
$\eta^{\alpha}_i$ and $\eta^{\bar{i}}_{\dot\alpha}$ for the coset representative $g_{\rm{old}}$ do not generate any compensators, and they act only as the shifts $\delta \psi^{\alpha}_i = \eta^{\alpha}_i$ and $\delta \theta^{\bar{i}}_{\dot\alpha}=\eta^{\bar{i}}_{\dot\alpha}$. 
Thus, the variation of the $Q$-exact term written above under this transformation is  given by
\begin{equation}
\delta_{\eta^{\alpha}_i, \eta^{\bar{i}}_{\dot\alpha}} Q \cdot \Omega =  Q \cdot \delta_{\eta^{\alpha}_i, \eta^{\bar{i}}_{\dot\alpha}} \Omega = Q \cdot (e^{-\frac{w-z}{2}} \lambda_- \eta) \, , 
\end{equation}
and it is nonzero. 
Inspecting the BRST transformations in Appendix B of \cite{VertexOld}, it is easy to verify 
that the BRST transformations of any of the untilded variables do not generate $\theta_-$. 
Thus, to cancel the variation above, it would be necessary 
to add terms linear in $\theta_-$,  
however, the latter would either not be $Q$-exact or would generated 
uncanceled contributions. Another argument uses $V_{\rm{old, dil}}$ in \eqref{olddilaton} written in terms 
of $\{\bar\lambda_+, \lambda_+ \}$. The only possibility to write it 
as $Q$-exact at lowest order in the fermionic variables is 
\begin{equation}
 Q \cdot \Omega^{\prime} \equiv Q \cdot ((\bar\lambda_+ \widetilde{\bar\theta}_+) + (\lambda_+ \widetilde{\theta}_+)) \, ,
 \label{TheOmegaQ}
\end{equation}
where $\widetilde{\bar\theta}_+ = e^{\frac{w+z}{2}}(\theta^{i}_{\dot\alpha}, \psi^{\alpha}_{\bar{i}})$ 
and $\widetilde\theta_+ = e^{\frac{w-z}{2}}(\theta^{i}_{\alpha}, \psi^{\dot\alpha}_{\bar{i}})$.

Note that  $Q \cdot \Omega^{\prime}$ contains, among other terms, the term  
 \begin{equation}
(\bar{\lambda}_+ \widetilde{\bar{\theta}}_+) (\lambda_-  \widetilde\theta_+) \, .
\end{equation}
One can see that this term cannot be removed by adding any chain of $Q$-exact terms without
producing terms with fewer $\theta$'s. 

Now, consider 
the states with arbitrary $n$ for $g_{\rm{old}}$. The vertex operators are of the form 
$V_n = e^{n(z+w)} V_{\rm{dil}} + \ldots$. The only way to remove 
this first term is to add, schematically, $Q \cdot  e^{n(z+w)} \lambda \theta$. 
The BRST operator produces terms proportional to $n$ when acting on the exponential, and this does not help with possible cancelations. 
Thus, the fact that the vertex operators are not $Q$-exact follows from an argument similar to the one given above for 
$V_{\rm{dil}}$, see \cite{VertexOld} for more details.  
Another argument that  
the vertex operators are not $Q$-exact is that, apart from the dilaton vertex operator, which is $Q$-exact in this form in the flat-space limit, all other vertex operators for any $n$ reduce, as expected, to states in the flat space cohomology
of \cite{Martins:2019wwi} by taking the flat space limit.

\subsubsection{Comments about $SU(4)$ covariance}
\label{Comments about $SU(4)$ covariance}

As reviewed in the Subsection \ref{Null polarization vectors}, the half-BPS operators  
of $\mathcal{N}=4$ SYM are polynomially parametrized by the null polarizations vectors $(y_0)_{IJ}$. 
This implies that they are in the symmetric traceless representation of $SO(6)$ and transform 
covariantly.  
In \cite{Mikhailov:2009rx}, it was proved that, unlike in flat space, there is 
a vertex operator representative 
in $AdS_5 \times S^5$ that transforms covariantly without the necessity of adding
$Q$-exact terms; 
see also \cite{OliveiraFerreira:2026rpm} 
for more recent discussions on the symmetry constraints on the allowed representations. 

The procedure described in Subsection   
\ref{su22 covariant vertex operators} for changing the variables from $g^{\prime}_{\text{new}}$ in \eqref{gprime} to the ones from 
$g_{\text{new}}$ in \eqref{CosetIdeal} introduces poles $(y_0 \cdot Y)$ at intermediate steps depending on the
vertex representative and the value of $n$.
The vertex operators constructed in this paper are already fully polynomial in the null polarization vectors $(y_0)_{IJ}$ for both $n < -7$ and 
$n = 0$. 
This follows because the maximum number of normalized polarizations $\dot{y}_0$ appearing in our vertices 
is seven, occurring in $V_{12}$, see the {\it{Mathematica}} ancillary file.   
For the others few values $n=-1, \ldots,-6$, we succeed in finding a polynomial representative up to the second order in the fermionic variables by adding $Q$-exact terms, see    
\eqref{TheV2covariant} for the explicit final expression for $V_{2}$ without poles for any $n$.   
For the higher orders in the fermionic variables, it is not immediate to find the $Q$-exact terms that remove all the poles 
and select the covariant vertex operators representative for the full vertex. We leave this for future work.

Recall that the polarizations $(y_0)_{IJ}$ select a particular $U(1)$ inside $SO(6)$. 
Notice that the fermionic variables of $g_{\text{new}}$ and
the covariant 
bosonic variables are charged with respect to this $U(1)$. 
The denominators $(y_0 \cdot Y)$ guarantee that all the terms of the vertex operators 
have the same $U(1)$ charge and the overall $(y_0 \cdot Y)^{-n}$. 
They should disappears from the computation of physical 
quantities such as string amplitudes. This is expected because 
$Q$-exact terms decouple from amplitude computations.  
We should mention that spurious poles on the $y$'s appear at intermediate stages also 
on  
twistor computations in $\mathcal{N}=4$ SYM, see \cite{Chicherin:2014uca}.  

The vertex operators using the variables of $g^{\prime}_{\text{new}}$ in \eqref{gprime} depend on 
$W, Z, \widehat{\Theta } _{\widetilde{R}}^{\widetilde{I}}, \widehat{\Psi }^{\widetilde{R}}_{\widetilde{I}}, \lambda _{\widetilde{R}}^{\widetilde{I}}, \lambda^{\widetilde{R}}_{\widetilde{I}}$, and it is written 
in $\mathfrak{su}_2^{\oplus 4}$-component notation. As discussed above, the procedure 
described in Section \ref{su22 covariant vertex operators} generates poles at intermediate stages depending 
on the vertex representative. 
It is not easy to find the gauge-fixed good vertex representative in $g^{\prime}_{\text{new}}$ 
which does not produce poles after introducing the null polarization vectors using our procedure.
Nevertheless, it is always possible by adding and subtracting terms to {\it{complete the indices}} and project the terms of the vertex operators into irreducible representations.  
Let us consider a simple example for the projection for $SO(5)$ under global $SO(6)$ transformations. 
It uses the following combination 
\begin{equation}
S^{\widetilde{I} \widetilde{J}} = (\lambda \lambda)^{\widetilde{I}\widetilde{J}} \, .
\end{equation}
By using the 6d Pauli matrizes in \eqref{sigma6matrices}, we define the variables 
\begin{equation}
y^{\mathsf{I}}= \frac{1}{4} (\sigma^{\mathsf{I}})_{\widetilde{I} \widetilde{J}} S^{\widetilde{J} \widetilde{I}} \, , 
\end{equation}
where $\mathsf{I}=1,\ldots,6$. 
Consider
the function $f = \sum_{\mathsf{I}=1}^3 y^{\mathsf{I}} y^{\mathsf{I}}$.
This function is projected into the irreducible representations of $SO(6)$ as 
\begin{equation}
f=(y^{\mathsf{1}})^2+(y^{\mathsf{2}})^2+(y^{\mathsf{3}})^2 = \underbrace{\frac{1}{2} \, y^2}_{\textbf{1}} + \underbrace{\frac{1}{2}\left(\sum_{\mathsf{I}=\mathsf{1}}^{\mathsf{3}} (y^{\mathsf{I}})^2 - \sum_{\mathsf{I}=\mathsf{4}}^{\mathsf{6}} (y^{\mathsf{I}})^2 \right)}_{\textbf{20}'} = \sum_{\lambda=\mathbf{1},\mathbf{15},\mathbf{20'}} \Pi^\lambda_{\mathsf{I} \mathsf{J},\mathsf{K} \mathsf{L}} \, \delta_{(3)}^{\mathsf{K} \mathsf{L}}  y^{\mathsf{I}} y^{\mathsf{J}}\,  \, , 
\end{equation}
where $\delta_{(3)}^{\mathsf{K} \mathsf{L}} = \mathrm{diag}(1, 1, 1, 0, 0, 0)^{\mathsf{K} \mathsf{L}}$ is a polarization vector. The projection into the representation $\bf{15}$ is zero for this particular choice of $f$. The relevant projectors appearing above are  
\begin{equation}
\Pi^\mathbf{1}_{\mathsf{I} \mathsf{J},\mathsf{K} \mathsf{L}} = \frac{1}{6} \delta_{\mathsf{I}\mathsf{J}} \delta_{\mathsf{K} \mathsf{L}} \, , \quad  \Pi^\mathbf{15}_{\mathsf{I} \mathsf{J},\mathsf{K} \mathsf{L}} =  \frac{1}{2} \left( \delta_{\mathsf{I} \mathsf{K}} \delta_{\mathsf{J} \mathsf{L}} - \delta_{\mathsf{I} \mathsf{L}} \delta_{\mathsf{J} \mathsf{K}} \right) \, , \quad 
\Pi^\mathbf{20}_{\mathsf{I} \mathsf{J},\mathsf{K} \mathsf{L}} =  \left( \delta_{\mathsf{I} \mathsf{K}} \delta_{\mathsf{J} \mathsf{L}} - \frac{1}{6} \delta_{\mathsf{I} \mathsf{J}}  \delta_{\mathsf{K} \mathsf{L}}\right) \, . 
\end{equation}

The decomposition of the vertex operators of $g^{\prime}_{\text{new}}$  into irreducible representations may help with finding 
the gauge fixing condition eliminating the poles of the terms $V^n_{i \ge 4}$ for $n=-1, \ldots,-6$. 
We leave further explorations for future work.
We recall that the vertex operators of this paper are obtained by super-reparametrizations and global 
$g\in SU(2,2)\times SU(4)$ rotations from the ones in
\cite{VertexOld}, which preserves twenty-four supersymmetries and satisfies $\Delta = J$.

\subsubsection{Applying supersymmetries transformations: the supermultiplet}
\label{Applying supersymmetries transformations: the supermultiplet}

As mentioned in the Section \ref{sectionreviewing}, we have obtained the vertex operators corresponding to the bottom state of the half-BPS supermultiplets. These states are annihilated by twenty-four supersymmetries. 
Given an object that preserves a set of supersymmetries, it is natural to look for the fermionic parameters that decouple from it. For example, F-terms in $\mathbb{R}^{4|4}$ can be described using chiral superspace $\mathbb{R}^{4|2}$, where the number of fermionic parameters is halved. However, it is important to consider the integrability of these fermionic parameters associated with the supersymmetries. Just as bosonic symmetries induce bosonic vector fields, and a set of integrable bosonic vector fields defines projections on the total space, the same principle applies to supervector fields and superspaces. In the example of $\mathbb{R}^{4|2}$ chiral superspace, the anticommutator of the fermionic vector fields associated with the projection of $\mathbb{R}^{4|4}$ down to $\mathbb{R}^{4|2}$ vanishes, so they are integrable and a projection exists. In the case of half-BPS vertex operators in $AdS_5\times S^5$, where the number of supersymmetries is twenty-four, the anticommutator clearly does not vanish. To define a set of integrable supervector fields for the half-BPS case, it is necessary to include the nine bosonic vector fields that appear in the anticommutator of the twenty-four fermionic ones.
Once the integrable set of supervector fields is established, the next step is to choose a parametrization that isolates the set of parameters that will be projected. When we parametrize the superspace via supercosets, this is done by bringing all the generators of the integrable set to the left of the supercoset element. 
This explains why the vertex operators at infinity of \cite{VertexOld} 
only depends on $\widetilde\theta^+$ using $g_{\rm{old}}$.
The same physical states are described by vertex operators depending 
on more fermionic coordinates using $g_{\rm{new}}$.   
   
The vertex operators of  \cite{VertexOld} are annihilated by all susy generators obeying $J - \Delta \geq 0$. 
The $J- \Delta = -1$ susy transformations acting nontrivially on the vertex operators are generated by $\eta^i_{\alpha} q^{\alpha}_i$ and $\eta^{\dot\alpha}_{\bar{i}} q^{\bar{i}}_{\dot\alpha}$. 
After the change of variables from $g_{\rm{old}} \rightarrow g^{\prime}_{\rm{new}}  \rightarrow g_{\rm{new}} $, the vertex operators with polarizations $(x_{\infty}, y_{\infty})$ are annihilated by the same set of supersymmetries as the ones in \cite{VertexOld}.  
In this Subsection, we will apply susy transformations on the $g_{\rm{new}}$
vertex operators. 
Applying susy transformations successively, the entire supermultiplet is generated. 
We use the results and notations of the Appendix \ref{psu224ideal}. 
First, we consider the transformations of the type $A$. The only variable 
transforming in this case is $\Psi$ and the transformation is
\begin{equation}
\delta_A \Psi^R_I = A^R_I \, .  
\label{typeA}
\end{equation}
It is easy to compute the transformation of the vertex operators under these shifts of $\Psi$. The component without fermionic
variables is invariant $\delta_A V_0$ = 0. 
Using the vertex representative $V_2$ of \eqref{v2},
one verifies 
that $A^R_I$ always appears contracted with $x_0$ and $y_0$ in the variation $\delta_A V_2$.
In particular, by taking $(x_0,y_0)$ to be $(x_{\infty}, y_{\infty})$,  
\begin{equation}
(x_{\infty})_{RS} A^S_{I} (y_{\infty})^{IJ}  = 
\begin{bmatrix}
   0&0\\0& \varepsilon_{\dot\beta\dot\alpha}
\end{bmatrix}
\begin{bmatrix}
   A^{\alpha}_i & A^{\alpha}_{\bar{i}} \\ A^{\dot\alpha}_{\bar{i}} & A^{\dot\alpha}_{\bar{i}}
\end{bmatrix}
\begin{bmatrix}
   0&0\\0& - \varepsilon^{\bar{i} \bar{j}} 
\end{bmatrix}
=
\begin{bmatrix}
   0&0\\0& - \varepsilon_{\dot\beta\dot\alpha} A^{\dot\alpha}_{\bar{i}}\varepsilon^{\bar{i} \bar{j}} 
\end{bmatrix} \, .
\label{xAy}
\end{equation}
We see that the only components of $A$ that do not annihilate the vertex operators for $(x_{\infty}, y_{\infty})$ are the ones corresponding to the susy generators of type $J- \Delta = -1$, as expected by construction. 
For general $(x_0,y_0)$, linear combinations of the susy generators annihilate the vertex operators.  
The combinations are obtained from the zeros of $(x_0)_{RS} A^S_{I} (y_0)^{IJ}$.    
It is not difficult to apply further susy transformations of type $A$ with different parameters or to consider 
the higher-order terms of the vertex operators in the fermionic expansion. We now proceed to analyse the transformations of type $B$ in the notation of the Appendix \ref{psu224ideal}. 
The fermions transform as 
\begin{equation}
\delta_B \Psi = - \frac{1}{2} [[B,\Psi],\Psi] + [[B,\Psi],\Psi] \, , \quad \quad  \delta_B \Theta = B + [[B,\Psi],\Theta] \, , 
\end{equation}
and we have written the $\Psi$ transformation in a convenient way. The second part of its transformation is a $SU(2,2) \times SU(4)$ rotation that will cancel in fully contracted expressions such as the vertex operators. By evaluating the commutator, one verifies
\begin{equation}
[B,\Psi] = [B^{I}_{R}{q}^{R}_{I},  \Psi^{S}_{J}{q}^{J}_{S}]
= - B^{I}_{R} \Psi^{S}_{J} (\delta^J_I M^R_S  +  \delta^R_S M^J_I) 
= - B^{I}_{R} \Psi^{S}_{I}  M^R_S  - B^{I}_{R} \Psi^{R}_{J}  M^J_I \, . 
\end{equation}
As a consequence of \eqref{eqBallpo}, the other covariant variables, such as the embedding coordinates $X_{RS}$, rotate with the same $[B,\Psi]$ rotation matrix, 
\begin{equation}
\delta_B \cdot X_{R S} =  B^I_R \Psi^T_I X_{T S} +
B^I_S \Psi^T_I X_{R T} - \frac{1}{2} (B^I_T \Psi^T_I) X_{R S} \, .  
\end{equation}
By defining rotated null polarization vectors $x_0^{\prime}$ and $y_0^{\prime}$, such that they compensate the $[B,\Psi]$ rotations, 
\begin{equation}
\begin{aligned}
& (x^{\prime}_0)_{RS} 
 = - B^I_R \Psi^T_I (x_0)_{T S} -
B^I_S \Psi^T_I (x_0)_{R T} + \frac{1}{2} (B^I_T \Psi^T_I) (x_0)_{R S} \, , \\
& (y^{\prime}_0)^{IJ} 
 = B^I_R \Psi^R_K (y_0)^{K J} +
B^J_R \Psi^R_K (y_0)^{I K} - \frac{1}{2} (B^I_T \Psi^T_I) (y_0)^{IJ} \, ,
\end{aligned} 
\end{equation}
the susy transformations of the type $B$ amount to 
\begin{equation}
\delta_B \Psi  \rightarrow  - \frac{1}{2} [[B,\Psi],\Psi]  \, , \quad \quad  \delta_B \Theta  \rightarrow  B \, , \quad \quad x_0, y_0  \rightarrow x^{\prime}_0, y^{\prime}_0 \, . 
\label{TheBtrans}
\end{equation} 
Thus, one has for $V_0$ given in \eqref{dilatoncovariant},  
\begin{equation}
\delta_B V_0 = n \left(\frac{x_0\cdot X}{y_0\cdot Y}\right)^{n-1} (x^{\prime}_0\cdot X - y^{\prime}_0\cdot Y)  \left[(\overline\Lambda X \overline \Lambda\ Y) - (\Lambda Y \Lambda X)\right] \,, 
\label{variationV0} 
  \end{equation}
and it is not difficult to find $\delta_B V_2$.  
For transformations of type $B$, it is possible to perform a similar analysis as done for type $A$ and find the linear combination of supersymmetries that annihilates the vertex operators for general $(x_0,y_0)$. 
The type $B$ transformations involve all variations in \eqref{TheBtrans}, unlike  
the type $A$ transformations, which only transform the fermions $\Psi$. 
The $\delta_B V_0$ above is cancelled by the corresponding terms of $\delta_B V_2$ linear 
in the fermionic variables.  
Considering the $V_2$ representative in \eqref{v2},
apart from the terms in the first line, all the $\Theta$'s are 
contracted with the null polarizations vectors. Setting 
$(x_{\infty},y_{\infty})$, all those terms, when $\delta_B \Theta = B$, are annihilated by the cases,     
\begin{equation}
(x_{\infty})^{RS} B_S^{I} (y_{\infty})_{IJ}  = 
\begin{bmatrix}
    \varepsilon^{\beta \alpha}
&0\\0& 0
\end{bmatrix}
\begin{bmatrix}
   B_{\alpha}^i & B_{\alpha}^{\bar{i}} \\ B_{\dot\alpha}^{\bar{i}} & B_{\dot\alpha}^{\bar{i}}
\end{bmatrix}
\begin{bmatrix}
  -  \varepsilon_{i j}  &0\\0& 0
\end{bmatrix}
=
\begin{bmatrix}
  -  \varepsilon^{\beta \alpha} B_{\alpha}^{i}\varepsilon_{i j} &0\\0& 0
\end{bmatrix} \, .
\label{xBy}
\end{equation}
The nonzero susy transformations are therefore generated by $B^i_{\alpha}$ and are of the type $J- \Delta = -1$, as expected. Moreover, the type $B$ variations of the terms in the first line of $V_2$ in \eqref{v2} linear in the fermions 
cancel the variations of $V_0$ in \eqref{variationV0} for the susy types $J- \Delta \neq -1$. 
To see this, 
one rearranges 
the term  
 $(B X \dot{x}_0 \Psi \dot{y}_0 Y)$ 
 to cancel \eqref{variationV0} for all 
 $B$'s different from $B^i_{\alpha}$. 
 
 Since $V_{i \geq 4}$ has many terms, we are not going to write here either $\delta_A V_{i \geq 4}$ or
 $\delta_B V_{i \geq 4}$ and show explicitly that the full vertex is annihilated by all the susy generators with $J- \Delta \neq -1$ when 
 the null polarizations are set to $(x_{\infty},y_{\infty})$.
 The necessary manipulations are easier to perform using {\it{Mathematica}}. 
 Using both the transformations \eqref{typeA} and \eqref{TheBtrans},
 it is not difficult to compute the entire supermultiplet by applying several susy transformations sequentially.  
  
\subsubsection{Flat space limit} 
\label{Flat space limit}

As mentioned in the Introduction, it is possible, for example, to organize an expansion of the 
string amplitudes in small curvatures. The leading term in 
this expansion is the flat space amplitude.  
Here we will consider the flat space limit of the vertex operators constructed in this paper in  unintegrated  and integrated form. We start by reviewing the flat space limit of the vertex operators of \cite{VertexOld} using $g_{\rm{old}}$.    
It was stablished in \cite{VertexOld}  
that the 
$g_{\rm{old}}$ 
vertices correctly reduce 
to the flat space ones of \cite{Martins:2019wwi}. In addition,  
the $AdS_5 \times S^5$ BRST transformations becomes the flat space BRST transformations. 
The procedure to take the limit is to organize the $\mathfrak{psu}(2,2|4)$ generators 
as 
\begin{equation}
M = \{ \widehat M^\alpha_{\beta}, \widehat M^{\dot{\alpha}}_{\dot{\beta}}, (P^{\alpha}_{\dot{\alpha}} + \epsilon_{\dot{\alpha} \dot{\beta}} \epsilon^{\alpha \beta} K^{\dot\beta}_{ \beta}), \widehat M^i_{j}, \widehat M^{\bar{i}}_{\bar{j}}, (P^{\bar{i}}_{\ i} + \epsilon_{ij} \epsilon^{\bar{i} \bar{j}} K^{j}_{\bar{j}})\},
\end{equation}
\[
P = \{K^{\dot\beta}_{\beta}, K^{j}_{\bar{j}}, \Delta, J\}, \qquad q = \{q^\alpha_i, q^i_{\dot{\alpha}}, q^{\bar{i}}_\alpha, q^{\dot{\alpha}}_i, q^i_\alpha, q^{\dot{\alpha}}_{\bar{i}}, q^{\bar{i}}_{\dot{\alpha}}, q^\alpha_i\} , 
\]
where $M$ are the generators of the isotropy group and both $P$ and $q$ parametrize $g_{\rm{old}}$. The 
$\mathfrak{psu}(2,2|4)$ superalgebra is schematically of the form in this notation  
\[
[M, M] \sim M, \quad [P, P] \sim M + P, \quad [M, P] \sim P + M,
\]
\[
[P, q] \sim q, \quad [M, q] \sim q, \quad [q, q] \sim P + M.
\]
Rescaling 
both the generators and the structure constants of the superalgebra appropriately by a parameter 
$\Lambda$, it is possible to take a limit $\Lambda \rightarrow 0$ by including new conserved currents 
in such a way that the 
$\mathfrak{psu}(2,2|4)$ algebra is deformed to the super-Poincaré algebra continuously. 
This limit corresponds to looking at small 
distances around the origin of the coset representative.  The precise scaling of the generators is 
\[
M \rightarrow M \, , \quad P \rightarrow \Lambda^{-1} P \, , \quad 
q \rightarrow \Lambda^{-\frac{1}{2}} q \, . 
\]
The rescaling above implies the scaling of all the momenta, in particular $n \rightarrow \Lambda^{-1} n$. 
A non-trivial limit is obtained by also 
redefining the variables as  
\[
z \to \Lambda z, \quad w \to \Lambda w, \quad x \to \Lambda x, \quad y \to \Lambda y, \quad \theta \to \Lambda^{\frac{1}{2}} \theta\, , \quad \psi \to \Lambda^{\frac{1}{2}} \psi \, . 
\]

The flat space vertex operators of \cite{Martins:2019wwi} are in a specific gauge. They only depend on the pure spinors $\bar{\lambda}_+$ of \eqref{lambdasso8} and on $\widetilde\theta_+$.
The vertex operators of \cite{VertexOld} reduce precisely to the expressions of \cite{Martins:2019wwi} in the flat space limit without any further simplification. 
The reduction of the vertex operators constructed in this paper for $(x_{\infty}, y_{\infty})$ will depend on more than eight fermionic parameters. They are related to 
the ones in \cite{Martins:2019wwi} by adding $Q$-exact terms. 
This is guarantee because they describe the same physical string states and the gauge in which the flat space vertex operators only depend on $\bar{\lambda}_+$ and on $\widetilde\theta_+$ is always reachable.
The vertex operators for general $(x_0,y_0)$ reduce to flat space vertex operators with different polarizations.

Taking the flat space limit of the integrated vertex operators is more involved. The procedure is as follows.   
Let us first review the flat space limit of the $AdS_5 \times S^5$ worldsheet action, which is the integrated dilaton vertex operator, see \cite{BerkovitsNewLimit} for more details. It is convenient 
to rewrite the terms of the action with two fermionic currents in its original first-order form plus the integral
of a Wess-Zumino-Witten three-form after an integration by parts,
\begin{equation}
\begin{aligned}
& \quad \quad \int d^2 z \,  \frac{3}{4} \eta_{\alpha \widehat\beta} J^{\widehat\beta} \overline{J}^{\alpha} - \frac{1}{4} \eta_{\alpha \widehat\beta} \overline{J}^{\alpha } J^{\widehat\beta}  \, \rightarrow  
 \,  \\
 \int d^2 z \; [ \overline{J}^{\alpha} d^{\prime}_{\alpha} + J^{\widehat\alpha} \widehat{d}^{\prime}_{\widehat\alpha} 
 & + \eta^{\alpha \widehat\beta}  d^{\prime}_{\alpha}  \widehat{d}^{\prime}_{\widehat\beta} 
+ \frac{1}{4} \int d \sigma_3 (\gamma_{c \alpha \beta} J^c \wedge J^{\alpha} \wedge J^{\beta} -
\gamma_{c \widehat\alpha \widehat\beta} J^c \wedge J^{\widehat\alpha} \wedge J^{\widehat\beta}) ] \, , 
\label{actionflatlimit}
\end{aligned}
\end{equation}
and the Maurer-Cartan identities were used. 
Notice that the first line is recovered by using the equations of motion for the variables $d^{\prime}_{\alpha}$ and $\widehat{d}^{\prime}_{\widehat{\alpha}}$.
The flat space limit is obtained by rescaling  
$\eta_{\alpha \widehat\alpha} \rightarrow (1/\Lambda) \eta_{\alpha \widehat\alpha}$ and 
$\eta^{\alpha \widehat\alpha} \rightarrow \Lambda \eta^{\alpha \widehat\alpha}$, and by taking $\Lambda \rightarrow 0$, which makes the term 
$d^{\prime}_{\alpha}  \widehat{d}^{\prime}_{\widehat\beta}$ above vanish. In addition, the currents truncate at linear order in $X$ and at quadratic order in 
$\theta$ in the same limit. Using the definitions    
\begin{align*}
d_\alpha = p_\alpha 
             + \frac{1}{2} (\gamma^{\underline{a}} \theta)_\alpha \bigg( \partial X_{\underline{a}} 
             - \frac{1}{4} (\theta \gamma_{\underline{a}} \partial \theta) \bigg) \, , \quad \quad \Pi^{\underline{a}} = \partial X^{\underline{a}} 
            + \frac{1}{2} (\theta \gamma^{\underline{a}} \partial \theta) 
            + \frac{1}{2} (\widehat{\theta} \gamma^{\underline{a}} \bar\partial \widehat\theta),
\end{align*}
one recovers the pure spinor flat space action. The integrated flat space vertex operators 
depend on the $d$'s. To get them as the flat space limit of the $AdS_5 \times S^5$ integrated vertex operators, it is necessary to rewrite some bilinear terms of the fermionic currents as depending on the $d$'s via the equations of motion of \eqref{actionflatlimit}.    
After rearranging the terms, one can show that the $AdS_5 \times S^5$ integrated vertex operators reduce to the flat space ones. 
In particular the term with $\eta^{\alpha \hat\alpha}$ in \eqref{TheUsuperfield} drops out.

\subsubsection{Plane-wave background}   \label{Plane-wave background} 

It is known that the plane-wave superalgebra can be obtained as a contraction of the $\mathfrak{psu}(2,2|4)$ superalgebra \cite{ppwaveII}. Notice that the plane-wave background also preserves thirty-two supersymmetries.  
The supergravity vertex operators on this background are discussed in 
\cite{ppwaveI} in terms of superfields. The contraction of the algebra is as follows. 
Define the combination of generators
\begin{equation}
E_+ = \Omega (\Delta + J) \, , \quad \quad 
E_- = \frac{1}{\Omega} (\Delta - J) \, , 
\end{equation}
where $\Omega$ is a parameter. Below, we will take $\Omega \rightarrow \infty$. 
The $E_-$ will be a central element of the resulting algebra and $E_+$ will act as an outer automorphism. 
 The boosts in these directions will also going to be rescaled as   
\begin{equation}
\widetilde{M}_{a \Delta} = \Omega \widehat{P}_{a}  \, , \quad  
\widetilde{M}_{a^{\prime} J} = \Omega \widehat{P}_{a^{\prime}} \, .   
\end{equation}
For the supercharges, it is convenient to break them in $SO(8)$ notation as 
\begin{equation}
q_{\alpha} = (q_a , q_{\dot{a}}) \, , \quad \quad 
q_{\widehat\alpha} = (\overline{q}_a , \overline{q}_{\dot{a}}) \, , 
\end{equation}
and define the rescaling 
\begin{equation}
S_{a} = \Omega^{\frac{1}{2}} q_a \, , \quad \overline{S}_{a} = \Omega^{\frac{1}{2}}  \overline{q}_a  \, , \quad  
S_{\dot{a}} =\Omega^{-\frac{1}{2}} q_{\dot{a}} \, , \quad  \overline{S}_{\dot{a}} =\overline{q}_{\dot{a}} \, . 
\end{equation}
The plane-wave superalgebra is obtained by substituting all the definitions above into the $\mathfrak{psu}(2,2|4)$ superalgebra \eqref{PSU algebra in SO} and taking the limit after decompositing the gamma matrices in $SO(8)$ notation, writing them in terms of the $SO(8)$ Pauli matrices $\sigma^{i}_{a \dot{a}}$ and their transpose $\sigma^{i}_{\dot{a} a}=(\sigma^{i}_{a \dot{a}})^T$, where $i =1, \ldots 8$. 
As discussed in \cite{ppwaveI}, the plane-wave superalgebra  has a $\mathbb{Z}_4$-graded subalgebra containing the generators
\begin{equation}
{\rm{Heisenberg \; superalgebra}}: \mathfrak{h}(8|8) \rtimes U(1) : \quad \{ \widehat{P}_{a} ,  \widehat{P}_{a^{\prime}}; S_{\dot{a}}; E_+, E_-, \widetilde{P}_{a}, \widetilde{P}_{a^{\prime}};   \bar{S}_{\dot{a}} \} ,   
\end{equation}    
which are the only generators appearing on the reduction of the $\mathfrak{psu}(2,2|4)$ Casimir and generating the supergravity supermultiplet. 

The Ramond-Ramond fluxes in the pure spinor formalism are organized in the 
superfield $P^{\alpha \widehat{\alpha}}$ appearing in the general pure spinor action \cite{DescentII}. Unlike the $AdS_5 \times S^5$ background, this superfield is not invertible in a plane-wave background (only one component of the self-dual five-form field strength is nonzero in this case) and the worldsheet variables $d_{\alpha}$ and $d_{\widehat\alpha}$ cannot be integrated out using their equations of motion. The BRST-like operator depends on these variables. 
In addition, the WZ term on the action cannot be written as a two-form as in the $AdS_5 \times S^5$ background. The BRST transformations in this background can be found in \cite{ppwaveI}. In particular, its action on a function of the coset element $g$ is the same as in 
\eqref{BRSTaction}.             

The unintegrated vertex operators are written only in terms of variables with zero conformal weight. They do not depend on the conformal weight one variables $\{d_{\alpha}, d_{\widehat\alpha}\}$. In the plane-wave background, the fermionic variables can be organized by their charges. It is consistent to assign positive charge to $(\theta^a, \bar\theta^a)$ and the unintegrated vertex operators are decomposed into terms of different charge. Let us call this charge the $S$-charge. It is argued in \cite{ppwaveI} that it is enough to compute the term of zero $S$-charge since the terms with nonzero $S$-charge can be obtained by application of global symmetries to it. The rescaling of the generators induces a rescaling 
of the variables such that the $AdS_5 \times S^5$ BRST transformations reduce to the plane-wave background when acting on the matter variables. Thus it is possible to take the limit of our vertex operators and eventually restrict to the set of zero $S$-charge to get the expansions of the superfields appearing in the plane-wave background. We leave the details of the computation for future work.

\subsubsection{Further comments about the gauge degrees of freedom}
\label{Further comments about the gauge degrees of freedom}

\begin{figure}[ht!]
\centering
\includegraphics{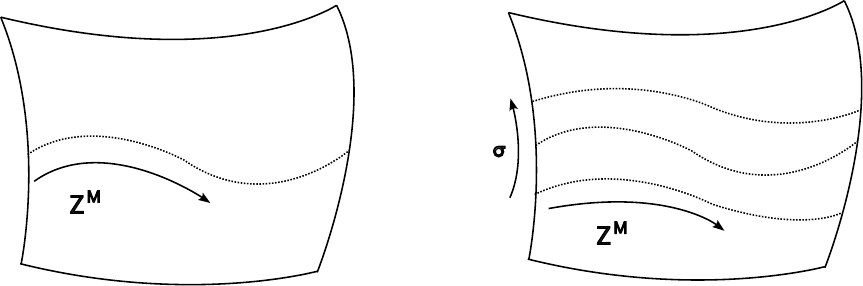}
\caption{Tubular neighborhood around a gauge slice. The $Z^M$ are gauge fixed worldsheet variables and the $\sigma$'s are the new coordinates introduced by the gauge transformation. }
\label{tubular}
\end{figure}

The vertex operators constructed in the Subsections \ref{vertexoperatorsfromchangingcoordinates} and \ref{recovering} are in a gauge-fixed form, i.e., 
they depend on the worldsheet variables parametrizing $g_{\rm{new}}$ of 
\eqref{CosetIdeal}. 
The gauge freedom was used to select a particular coset representative
  \begin{equation}
  g = g(Z) h \, ,
  \end{equation}
where $Z$ is the set of variables. As discussed in the Subsection \ref{Covariant superfields},
this gauge-fixing procedure can be reversed by applying a gauge transformation generated by $h = e^\sigma$ and then promoting the gauge parameters 
to be \emph{new degrees of freedom}.
This is precisely what was done in  \cite{BerkovitsCovariant,BerkovitsNewLimit}.
In particular,  
the action was written in a manifestly gauge invariant way by using 
$USp(2,2) \times USp(4)$ connections $\mathcal{A}^{\underline{ab}}, \bar{\mathcal{A}}^{\underline{ab}}$. These connections can be eventually integrated out to recover the usual worldsheet action of \eqref{action}.

In practical terms, these gauge transformations only ``rotate" the pure spinor variables $h^{-1}\cdot\lambda$, and so the gauge degrees of freedom $\sigma$ will only appear in the combination
\begin{equation}
(\Lambda_h)_I^R = V_I^{\widetilde{I}} h_{\widetilde I}^{\widetilde J} \lambda_{\widetilde{J}}^{\widetilde{R}} (h^{-1})^{\widetilde S}_{\widetilde R} (U^{-1})_{\widetilde{S}}^R \, , \quad \quad \quad 
(\bar\Lambda_h)^I_R = U_R^{\widetilde{R}} h_{\widetilde R}^{\widetilde S} \lambda^{\widetilde{I}}_{\widetilde{S}} (h^{-1})^{\widetilde J}_{\widetilde I} (V^{-1})_{\widetilde{J}}^I \, ,
\end{equation}
which follows from the definition of $\Lambda^R_I$ and $\overline\Lambda{}^I_S$ of   
\eqref{defLambda}. 
Furthermore, we can also trivially rewrite the embedding variables as
  \begin{equation}
  (X_h)_{RS} = U_R^{\widetilde R}\ h_{\widetilde R}^{\widetilde T}\ \Omega_{\widetilde T \widetilde U}\ h^{\widetilde U}_{\widetilde S}\ U^{\widetilde S}_{S} \quad ,\qquad (Y_h)_{IJ} = V_I^{\widetilde I}\ h^{\widetilde K}_{\widetilde I}\ \Omega_{\widetilde K\widetilde L}\ h^{\widetilde L}_{\widetilde J}\ V_J^{\widetilde J} \, . 
  \end{equation}

Geometrically, this procedure has the following interpretation. The vertex operators  
are only defined on a \emph{gauge slice} parametrized by the worldsheet variables $Z^M$. Then, by applying a gauge transformation, we transport the vertex operator everywhere in a tubular neighborhood of the gauge slice. This tubular neighborhood is now parametrized by $Z^M$ and the new degrees of freedom $\sigma$, 
see Figure \ref{tubular}.
The previous construction applied to 
our 
vertex operators leads exactly to the same expression for the vertices, where now the pure spinors $(\Lambda_h)^R_I$, $(\overline\Lambda_h)^I_R$ and embedding coordinates $(X_h)_{RS}$ and $(Y_h)_{IJ}$ are gauge-invariant objects containing the new degrees of freedom $\sigma$.

\subsubsection{Primary vertex operators}
\label{Primary vertex operators}

The unintegrated vertex operators can be gauge-fixed to be conformal primary fields with
zero worldsheet conformal dimension. 
This is equivalent to the vanishing of the double poles in the OPE's of the vertex operators and the energy-momentum tensor. 
In a flat background, computing this double-pole contribution is straightforward. Suppose that the massless vertex operator is in the gauge $V^{\rm{flat}} = \lambda^{\alpha}  \lambda^{\widehat\alpha} A^{\rm{flat}}_{\alpha \widehat\alpha}$. The condition of vanishing double poles is equivalent to 
$\partial^m \partial_m A^{\rm{flat}}_{\alpha \widehat\alpha} =0$.
In \cite{BerkovitsChandia}, the condition for the $AdS$ vertices to be conformal primary was analysed.
In the gauge \eqref{VonlyAAds}, 
the condition obtained due to the structure of the energy-momentum tensor is  
$\nabla^A \nabla_A A_{\alpha \widehat\alpha}=0$, here $A = \{ [\underline{a} \underline{b}], \underline \alpha, \underline{a}, \underline{\widehat{\alpha}}\}$.
If the superfields do not immediately satisfy this condition, it is always possible to modify them by adding $Q$-exact terms
such that the condition is satisfied.    
This condition can be derived analogously to the flat background case, by taking the leading-order OPE's of the $AdS$ energy-momentum tensor and the vertex operators. 
However, the computation is tedious. As argued in  \cite{BerkovitsChandia}, another way of deriving this condition is by analysing the $AdS$ isometries and imposing that the vertex operators have the correct flat space limit. Similar arguments can be applied to the vertex operators in the gauge \eqref{vertex10d}.
     
\pagebreak

\section{Discussion}
\label{Discussion}

In this paper, we have constructed multiple half-BPS vertex operators, both in unintegrated and integrated form, and in both the pure spinor and B-RNS-GSS formalisms. Their $SU(2,2)\times SU(4)$ transformations are now straightforward due to our choice of superspace parametrization. In fact, 
their transformations under $SU(2,2)\times SU(4)$ amount to changing the null polarization vectors $(x_0, y_0)$, 
as explained in Subsection \ref{recovering}. 
We have focused on the unintegrated vertex operators of the pure spinor formalism, since the B-RNS-GSS vertex operators and the pure spinor integrated vertex operators are obtained by acting with linear combinations of covariant derivatives on the unintegrated pure spinor ones. Our superspace parametrization is related to that used in \cite{VertexOld} by a reparametrization laid out in Subsection \ref{reparametrization old to new}.

Our main motivation for this work is the worldsheet computation of string scattering amplitudes in the $AdS_5\times S^5$ background. To compute such amplitudes, it is necessary to have vertex operators for each scattering state, together with a prescription that specifies how the zero modes are to be treated. For the bottom states of the supergravity multiplet, our work provides the required vertex operators. Moreover, 
as discussed in the Section \ref{The ideal coset vertices},
the vertex operators for the other states of the supergravity multiplet can be obtained through $PSU(2,2|4)$ transformations of our vertex operator, which are now straightforward due to our choice of parametrization. The next step is to establish the amplitude prescription, which we are currently undertaking in ongoing work \cite{workprogress}.
It should be mentioned that progress in computing amplitudes in other curved spaces was made recently, see \cite{TeseCassiano} for the $AdS_3$ case. However, the background considered was supported only 
by NS-NS fluxes, which admit left-right factorization.  

As mentioned in the Introduction, there are many CFT predictions for these scattering amplitudes. For two and three half-BPS states, the amplitudes are protected by supersymmetry and reduce to collections of propagators. Thus, once the amplitude prescription is established, the first step is to compute these protected amplitudes. The next step is the computation of four-point amplitudes, which can then be compared with non-trivial CFT predictions such as those in \cite{AldayHansen}. With the four-point amplitudes in hand, one can also study their factorization into poles to extract the massive spectrum.

Another application would be the computation, on the holographic dual side, of correlators between supersymmetric Wilson loops and half-BPS local operators  \cite{Giombi:2012ep}. Although attempts to apply supersymmetric localization to these computations using the pure spinor formalism have been made in the literature  \cite{LucasWilson} but yielded negative results, we believe that the approach taken in  \cite{LucasWilson} was naive. It considered the localizing equations to be simply the annihilation of all fermions by the sum of a supercharge and the BRST charge, while ignoring the choice of localizing fermion and the constraint that it must be annihilated by the square of the fermionic localizing charge. For example, \cite{LucasWilson} concludes that the supercharge must be such that its corresponding Killing spinor becomes a pure spinor somewhere inside $AdS_5 \times S^5$. We believe that this restriction is an artifact of the naive procedure of obtaining the supersymmetric localizing equations directly from the annihilation of the fermions by the localizing fermionic charge  \cite{LucasWilson2}.

In this work, we have only considered half-BPS vertex operators. It will be very interesting to construct more general vertex operators. The dimensions of the half-BPS operators are independent of the AdS radius, and the BRST operator, when acting on them, is 
just the covariant derivatives acting on the zero modes of the worldsheet variables. 
For constructing massive vertex operators, it is clear that $\alpha^{\prime}$ corrections need to be taken into account. In other words, it is necessary to evaluate the OPE's of the worldsheet variables. Unlike flat-space, the worldsheet theory in $AdS_5 \times S^5$  is interacting
and the OPE's could receive $\alpha^{\prime}$ corrections to all orders. 
Solving the cohomology at the massive level directly needs new strategies. 
One way would be to take the OPE's of the vertex operators constructed in this work.
In flat space, the DDF operators in the pure spinor formalism were constructed in \cite{DDFspectrum}.

\section*{Acknowledgement}

We would like to thank Luis Alday, Nathan Berkovits, Eggon Galvão, Andrei Mikhailov, Ulisses Portugal, Tobias Hansen, Luis Ypanaqué 
for very useful discussions.
We would like to thank Tobias Hansen for sharing with us powerful tensor structure codes. 
We would like to thank IIP, where part of this work was done. 
We would like to thank Serrapilheira Institute Grant (Serra–R-2012-38185), 
FAPESP Grants: 2021/09003-9 and 2018/07834-8, FUNPEC/UFRN/FRONTEIRAS DA CIÊNCIA, and 
Simons Foundation Grant (1023171 - RC) for financial support. 
TF thanks the CNRS for support via the IRP project NP-Strong.

\paragraph {AI disclosure:} The authors have used several versions of both ChatGPT and Claude  
to assist with the paper presentation and writing.  
The authors take full responsibility for its results. 

\appendix 

\section{The $\mathfrak{gl}(4|4)$ and $\mathfrak{psl}(4|4)$  superalgebras}
\label{apsupermatrices}

In this Appendix, we fix our conventions for the $\mathfrak{gl}(4|4)$ and $\mathfrak{psl}(4|4)$  Lie-superalgebras.
Note that only the former admits a supermatrix representation.

\subsection{$\mathfrak{gl}(4|4)$}

The superalgebra $\mathfrak{gl}(4|4)$ has 32 bosonic and 32 fermionic generators.  
The generators are organized in a supermatrix
\begin{equation}
M^A_B  =
\begin{bmatrix}
 M^{R}_{S}&{q}^{R}_{J}\\{q}^{I}_{S}&M^{I}_{J} 
\end{bmatrix} \, , 
\label{gl44}
\end{equation}
where the indices $R,S = 1\, , \ldots, 4$ are bosonic and the indices $I,J = 1\, , \ldots, 4$ are fermionic. 
The $\mathfrak{gl}(4,4)$ superalgebra contains a bosonic $\mathfrak{gl}(4)\oplus\mathfrak{gl}(4)$ subalgebra
\begin{equation}
[M^{I}_{J},M^{K}_{L}]=\delta^{K}_{J} M^{I}_{L} - \delta^{I}_{L} M^{K}_{J}\,,\qquad [M^{R}_{S},M^{T}_{U}]=\delta^{T}_{S} M^{R}_{U} - \delta^{R}_{U} M^{T}_{S}\,.
\end{equation}
The fermionic generators are in the (anti-)fundamental with respect to $\mathfrak{gl}(4)\oplus\mathfrak{gl}(4)$, i.e.  
\begin{equation}
[M^{R}_{S},{q}^{I}_{T}]=\delta^{R}_{T}{q}^{I}_{S}\,, \quad [M^{R}_{S},{q}^{T}_{I}]=-\delta^{T}_{S}{q}^{R}_{I}\, , \quad  
[M^{I}_{J},{q}^{K}_{R}]=\delta^{K}_{J}{q}^{I}_{R}\,, \quad [M^{I}_{J},{q}^{R}_{K}]=-\delta^{I}_{K}{q}^{R}_{J}\,.    
\end{equation}
Finally, the anti-commutators of the fermionic generators are given by 
\begin{equation}
\{{q}^{I}_{S},{q}^{R}_{J}\}=\delta^{I}_{J} M^{R}_{S} + \delta^{R}_{S} M^{I}_{J}.   
\end{equation}

As in the main text, the indices $I$ and $R$ split as 
$I = \{i
,  \overline i\}$ and     
$R = \{\alpha ,  
\dot \alpha
\}$. 
The matrix $M^A_B$ of \eqref{gl44} takes the form
\begin{equation}
M^A_B  =
\begin{bmatrix}
M^{\alpha}_{\beta} & P^{\alpha}_{\dot\beta} & q_j^{\alpha} & q_{\bar{j}}^{\alpha} \\
K^{\dot\alpha}_{\beta} & M^{\dot{\alpha}}_{\dot{\beta}} & q_j^{\dot{\alpha}} & q_{\bar{j}}^{\dot\alpha} \\
q_{\beta}^i & q_{\dot\beta}^i & M^{i}_j & K_{\bar{j}}^i \\
q_{\beta}^{\bar{i}} & q_{\dot\beta}^{\bar{i}} & P_{j}^{\bar{i}} & M^{\bar{i}}_{\bar{j}}  
\end{bmatrix} \, .
\label{gl44}
\end{equation}

In the expression above, the $P$'s are the translations and the $K$'s are the conformal boosts.
Next, we isolate the $U(1)_{\Delta}\times U(1)_{J}\times U(1)_{\mathbb{B}}$ generators and central charge $\mathbb{C}$, from the block diagonals
\begin{equation} 
\begin{aligned}
& M^{\alpha}_{\beta} = \widehat{M}^{\alpha}_{\beta} + \frac{1}{2} \delta^{\alpha}_{\beta} \Delta
+ \frac{1}{2} \delta^{\alpha}_{\beta} \mathbb{B} +\delta^{\alpha}_{\beta} \mathbb{C} \, , \quad \quad M^{\dot\alpha}_{\dot\beta}  = \widehat{M}^{\dot\alpha}_{\dot\beta} - \frac{1}{2} \delta^{\dot\alpha}_{\dot\beta} \Delta
+ \frac{1}{2} \delta^{\dot\alpha}_{\dot\beta} \mathbb{B} +  \delta^{\dot\alpha}_{\dot\beta}  \mathbb{C} \, , \\
& M^i_j =\widehat{M}^{i}_{j} + \frac{1}{2} \delta^{i}_{j} J
- \frac{1}{2} \delta^{i}_{j} \mathbb{B} +  \delta^{i}_{j} \mathbb{C} \, , \quad \quad \quad 
M^{\bar{i}}_{\bar{j}} =\widehat{M}^{\bar{i}}_{\bar{j}} - \frac{1}{2} \delta^{\bar{i}}_{\bar{j}} J
- \frac{1}{2} \delta^{\bar{i}}_{\bar{j}} \mathbb{B} +  \delta^{\bar{i}}_{\bar{j}} \mathbb{C}\, ,
\end{aligned} 
\label{TheHattedGenerators} 
\end{equation} 
where the $\widehat{M}$'s are the Lorentz generators, $\Delta$ is the dilatation generator and 
$J$ measures the $R$-charge with respect to a particular $U(1)$. $\mathbb{B}$ 
is 
the hypercharge. 
$\mathbb{B}$ and $\mathbb{C}$ 
are not generators of the $\mathfrak{psu}(4|4)$ superalgebra. The generator $\mathbb{C}$ 
commutes with all other generators, and
\begin{equation}
[ \Delta, M^A_B ] = C_{\Delta}(M^A_B) M^A_B \, , \quad [ J, M^A_B ] = C_{J}(M^A_B) M^A_B \, , 
\quad [ \mathbb{B}, M^A_B ] = C_{\mathbb{B}}(M^A_B) M^A_B \, , 
\end{equation}
with the charges 
\begin{table}[H]
\centering
\begingroup
\setlength{\tabcolsep}{8pt} 
\renewcommand{\arraystretch}{1.5}
\begin{tabular}{ccccccccc}
 & $K^{\dot{\alpha}}_{\alpha}$ & $P^{\alpha}_{\dot{\alpha}}$ & $K^i_{\bar{i}}$ & $P^{\bar{i}}_i$ & $(q^i_{\alpha}, q^{\dot\alpha}_{\bar{i}})$ & $(q^{\bar{i}}_{\dot\alpha},q^{\alpha}_i)$ & $(q^i_{\dot{\alpha}}, q^{\alpha}_{\bar{i}} ) $ & $ (q^{\bar{i}}_{\alpha}, q^{\dot\alpha}_i) $ \\ 
 \hline 
 $C_{J}:$ &  0 & 0 & 1 & -1 & $ \frac{1}{2}$ & $- \frac{1}{2} $ & $\frac{1}{2}$ & $-\frac{1}{2}$ \\
 $C_{\Delta}:$ &  -1 & 1 & 0 & 0 & $- \frac{1}{2}$ & $ \frac{1}{2} $ & $\frac{1}{2}$ & $- \frac{1}{2}$   \\
 $C_{\mathbb{B}}:$ &  0 & 0 & 0 & 0 & $ (-\frac{1}{2},\frac{1}{2}) $ & $(-\frac{1}{2},\frac{1}{2})$ & $(-\frac{1}{2},\frac{1}{2})$ & $(-\frac{1}{2},\frac{1}{2})$
\end{tabular}
\endgroup
\caption{The charge of the generators under $\Delta$, $J$ and $\mathbb{B}$. The
generators not appearing in the first row have charge zero with respect to all the  $\Delta$, $J$ and $\mathbb{B}$.} 
 \label{tab:ChargesDJ}
\end{table}

\subsection{$\mathfrak{psl}(4|4)$} 
\label{psl44su22}

The $\mathfrak{sl}(4|4)$ subalgebra is obtained from the  
 $\mathfrak{gl}(4|4)$ superalgebra by imposing a constraint on the generators. 
 The matrices of $\mathfrak{sl}(4|4)$ have vanishing supertraces, 
 \begin{equation}
{\rm{Str}} \, M =  M^{R}_{R} - M^I_I =0\,,    
\end{equation}
which eliminates $\mathbb{B}$, see \eqref{TheHattedGenerators}. The $\mathfrak{psl}(4|4)$ superalgebra is obtained from the $\mathfrak{sl}(4|4)$ superalgebra by gauging
\begin{equation}
g\cong \ell g\,, 
\end{equation}
such that at the level of generators, one has
\begin{equation}
M^{A}_{B}\cong M^{A}_{B}+\delta^{A}_{B}\mathbb{C}\,,   
\end{equation}
which eliminates $\mathbb{C}$. Because of the gauging, $\mathfrak{psl}(4|4)$ is not an algebra of constrained supermatrices.
For the reader's convenience, we are going to write all the nonzero $\mathfrak{psl}(4|4)$ commutators below in our conventions, 
\begin{equation}
\label{firstlinecommutator}
\begin{aligned}
[\Delta, P^{\alpha}_{\dot\alpha}]= P^{\alpha}_{\dot\alpha}\,,\qquad & [\Delta,K^{\dot\alpha}_{\alpha}]= -K^{\dot\alpha}_{\alpha}\,, \\
 [\Delta,{q}^{R}_{I}]=\frac{1}{2}\delta^{R}_{\alpha}{q}^{\alpha}_{I}-\frac{1}{2}\delta^{R}_{\dot\alpha}{q}^{\dot\alpha}_{I}\,,\qquad & [\Delta,{q}^{I}_{R}]=-\frac{1}{2}\delta^{\alpha}_{R}{q}^{I}_{\alpha}+\frac{1}{2}\delta^{\dot\alpha}_{R}{q}^{I}_{\dot\alpha}\,, \\
[J,P^{\overline i}_{i}]= -P^{\overline i}_{i}\,,\qquad  & [J, K^{i}_{\overline i}]=K^{i}_{\overline i}\,, \\
[J,{q}^{I}_{R}]=\frac{1}{2}\delta^{I}_{i}{q}^{i}_{R}-\frac{1}{2}\delta^{I}_{\overline i}{q}^{\overline i}_{R}\,,\qquad & [J,{q}^{R}_{I}]=-\frac{1}{2}\delta^{i}_{I}{q}^{R}_{i}+\frac{1}{2}\delta^{\overline i}_{I}{q}^{R}_{\overline i}\, .
\end{aligned} 
\end{equation}
The commutators involving the generators $\widehat{M}^R_S$ are 
\begin{equation}
\begin{aligned}
[\widehat  M^{\alpha}_{\beta},\widehat  M^{\gamma}_{\delta}]=\delta^{\gamma}_{\beta}\widehat  M^{\alpha}_{\delta}-\delta^{\alpha}_{\delta}\widehat  M^{\gamma}_{\beta}\,,\qquad & [\widehat  M^{\dot\alpha}_{\dot\beta},\widehat  M^{\dot\gamma}_{\dot\delta}]=\delta^{\dot\gamma}_{\dot\beta}\widehat  M^{\dot\alpha}_{\dot\delta}-\delta^{\dot\alpha}_{\dot\delta}\widehat  M^{\dot\gamma}_{\dot\beta}    \,, \\
[\widehat  M^{\alpha}_{\beta},{q}^{R}_{I}]=\delta^{R}_{\beta}{q}^{\alpha}_{I}-\frac{1}{2}\delta^{\alpha}_{\beta}{q}^{R}_{I
}\,,  \qquad  & [\widehat  M^{\alpha}_{\beta},{q}^{I}_{R}]=-\delta^{\alpha}_{R}{q}^{I}_{\beta}+\frac{1}{2}\delta^{\alpha}_{\beta}{q}^{I}_{R}\,, \\
[\widehat  M^{\dot\alpha}_{\dot\beta},{q}^{R}_{I}]=\delta^{R}_{\dot\beta}{q}^{\dot\alpha}_{I}-\frac{1}{2}\delta^{\dot\alpha}_{\dot\beta}{q}^{R}_{I}  \,,\qquad & [\widehat  M^{\dot\alpha}_{\dot\beta},{q}^{I}_{R}]=-\delta^{\dot\alpha}_{R}{q}^{I}_{\dot\beta}+\frac{1}{2}\delta^{\dot\alpha}_{\dot\beta}{q}^{I}_{R}\,, \\
[\widehat  M^{\alpha}_{\beta},P^{\gamma}_{\dot\gamma}]=\delta^{\gamma}_{\beta}P^{\alpha}_{\dot\gamma}-\frac{1}{2}\delta^{\alpha}_{\beta}P^{\gamma}_{\dot\gamma}\,,\qquad   &  [\widehat  M^{\dot \alpha}_{\dot \beta},P^{\gamma}_{\dot\gamma}]=-\delta^{\dot\alpha}_{\dot\gamma}P^{\gamma}_{\dot\beta}+\frac{1}{2}\delta^{\dot \alpha}_{\dot \beta}P^{\gamma}_{\dot\gamma} \,,
\\
[\widehat  M^{\dot\alpha}_{\dot\beta},K^{\dot\gamma}_{\gamma}]=\delta^{\dot\gamma}_{\dot\beta}K^{\dot\alpha}_{\gamma}-\frac{1}{2}\delta^{\dot\alpha}_{\dot\beta}K^{\dot\gamma}_{\gamma}\,,\qquad    & [\widehat  M^{ \alpha}_{ \beta},K^{\dot\gamma}_{\gamma}]=-\delta^{\alpha}_{\gamma}K^{\dot\gamma}_{\beta} +\frac{1}{2}\delta^{ \alpha}_{ \beta}K^{\dot\gamma}_{\gamma}\,,
\end{aligned}
\end{equation}
and the ones involving the generators $\widehat{M}^I_J$ are 
\begin{equation}
\begin{aligned} 
[\widehat  M^{i}_{j},\widehat  M^{k}_{l}]=\delta^{k}_{j}\widehat  M^{i}_{l}-\delta^{i}_{l}\widehat  M^{k}_{j}\,,\qquad & [\widehat  M^{\bar i}_{\bar j},\widehat  M^{\bar k}_{\bar l}]=\delta^{\bar k}_{\bar j}\widehat  M^{\bar i}_{\bar l}-\delta^{\bar i}_{\bar l}\widehat  M^{\bar k}_{\bar j} \, ,    \\
[\widehat  M^{i}_{j},{q}^{I}_{R}]=\delta^{I}_{j}{q}^{i}_{R}-\frac{1}{2}\delta^{i}_{j}{q}^{I}_{R
}    \,,\qquad &  [\widehat  M^{i}_{j},{q}^{R}_{I}]=-\delta^{i}_{I}{q}^{R}_{j}+\frac{1}{2}\delta^{i}_{j}{q}^{R}_{I}\,, \\
[\widehat  M^{\bar i}_{\bar j},{q}^{I}_{R}]=\delta^{I}_{\bar j}{q}^{\bar i}_{R}-\frac{1}{2}\delta^{\bar i}_{\bar j}{q}^{I}_{R} \,,\qquad & [\widehat  M^{\bar i}_{\bar j},{q}^{R}_{I}]=-\delta^{\bar i}_{I}{q}^{R}_{\bar j}+\frac{1}{2}\delta^{\bar i}_{\bar j}{q}^{R}_{I}\,,   
\\
[\widehat  M^{i}_{ j},K^{k}_{\bar k}]=\delta^{k}_{ j}K^{ i}_{\bar k}-\frac{1}{2}\delta^{ i}_{ j}K^{k}_{\bar k}\,,\qquad   
& [\widehat  M^{\bar   i}_{\bar   j},K^{k}_{\bar k}]=-\delta^{\bar  i}_{\bar k}K^{k}_{\bar  j}+\frac{1}{2}\delta^{\bar   i}_{\bar   j}K^{k}_{\bar k} \, , \\
[\widehat  M^{\bar  i}_{\bar  j},P^{\bar k}_{k}]=\delta^{\bar k}_{\bar  j}P^{\bar  i}_{k}-\frac{1}{2}\delta^{\bar  i}_{\bar  j}P^{\bar k}_{k}\,,\qquad   &  [\widehat  M^{  i}_{  j},P^{\widehat k}_{k}]=-\delta^{ i}_{k}P^{\widehat k}_{ j} +\frac{1}{2}\delta^{  i}_{  j}P^{\widehat k}_{k} \, . 
\end{aligned} 
\end{equation}
The remaining nonzero commutators involving bosonic generators are 
\begin{equation}
\begin{aligned} 
& [K^{\dot\alpha}_{\beta},q^{R}_{I}]=\delta^{R}_{\beta}q^{\dot\alpha}_{I}\,,\qquad [K^{\dot\alpha}_{\beta},q^{I}_{R}]=-\delta^{\dot\alpha}_{R}q^{I}_{\beta}\,, \\
& [P^{\alpha}_{\dot\beta},q^{R}_{I}]=\delta^{R}_{\dot\beta}q^{\alpha}_{I}\,,\qquad [P^{\alpha}_{\dot\beta},q^{I}_{R}]=-\delta^{\alpha}_{R}q^{I}_{\dot\beta}\,,    \\
& [K^{\dot\alpha}_{\alpha},P^{\beta}_{\dot\beta}]  = \delta^{\beta}_{\alpha} \widehat M^{\dot\alpha}_{\dot\beta} - \delta^{\dot\alpha}_{\dot\beta}\widehat M^{\beta}_{\alpha} - \delta^{\dot\alpha}_{\dot\beta}\delta^{\beta}_{\alpha} \Delta \, ,     
\label{TheKPqalgebra}
\end{aligned}
\end{equation}
and 
\begin{equation}
\begin{aligned} 
& [K^{i}_{\bar j},q^{I}_{R}]=\delta^{I}_{\bar j}q^{i}_{R}\,,\qquad [K^{i}_{\bar j},q^{R}_{I}]=-\delta^{i}_{I}q^{R}_{\bar j}\,,    \\
& [P^{\bar i}_{j},q^{I}_{R}]=\delta^{I}_{j}q^{\bar i}_{R}\,,\qquad [P^{\bar i}_{j},q^{R}_{I}]=-\delta^{\bar i}_{I}q^{R}_{j}\,,  \\  
& [K^{i}_{\bar i},P^{\bar j}_{j}]=\delta^{\bar j}_{\bar i} \widehat M^{i}_{j} - \delta^{i}_{j} \widehat M^{\bar j}_{\bar i} + \delta^{\bar j}_{\bar i}\delta^{i}_{j} J \, .     
\label{TheKPqalgebraI}
\end{aligned}
\end{equation}
Finally, the anticommutators of the fermionic generators are given by  
\begin{equation}
\begin{aligned} 
\label{algebra}
& \qquad \qquad \qquad \qquad \qquad \qquad \qquad   \{q^{R}_{J},q^{I}_{S}\}=   \\
& =\delta^{I}_{J}\left[\delta^{R}_{\alpha}\delta_{S}^{\beta}\left(\widehat  M^{\alpha}_{\beta}+\frac{1}{2}\Delta\delta^{\alpha}_{\beta}\right)  +\delta^{R}_{\dot\alpha}\delta_{S}^{\dot\beta}\left(\widehat  M^{\dot\alpha}_{\dot\beta}-\frac{1}{2}\Delta \delta^{\dot\alpha}_{\dot\beta}\right)+\delta^{R}_{\dot\alpha}\delta_{S}^{\beta} K^{\dot\alpha}_{\beta}+\delta^{R}_{\alpha}\delta_{S}^{\dot\beta} P^{\alpha}_{\dot\beta}\right]  \\   
& +\delta^{R}_{S}\left[\delta^{I}_{i}\delta_{J}^{j}\left(\widehat  M^{i}_{j}+\frac{1}{2}J\delta^{i}_{j}\right)  +\delta^{I}_{\bar i}\delta_{J}^{\bar j}\left(\widehat  M^{\bar i}_{\bar j}-\frac{1}{2}J \delta^{\bar i}_{\bar \beta}\right)+\delta^{I}_{\dot\alpha}\delta_{J}^{j} P^{\bar i}_{j}+\delta^{I}_{i}\delta_{J}^{\bar j} K^{i}_{\bar j}\right] \, . 
\end{aligned} 
\end{equation}

In our conventions the isotropy group $USp(2,2) \times USp(4)$ of the coset \eqref{supercoset} is parametrized by the following 
combination of generators ($\varepsilon_{12}=\varepsilon^{12}=1$), 
\begin{equation}
\{ \widehat{M}^{\alpha}_{\beta} , \, \widehat{M}^{\dot\alpha}_{\dot\beta}, \, (P^{\alpha}_{\dot\alpha} +\varepsilon_{\dot\alpha \dot\beta} \varepsilon^{\alpha \beta} 
K_{\beta}^{\dot\beta})  \}  \; \times \; \{ \widehat{M}^{i}_{j} , \, \widehat{M}^{\bar{i}}_{\bar{j}}, \, (P^{\bar{i}}_{i} +\varepsilon_{i j} \varepsilon^{\bar{i} \bar{j}} 
K_{\bar{j}}^{j})  \} \, .
\end{equation} 
Removing the generators of the isotropy group, the coset locally retains ten bosonic generators, which 
is the correct dimension of the bosonic subspace of $AdS_5 \times S^5$.

\subsection{The $\mathfrak{psu}(4|4)$ superalgebra in 10d notation}

In this subsection, the $\mathfrak{psu}(4|4)$ superalgebra is presented in 10d notation. Our conventions differ by a few signs and factors of $i$ from those of \cite{IntegratedVertex}. This change of notation is important for writing the integrated vertex operators compactly. 
Recall that the $SO(10)$ vector indices $\underline{a}$ are split into $AdS_5$ indices $a$ and $S^5$ indices $a'$. The bosonic generators in 10d notation are denoted as $\{\widetilde{M}_{ab}, \widetilde{M}_{a^{\prime}b^{\prime}}, \widetilde{P}_{\underline{a}} \}$. The fermionic ones are $\{q_{\alpha}, q_{\widehat\alpha} \}$. The algebra is given by  
\begin{equation}
\label{PSU algebra in SO}
[\widetilde{M}_{\underline{ab}},\widetilde{M}_{\underline{cd}}]=\eta_{\underline{a}[\underline{c}}\widetilde{M}_{\underline{d}]\underline{b}}-\eta_{\underline{b}[\underline{c}} \widetilde{M}_{\underline{d}]\underline{a}}\,, \qquad [\widetilde{P}_{a},\widetilde{P}_{b}]=\widetilde{M}_{ab}\,,\qquad [\widetilde{P}_{a'},\widetilde{P}_{b'}]=-\widetilde{M}_{a'b'}\,,   
\end{equation}
\begin{equation*}
[\widetilde{P}_{\underline{a}},\widetilde{M}_{\underline{bc}}]= \eta_{\underline{a}[\underline{b}}\widetilde{P}_{\underline{c}]}    \,,\qquad [\widetilde{M}_{\underline{ab}},q_{\alpha}]=-\frac{1}{2}\gamma_{\underline{ab}\alpha}\,^{\beta}q_{\beta}\,,\qquad [\widetilde{M}_{\underline{ab}},q_{\widehat\alpha}]=-\frac{1}{2}\gamma_{\underline{ab}\widehat\alpha}\,^{\widehat\beta}q_{\widehat\beta}\,,
\end{equation*}
\begin{equation*}
[\widetilde{P}_{\underline{a}},q_{\alpha}]=\frac{i}{2}(\gamma_{\underline{a}})_{\alpha \beta} \eta^{\beta \widehat\beta}q_{\widehat\beta}\,,\qquad [\widetilde{P}_{\underline{a}},q_{\widehat\alpha}]=-\frac{i}{2}(\gamma_{\underline{a}})_{\widehat\alpha \widehat\beta} \eta^{\beta \widehat\beta}q_{\beta}\,,    
\end{equation*}
\begin{equation*}
\{q_{\alpha},q_{\beta}\}=\gamma^{\underline{a}}_{\alpha\beta}\widetilde{P}_{\underline{a}}\,,\qquad \{q_{\widehat\alpha},q_{\widehat\beta}\}=\gamma^{\underline{a}}_{\widehat\alpha\widehat\beta}\widetilde{P}_{\underline{a}} \,,   
\end{equation*}
\begin{equation*}
\{q_{\alpha},q_{\widehat\alpha}\}=-\frac{i}{2}\left[(\gamma^{ab})_{\alpha}\,^{\beta} \eta_{\beta \widehat\alpha}  \widetilde{M}_{ab}-(\gamma^{a'b'})_{\alpha}\,^{\beta} \eta_{\beta \widehat\alpha}  \widetilde{M}_{a'b'}\right] 
\equiv - \frac{i}{2} \left[ (\gamma^{[ab]}\eta)_{\alpha\widehat\alpha}\widetilde{M}_{[ab]} \right]\,,  
\end{equation*}
where 
$\eta_{\alpha \widehat\beta} = -\gamma^{12345}_{\alpha \widehat\beta} = - \eta_{\widehat\beta \alpha}$ and 
\begin{equation}
(\gamma_{\underline{a} \underline{b}})_{\alpha}\,^{\beta} = \frac{1}{2} (\gamma_{\underline{a} \alpha \gamma} \gamma_{\underline{b}}^{\gamma \beta} - \gamma_{\underline{b} \alpha \gamma} \gamma_{\underline{a}}^{\gamma \beta}) \, , 
\end{equation}
and similarly for the hatted indices. The expression for $\eta^{\alpha\widehat{\beta}} = - \eta^{\widehat{\beta} \alpha}$ can be derived from  
\begin{equation}
\label{propertieseta} 
\eta^{\alpha\widehat{\beta}} \,\eta_{\widehat{\beta}\rho} \;=\; \delta^\alpha_{\rho},
\quad \quad 
\eta^{\widehat{\alpha}\beta} \,\eta_{\beta\widehat{\gamma}} \;=\; \delta^{\widehat{\alpha}}_{\widehat{\gamma}}.
\end{equation}
Recall that the gamma matrices above satisfy the algebra 
\begin{equation}
(\gamma^{\underline{a}})_{\alpha \beta} (\gamma^{\underline{b}})^{\beta \gamma} +
(\gamma^{\underline{b}})_{\alpha \beta} (\gamma^{\underline{a}})^{\beta \gamma} = 2 \eta^{\underline{ab}} \delta^{\gamma} _{\alpha} \, . 
\end{equation}
Notice that the gamma matrices with hatted and unhatted indices are related as follows
\begin{equation}
\begin{aligned}
&(\gamma^{a})^{\alpha \beta} =  \, \eta^{\alpha \widehat\alpha}  \eta^{\beta \widehat\beta} 
(\gamma^{a})_{\widehat\alpha \widehat\beta} \, , \quad  \quad \, 
(\gamma^{a})_{\alpha \beta} =  \, \eta_{\alpha \widehat\alpha}  \eta_{\beta \widehat\beta} 
(\gamma^{a})^{\widehat\alpha \widehat\beta} \, , \\
&(\gamma^{a'})^{\alpha \beta} =  \, - \eta^{\alpha \widehat\alpha}  \eta^{\beta \widehat\beta} 
(\gamma^{a'})_{\widehat\alpha \widehat\beta} \, , \quad  
(\gamma^{a'})_{\alpha \beta} =  \, - \eta_{\alpha \widehat\alpha}  \eta_{\beta \widehat\beta} 
(\gamma^{a'})^{\widehat\alpha \widehat\beta} \, . 
\label{gammaAdSS}
\end{aligned}
\end{equation}
This implies that the last commutator of \eqref{PSU algebra in SO} can also be written in terms of gamma matrices with hatted
indices by using    
\begin{equation}
(\gamma^{\underline{ab}})_{\beta}\,^{\alpha} \eta_{\alpha \widehat{\alpha}} 
= -  (\gamma^{\underline{ab}})_{\widehat\alpha}\,^{\widehat\beta}   \eta_{\beta \widehat{\beta}} \, , 
\end{equation}
which follows from applying \eqref{gammaAdSS}. 
The gamma matrices satisfy several identities.
Many of them were used in the Subsection \ref{The integrated vertex operator} for establishing the descent equations. In addition, they are useful for establishing  the super Jacobi identities   
\begin{equation}
(-1)^{ A C} [T_A,[T_B,T_C]] + 
(-1)^{ A B} [T_B,[T_C,T_A]] +
(-1)^{ C B} [T_C,[T_A,T_B]]  = 0 \, . 
\end{equation}
For example,  
\begin{equation}
\{ Q_{\alpha}, \{ Q_{\beta}, Q_{\gamma} \} \} 
+ \{ Q_{\beta}, \{ Q_{\gamma}, Q_{\alpha} \} \} +
\{ Q_{\gamma}, \{ Q_{\alpha}, Q_{\beta} \} \} =0 \, , 
\end{equation}
follows from 
\begin{equation}
\gamma^{ \underline{a}}_{( \alpha \beta} (\gamma_{\underline{a}})_{\gamma) \delta} =0 \, . 
\end{equation}
Moreover 
\begin{equation}
\{ Q_{\alpha} ,\{ Q_{\beta}, Q_{\widehat\beta} \} \} + \{ Q_{\beta}, \{ Q_{\alpha}, Q_{\widehat\beta} \} \}+\{ Q_{\widehat\beta}, \{ Q_{\beta}, Q_{\alpha} \} \} =0 \, , 
\end{equation}
is a consequence of 
\begin{equation}
2  (\gamma_{a})_{\alpha \beta} (\gamma^{a})^{\nu \rho}- 2  (\gamma_{a^{\prime}})_{\alpha \beta} (\gamma^{a^{\prime}})^{\nu \rho} + (\gamma^{[ab]})_{(\beta}\,^{\nu}(\gamma_{[ab]})_{\alpha)}\,^{\rho} =0 \, . 
\end{equation}

We end this Subsection by listing a few more useful gamma matrices identities 

\begin{equation}
\gamma^{a_1 \ldots a_p}_{\alpha \beta} \gamma_{b_1 \ldots b_p}^{\alpha \beta} = 16  p! \delta^{a_1 \ldots a_p}_{b_1 \ldots b_p} \, , \quad \quad 
\gamma^{a_1 \ldots a_p \, \alpha}\,_{\beta} \gamma_{b_1 \ldots b_p}\,^{\beta}\,_{\alpha} = 16  p! \delta^{a_1 \ldots a_p}_{b_1 \ldots b_p} \, , 
\end{equation}
and 
\begin{equation}
\gamma_{ab} \gamma_{\underline{cdefg}} \gamma^{ab} - \gamma_{a^{\prime}b^{\prime}} \gamma_{\underline{cdefg}} \gamma^{a^{\prime}b^{\prime}}=0 \, . 
\end{equation}

\subsection{Relating the notations: 6d Pauli Matrices}
\label{6dPauliMatrices}

In order to relate the two notations of the $\mathfrak{psu}(4|4)$ superalgebra given in the two previous Subsections, the
6d Pauli Matrices will be used. Here, we set our conventions for these matrices. 
The matrices are given by  
 \begin{equation}
\sigma_{RS}^1 =
\left(\begin{tabular}{cccc}
0 & 0 & 1 & $0$ \\
0 & 0 & $0$ & 1 \\
-1 & $0$ & 0 & 0 \\
$0$ & -1 & 0 & 0
\end{tabular} \right) \, ,  \quad
\sigma_{RS}^2 =
\left( \begin{tabular}{cccc}
0 & 0 & $0$ & $i$ \\
0 & 0 & $i$ & $0$ \\
$0$ & -$i$ & 0 & 0 \\
-$i$ & $0$ & 0 & 0
\end{tabular} \right) \, ,
\label{sigma6matrices}
\end{equation}
\[
\sigma_{RS}^3 =
\left(\begin{tabular}{cccc}
0 & $0$ & 0 & 1 \\
$0$ & 0 & -1 & 0 \\
0 & 1 & 0 & $0$ \\
-1 & 0 & $0$ & 0
\end{tabular} \right) \, , \quad
\sigma_{RS}^4 =
\left( \begin{tabular}{cccc}
0 & 0 & $i$ & $0$ \\
0 & 0 & 0 & -$i$ \\
-$i$ & $0$ & 0 & 0 \\
0 & $i$ & 0 & 0
\end{tabular} \right) \, ,
\]
\[
\sigma_{RS}^5 =
\left( \begin{tabular}{cccc}
0 & $i$ & 0 & 0 \\
-$i$ & 0 & 0 & 0 \\
0 & 0 & 0 & $i$ \\
0 & 0 & -$i$ & 0
\end{tabular} \right) \, , \quad
\sigma_{RS}^6 =
\left( \begin{tabular}{cccc}
0 & 1 & 0 & 0 \\
$-1$ & 0 & 0 & 0 \\
0 & 0 & 0 & $-1$ \\
0 & 0 & 1 & 0
\end{tabular} \right) \, .
\]
Notice that $\sigma^{\mathsf{I}}_{RS} = - \sigma^{\mathsf{I}}_{SR}$. The matrices above are so named because the 6d eight by eight gamma matrices 
in the Weyl basis are written in terms of them as 
\begin{equation}
\Gamma^{\mathsf{I}} = \left( \begin{tabular}{cc}
0 &  $\sigma_{RS}^{\mathsf{I}}$ \\
$\sigma^{\mathsf{I}RS}$ & 0  \\
\end{tabular} \right) \, , 
\end{equation}
with $\mathsf{I},\mathsf{J}=1, \ldots,6$. The matrices with indices $R,S$ above are related to the previous ones as ($\varepsilon_{1234}=\varepsilon^{1234}=1$), 
\begin{equation}
\sigma^{\mathsf{I}RS} = \frac{1}{2} \varepsilon^{RSTU} \sigma^{\mathsf{I}}_{TU} \, , \quad \quad 
\sigma^{\mathsf{I}}_{RS} = \frac{1}{2} \varepsilon_{RSTU} \sigma^{\mathsf{I}TU} \, . 
\end{equation} 
The gamma matrices obey the Clifford algebra $\{ \Gamma^\mathsf{I}, \Gamma^\mathsf{J} \} = 2 \delta^{\mathsf{I} \mathsf{J}}$ and this is equivalent to 
\begin{equation}
\sigma^{\mathsf{I} RS} \sigma^{\mathsf{J}}_{ST} + \sigma^{\mathsf{J} RS} \sigma^{\mathsf{I}}_{ST} = 2 \delta^R_T  \delta^{\mathsf{I} \mathsf{J}} \,.  
\end{equation}
   
Using the notation 
\[
(\sigma^{\mu})_{\alpha \dot\beta} = (\mathbbold{1}, i \sigma^{i}_{\alpha \dot\beta}) \, , \quad {\rm{and}} \quad
(\sigma^{\mu})^{\dot\beta \alpha} = \varepsilon^{\dot\beta \dot\gamma} \varepsilon^{\alpha \gamma} \sigma^{\mu}_{\gamma \dot\gamma} \, , 
\]
where $\sigma^{i}_{\alpha \dot\beta}$ are the Pauli Matrices with $i=1,2,3$. The matrices of \eqref{sigma6matrices} can be written in block matrix notation as 
\begin{equation}
\sigma^{\mu}_{RS} =\left( \begin{tabular}{cc}
0 & $\sigma^{\mu}_{\alpha \dot\beta}$  \\
$-(\sigma^{\mu}_{\alpha \dot\beta})^T$ & 0 \\
\end{tabular} \right) \, , \quad 
\sigma^{5}_{RS} = i \left( \begin{tabular}{cc}
$\varepsilon_{\alpha \beta}$  & 0  \\
0 & $\varepsilon_{\dot\alpha \dot\beta}$ \\
\end{tabular} \right) \, , \quad 
\sigma^{6}_{RS} =\left( \begin{tabular}{cc}
$\varepsilon_{\alpha \beta}$ & 0  \\
0 & -$\varepsilon_{\dot\alpha \dot\beta}$  \\
\end{tabular} \right) \, . 
\label{sigma6d}
\end{equation}
 
\subsection{Relating the notations: bosonic generators}
The map between the bosonic generators of the $\mathfrak{psu}(4|4)$ algebra is straightforward in the $\mathfrak{gl}(4|4)$ notation of \eqref{gl44}, 
\begin{equation}
\label{10dtoSU22}
\widetilde{P}_{a} = \frac{1}{4} \delta_{a}^{[RS]}(M \pm \Delta)^{T}\,_{[R}\sigma^6_{S] T}\,,\qquad \widetilde{P}_{a'}= \frac{i}{4}  \delta_{a'}^{[IJ]}(M \pm J)^{K}\,_{[I}\sigma^6_{J]K}\,,    
\end{equation}
\begin{equation*}
\widetilde{M}_{[ab]}= \frac{1}{4} \delta_{[ab]}^{(RS)}( M \pm \Delta)^{T}\,_{(R}\sigma^6_{S)T}\,,\qquad    \widetilde{M}_{[a'b']}= \frac{1}{4} \delta_{[a'b']}^{(IJ)}( M \pm J)^{K}\,_{(I}\sigma^6_{J)K}\,,
\end{equation*}
where $[ ,]$ means anti-symmetrization and $(,)$ means symmetrization, 
and $\sigma^6_{RS}$ was defined in \eqref{sigma6d} together
with the other 6d Pauli Matrices used to change the representations. 
The signs $\pm$ follow from \eqref{TheHattedGenerators}. The notation $M \pm \Delta$ 
corresponds to the first two terms on the right side of \eqref{TheHattedGenerators}.
The indices above are split as $R,S,T=(\alpha,\dot\alpha)$ and $I,J,K=(i,\overline i)$. 
First, let us focus on the $\mathfrak{usp}(2|2)$ subalgebra.
The right-hand side of \eqref{10dtoSU22} decomposes as 
\begin{equation}
(M \pm \Delta)^{R}\,_{[\alpha}\sigma^6_{\beta]R}=-  \Delta\varepsilon_{\alpha\beta}\,,\qquad (M \pm \Delta)^{R}\,_{[\dot\alpha}\sigma^6_{\dot\beta]R}=-  \Delta\varepsilon_{\dot\alpha\dot\beta}\,,    
\end{equation}
\begin{equation*}
( M \pm \Delta)^{R}\,_{[\dot\alpha}\sigma^6_{\beta]R}= \frac{1}{2}(P^{\gamma}_{\dot\alpha}\varepsilon_{\beta\gamma} + K^{\dot\gamma}_{\beta}\varepsilon_{\dot\alpha\dot\gamma})\,,\qquad ( M \pm \Delta)^{R}\,_{[\alpha}\sigma^6_{\dot\beta]R}= -\frac{1}{2}(P^{\gamma}_{\dot\beta}\varepsilon_{\alpha\gamma}+K^{\dot\gamma}_{\alpha}\varepsilon_{\dot\beta\dot\gamma})\,,   
\end{equation*}
and
\begin{equation}
( M \pm \Delta)^{R}\,_{(\alpha}\sigma^6_{\beta)R}=  \widehat{M}^{\gamma}_{(\alpha}\varepsilon_{\beta)\gamma}\,,\qquad ( M \pm \Delta)^{R}\,_{(\dot\alpha}\sigma^6_{\dot\beta)R}= - \widehat{M}^{\dot\gamma}_{(\dot\alpha}\varepsilon_{\dot\beta)\dot\gamma}\,,    
\end{equation}
\begin{equation*}
( M \pm \Delta)^{R}\,_{(\dot\alpha}\sigma^6_{\beta)R}=\frac{1}{2}(P^{\gamma}_{\dot\alpha}\varepsilon_{\beta\gamma}-K^{\dot\gamma}_{\beta}\varepsilon_{\dot\alpha\dot\gamma})\,.
\end{equation*}
Thus one has the following relations between generators
\begin{equation}
\widetilde{P}_{\alpha\dot\alpha}= -\frac{1}{2} (P^{\gamma}_{\dot\alpha} \varepsilon_{\alpha\gamma}+ K^{\dot\gamma}_{\alpha} \varepsilon_{\dot\alpha\dot\gamma})  \,,\qquad \widetilde{P}_{5}=- \Delta \,,
\label{widenonwide}
\end{equation}
\begin{equation*}
4 \widetilde{M}_{\alpha\dot\alpha\beta\dot\beta}= - \varepsilon_{\alpha\beta} \widehat{M}^{\dot\gamma}\,_{(\dot\alpha}\varepsilon_{\dot\beta)\dot\gamma}-  \varepsilon_{\dot\alpha \dot\beta} \widehat{M}^{\gamma}\,_{(\alpha}\varepsilon_{\beta)\gamma}  \,,\qquad \widetilde{M}_{5\alpha\dot\alpha}= \frac{1}{2} (P^{\gamma}_{\dot\alpha}\varepsilon_{\alpha\gamma}-K^{\dot\gamma}_{\alpha}\varepsilon_{\dot\alpha\dot\gamma})\, .
\end{equation*}

The $\mathfrak{usp}(2|2)$ subalgebra of the algebra \eqref{PSU algebra in SO} has to be compared with 
the algebra given in Subsection \ref{psl44su22}. 
The commutators are 
\begin{equation}
[\widetilde{P}_{a},\widetilde{P}_{b}]=\widetilde{M}_{ab}\ \; \rightarrow \quad [\widetilde{P}_{5},\widetilde{P}_{\alpha\dot\alpha}]=\widetilde{M}_{5\alpha\dot\alpha} \quad {\rm{and}} 
\quad
[\widetilde{P}_{\alpha\dot\alpha},\widetilde{P}_{\beta\dot\beta}]=\widetilde{M}_{\alpha\dot\alpha\beta\dot\beta}\,.  
\end{equation}
Let us work out the first commutator on the right hand side in detail
\begin{equation}
[\widetilde{P}_{5},\widetilde{P}_{\alpha\dot\alpha}]=\widetilde{M}_{5\alpha\dot\alpha} \; \rightarrow \; 
[ \Delta, \frac{1}{2} (P^{\gamma}_{\dot\alpha} \varepsilon_{\alpha\gamma}+ K^{\dot\gamma}_{\alpha} \varepsilon_{\dot\alpha\dot\gamma})] = \frac{1}{2} (P^{\gamma}_{\dot\alpha}\varepsilon_{\beta\gamma}-K^{\dot\gamma}_{\beta}\varepsilon_{\dot\alpha\dot\gamma}) \, , 
\end{equation}
which implies 
\begin{equation}
[ \Delta, P^{\gamma}_{\dot\alpha}] = P^{\gamma}_{\dot\alpha} \, , \quad \quad 
[ \Delta, K_{\gamma}^{\dot\alpha}] = - K_{\gamma}^{\dot\alpha} \, .  
\end{equation}
The commutators above corresponds to the first line of \eqref{firstlinecommutator}.
In addition, the commutator $[\widetilde{P}_{\underline{a}},\widetilde{M}_{\underline{bc}}]=\eta_{\underline{a}[\underline{b}}\widetilde{P}_{\underline{c}]}$ gives 
($\eta_{5 5} = 1$)

\begin{equation}
\label{TheAlgebra}
 [\widetilde{P}_{5},\widetilde{M}_{5\alpha\dot\alpha}]= \widetilde{P}_{\alpha\dot\alpha}\,,\qquad [\widetilde{P}_{\beta\dot\beta},\widetilde{M}_{5\alpha\dot\alpha}]=- \eta_{\beta\dot\beta\alpha\dot\alpha} \widetilde{P}_{5} \,,  
\end{equation}
\begin{equation*}
[\widetilde{P}_{\gamma\dot\gamma},\widetilde{M}_{\alpha\dot\alpha\beta\dot\beta}]= \eta_{\gamma\dot\gamma\alpha\dot\alpha}\widetilde{P}_{\beta\dot\beta} -
\eta_{\gamma\dot\gamma\beta\dot\beta}\widetilde{P}_{\alpha\dot\alpha}  \,,  
\end{equation*}
and 
\begin{equation}
\eta_{\alpha \dot\alpha \beta \dot\beta} =  -2 \varepsilon_{\alpha \beta} 
\varepsilon_{\dot\alpha \dot\beta} \, .  
\end{equation}
Finally for the commutators involving only the $\widetilde{M}$ generators, one has   
\begin{equation}
\label{lastone}
[\widetilde M_{\alpha\dot\alpha\beta\dot\beta}, \widetilde M_{\gamma\dot\gamma\delta\dot\delta}]=    
\end{equation}
\begin{equation*}
= \eta_{\alpha\dot\alpha\gamma\dot\gamma} \widetilde M_{\delta\dot\delta\beta\dot\beta}
- \eta_{\alpha\dot\alpha\delta\dot\delta} \widetilde M_{\gamma\dot\gamma\beta\dot\beta}
-\eta_{\beta\dot\beta\gamma\dot\gamma} \widetilde M_{\delta\dot\delta\alpha\dot\alpha}
+\eta_{\beta\dot\beta\delta\dot\delta} \widetilde M_{\gamma\dot\gamma\alpha\dot\alpha} \, . 
\end{equation*}
There are also commutators involving $\widetilde{M}_{5\alpha\dot\alpha}$ that we will not write explicitly. By substituting the relations \eqref{widenonwide} into \eqref{TheAlgebra}
and \eqref{lastone}, the algebra given in Subsection  \ref{psl44su22} is recovered. 
The indices $a$ can also be split as  $a = \{ \mu, 5 \}$ where $\mu =1,\ldots,4$ is a 4d vector index. The two ways of splitting the indices are related by contracting with 
the 4d Pauli matrices
\begin{equation}
\widetilde{P}_{\mu}= i \sigma_{\mu}^{\dot\alpha \alpha} \widetilde{P}_{\alpha \dot\alpha}  \,,\qquad \widetilde{P}_{5}= - \Delta \,,
\end{equation}
\begin{equation*}
\widetilde{M}_{\mu \nu}= \frac{1}{2} \widehat{M}^{\alpha}_{\beta} (\sigma_{\mu \nu})^{\beta}_{\alpha} -
\frac{1}{2} \widehat{M}^{\dot\alpha}_{\dot\beta} (\sigma_{\mu \nu})^{\dot\beta}_{\dot\alpha}  \, \quad \quad \widetilde{M}_{5 \mu}= i \sigma_{\mu}^{\dot\alpha \alpha} \widetilde{M}_{5 \alpha \dot\alpha}\, .
\end{equation*}

For the $\mathfrak{usp}(4)$ generators, the right hand side of the \eqref{10dtoSU22} decomposes as 
\begin{equation}
(M \pm J)^{I}\,_{[i}\varepsilon_{j]R} = - i J\varepsilon_{ij}\,,\qquad (M \pm J)^{I}\,_{[\overline i}\varepsilon_{\overline j]I}=- i J\varepsilon_{\overline{ij}}    \,,
\end{equation}
\begin{equation*}
(M \pm J)^{I}\,_{[\overline i}\varepsilon_{ j]I}= \frac{i}{2}(K^{k}_{\overline i}\varepsilon_{jk}+P^{\overline k}_{j}\varepsilon_{\overline i\overline k})\,,\qquad (M \pm J)^{I}\,_{[ i}\varepsilon_{\overline j]I}= -\frac{i}{2}(P^{\overline k}_{i}\varepsilon_{\overline j\overline k}+K^{k}_{\overline j}\varepsilon_{ik})\,,    
\end{equation*}
and 
\begin{equation}
(M \pm J)^{I}\,_{(i}\varepsilon_{j)R} = \widehat M^{k}\,_{(i}\varepsilon_{j)k}\,,\qquad (M \pm J)^{I}\,_{(\overline i}\varepsilon_{\overline j)I}=-\widehat M^{\overline k}\,_{(\overline{i}}\varepsilon_{\overline j)\overline k}    \,,
\end{equation}
\begin{equation*}
(M \pm J)^{I}\,_{(\overline i}\varepsilon_{ j)I}= \frac{1}{2}(K^{k}_{\overline i}\varepsilon_{jk}-P^{\overline k}_{j}\varepsilon_{\overline i\overline k})\,.
\end{equation*}

It is possible, following the same steps as those for the $\mathfrak{usp}(2|2)$ algebra, to show that the relations above are a map between the $\mathfrak{usp}(4)$ algebra in the two different notations.    
Similarly to the previous case, splitting the indices $a^{\prime} =\{\mu', 5'\}$ enables one to 
write for 
the generators 
\begin{equation}
\widetilde{P}_{5'}= - i J \,,\qquad \widetilde{P}_{\mu'}=  \sigma_{\mu'}^{i \bar{i}} (K^j_{\bar{i}} \varepsilon_{i j}+P_i^{\bar{j}} \varepsilon_{\bar{i} \bar{j}} )  \,,
\end{equation}
\begin{equation*}
\widetilde M_{\mu',\nu'}= \frac{1}{2} \widehat M^{i}_{j} (\sigma_{\mu' \nu'})^j_i 
- \frac{1}{2} \widehat M^{\bar{i}}_{\bar{j}} (\sigma_{\mu' \nu'})^{\bar{j}}_{\bar{i}} \,,\qquad \widetilde M_{5', \mu'}=i \sigma_{\mu'}^{i \bar{i}} (K^j_{\bar{i}} \varepsilon_{i j} -P_i^{\bar{j}} \varepsilon_{\bar{i} \bar{j}} ) \,.    
\end{equation*}

\subsection{Relating the notations: fermionic generators}

First, we decompose the $SO(10)$ spinors into representations of an $ SO(4)\times SO(4) \times SO(2)$ subgroup, where the first $SO(4)\cong SU(2)_{\alpha}\times SU(2)_{\dot\alpha}$ corresponds to rotations of the $\mu$ directions, while the second $SO(4)\cong SU(2)_{i}\times SU(2)_{\overline i}$ corresponds to rotations of the $\mu'$ directions. The $SO(2)$ rotates the plane spanned by the generators $\Delta$ and $J$. 
The decomposition of the spinorial index $\underline\alpha = 1, \ldots,16$ is given by
\begin{equation}
\underline\alpha =\begin{matrix}
++|++|+,& --|--|+,& ++|--|+,& --|++|+\\
++|+-|-,& ++|-+|-,& --|+-|-,& --|-+|-\\
+-|++|-,& -+|++|-,& +-|--|-,& -+|--|-\\
-+|-+|+,& +-|-+|+,& +-|+-|+,& -+|+-|+\\
\end{matrix}
\end{equation}
\begin{equation*}
\alpha = ++|,--|  ,\qquad  \dot\alpha = +-|,-+| ,\qquad  i=|++,|-- ,\qquad  \overline i =|+-,|-+ \, . 
\end{equation*}
where indices can be raised and lowered by contracting with $\varepsilon_{ij}$, $\varepsilon_{\overline i\overline j}$, $\varepsilon_{\alpha\beta}$ and $\varepsilon_{\dot\alpha\dot\beta}$. We follow the conventions of \cite{VertexOld}, where the supercharges appearing in \eqref{PSU algebra in SO} are given by
\begin{equation}
q_{\underline\alpha}=( q_{L \, \alpha}^{\; \; \;  i}, q_{L \, \dot\alpha}^{\; \; \; \overline i}, q_{L \, \alpha}^{\; \; \; \overline i}, q_{L \, \dot\alpha}^{\; \; \;  i}) \,,\qquad q_{\widehat{\underline\alpha}}=( q_{R \, \alpha}^{i}, q_{R \, \dot\alpha}^{\overline i}, q_{R \, \alpha}^{\overline i}, q_{R \, \dot\alpha}^{i})    \,.
\end{equation}
where
\begin{equation}
 q^{i}_{L \, \alpha} = q^{i}_{\alpha}+\varepsilon^{ij}\varepsilon_{\alpha\beta}q^{\beta}_{j} \,,\qquad  q^{\overline i}_{L \, \dot\alpha} = q^{\overline i}_{\dot\alpha}+\varepsilon^{\overline i\overline j}\varepsilon_{\dot\alpha\dot\beta}q^{\dot\beta}_{\overline j} \,,   
\label{relatingthesusys} 
\end{equation}
\begin{equation*}
 q^{\overline i}_{L \, \alpha} = q^{\overline i}_{\alpha}-\varepsilon^{\overline i\overline j}\varepsilon_{\alpha\beta}q^{\beta}_{\overline j} \,,\qquad  q^{i}_{L \, \dot\alpha} = q^{i}_{\dot\alpha}-\varepsilon^{ij}\varepsilon_{\dot\alpha\dot\beta}q^{\dot\beta}_{j} \,,   
\end{equation*}
\begin{equation*}
i q^{i}_{R \, \alpha} = q^{i}_{\alpha}-\varepsilon^{ij}\varepsilon_{\alpha\beta}q^{\beta}_{j} \,,\qquad i q^{\overline i}_{R \, \dot\alpha} = q^{\overline i}_{\dot\alpha}-\varepsilon^{\overline i\overline j}\varepsilon_{\dot\alpha\dot\beta}q^{\dot\beta}_{\overline j}\,,    
\end{equation*}
\begin{equation*}
i  q^{\overline i}_{R \, \alpha} = q^{\overline i}_{\alpha}+\varepsilon^{\overline i\overline j}\varepsilon_{\alpha\beta}q^{\beta}_{\overline j} \,,\qquad i  q^{i}_{R \, \dot\alpha} = q^{i}_{\dot\alpha}+\varepsilon^{ij}\varepsilon_{\dot\alpha\dot\beta}q^{\dot\beta}_{j}  \,.  
\end{equation*}
The $\mathfrak{psu}(4|4)$ superalgebra in 10d notation of \eqref{PSU algebra in SO}
 involves the $\gamma^{\underline{a}}_{\alpha \beta}$ matrices. 
By $SO(4)\times SO(4)$ equivariance alone, one concludes that the nonzero components of those matrices are given by
\begin{equation}
\gamma^{\mu}\,_{\dot\alpha}^{i}\,_{\beta}^{j}=c_1\,\varepsilon^{ij}\sigma^{\mu}_{\beta\dot\alpha}=\gamma^{\mu}\,_{\beta}^{j}\,_{\dot\alpha}^{i}    \,,\qquad \gamma^{\mu}\,_{\dot\alpha}^{\overline i}\,_{\beta}^{\overline j}=c_2\,\varepsilon^{\overline i\overline j}\sigma^{\mu}_{\beta\dot\alpha}=\gamma^{\mu}\,_{\beta}^{\overline j}\,_{\dot\alpha}^{\overline i}\,,   
\end{equation}
\begin{equation*}
\gamma^{5}\,_{\alpha}^{i}\,_{\beta}^{j}=c_3\,\varepsilon^{ij}\varepsilon_{\alpha\beta}   \,,\qquad \gamma^{5}\,_{\dot\alpha}^{\overline i}\,_{\dot\beta}^{\overline j}=c_4\,\varepsilon^{\overline i\overline j}\varepsilon_{\dot\alpha \dot\beta}\,,\qquad \gamma^{5}\,_{\alpha}^{\overline i}\,_{\beta}^{\overline j}=c_5\,\varepsilon^{\overline i\overline j}\varepsilon_{\alpha\beta}   \,,\qquad \gamma^{5}\,_{\dot\alpha}^{ i}\,_{\dot\beta}^{ j}=c_6\,\varepsilon^{ i j}\varepsilon_{\dot\alpha \dot\beta}\,,
\end{equation*}
\begin{equation*}
\gamma^{\mu'}\,_{\alpha}^{\overline i}\,_{\beta}^{j}=c_7\,\varepsilon_{\alpha\beta}\sigma^{\mu'\overline ij}=\gamma^{\mu'}\,_{\beta}^{j}\,_{\alpha}^{\overline i}    \,,\qquad \gamma^{\mu'}\,_{\dot\alpha}^{\overline i}\,_{\dot\beta}^{ j}=c_8\,\varepsilon_{\dot\alpha\dot\beta}\sigma^{\mu'\overline ij}=\gamma^{\mu'}\,_{\dot\beta}^{ j}\,_{\dot\alpha}^{\overline i}\,,     
\end{equation*}
\begin{equation*}
\gamma^{5'}\,_{\alpha}^{i}\,_{\beta}^{j}=c_9\,\varepsilon^{ij}\varepsilon_{\alpha\beta}   \,,\qquad \gamma^{5'}\,_{\dot\alpha}^{\overline i}\,_{\dot\beta}^{\overline j}=c_{10}\,\varepsilon^{\overline i\overline j}\varepsilon_{\dot\alpha \dot\beta}\,,\qquad \gamma^{5'}\,_{\alpha}^{\overline i}\,_{\beta}^{\overline j}=c_{11}\,\varepsilon^{\overline i\overline j}\varepsilon_{\alpha\beta}   \,,\qquad \gamma^{5'}\,_{\dot\alpha}^{ i}\,_{\dot\beta}^{ j}=c_{12}\,\varepsilon^{ i j}\varepsilon_{\dot\alpha \dot\beta}\,,
\end{equation*}
where the pre-factors $c_1$, $c_2$, $c_3$, $c_4$, $c_5$, $c_6$, $c_7$, $c_8$, $c_9$, $c_{10}$, $c_{11}$, and $c_{12}$, are fixed by comparing \eqref{algebra} with \eqref{PSU algebra in SO}.
We have
\begin{equation}
c_{1}=-i \,,\qquad c_{2}=i \,,\qquad c_{3}=-1 \,,\qquad c_{4}=1\,,\qquad c_{5}=1 \,,\qquad c_{6}=  -1 \,, 
\end{equation}
\begin{equation*}
c_{7}=1 \,,\qquad c_{8}=-1 \,,\qquad c_{9}= i \,,\qquad c_{10}= -i\,,\qquad c_{11}= i\,,\qquad c_{12}= -i \,.  
\end{equation*}
In other words, 
\begin{equation}
\{q_{L \,  \alpha}^{\; \; \; i},q_{L \,  \beta}^{\; \; \; j}\}=(\Delta+J)\varepsilon^{ij}\varepsilon_{\alpha\beta}\,,\qquad \{q_{L \,  \alpha}^{\; \; \; i},q_{L \,  \dot\beta}^{\; \; \; \overline j}\}=0\,,  
\end{equation}
\begin{equation*}
 \{q_{L \, \alpha}^{\; \; \; i},q_{L \, \dot\beta}^{\; \; \; j}\}=\varepsilon^{ij}(\varepsilon_{\dot\beta\dot\gamma}K^{\dot\gamma}_{\alpha}+\varepsilon_{\alpha\gamma}P^{\gamma}_{\dot\beta}) \,,\qquad \{q_{L \, \alpha}^{\; \; \;  i},q_{L \, \beta}^{\; \; \; \overline j}\}=\varepsilon_{\alpha\beta}(\varepsilon^{ik}P^{\overline j}_{k}+\varepsilon^{\overline j\overline k}K^{i}_{\overline k})\,,    
\end{equation*}
\begin{equation*}
\{q_{L \, \dot\alpha}^{\; \; \; \overline i},q_{L \, \dot\beta}^{\; \; \; \overline j}\}=-(\Delta+J)\varepsilon^{\overline i\overline j}\varepsilon_{\dot\alpha\dot\beta}    \,,\qquad \{q_{ L \, \dot\alpha}^{\; \; \; \overline i},q_{ L \, \dot\beta}^{\; \; \; j}\}=\varepsilon_{\dot\alpha\dot\beta}(\varepsilon^{jk}P^{\overline i}_{k}+\varepsilon^{\overline i\overline k}K^{j}_{\overline k})\,,
\end{equation*}
\begin{equation*}
\{q_{L \,  \dot\alpha}^{\; \; \; \overline i},q_{L \,  \beta}^{\; \; \; \overline j}\}=\varepsilon^{\overline i\overline j}(\varepsilon_{\beta\gamma}P^{\gamma}_{\dot\alpha}+\varepsilon_{\dot\alpha\dot\gamma}K^{\dot\gamma}_{\beta})\,,  \qquad \{q_{L \,  \dot\alpha}^{\; \; \; i},q_{L \,  \dot\beta}^{\; \; \; j}\}=(\Delta-J)\varepsilon_{\dot\alpha\dot\beta}\varepsilon^{ij}\,,  
\end{equation*}
\begin{equation*}
\{q_{L \, \dot\alpha}^{\; \; \; i},q^{\; \; \; \overline j}_{L \, \beta}\}=0\,,\qquad \{q^{\; \; \;  \overline i}_{L \, \alpha},q_{L \, \beta}^{\; \; \;  \overline j}\}=-(\Delta-J)\varepsilon_{\alpha\beta}\varepsilon^{\overline i\overline j}\,.    
\end{equation*}

Similarly, it is possible to obtain the nonzero components of the matrices  $\gamma^{\underline{a}}_{\widehat\alpha \widehat\beta}$. The result is $\widehat{c}_i = c_i$ for any $i$, i.e.,
the matrices are numerically equal. To obtain the Ramond-Ramond bi-spinor $\eta^{\alpha \widehat\beta}$ one can extract $\frac{i}{2}(\gamma^{\underline{a}}\eta)_{\alpha}\,^{\widehat\beta}$ from \eqref{PSU algebra in SO}, by comparing with the results obtained from \eqref{TheKPqalgebra} and \eqref{TheKPqalgebraI} written using the combination of generators of \eqref{relatingthesusys} and the map of the bosonic generators of the previous Subsection:
\begin{equation*}
[P_5,q^{\; \; \;  i}_{L \, \alpha}]= \frac{i}{2} q^{\; \; \;  i}_{R \, \alpha}\,,\qquad [P_5,q^{\; \; \;  \overline i}_{L \, \dot\alpha}]=- \frac{i}{2}  q^{\; \; \;  \overline i}_{R \, \dot\alpha}\,,\qquad [P_5,q^{\; \; \;  i}_{L \, \dot\alpha}]= - \frac{i}{2} q^{\; \; \; i}_{R \, \dot\alpha}\,,\qquad [P_5,q^{\; \; \; \overline i}_{L \, \alpha}]= \frac{i}{2} q^{\; \; \;  \overline i}_{R \, \alpha}\,,    
\end{equation*}
\begin{equation*}
[P_{\mu},q^{\; \; \;  i}_{L \, \alpha}]= \frac{1}{2} \varepsilon_{\alpha \beta} \sigma_{\mu}^{\dot\beta \beta} q^{\; \; \;  i}_{L \, \dot\beta}\,,\qquad
[P_{\mu},q^{\; \; \;  \bar{i}}_{L \, \dot\alpha}]= \frac{1}{2} \varepsilon_{\dot\alpha \dot\beta} \sigma_{\mu}^{\dot\beta \beta} q^{\; \; \;  \bar{i}}_{L \, \beta} \, , 
\end{equation*}
\begin{equation*}
[P_{\mu},q^{\; \; \;  i}_{L \, \dot\alpha}]= \frac{1}{2} \varepsilon_{\dot\alpha \dot\beta} \sigma_{\mu}^{\dot\beta \beta} q^{\; \; \;  i}_{L \, \beta} \,,\qquad
[P_{\mu},q^{\; \; \;  \bar{i}}_{L \, \alpha}]= \frac{1}{2} \varepsilon_{\alpha \beta} \sigma_{\mu}^{\dot\beta \beta} q^{\; \; \;  \bar{i}}_{L \, \dot\beta} \, . 
\end{equation*}
and 
\begin{equation*}
[P_{5^{\prime}},q^{\; \; \;  i}_{L \, \alpha}]= \frac{1}{2} q^{\; \; \;  i}_{R  \, \alpha}\,,\qquad [P_{5^{\prime}},q^{\; \; \;  \overline i}_{L \, \dot\alpha}]=- \frac{1}{2} q^{\; \; \;  \overline i}_{R \, \dot\alpha}\,,\qquad [P_{5^{\prime}},q^{\; \; \;  i}_{L \, \dot\alpha}]=  \frac{1}{2}  q^{\; \; \;  i}_{R \, \dot\alpha}\,,\qquad [P_{5^{\prime}},q^{\; \; \;  \overline i}_{L \, \alpha}]= -\frac{1}{2}  q^{\; \; \;  \overline i}_{R \, \alpha}\,,    
\end{equation*}
\begin{equation*}
[P^{\mu^{\prime}},q^{\; \; \;  i}_{L \, \alpha}]= \frac{i}{2} \varepsilon^{i j} \sigma^{\mu^{\prime}}_{i \bar{i}} q^{\; \; \; \bar{i}}_{L \, \alpha}\,,\qquad
[P^{\mu^{\prime}},q^{\; \; \;  \bar{i}}_{L \, \dot\alpha}]= \frac{i}{2} \varepsilon^{\bar{i} \bar{j}} \sigma^{\mu^{\prime}}_{i \bar{j}} q^{\; \; \;  i}_{L \, \dot\alpha} \, , 
\end{equation*}
\begin{equation*}
[P^{\mu^{\prime}},q^{\; \; \; i}_{L \, \dot\alpha}]=\frac{i}{2} \varepsilon^{i j} \sigma^{\mu^{\prime}}_{j \bar{i}} q^{\; \; \; \bar{i}}_{L \, \dot\alpha} \,,\qquad
[P^{\mu^{\prime}},q^{\; \; \; \bar{i}}_{L \, \alpha}]=\frac{i}{2} \varepsilon^{\bar{i} \bar{j}} \sigma^{\mu^{\prime}}_{i \bar{j}} q^{\; \; \; i}_{L \, \alpha} \, . 
\end{equation*}
The result is   
\begin{equation}
\label{etasymbol}
\eta^{\alpha \beta}_{i j} =  \varepsilon_{i j} \varepsilon^{\alpha \beta} \, , \quad
\eta^{\dot\alpha \dot\beta}_{i j} =  - \varepsilon_{i j} \varepsilon^{\dot\alpha \dot\beta} \, , \quad
\eta^{\alpha \beta}_{\bar{i} \bar{j}} =  - \varepsilon_{\bar{i} \bar{j}} \varepsilon^{\alpha \beta} \, , \quad
\eta^{\dot\alpha \dot\beta}_{\bar{i} \bar{j}} =  \varepsilon_{\bar{i} \bar{j}} \varepsilon^{\dot\alpha \dot\beta} \, .
\end{equation}

It is possible to use the expressions above and check that the commutator
$[\widetilde{P}_{\underline{a}},q_{\widehat\alpha}]$ of \eqref{PSU algebra in SO} which has $
-\frac{i}{2}(\gamma_{a}\eta)_{\widehat\alpha}\,^{\beta}    
$
is also satisfied using the decomposition above. The missing elements are matrices with indices $\{ \gamma_{a}^{\alpha \beta}, \gamma_{a}^{\widehat\alpha \widehat\beta} \}$, which are
necessary for the commutators  
of $\widetilde{M}_{\underline{ab}}$ and $q_{\alpha}$ for example. 
One way of fixing them is either by using the relation
\begin{equation}
\gamma^M_{\alpha \beta} \gamma^{N \beta \gamma} +
\gamma^N_{\alpha \beta} \gamma^{M \beta \gamma} = 2 \eta^{M N} \delta^{\gamma}_{\alpha} \, , 
\end{equation}
or the algebra. One concludes that 
\begin{equation}
\gamma^{\mu}\,^{\dot\alpha}_{i}\,^{\beta}_{j}=d_1\,\varepsilon_{ij}\sigma^{\mu \dot\alpha  \beta}=\gamma^{\mu}\,^{\beta}_{j}\,^{\dot\alpha}_{i}    \,,\qquad \gamma^{\mu}\,^{\dot\alpha}_{\overline i}\,^{\beta}_{\overline j}=d_2\,\varepsilon_{\overline i\overline j}\sigma^{\mu \dot\alpha \beta}=\gamma^{\mu}\,^{\beta}_{\overline j}\,^{\dot\alpha}_{\overline i}\,,   
\end{equation}
\begin{equation*}
\gamma^{5}\,^{\alpha}_{i}\,^{\beta}_{j}=d_3\,\varepsilon_{ij}\varepsilon^{\alpha\beta}   \,,\qquad \gamma^{5}\,^{\dot\alpha}_{\overline i}\,^{\dot\beta}_{\overline j}=d_4\,\varepsilon_{\overline i\overline j}\varepsilon^{\dot\alpha \dot\beta}\,,\qquad \gamma^{5}\,^{\alpha}_{\overline i}\,^{\beta}_{\overline j}=d_5\,\varepsilon_{\overline i\overline j}\varepsilon^{\alpha\beta}   \,,\qquad \gamma^{5}\,^{\dot\alpha}_{ i}\,^{\dot\beta}_{ j}=d_6\,\varepsilon_{ i j}\varepsilon^{\dot\alpha \dot\beta}\,,
\end{equation*}
\begin{equation*}
\gamma^{\mu'}\,^{\alpha}_{\overline i}\,^{\beta}_{j}=d_7\,\varepsilon^{\alpha\beta}\sigma^{\mu'}_{\overline ij}=\gamma^{\mu'}\,^{\beta}_{j}\,^{\alpha}_{\overline i}    \,,\qquad \gamma^{\mu'}\,^{\dot\alpha}_{\overline i}\,^{\dot\beta}_{ j}=d_8\,\varepsilon^{\dot\alpha\dot\beta}\sigma^{\mu'}_{\overline ij}=\gamma^{\mu'}\,^{\dot\beta}_{ j}\,^{\dot\alpha}_{\overline i}\,,     
\end{equation*}
\begin{equation*}
\gamma^{5'}\,^{\alpha}_{i}\,^{\beta}_{j}=d_9\,\varepsilon_{ij}\varepsilon^{\alpha\beta}   \,,\qquad \gamma^{5'}\,^{\dot\alpha}_{\overline i}\,^{\dot\beta}_{\overline j}=d_{10}\,\varepsilon_{\overline i\overline j}\varepsilon^{\dot\alpha \dot\beta}\,,\qquad \gamma^{5'}\,^{\alpha}_{\overline i}\,^{\beta}_{\overline j}=d_{11}\,\varepsilon_{\overline i\overline j}\varepsilon^{\alpha\beta}   \,,\qquad \gamma^{5'}\,^{\dot\alpha}_{ i}\,^{\dot\beta}_{ j}=d_{12}\,\varepsilon_{ i j}\varepsilon^{\dot\alpha \dot\beta}\,,
\end{equation*}
where 
\begin{equation}
d_{1}=i \,,\qquad d_{2}=-i \,,\qquad d_{3}=-1 \,,\qquad d_{4}=1\,,\qquad d_{5}=1 \,,\qquad d_{6}= -1 \,, 
\end{equation}
\begin{equation*}
d_{7}=1 \,,\qquad d_{8}=-1 \,,\qquad d_{9}= -i \,,\qquad d_{10}= i\,,\qquad d_{11}= -i\,,\qquad d_{12}= i \,.  
\end{equation*}
A similar calculation gives $\widehat{d}_i = d_i$ for all $i$. 
It is also useful to compute $\eta^{\widehat\alpha \alpha}$ and 
$\eta_{\alpha \widehat\alpha} = - \eta_{\widehat\alpha \alpha}$.
The easiest way is to use the properties \eqref{propertieseta}.
Since $\eta^{\widehat\alpha \alpha}= - \eta^{\alpha \widehat\alpha}$, 
the value of this symbol is obtained by multiplying \eqref{etasymbol} 
by minus one. 
For the symbol $\eta_{\alpha \widehat\alpha}$, one writes 
the decomposition 
\begin{equation}
\label{etasymboldown}
\eta_{\alpha \beta}^{i j} = k_1 \, i  \varepsilon^{i j} \varepsilon_{\alpha \beta} \, , \quad
\eta_{\dot\alpha \dot\beta}^{i j} = k_2  \, i \varepsilon^{i j} \varepsilon_{\dot\alpha \dot\beta} \, , \quad
\eta_{\alpha \beta}^{\bar{i} \bar{j}} =  k_3 \, i \varepsilon^{\bar{i} \bar{j}} \varepsilon_{\alpha \beta} \, , \quad
\eta_{\dot\alpha \dot\beta}^{\bar{i} \bar{j}} = k_4 \,  i \varepsilon^{\bar{i} \bar{j}} \varepsilon_{\dot\alpha \dot\beta} \, .
\end{equation}
and \eqref{propertieseta} fixes the constants to be  
\begin{equation}
k_1 = i \, , \quad  k_2 = - i \, , \quad k_3 = - i \, , \quad  k_4 = i \, . 
\end{equation}

\section{Superfields of the integrated vertex operators}
\label{Integrated vertex operators superfields}

Here, we are going to give only the defining equations for the superfields $\mathcal{V}$ and $\overline{\mathcal{V}}$ 
appearing in the integrated vertex operators of \eqref{TheUsuperfield}. The proof of the full descent equations follows by applying covariant derivatives on these superfields.  In the equations below, by setting all the $B$'s to zero, the result of \cite{IntegratedVertex} is recovered.  
We omit the underline on the Greek indices. 
The equations fixing $\mathcal{V}_{\underline{ab}}$ are    
\begin{equation}
\begin{aligned}
& \gamma^{\underline{b}}_{\widehat\alpha \widehat\beta} \, \mathcal{V}_{\underline{ab}} =
\nabla_{(\widehat\alpha} A_{\widehat\beta) \underline{a}}
- \frac{i}{2} (\gamma_{\underline{a}})_{(\widehat\alpha | \widehat\gamma} \eta^{\gamma \widehat\gamma} A_{\widehat\beta) \gamma} - 2 \nabla_{\underline{a}} B_{\widehat\alpha \widehat\beta} \, ,  \\ 
& \gamma^{\underline{b}}_{\alpha \beta} \mathcal{V}_{\underline{b} \underline{a}} 
= - \nabla_{(\alpha} A_{\beta) \underline{a}} - \frac{i}{2} (\gamma_{\underline{a}})_{\gamma (\alpha} \eta^{\gamma \widehat\gamma} A_{\beta) \widehat\gamma} 
- 2 \nabla_{\underline{a}} B_{\alpha \beta} \, .
\end{aligned} 
\end{equation}
The superfields $\overline{\mathcal{V}}_{\underline{a}  \alpha}$ and $\mathcal{V}_{\underline{a} \widehat{\alpha}}$ are obtained from  
\begin{equation}
\begin{aligned}
& \gamma^{\underline{a}}_{\beta \gamma} \overline{\mathcal{V}}_{\underline{a}  \alpha}= 
(\gamma^{\underline{a}})_{\alpha (\beta} A_{\gamma) \underline{a}} 
- 2 \eta_{\alpha \widehat{\gamma}} \nabla_{(\beta} W_{\gamma)}^{\widehat{\gamma}} 
- 2 \nabla_{\alpha} B_{\beta \gamma} \, ,  \\
& \gamma^{\underline{a}}_{\widehat\beta \widehat\gamma} \mathcal{V}_{\underline{a} \widehat{\alpha}} = - (\gamma^{\underline{a}})_{\widehat{\alpha} (\widehat{\beta}} A_{\underline{a} \widehat{\gamma})} - 2 \eta_{\gamma \widehat{\alpha}} \nabla_{(\widehat\beta} W_{\widehat\gamma)}^{\gamma} + 2 \nabla_{\widehat{\alpha}} B_{\widehat{\beta}\widehat{\gamma}} \, . 
\end{aligned} 
\end{equation}
The superfield $\mathcal{V}_{\beta \widehat{\gamma}}$ obeys   
\begin{equation}
\begin{aligned}
& \frac{i}{2} (\gamma_{\underline{a}})_{\alpha \gamma} \eta^{\gamma \widehat{\gamma}} \,\mathcal{V}_{\beta \widehat{\gamma}} 
+ \frac{i}{2} (\gamma_{\underline{a}})_{\beta \gamma} \eta^{\gamma \widehat{\gamma}} A_{\alpha \widehat{\gamma}}
- \gamma^{\underline{b}}_{\alpha \beta} \mathcal{V}_{\underline{ab}}
- \nabla_{\alpha} \overline{\mathcal{V}}_{\underline{a}\beta}
- 2 \eta_{\beta \widehat{\gamma}} \nabla_{\underline{a}} W_{\alpha}^{\widehat{\gamma}}
+ 2 \nabla_{\beta} B_{\alpha \underline{a}}= 0 \, ,  \\
& \frac{i}{2} (\gamma_{\underline{a}})_{\widehat{\alpha} \widehat{\gamma}} \eta^{\gamma \widehat{\gamma}} \mathcal{V}_{\gamma \widehat{\beta}}
+ \frac{i}{2} (\gamma_{\underline{a}})_{\widehat{\beta} \widehat{\gamma}} \eta^{\gamma \widehat{\gamma}} A_{\gamma \widehat{\alpha}}
- (\gamma^{\underline{b}})_{\widehat{\alpha} \widehat{\beta}} \mathcal{V}_{\underline{ba}}
- \nabla_{\widehat{\alpha}} \mathcal{V}_{\underline{a} \widehat{\beta}} - 2 \eta_{\gamma \widehat\alpha} \nabla_{\underline{a}} W^{\gamma}_{\widehat\beta}
- 2 \nabla_{\widehat\alpha} B_{\underline{a} \widehat\beta} =0 \, . 
\end{aligned} 
\end{equation}
The superfield $\mathcal{V}_{\underline{a} \underline{bc}}$ is determined by 
\begin{equation}
\begin{aligned}
\frac{1}{4} (\gamma^{\underline{bc}})_{\beta}\,^{\alpha} \eta_{\alpha \widehat{\alpha}} \mathcal{V}_{\underline{a} \underline{bc}} + \nabla_{\beta} A_{\underline{a} \widehat{\alpha}}
- \nabla_{\widehat{\alpha}} \mathcal{V}_{\underline{a} \beta} + (\gamma_{\underline{a}})_{\widehat{\alpha} \widehat{\beta}}\eta^{\gamma \widehat{\beta}}
\eta_{\beta \widehat{\gamma}} W^{\widehat{\gamma}}_{\gamma} - 4 
\eta_{\beta \widehat{\gamma}} \nabla_{\underline{a}} W^{\widehat{\gamma}}_{\widehat{\alpha}} 
+ i B_{\widehat\alpha \widehat\beta} \eta^{\gamma \widehat\beta}(\gamma_{\underline{a}})_{\gamma \beta}=0 \, ,   \\
\frac{1}{4} (\gamma^{\underline{bc}})_{\beta}\,^{\gamma} \eta_{\gamma \widehat\alpha} \mathcal{V}_{\underline{abc}} 
- \nabla_{\beta} A_{\underline{a} \widehat\alpha} - \nabla_{\beta} \overline{\mathcal{V}}_{\underline{a} \widehat{\alpha}}
- (\gamma_{\underline{a}})_{\alpha \beta} \eta^{\alpha \widehat\beta}\eta_{\gamma \widehat\alpha} W^{\gamma}_{\widehat\beta} - 4  \eta_{\gamma \widehat\alpha} \nabla_{\underline{a}} W^{\gamma}_{\beta}+ i B_{\alpha \beta} \eta^{\alpha \widehat\beta}(\gamma_{\underline{a}})_{\widehat\alpha \widehat\beta}=0 \, .   
\end{aligned} 
\end{equation}
The superfields $\overline{\mathcal{V}}_{\underline{ab} \gamma}$ and $\mathcal{V}_{\underline{ab} \widehat\beta}$ follows from  
\begin{equation}
\begin{aligned}
\frac{1}{4} (\gamma^{\underline{ab}})_{\alpha}\,^{\beta} \eta_{\beta \widehat\beta} \overline{\mathcal{V}}_{\underline{ab} \gamma} 
- 
\nabla_{\alpha} \overline{\mathcal{V}}_{\gamma \widehat\beta} %
- (\gamma^{\underline{a}})_{\alpha \gamma} \mathcal{V}_{\underline{a} \widehat\beta} + 2 i  \eta_{\gamma \widehat{\gamma}} \nabla_{\widehat{\beta}} W^{\widehat{\gamma}}_{\alpha}  - 4 i \eta_{\beta \widehat{\beta}} \nabla_{\gamma}  W^{\beta}_{\alpha}=0 \, ,  \\
\frac{1}{2} 
(\gamma^{\underline{ab}})_{\alpha}\,^{\beta} \eta_{\beta \widehat\alpha} 
\mathcal{V}_{\underline{ab} \widehat\beta}+ \nabla_{\widehat\alpha} \overline{\mathcal{V}}_{\alpha \widehat\beta}
+
(\gamma^{\underline{a}})_{\widehat\alpha \widehat\beta} \overline{\mathcal{V}}_{\underline{a} \alpha}
- 2 i  \eta_{\beta \widehat{\beta}} \nabla_{\alpha} W^{\beta}_{\widehat{\alpha}} 
+ 4 i  \eta_{\alpha \widehat{\gamma}} \nabla_{\widehat{\alpha}} W^{\widehat{\gamma}}_{\widehat{\beta}}=0 \, . 
\end{aligned} 
\end{equation}
Finally, the superfield $\mathcal{V}_{\underline{cd}\underline{ab}}$ solves the following two equations  
\begin{equation}
\begin{aligned}
\left[\frac{1}{8}(\gamma^{\underline{cd}}\eta)_{\alpha\widehat\beta}\mathcal{V}_{\underline{cd}\underline{ab}} - i \nabla_{\widehat\beta}(F_{\alpha\underline{ab}} + 2 i B_{\underline{a} \widehat{\delta}}\eta^{\gamma \widehat{\delta}}(\gamma_{\underline{b}})_{\gamma \alpha}) + \frac{1}{2}\nabla_{\alpha}\mathcal{V}_{\underline{ab}\widehat\beta}\right] \overline N^{\underline{ab}} \\
+\frac{i}{2}\overline{N}^{[ab]}\left[(\gamma_{[ab]})_{\alpha}\,^{\gamma}\overline{\mathcal{V}}_{\gamma\widehat\beta}+(\gamma_{[ab]})_{\widehat\beta}\,^{\widehat\gamma}A_{\alpha\widehat\gamma} - 4 i (\gamma_{[ab]})_{\widehat\beta}\,^{\widehat\gamma}\eta_{\gamma \widehat\gamma} W^{\gamma}_{\beta} \right]=0\,,   
\end{aligned} 
\end{equation}

\begin{equation*}
\begin{aligned}
\left[- \frac{1}{8}(\gamma^{\underline{cd}}\eta)_{\widehat\alpha\beta}\mathcal{V}_{\underline{ab}\underline{cd}} - i \nabla_{\beta}(F_{\underline{ab}\widehat\alpha} - 2 i  B_{\underline{a} \beta} \eta^{\beta \widehat\gamma} (\gamma_{\underline{b}})_{\widehat\alpha \widehat{\gamma}})-\frac{1}{2} \nabla_{\widehat\alpha}\overline{\mathcal{V}}_{\underline{ab}\beta}\right] N^{\underline{ab}} \\ -\frac{i}{2} N^{[ab]}\left[(\gamma_{[ab]})_{\widehat\alpha}\,^{\widehat\gamma}\overline{\mathcal{V}}_{\beta\widehat\gamma}-(\gamma_{[ab]})_{\beta}\,^{\gamma}A_{\gamma\widehat\alpha} + 4 i (\gamma_{[ab]})_{\beta}\,^{\gamma} \eta_{\gamma \widehat{\beta}} W^{\widehat{\beta}}_{\widehat\alpha} \right]=0\,.   
\end{aligned}
\end{equation*}

\section{The infinitesimal vertices} 
\label{InfinitesimalVertices} 

In this Appendix, we give part of the result of translating the vertices 
by the infinitesimal parameters $\eta^{\dot\beta}_{\beta}$. 
Precisely, we give the highest order in $n$ for each order in the $\theta$ expansion of the vertices.  
By setting $\eta^{\dot\beta}_{\beta}=0$, one recovers the vertex operators 
given in the Appendix \ref{DilatonOldCoset}. The infinitesimally translated vertices $V_{inf}$ 
have a $\theta$ expansion up to order $\theta^{10}$, 
\begin{equation}
V_{inf} = V_{inf,0} + V_{inf,2}+V_{inf,4}+V_{inf,6}+ V_{inf,8}+V_{inf,10} \, ,
\end{equation}
where the subscript means the order in $\theta$ of the expansion. Using the definition 
\begin{equation}
V_{\rm{dilaton}} = (\bar{\lambda}_+ \bar{\lambda}_+) + (\lambda_+ \lambda_+) \, ,  
\end{equation} 
one has at order $\theta^0$,
\[
 V^n_{inf,0}  = - e^{n (w+z)} n V_{\rm{dilaton}} \eta _{\beta }^{\dot{\beta }} \widetilde{x}_{\dot{\beta }}^{\beta } \, . 
\]
where $\widetilde{x}^{\alpha}_{\dot\alpha} \equiv e^{-z} x^{\alpha}_{\dot\alpha}$. At order $\theta^2$, there are contributions for $n^2$ and $n$. 
Here we write down only the $n^2$ piece  
and the tilded fermions appearing below were defined in \eqref{tildefermions}.
 One has  
\begin{equation}
\begin{aligned}
V^{n^2}_{I,2} & = e^{n (w+z)}  n^2 (\eta \cdot \widetilde{x}) 
 \widetilde{\theta }_{\alpha }^i \widetilde{\psi }_{\bar{i}}^{\dot{\alpha }}
 ( \varepsilon ^{\beta  \alpha} \varepsilon _{j i}  \lambda _{\beta }^j \lambda _{\dot{\alpha }}^{\bar{i}} 
 -\varepsilon^{\bar{j} \bar{i}}  \varepsilon _{\dot{\beta } \dot{\alpha }} \lambda _i^{\alpha } \lambda _{\bar{j}}^{\dot{\beta }}) \\
 & - 2 e^{n (w+z)}  n^2 (\eta \cdot \widetilde{x}) 
 \widetilde{\theta }_{\alpha }^i \widetilde{\psi }_{\bar{i}}^{\dot{\alpha }}
 (  \varepsilon ^{\beta  \alpha } \varepsilon _{j i}  \lambda _{\beta }^{\bar{i}} \lambda _{\dot{\alpha }}^j  +
  \varepsilon ^{\bar{j} \bar{i}}  \varepsilon _{\dot{\beta } \dot{\alpha }} \lambda _i^{\dot{\beta }} \lambda _{\bar{j}}^{\alpha }) \\ 
& + e^{n (w+z)}  n^2 (\eta \cdot \widetilde{x})    
\widetilde{\psi }_{\bar{i}}^{\dot{\alpha }} \widetilde{\psi }_{\bar{j}}^{\dot{\beta }} (
 \varepsilon ^{\bar{i} \bar{k}} \varepsilon _{\dot{\beta } \dot{\gamma }} \lambda _i^{\dot{\gamma }} \lambda _{\dot{\alpha }}^i   
 - \varepsilon ^{\bar{j} \bar{k}}  \varepsilon _{\dot{\beta } \dot{\alpha }} \lambda _{\alpha }^{\bar{i}} \lambda _{\bar{k}}^{\alpha }  
)  \\
& - e^{n (w+z)}  n^2 (\eta \cdot \widetilde{x}) 
 \widetilde{\theta }_{\alpha }^i \widetilde{\theta }_{\beta }^j
 ( \varepsilon ^{\gamma  \alpha } \varepsilon _{j i}  \lambda _{\gamma }^{\bar{i}} \lambda _{\bar{i}}^{\beta }   - 
  \varepsilon ^{\alpha  \beta } \varepsilon _{k i} \lambda _j^{\dot{\alpha }} \lambda _{\dot{\alpha }}^k) \, \\
& + \frac{1}{2}  e^{n (w+z)}  n^2 (\eta \cdot \widetilde{x}) (\widetilde\theta^i_{\alpha} \widetilde\theta^j_{\beta} \varepsilon^{\alpha \beta} \varepsilon_{ik} \lambda^k_{\gamma} \lambda^{\gamma}_j 
+ \widetilde\psi^{\dot\alpha}_{\bar{i}} \widetilde\psi^{\dot\beta}_{\bar{j}} \varepsilon^{\bar{i} \bar{j}} \varepsilon_{\dot\alpha \dot\gamma} \lambda^{\bar{k}}_{\dot\beta} \lambda^{\dot\gamma}_{\bar{k}}) \\
& - \frac{1}{2}  e^{n (w+z)}  n^2 (\eta \cdot \widetilde{x}) (\widetilde\theta^i_{\alpha} \widetilde\theta^j_{\beta} \varepsilon_{i j} \varepsilon^{\gamma \alpha} \lambda_{\gamma}^k \lambda_k^{\beta} + 
\widetilde\psi^{\dot\alpha}_{\bar{i}} \widetilde\psi^{\dot\beta}_{\bar{k}} \varepsilon_{\dot\alpha \dot\beta} \varepsilon^{\bar{j} \bar{i}} \lambda^{\bar{k}}_{\dot\gamma} \lambda^{\dot\gamma}_{\bar{j}}) \, . 
\end{aligned}
\end{equation}
At order $\theta^4$ and $n^3$, one has 
\begin{equation}
\begin{aligned}
V^{n^3}_{I,4} & = \frac{1}{24} e^{n (w+z)} n^3 (\eta \cdot \widetilde{x}) (\widetilde\theta^4 + \widetilde\psi^4 ) (\lambda^{\bar{i}}_{\alpha} \varepsilon^{\alpha \beta} \varepsilon_{\bar{i} \bar{j}} \lambda^{\bar{j}}_{\beta} + \lambda_{i}^{\dot\alpha} \varepsilon_{\dot\alpha \dot\beta} \varepsilon^{i j} \lambda_{j}^{\dot\beta} - \lambda^{\alpha}_{\bar{i}} \varepsilon_{\alpha \beta} \varepsilon^{\bar{i} \bar{j}}
\lambda^{\beta}_{\bar{j}} - 
\lambda^{i}_{\dot{\alpha}} \varepsilon_{\dot\alpha \dot\beta} \varepsilon^{i j}
\lambda^{j}_{\dot\beta})  \\
& - \frac{1}{2} e^{n (w+z)} n^3
(\eta \cdot \widetilde{x}) \widetilde{\theta }_{\alpha }^i \widetilde{\theta }_{\gamma }^j \widetilde{\psi }_{\bar{i}}^{\dot{\gamma }} \widetilde{\psi }_{\bar{j}}^{\dot{\alpha }}
(\varepsilon ^{\alpha  \gamma} \varepsilon^{\bar{i} \bar{j}} \varepsilon _{k j} \varepsilon _{i s} \lambda _{\dot{\alpha }}^k \lambda _{\dot{\gamma }}^s + \varepsilon ^{\delta  \gamma}  \varepsilon^{\alpha  \eta } \varepsilon _{j i}  \varepsilon _{\dot{\gamma } \dot{\alpha }} \lambda _{\delta }^{\bar{i}} \lambda _{\eta }^{\bar{j}} ) \\
& + \frac{1}{2} e^{n (w+z)} n^3 (\eta \cdot \widetilde{x}) \widetilde{\theta }_{\alpha }^i \widetilde{\theta }_{\gamma }^j \widetilde{\psi }_{\bar{i}}^{\dot{\delta }} \widetilde{\psi }_{\bar{j}}^{\dot{\gamma }} (\varepsilon ^{\bar{k} \bar{j}} \varepsilon^{\bar{i} \bar{l}} \varepsilon _{i j}\varepsilon _{\dot{\gamma } \dot{\delta }} \lambda _{\bar{k}}^{\alpha } \lambda _{\bar{l}}^{\gamma}  
+ \varepsilon ^{\alpha  \gamma} \varepsilon^{\bar{i} \bar{j}} \varepsilon _{\dot{\delta } \dot{\gamma }} \varepsilon _{\dot{\alpha } \dot{\rho }} \lambda _j^{\dot{\alpha }} \lambda _i^{\dot{\rho}}    ) \\
& 
+ \frac{2}{3} e^{n (w+z)} n^3 (\eta \cdot \widetilde{x}) 
\widetilde{\theta }_{\alpha }^k \widetilde{\theta }_{\gamma }^l \widetilde{\theta }_{\delta }^j \widetilde{\psi }_{\bar{j}}^{\dot{\alpha }} (
 \varepsilon ^{\gamma  \delta} \varepsilon^{\bar{i} \bar{j}} \varepsilon _{i j} \varepsilon _{k l} \lambda _{\bar{i}}^{\alpha } \lambda _{\dot{\alpha }}^i 
+ \varepsilon ^{\alpha  \gamma} \varepsilon^{\rho  \delta} \varepsilon _{j k}  \varepsilon _{\dot{\gamma } \dot{\alpha }} \lambda _l^{\dot{\gamma }} \lambda _{\rho }^{\bar{j}})\\
&-\frac{2}{3} e^{n (w+z)} n^3 (\eta \cdot \widetilde{x}) \widetilde{\theta }_{\alpha }^j \widetilde{\psi }_{\bar{j}}^{\dot{\delta }} \widetilde{\psi }_{\bar{k}}^{\dot{\alpha }} \widetilde{\psi }_{\bar{l}}^{\dot{\gamma }}   (\varepsilon ^{\bar{j} \bar{k}} \varepsilon^{\bar{i} \bar{l}} \varepsilon _{i j}  \varepsilon _{\dot{\gamma } \dot{\delta }} \lambda _{\bar{i}}^{\alpha } \lambda _{\dot{\alpha }}^i 
+ \varepsilon ^{\gamma  \alpha} \varepsilon^{\bar{k} \bar{l}}  \varepsilon _{\dot{\delta } \dot{\alpha }} \varepsilon _{\dot{\rho } \dot{\gamma }} \lambda _j^{\dot{\rho }} \lambda _{\gamma }^{\bar{j}}) \,. 
\end{aligned} 
\end{equation}
In the expressions above, we have used the definitions 
\begin{equation}
\widetilde\theta^4  = 12 \, \theta^1_{1} \theta^1_{2} \theta^2_{1} \theta^2_{2} \, ,  \quad \quad
\widetilde\psi^4   = 12 \, \psi^{\dot{1}}_{\bar{1}} \psi^{\dot{1}}_{\bar{2}} \psi^{\dot{2}}_{\bar{1}} \psi^{\dot{2}}_{\bar{2}} \, .
\end{equation}
At order $\theta^6$, the contribution is  
\begin{equation}
\begin{aligned}
V^{n^4}_{I,6}  & =  \frac{1}{12} e^{n (w+z)} n^4 \widetilde\theta^4  (\eta \cdot \widetilde{x})\widetilde{\psi }_{\bar{i}}^{\dot{\alpha }} \widetilde{\psi }_{\bar{j}}^{\dot{\gamma }}  (\varepsilon ^{\bar{j} \bar{k}} \varepsilon _{ \dot{\alpha } \dot{\gamma }} \lambda _{\alpha }^{\bar{i}} \lambda _{\bar{k}}^{\alpha }  + \varepsilon ^{\bar{i} \bar{j}} \varepsilon _{\dot{\delta } \dot{\gamma }} \lambda _i^{\dot{\delta }} \lambda _{\dot{\alpha }}^i )\\
& -\frac{1}{12} e^{n (w+z)} n^4  \widetilde\psi^4  (\eta \cdot \widetilde{x})   \widetilde{\theta }_{\alpha }^i \widetilde{\theta }_{\gamma }^j (\varepsilon ^{ \delta  \gamma } \varepsilon _{i j}  \lambda _{\delta }^{\bar{i}} \lambda _{\bar{i}}^{\alpha } 
+ \varepsilon ^{\alpha  \gamma} \varepsilon _{k i} \lambda _j^{\dot{\alpha }} \lambda _{\dot{\alpha }}^k ) \\
& +
\frac{2}{9} e^{n (w+z)} n^4  (\eta \cdot \widetilde{x}) \widetilde{\theta }_{\alpha }^i \widetilde{\theta }_{\gamma }^j \widetilde{\theta }_{\delta }^k \widetilde{\psi }_{\bar{i}}^{\dot{\alpha }} \widetilde{\psi }_{\bar{j}}^{\dot{\gamma }} \widetilde{\psi }_{\bar{k}}^{\dot{\eta }}(
\varepsilon ^{\alpha  \gamma} \varepsilon^{\eta  \delta} \varepsilon^{\bar{j} \bar{k}} \varepsilon _{i j} \varepsilon _{k l} \varepsilon _{\dot{\alpha } \dot{\gamma }} \lambda _{\eta }^{\bar{i}} \lambda _{\dot{\eta }}^l  
-
\varepsilon ^{\gamma  \alpha} \varepsilon^{\bar{i} \bar{j}} \varepsilon^{\bar{l} \bar{k}} \varepsilon _{i k}  \varepsilon _{\dot{\eta } \dot{\alpha }} \varepsilon _{\dot{\delta } \dot{\gamma }} \lambda _k^{\dot{\delta }} \lambda _{\bar{l}}^{\delta } ) \,. 
\end{aligned}
\end{equation}
Finally, the two remaining terms are 
\begin{equation}
V^{n^5}_{I,8} = \frac{1}{144} e^{n (w+z)} n^5 \widetilde\theta^4 \widetilde\psi^4 (\eta \cdot \widetilde{x})( \lambda^{\bar{i}}_{\alpha} \varepsilon^{\alpha \beta} \varepsilon_{\bar{i} \bar{j}} \lambda^{\bar{j}}_{\beta} -\lambda_{i}^{\dot\alpha} \varepsilon_{\dot\alpha \dot\beta} \varepsilon^{i j} \lambda_{j}^{\dot\beta}) \, ,
\end{equation}
and 
\begin{equation}
V^{n^4}_{I,10} = 
\frac{1}{72} e^{-z+n (w+z)} n^4 \widetilde\theta^4 \widetilde\psi^4 
\widetilde{\theta }_{\dot{\alpha }}^i  \eta _{\alpha }^{\dot{\alpha }}  \widetilde{\psi }_{\bar{i}}^{\alpha }(
\varepsilon ^{\bar{i} \bar{j}} \varepsilon _{\dot{\beta } \dot{\gamma }} \lambda _i^{\dot{\beta }} \lambda _{\bar{j}}^{\dot{\gamma }} - \varepsilon ^{\beta  \gamma} \varepsilon _{j i}  \lambda _{\beta }^j \lambda _{\gamma }^{\bar{i}}
) \, . 
\end{equation}

\section{Finite transformations in $g_{\rm{old}}$}
\label{finiteOldcoset}

In this Appendix, we collect more results for the finite translations of the variables of $g_{\rm{old}}$.The transformations of the variables admit a fermionic expansion, and 
the subscripts below indicate the fermionic order in the expansion.    
Here, only the transformations up to order $\theta^2$ are shown, and 
we use the definitions \eqref{fewdefinitions} and   \eqref{OmegaFactor}.

\paragraph{Order $\theta$ (Space-time translations)}
  \begin{align}
  (\theta_1)^\alpha_i & = \frac{1}{\Omega} \left( ((1+x\eta)\delta^\alpha_\beta - (x\eta)^\alpha_\beta) \theta^\beta_i - e^{-2z} (\ve\ve\eta)^\alpha_\bd \theta^\bd_i\right) \, , \\
 (\theta_1)^\ad_i & = -\eta^\ad_\beta \theta^\beta_i + (\delta^\ad_\bd + (\eta x)^\ad_\bd) \theta^\bd_i \, , \nonumber \\
 (\theta_1)^\alpha_\ib & = \frac{1}{\Omega} \left( ((1+x\eta)\delta^\alpha_\beta - (x\eta)^\alpha_\beta) \theta^\beta_\ib - e^{-2z} (\ve\ve\eta)^\alpha_\bd \theta^\bd_\ib\right) \, , \nonumber \\
 (\theta_1)^\ad_\ib & = -\eta^\ad_\beta \theta^\beta_\ib + (\delta^\ad_\bd + (\eta x)^\ad_\bd ) \theta^\bd_\ib \, , \nonumber \\
 (\theta_1)^i_\alpha & = \theta^i_\beta (\delta^\beta_\alpha + (x\eta)^\beta_\alpha) + \theta^i_\bd \eta^\bd_\alpha \, , \nonumber \\
 (\theta_1)^i_\ad & = \frac{1}{\Omega} \left( \theta^i_\bd ( \delta^\bd_\ad (1+\eta x) - (\eta x)^\bd_\ad ) + e^{-2z} \theta^i_\beta (\ve\ve\eta)^\beta_\ad \right) \, , \nonumber  \\  
 (\theta_1)^\ib_\alpha & = \theta^\ib_\beta (\delta^\beta_\alpha + (x\eta)^\beta_\alpha ) + \theta^\ib_\bd \eta^\bd_\alpha \, , \nonumber  \\
(\theta_1)^\ib_\ad & = \frac{1}{\Omega} \left( \theta^\ib_\bd (\delta^\bd_\ad (1+\eta x) - (\eta x)^\bd_\ad ) + e^{-2z} \theta^\ib_\beta (\ve\ve\eta)^\beta_\ad \right) \, . \nonumber 
  \end{align}
  
\paragraph{Order $\theta$ ($R$-charge translations)}
  \begin{align}
  (\theta_1)^\alpha_i & = \frac{1}{\overline\Omega} 
  \left(\theta^\alpha_j  (\delta^j_i (1 + \overline{x\eta}) - (\eta x)^j_i ) + e^{2w} \theta^\a_\jb (\ve\ve \eta)^\jb_i \right) \, ,  \\
  (\theta_1)^\ad_i & = \frac{1}{\overline\Omega} \left( \theta^\ad_j (\delta^j_i (1 + \overline{x\eta}) - (\eta x)^j_i ) + e^{2w} \theta^\ad_\jb (\ve\ve \eta)^\jb_i \right) \, , \nonumber \\
   (\theta_1)^\a_\ib & = \theta^\a_j \eta^j_\ib + \theta^\a_\jb \left( \delta^\jb_\ib + (x \eta)^\jb_\ib \right) \, , \nonumber \\
(\theta_1)^\ad_\ib & = \theta^\ad_j \eta^j_\ib + \theta^\ad_\jb \left( \delta^\jb_\ib + (x \eta)^\jb_\ib \right) \, , \nonumber  \\
 (\theta_1)^i_\a & = \left(\delta^i_j + (\eta x)^i_j \right) \theta^j_\a - \eta^i_\jb \theta^\jb_\a \, , \nonumber  \\
 (\theta_1)^i_\ad & =  \left(\delta^i_j + (\eta x)^i_j \right) \theta^j_\ad - \eta^i_\jb \theta^\jb_\ad \, , \nonumber \\
(\theta_1)^\ib_\a & =  \frac{1}{\overline\Omega} \left( (\delta^\ib_\jb(1+\overline{x\eta}) - (x \eta)^\ib_\jb)\theta^\jb_\a + e^{2w} (\ve\ve\eta)^\ib_j \theta^j_\a \right) \, , \nonumber \\
(\theta_1)^\ib_\ad & = \frac{1}{\overline\Omega} \left( (\delta^\ib_\jb(1+\overline{x\eta}) - (x \eta)^\ib_\jb)\theta^\jb_\ad + e^{2w} (\ve\ve\eta)^\ib_j \theta^j_\ad \right) \, . 
\end{align}    
  
\paragraph{Order $\theta^2$ (Space-time translations)}
  \begin{align}
  e^{z_0}z_2 & = 0 \, , \\
  e^{w_0}w_2 & = 0 \, , \nonumber  \\
  (x_2)^\a_\ad & = 0 \, , \nonumber \\
  (x_2)^\ib_i & = \frac{1}{e^{2z}\Omega}\Bigg[ \left( -\theta^\ib_\a\theta^\a_i + \theta^\ib_\ad\theta^\ad_i \right)\frac{(\eta\cdot\eta)}{2} + \theta^\ib_\a\theta^\ad_i\left( (\ve\ve\eta)^\a_\ad + x^\a_\ad\frac{(\eta\cdot\eta)}{2} \right) \nonumber \\
& \qquad -e^{2z}\theta^\ib_\a\theta^\ad_i\left( \eta^\ad_\a + (\ve\ve x)^\ad_\a \frac{(\eta\cdot\eta)}{2} \right) + e^{2z+2w} (\ve\ve\theta)_i^\ad (\ve\ve\theta)^\ib_\a\left( (\ve\ve\eta)^\a_\ad + x^\a_\ad \frac{(\eta\cdot\eta)}{2} \right) \nonumber \\
& \qquad + e^{2w} \left( (\ve\ve\theta)^\a_i (\ve\ve\theta)_\a^\ib - (\ve\ve\theta)^\ad_i (\ve\ve\theta)_\ad^\ib \right) \frac{(\eta\cdot\eta)}{2} - e^{2w}(\ve\ve\theta)^\a_i(\ve\ve\theta)^\ib_\ad \left( \eta^\ad_\a + (\ve\ve x)^\ad_\a \frac{(\eta\cdot\eta)}{2} \right) \Bigg] \, . \nonumber 
  \end{align}
  
  \paragraph{Order $\theta^2$ ($R$-charge translations)}
  \begin{align}
  e^{z_0}z_2 & = - \frac{1}{2\overline\Omega} \Bigg[ e^z \left( \theta^\ib_\ad \theta^\ad_i - \theta^\ib_\a \theta^\a_i \right) \left( \eta^i_\ib + (\ve\ve x)^i_\ib \frac{(\bar\eta \cdot \bar\eta)}{2} \right) + e^{z+2w}\left( \theta^i_\a\theta^\a_\ib - \theta^i_\ad\theta^\ad_\ib \right)\left( (\ve\ve\eta)^\ib_i + x^\ib_i \frac{(\bar\eta \cdot \bar\eta)}{2} \right) \nonumber \\
& \qquad + e^{z+2w}\left( \theta^i_\a\theta^\a_i - \theta^i_\ad\theta^\ad_i - \theta^\ib_\a\theta^\a_\ib + \theta^\ib_\ad\theta^\ad_\ib \right)\frac{(\bar\eta \cdot \bar\eta)}{2} \Bigg] \, , \\
 e^{w_0}w_2 & = \frac{1}{2\overline\Omega^2} \Bigg[ e^{3w}\left( \theta^i_\ad\theta^\ad_i + \theta^i_\a \theta^\a_i + \theta^\ib_\ad \theta^\ad_\ib + \theta^\ib_\a \theta^\a_\ib \right) \frac{(\bar\eta \cdot \bar\eta)}{2} + e^{3w} \left( \theta^i_\a\theta^\a_\ib - \theta^i_\ad\theta^\ad_\ib \right) \left( (\ve\ve\eta)^\ib_i + x^\ib_i \frac{(\bar\eta \cdot \bar\eta)}{2} \right) \nonumber \\
& \qquad - e^w \left( \theta^\ib_\a \theta^\a_i - \theta^\ib_\ad \theta^\ad_i \right)\left(\eta^i_\ib +(\ve\ve x)^i_\ib \frac{(\bar\eta \cdot \bar\eta)}{2} \right) \Bigg] \, , \nonumber \\
 (x_2)^\a_\ad & = -\frac{1}{e^{2z}\overline\Omega} \Bigg[ (\ve\ve\theta)^\a_\ib (\ve\ve\theta)_\ad^i \left( (\ve\ve\eta)^\ib_i + x^\ib_i \frac{(\bar\eta \cdot \bar\eta)}{2} \right)  - e^{2z} \theta^\ib_\ad\theta^\a_i \left( \eta^i_\ib + (\ve\ve x)^i_\ib \frac{(\bar\eta \cdot \bar\eta)}{2} \right) \nonumber \\
& \qquad - e^{2w} (\ve\ve\theta)^\a_i (\ve\ve\theta)^\ib_\ad \left( \eta^i_\ib + (\ve\ve x)^i_\ib\frac{(\bar\eta \cdot \bar\eta)}{2} \right) + e^{2w} \left( (\ve\ve\theta)^\a_\ib(\ve\ve\theta)_\ad^\ib - (\ve\ve\theta)^\a_i(\ve\ve\theta)_\ad^i \right) \frac{(\bar\eta \cdot \bar\eta)}{2} \nonumber \\
& \qquad + e^{2z+2w} \left( \theta^i_\ad\theta^\a_i - \theta^\ib_\ad\theta^\a_\ib \right) \frac{(\bar\eta \cdot \bar\eta)}{2} + e^{2z+2w} \theta^i_\ad\theta^\a_\ib \left( (\ve\ve\eta)^\ib_i + x^\ib_i \frac{(\bar\eta \cdot \bar\eta)}{2} \right) \Bigg] \, , \nonumber \\
(x_2)^\ib_i & = -\frac{e^{2w}}{\overline\Omega^2} \left( \theta^j_\ad\theta^\ad_j + \theta^\jb_\a\theta^\a_\jb \right) \left( (1+ \overline{x\eta})(\ve\ve\eta)^\ib_i - x^\ib_i\frac{(\bar\eta \cdot \bar\eta)}{2}  \right) \, . \nonumber 
  \end{align}

\section{The $g_{\rm{new}}$ BRST transformations} 
\label{BRSTideal}

In this Appendix, we present explicit formulas for the BRST transformations of the variables in $g_{\rm{new}}$, along with their derivations. 
The transformations are expressed using both superalgebra commutators and $\mathfrak{su}_2^{\oplus 4}$-component notation. 
We also extract and display the fermionic covariant derivatives arising from these transformations. Finally, we discuss the general transformation of the BRST operator under arbitrary coset reparametrizations, with the transformation $g_{\rm{old}} \leftrightarrow g_{\rm{new}}$ serving as a concrete example.

\subsection{BRST transformations of $g_{\rm{new}}$: derivation}
\label{BRST transformation new}

We changed the variables of the vertex operators from \cite{VertexOld} to the variables of $g_{\rm{new}}$ to facilitate the action of global transformations. After this %
reparametrization, the vertex operators lie in the cohomology of the BRST operator $Q_n$ acting on the variables of $g_{\rm{new}}$. In this Subsection, we derive the BRST transformations of the new variables. One approach is to use the general formulas presented in Subsection E.4 below. Alternatively, it is instructive to solve the following defining equation
\begin{equation}
Q_{n} \cdot g_{\rm{new}}=g_{\rm{new}} \,\left(\lambda_{A}+\lambda_{B}+\Sigma_{n}\right)\,,    
\label{BRSTnewvariables}
\end{equation}
where
\begin{equation}
\lambda_{A}=\lambda^{R}_{I}q^{I}_{R}\,,\qquad \lambda_{B}=\lambda^{I}_{R}q^{R}_{I}\,,\qquad \Sigma_{n}\in \mathfrak{usp}(2,2)\oplus\mathfrak{usp}(4) \, .    
\label{defllsigma} 
\end{equation}

To solve \eqref{BRSTnewvariables} and determine the BRST transformations, it suffices to compute the contribution of the pure spinor variable $\lambda_A$. The contribution of $\lambda_B$ to the transformations is obtained by suitable replacements in the formulas below. The gauge-invariant pure spinors 
$\Lambda^R_I$ and $\bar\Lambda^I_R$ of \eqref{defLambda} appear naturally in the manipulations. One has
 \begin{equation}
 \begin{aligned}
g_{\rm{new}} \,\lambda_{A} & = e^{\Psi} e^{\Theta} U V  \lambda_{A}= e^{\Psi} e^{\Theta}  \Lambda U V  \\
& = e^{\Psi} \left( \Lambda + [\Theta, \Lambda] + \frac{1}{2} [\Theta,[\Theta,\Lambda]] \right) e^{\Theta} U V \, ,
\end{aligned} 
\end{equation}
 where $\Lambda$ without free indices is obtained by contracting 
$\Lambda^R_I$ with fermionic generators.
The first term inside the parentheses  contributes to $Q_n \cdot \Psi$ and the last one
contributes to $Q_n \cdot \Theta$.  It is convenient to rewrite the middle term containing  $m_Q = [\Theta, \Lambda] \in \mathfrak{su}(2,2)\oplus\mathfrak{su}(4)$ 
as  
\begin{equation}
e^{\Psi}m_{Q}\,e^{\Theta}UV=e^{\Psi}e^{\Theta}(m_{Q}+[m_{Q},\Theta])U V \,,  
\label{mQ}
\end{equation}
and the term containing $[m_{Q},\Theta]$ also contributes to $Q_n \cdot \Theta$. We are left with a purely bosonic computation that determines the gauge compensators and the transformations of the bosonic variables 
\begin{equation}
m_{Q}\,UV(1+\Sigma_{n})= e^{X+ Q_n \cdot X} e^{Z +Q_n \cdot Z}
e^{Y+ Q_n \cdot Y} e^{W + Q_n \cdot W} \, .
\end{equation}
The equation above can be solved by the same manipulations as
the ones of the Appendix \ref{psu224ideal} by replacing 
$[B,\Psi]$ there with $m_{Q}$ in \eqref{eqBallpo}.   
The resulting 
BRST transformations starting from the fermionic variables ones are given by
\begin{equation}
Q_n \cdot \Psi^S_I = \Lambda^S_I \, ,
\label{BRSTPsi}
\end{equation}
and  
\begin{equation}
Q_n \cdot \Theta^I_R = \bar\Lambda^I_R - \Theta^I_S \Lambda^S_J \Theta^J_R \, .
\label{BRSTTheta}
\end{equation}

To write down the bosonic transformations, we decompose $m_Q  \in \mathfrak{su}(2,2)\oplus\mathfrak{su}(4)$ as     
\begin{equation}
m_{\text{Q}}\,
= m_{\text{Q}}^{\Delta}+m_{\text{Q}}^{J}+m_{\text{Q}}^{P}
+m_{\text{Q}}^{K}+m_{\text{Q}}^{\widehat{M}} \, , 
\label{mq}
\end{equation}
where the superscript generators indicates the contributions linear in the respective generators. 
The $Z$ and $W$ transformations are
\begin{equation}
Q_n \cdot Z  =   m_{Q}^{\Delta}+ [m_{\text{Q}}^{P^{\sigma}_{\dot\sigma}},X^{\alpha}_{\dot\alpha} K^{\dot\alpha}_{\alpha}]^{\Delta}\,, \quad \quad
Q_n \cdot W  =   m_{Q}^{J}+[m_{\text{Q}}^{P^{\bar{k}}_{k}},Y^{\bar{i}}_i K^{i}_{\bar{i}}]^{J}\, , 
\end{equation}
and for the $X$'s and $Y$'s, one has 
\begin{equation}
\begin{aligned}
(Q_n \cdot X^{\alpha}_{\dot\alpha}) K^{\dot\alpha}_{\alpha} & = 
-e^{2 Z} (m_Q)^{\dot\alpha}_{\alpha} \epsilon^{\alpha \beta} \epsilon_{\dot\alpha \dot\beta} K^{\dot\beta}_{\beta} + m_Q^{K^{\dot\alpha}_{\alpha}} + [m_Q^{\widehat{M}^{\alpha}_{\alpha}}+m_Q^{\widehat{M}^{\dot\alpha}_{\dot\alpha}} + m_Q^{\Delta},X^{\beta}_{\dot\beta} K^{\dot\beta}_{\beta}] \\ 
& + \frac{1}{2} [ [m_Q^{K^{\dot\alpha}_{\alpha}},X^{\beta}_{\dot\beta} K^{\dot\beta}_{\beta}],X^{\gamma}_{\dot\gamma} K^{\dot\gamma}_{\gamma}] \, ,    \\
(Q_n \cdot Y^{\bar{i}}_{i}) K^{i}_{\bar{i}} & = -e^{- 2 W} (m_Q)^{i}_{\bar{i}} \epsilon^{\bar{i} \bar{j}} \epsilon_{i j} M^{j}_{\bar{i}} + m_Q^{K^{i}_{\bar{i}}} + [m_Q^{\widehat{M}^{i}_{i}}+m_Q^{\widehat{M}^{\bar{i}}_{\bar{i}}} + m_Q^{J},Y^{\bar{i}}_{i} K^{i}_{\bar{i}}] \\
& + \frac{1}{2} [ [m_Q^{K^{i}_{\bar{i}}},Y^{\bar{j}}_{j} K^{j}_{\bar{j}}],Y^{\bar{k}}_{k} K^{k}_{\bar{k}}] \, .
\end{aligned} 
\end{equation}
Finally, the compensators $\Sigma$'s are given by   
\begin{equation}
\begin{aligned}
\Sigma_{Q_n}|^{\mathfrak{usp}(2|2)} & = - m_{Q}^{\widehat{M}^{\alpha}_{\alpha}} - m_{Q}^{\widehat{M}^{\dot{\alpha}}_{\dot{\alpha}}} - [m_{Q}^{P^{\alpha}_{\dot\alpha}}, X^{\beta}_{\dot\beta} K^{\dot\beta}_{\beta}] - 
e^{-Z}(m_{Q})^{\dot\alpha}_{\alpha}\left(P^{\alpha}_{\dot\alpha}+\varepsilon_{\dot\alpha\dot\beta}\varepsilon^{\alpha\beta}K^{\dot\beta}_{\beta}\right) \, ,  \\
\Sigma_{Q_n}|^{\mathfrak{usp}(4)} & = - m_{Q}^{\widehat{M}^i_i} -m_{Q}^{\widehat{M}^{\bar{i}}_{\bar{i}}}- [m_{Q}^{P^{\bar{i}}_i}, Y^{\bar{i}}_i K^{i}_{\bar{i}}] 
-e^{W}(m_{Q})^{i}_{\overline i}\left(P^{\overline i}_{i}+\varepsilon_{ij}\varepsilon^{\overline i\overline j}K^{j}_{\overline j}\right) \,.   
\end{aligned} 
\end{equation}

For several applications, such as the construction of integrated vertex operators, knowledge of the covariant derivatives is required. Recall that the BRST operator can also be written in terms of the fermionic covariant derivatives as
\begin{equation}
Q_n = \lambda^I_{R} \nabla^{R}_I +\lambda_I^{R} \nabla_{R}^I \, .
\end{equation}
Thus, the fermionic covariant derivatives can be read off from the transformations derived above. Their explicit expressions are given in Subsection
\ref{covariantderivativesfermionic}. 

\subsection{BRST transformations in $\mathfrak{su}_2^{\oplus 4}$-component notation}

In the expressions below, the following definitions are used
\begin{equation}
\begin{aligned}
& (\widetilde\lambda^i_{\alpha}, \widetilde\lambda^{\dot\alpha}_{\bar{i}}) = e^{\frac{Z-W}{2}}  (\lambda^i_{\alpha}, \lambda^{\dot\alpha}_{\bar{i}}) \, , \quad & (\widetilde\lambda^{\bar{i}}_{\dot\alpha}, \widetilde\lambda^{\alpha}_i) = e^{\frac{W-Z}{2}}  (\lambda^{\bar{i}}_{\dot\alpha}, \lambda^{\alpha}_i) \, , \\
& (\widetilde\lambda^i_{\dot{\alpha}} , \widetilde\lambda^{\alpha}_{\bar{i}}) =  
e^{-\frac{Z+W}{2}} (\lambda^i_{\dot{\alpha}} , \lambda^{\alpha}_{\bar{i}}) \, ,  \quad 
& (\widetilde\lambda^{\bar{i}}_{\alpha}, \widetilde\lambda^{\dot{\alpha}}_i) = e^{\frac{Z+W}{2}} (\lambda^{\bar{i}}_{\alpha}, \lambda^{\dot{\alpha}}_i)  \, . 
\end{aligned} 
\end{equation}
We first consider the transformations of the fermionic coordinates. These are given for the $\Psi$'s by 
\begin{equation}
\begin{aligned}
& Q_{n} \cdot \Psi^{\dot\alpha}_{\bar{i}} = 
\widetilde\lambda^{\dot\alpha}_{\bar{i}} \, ,  \\
& Q_{n} \cdot \Psi^{\alpha}_i = 
\widetilde\lambda^{\alpha}_i + 
Y^{\bar{j}}_i \widetilde\lambda^{\alpha}_{\bar{j}} -  
X^{\alpha}_{\dot\alpha} \widetilde\lambda^{\dot\alpha}_ i
- 
Y^{\bar{j}}_i X^{\alpha}_{\dot\alpha} \widetilde\lambda^{\dot\alpha}_{\bar{j}} \, ,\\
& Q_{n} \cdot \Psi^{\alpha}_{\bar{i}} =  
\widetilde\lambda^{\alpha}_{\bar{i}} -
X^{\alpha}_{\dot\beta} \widetilde\lambda^{\dot\beta}_{\bar{i}} \, , \quad \quad 
 Q_{n} \cdot \Psi^{\dot\alpha}_i =  
 \widetilde\lambda^{\dot\alpha}_i + 
 Y^{\bar{j}}_i \widetilde\lambda^{\dot\alpha}_{\bar{j}} \, , 
\end{aligned}
\end{equation}
and for the $\Theta$'s,
\begin{equation}
\begin{aligned}
Q_{n} \cdot \Theta^i_{\alpha} & = \widetilde\lambda^i_{\alpha} - \Theta^i_{\beta} \Theta^{\bar{j}}_{\alpha} \widetilde\lambda^{\beta}_{\bar{j}} + \Theta^j_{\alpha} \Theta^i_{\beta} \widetilde\lambda^{\beta}_{j} - \Theta^i_{\dot\alpha} \Theta^{\bar{j}}_{\alpha} \widetilde\lambda^{\dot\alpha}_{\bar{j}} + \Theta^j_{\alpha} \Theta^i_{\dot\alpha} \widetilde\lambda^{\dot\alpha}_j \\
 & +  \Theta^j_{\alpha} \Theta^i_{\beta} Y^{\bar{i}}_j \widetilde\lambda^{\beta}_{\bar{i}} - \Theta^j_{\alpha} \Theta^i_{\beta} X^{\beta}_{\dot\beta} \widetilde\lambda^{\dot\beta}_j+ \Theta^i_{\beta} \Theta^{\bar{i}}_{\alpha}  X^{\beta}_{\dot\beta} \widetilde\lambda^{\dot\beta}_{\bar{i}} + \Theta^j_{\alpha} \Theta^i_{\dot\alpha} Y^{\bar{i}}_j 
\widetilde\lambda^{\dot\alpha}_{\bar{i}} - \Theta^j_{\alpha} \Theta^i_{\beta}  Y_j^{\bar{i}} X^{\beta} _{\dot\alpha} \widetilde\lambda^{\dot\alpha}_{\bar{i}} \, , \\
Q_{n} \cdot \Theta^i_{\dot\alpha} & = \widetilde\lambda^i_{\dot\alpha}+ X^{\beta}_{\dot\alpha} \widetilde\lambda^i_{\beta}  - \Theta^i_{\beta} \Theta^{\bar{i}}_{\dot\alpha} \widetilde\lambda^{\beta}_{\bar{i}} 
+ \Theta^j_{\dot\alpha} \Theta^i_{\beta} \widetilde\lambda^{\beta}_{j} - \Theta^i_{\dot\beta} \Theta^{\bar{j}}_{\dot\alpha} \widetilde\lambda^{\dot\beta}_{\bar{j}} +
\Theta^j_{\dot\alpha} \Theta^i_{\dot\beta} \widetilde\lambda^{\dot\beta}_j  \\
& + \Theta^j_{\dot\alpha} \Theta^i_{\beta} Y^{\bar{i}}_j \widetilde\lambda^{\beta}_{\bar{i}} -
\Theta^j_{\dot\alpha} \Theta^i_{\beta} X^{\beta}_{\dot\beta} \widetilde\lambda^{\dot\beta}_j + \Theta^i_{\beta} \Theta^{\bar{i}}_{\dot\alpha} X^{\beta}_{\dot\beta} \widetilde\lambda^{\dot\beta}_{\bar{i}} + \Theta^j_{\dot\alpha} \Theta^i_{\dot\beta} Y^{\bar{j}}_j \widetilde\lambda^{\dot\beta}_{\bar{j}} - \Theta^j_{\dot\alpha} \Theta^i_{\beta} Y^{\bar{i}}_j X^{\beta}_{\dot\beta} \widetilde\lambda^{\dot\beta}_{\bar{i}} \, , \\     
Q_{n} \cdot \Theta^{\bar{i}}_{\alpha} &=  \widetilde\lambda^{\bar{i}}_{\alpha} - Y^{\bar{i}}_j \widetilde\lambda^j_{\alpha}+ \Theta^j_{\alpha} \Theta^{\bar{i}}_{\beta} \widetilde\lambda^{\beta}_j + \Theta^{\bar{j}}_{\alpha} \Theta^{\bar{i}}_{\beta} \widetilde\lambda^{\beta}_{\bar{j}} + \Theta^j_{\alpha} \Theta^{\bar{i}}_{\dot\alpha} \widetilde\lambda^{\dot\alpha}_j  + \Theta^{\bar{j}}_{\alpha} \Theta^{\bar{i}}_{\dot\alpha} \widetilde\lambda^{\dot\alpha}_{\bar{j}} \\
&  + \Theta^j_{\alpha} \Theta^{\bar{i}}_{\beta} Y^{\bar{j}}_j \widetilde\lambda^{\beta}_{\bar{j}}    
- \Theta^j_{\alpha} \Theta^{\bar{j}}_{\beta} X^{\beta}_{\dot\beta} \widetilde\lambda^{\dot\beta}_j - \Theta^{\bar{j}}_{\alpha} \Theta^{\bar{i}}_{\beta} X^{\beta}_{\dot\beta} \widetilde\lambda^{\dot\beta}_{\bar{j}} + \Theta^j_{\alpha} \Theta^{\bar{i}}_{\dot\alpha} Y^{\bar{k}}_j \widetilde\lambda^{\dot\alpha}_{\bar{k}} - \Theta^j_{\alpha} \Theta^{\bar{i}}_{\beta} Y_j^{\bar{k}} X^{\beta}_{\dot\beta} \widetilde\lambda^{\dot\beta}_{\bar{k}} \, ,  \\
Q_{n} \cdot \Theta^{\bar{i}}_{\dot\alpha}  & = \widetilde\lambda^{\bar{i}}_{\dot\alpha} + \Theta^j_{\dot\alpha} \Theta^{\bar{i}}_{\beta} \widetilde\lambda^{\beta}_j  + \Theta^{\bar{j}}_{\dot\alpha} \Theta^{\bar{i}}_{\beta} \widetilde\lambda^{\beta}_{\bar{j}} + \Theta^j_{\dot\alpha} \Theta^{\bar{i}}_{\dot\beta} \widetilde\lambda^{\dot\beta}_j + \Theta^{\bar{j}}_{\dot\alpha} \Theta^{\bar{i}}_{\dot\beta} \widetilde\lambda^{\dot\beta}_{\bar{j}}       \\
& - Y^{\bar{i}}_j \widetilde\lambda^j_{\dot\alpha} + X^{\beta}_{\dot\alpha} \widetilde\lambda^{\bar{i}}_{\beta} + \Theta^j_{\dot\alpha} \Theta^{\bar{i}}_{\beta} Y^{\bar{j}}_j \widetilde\lambda^{\beta}_{\bar{j}}  - \Theta^j_{\dot\alpha} \Theta^{\bar{i}}_{\beta} X^{\beta}_{\dot\beta} \widetilde\lambda^{\dot\beta}_j - \Theta^{\bar{j}}_{\dot\alpha} \Theta^{\bar{i}}_{\beta}  X^{\beta}_{\dot\beta} \widetilde\lambda^{\dot\beta}_{\bar{j}} + \Theta^k_{\dot\alpha} \Theta^{\bar{i}}_{\dot\beta} Y^{\bar{j}}_k \widetilde\lambda^{\dot\beta}_{\bar{j}}  \\
& - Y^{\bar{i}}_j X^{\beta}_{\dot\alpha} \widetilde\lambda^j_{\beta} - \Theta^i_{\dot\alpha} \Theta^{\bar{i}}_{\beta} Y^{\bar{k}}_i X^{\beta}_{\dot\beta} \widetilde\lambda^{\dot\beta}_{\bar{k}}  \, . 
\end{aligned}
\end{equation}
The transformations of the bosonic variables are 
\begin{equation}
\begin{aligned}
2 Q_{n}  \cdot Z & = \widetilde\lambda^{\alpha}_i \Theta^{i}_{\alpha} +  \widetilde\lambda^{\alpha}_{\bar{i}} \Theta^{\bar{i}}_{\alpha} - \widetilde\lambda^{\dot\alpha}_i \Theta^i_{\dot\alpha}-\widetilde\lambda^{\dot\alpha}_{\bar{i}} \Theta^{\bar{i}}_{\dot\alpha}  \\
& + \widetilde\lambda^{\alpha}_{\bar{i}} Y^{\bar{i}}_{i} \Theta^i_{\alpha} + \widetilde\lambda^{\dot\alpha}_{i} X^{\alpha}_{\dot\alpha} \Theta^i_{\alpha}   
+ \widetilde\lambda^{\dot\alpha}_{\bar{i}} X_{\dot\alpha}^{\alpha} \Theta^{\bar{i}}_{\alpha} - \widetilde\lambda^{\dot\alpha}_{\bar{i}} Y^{\bar{i}}_i \Theta^i_{\dot\alpha} 
+ X^{\alpha}_{\dot\alpha} Y^{\bar{i}}_i \Theta^i_{\alpha} \widetilde\lambda^{\dot\alpha}_{\bar{i}} \, , \\
2 Q_{n} \cdot W & = \widetilde\lambda^{\alpha}_i \Theta^{i}_{\alpha} - \widetilde\lambda^{\alpha}_{\bar{i}} \Theta^{\bar{i}}_{\alpha} + \widetilde\lambda^{\dot\alpha}_i \Theta^i_{\dot\alpha} - \widetilde\lambda^{\dot\alpha}_{\bar{i}} \Theta^{\bar{i}}_{\dot\alpha}    \\
& - \widetilde\lambda^{\alpha}_{\bar{i}} Y^{\bar{i}}_i \Theta^i_{\alpha} - \widetilde\lambda^{\dot\alpha}_i X^{\alpha}_{\dot\alpha} \Theta^i_{\alpha}  
+ \widetilde\lambda^{\dot\alpha}_{\bar{i}} X^{\alpha}_{\dot\alpha} \Theta^{\bar{i}}_{\alpha} - \widetilde\lambda^{\dot\alpha}_{\bar{i}} Y^{\bar{i}}_{i} \Theta^{i}_{\dot\alpha}  
+ X^{\alpha}_{\dot\alpha} Y^{\bar{i}}_{i} \Theta^i_{\alpha} \widetilde\lambda^{\dot\alpha}_{\bar{i}}  \, ,
\end{aligned}
\end{equation}
and 
\begin{equation}
\begin{aligned}
Q_{n} \cdot X^{\alpha}_{\dot\alpha} & = \widetilde\lambda^{\alpha}_i \Theta^i_{\dot\alpha} + \widetilde\lambda^{\alpha}_{\bar{i}} \Theta^{\bar{i}}_{\dot\alpha}   - \epsilon_{\dot\alpha \dot\beta} \epsilon^{\alpha \beta} \Theta^j_{\beta} \widetilde\lambda^{\dot\beta}_{j}  - \epsilon_{\dot\alpha \dot\beta} \epsilon^{\alpha \beta} \Theta_{\beta}^{\bar{i}} \widetilde\lambda_{\bar{i}}^{\dot\beta} \\
& - X^{\beta}_{\dot\alpha} \Theta^j_{\beta} \widetilde\lambda^{\alpha}_j + Y^{\bar{i}}_j \Theta^j_{\dot\alpha} \widetilde\lambda^{\alpha}_{\bar{i}}  - X^{\beta}_{\dot\alpha} \Theta^{\bar{i}}_{\beta} \widetilde\lambda^{\alpha}_{\bar{i}}   - \epsilon_{\dot\alpha \dot\beta} \epsilon^{\alpha \beta} Y^{\bar{j}}_j \Theta^j_{\beta} \widetilde\lambda^{\dot\beta}_{\bar{j}} - Y^{\bar{i}}_i X^{\beta}_{\dot\alpha} \Theta^i_{\beta} \widetilde\lambda^{\alpha}_{\bar{i}} \, ,  \\
Q_{n} \cdot  Y^{\bar{i}}_{i} & = \widetilde\lambda^{\beta}_i  \Theta^{\bar{i}}_{\beta} +  \widetilde\lambda^{\dot\alpha}_i \Theta^{\bar{i}}_{\dot\alpha}   - \epsilon_{i j} \epsilon^{\bar{i} \bar{j}} \Theta^j_{\beta} \widetilde\lambda^{\beta}_{\bar{j}} - \epsilon_{i j} \epsilon^{\bar{i} \bar{j}} \Theta^j_{\dot\beta} \widetilde\lambda^{\dot\beta}_{\bar{j}}\\
& + Y^{\bar{i}}_j \Theta^j_{\beta} \widetilde\lambda^{\beta}_i  + Y^{\bar{i}}_j \Theta^j_{\dot\alpha} \widetilde\lambda^{\dot\alpha}_i -
X^{\beta}_{\dot\beta} \Theta^{\bar{i}}_{\beta} \widetilde\lambda^{\dot\beta}_i + \epsilon_{i j} \epsilon^{\bar{i} \bar{j}} X^{\beta}_{\dot\beta} \Theta^j_{\beta} \widetilde\lambda^{\dot\beta}_{\bar{j}}  - Y^{\bar{i}}_j X^{\beta}_{\dot\beta} \Theta^j_{\beta} \widetilde\lambda^{\dot\beta}_i   \, . 
\end{aligned}
\end{equation}
Finally, the compensators are given by  
\begin{equation}
\begin{aligned}
& \Sigma^{\dot\alpha}_{\dot\beta} = - \widetilde\lambda^{\dot\alpha}_i \Theta^i_{\dot\beta} - \widetilde\lambda^{\dot\alpha}_{\bar{i}} \Theta^{\bar{i}}_{\dot\beta} 
+ \widetilde\lambda^{\dot\alpha}_{i} X^{\alpha}_{\dot\beta} \Theta^{i}_{\alpha} - \widetilde\lambda^{\dot\alpha}_{\bar{i}} Y^{\bar{i}}_i \Theta^i_{\dot\beta}  
+ \widetilde\lambda^{\dot\alpha}_{\bar{i}} X^{\alpha}_{\dot\beta} \Theta^{\bar{i}}_{\alpha} + X^{\alpha}_{\dot\beta} Y^{\bar{i}}_i \Theta^i_{\alpha} \widetilde\lambda^{\dot\alpha}_{\bar{i}} \, ,  \\
& \Sigma^{\alpha}_{\beta}  = - \widetilde\lambda^{\alpha}_i \Theta^i_{\beta} - \widetilde\lambda^{\alpha}_{\bar{i}} \Theta^{\bar{i}}_{\beta} - \widetilde\lambda^{\alpha}_{\bar{i}} Y^{\bar{i}}_i \Theta^i_{\beta} \, , \\
& \Sigma^i_j = - \widetilde\lambda^{\alpha}_j \Theta^{i}_{\alpha}  - \widetilde\lambda^{\dot\alpha}_j \Theta^i_{\dot\alpha} + \widetilde\lambda^{\dot\alpha}_j X^{\alpha}_{\dot\alpha} \Theta^i_{\alpha} \, , \\
& \Sigma^{\bar{i}}_{\bar{j}} = - \widetilde\lambda^{\alpha}_{\bar{j}} \Theta^{\bar{i}}_{\alpha} - \widetilde\lambda^{\dot\alpha}_{\bar{j}} \Theta^{\bar{i}}_{\dot\alpha} - \widetilde\lambda^{\alpha}_{\bar{j}} Y^{\bar{i}}_i \Theta^i_{\alpha} + \widetilde\lambda^{\dot\alpha}_{\bar{j}} X^{\alpha}_{\dot\alpha} \Theta^{\bar{i}}_{\alpha} -
\widetilde\lambda^{\dot\alpha}_{\bar{j}} Y^{\bar{i}}_i \Theta^i_{\dot\alpha} + X^{\alpha}_{\dot\alpha} Y^{\bar{i}}_i \Theta^i_{\alpha} \widetilde\lambda^{\dot\alpha}_{\bar{j}} \, , 
\end{aligned} 
\end{equation}
and the remaining ones are 
\begin{equation}
\begin{aligned}
e^Z \Sigma^{\dot\alpha}_{\alpha} & = - \widetilde\lambda^{\dot\alpha}_{i} \Theta^{i}_{\alpha} - \widetilde\lambda^{\dot\alpha}_{\bar{i}} \Theta^{\bar{i}}_{\alpha} - \widetilde\lambda^{\dot\alpha}_{\bar{i}} Y^{\bar{i}}_i \Theta^i_{\alpha} \, , \quad  & \Sigma^{\alpha}_{\dot\alpha} =\epsilon^{\alpha \beta} \epsilon_{\dot\alpha \dot\beta} \Sigma_{\beta}^{\dot\beta} \, , \\
e^{-W} \Sigma^{i}_{\bar{i}}  & = - \widetilde\lambda^{\alpha}_{\bar{i}} \Theta^i_{\alpha}  - \widetilde\lambda^{\dot\alpha}_{\bar{i}} \Theta^i_{\dot\alpha} 
+ \widetilde\lambda^{\dot\alpha}_{\bar{i}} X^{\alpha}_{\dot\alpha} \Theta^i_{\alpha} \, , \quad   & \Sigma^{\bar{i}}_{i} =\epsilon^{\bar{i} \bar{j}} \epsilon_{i j} \Sigma^{j}_{\bar{j}} \, .
 \end{aligned} 
\end{equation}

\subsection{Covariant Derivatives}  
\label{covariantderivativesfermionic}

Using the BRST transformations of the previous section, the fermionic covariant derivatives can be read off directly. We write them explicitly, as they are also needed to obtain the integrated vertex operators. We first write the contribution containing the derivatives of the fermionic variables.
From \eqref{BRSTPsi} and \eqref{BRSTTheta}, one has
\begin{equation}
\begin{aligned} 
\lambda^I_S \nabla_I^S \Big|_{\rm{fermionic}}  & =  \lambda^{\widetilde{I}}_{\widetilde{S}} (V^{-1})^J_{\widetilde{I}} U_{R}^{\widetilde{S}} \frac{\partial}{\partial \Theta_R^J}\, , \quad \quad \\
\lambda_I^S \nabla^I_S \Big|_{\rm{fermionic}} & =  \lambda_{\widetilde{I}}^{\widetilde{S}} V_J^{\widetilde{I}} (U^{-1})^{R}_{\widetilde{S}} \frac{\partial}{\partial \Psi^R_J} - \Theta^I_S (U^{-1})^S_{\widetilde{S}} \lambda^{\widetilde{S}}_{\widetilde{J}} V_J^{\widetilde{J}}\Theta^J_R 
\frac{\partial}{\partial \Theta^I_R} \, . 
\end{aligned} 
\end{equation}
The BRST operator also acts on the pure spinor variables due to the compensating gauge transformations. The pure spinor transformations can be written as functions of the generators given below, whose actions are
\begin{equation}
\dot{M}^{A}_B \cdot \lambda^C = \delta^C_B \lambda^A - \frac{1}{2} \delta^A_B \lambda^C \, , \quad \quad 
\dot{M}^{A}_B \cdot \lambda_C = - \delta_C^A \lambda_B + \frac{1}{2} \delta^A_B \lambda_C \, .
\end{equation}
Then one has 
\begin{equation}
\begin{aligned}
\lambda^{ \alpha}_{i} \nabla_{\alpha}^{i} \Big|_{\rm{bosonic}}& =  \widetilde\lambda^{\alpha}_i \Theta^i_{\dot\alpha} 
\frac{\partial}{\partial X^{\alpha}_{\dot\alpha}}
- X^{\beta}_{\dot\alpha} \Theta^j_{\beta} \widetilde\lambda^{\alpha}_j \frac{\partial}{\partial X^{\alpha}_{\dot\alpha}}+\Theta^{\bar{i}}_{\beta} \widetilde\lambda^{\beta}_i \frac{\partial}{\partial Y^{\bar{i}}_{i}}+ Y^{\bar{i}}_j \Theta^j_{\beta} \widetilde\lambda^{\beta}_i \frac{\partial}{\partial Y^{\bar{i}}_{i}} \\
& + \frac{1}{2} \widetilde\lambda^{\alpha}_i \Theta^{i}_{\alpha}  (\frac{\partial}{\partial Z} + \frac{\partial}{\partial W}) - \widetilde\lambda^{\alpha}_i \Theta^i_{\beta} \dot{M}^{\beta}_{\alpha}
- \widetilde\lambda^{\alpha}_j \Theta^{i}_{\alpha} \dot{M}^{j}_{i} \, ,  \\ 
\lambda^{\dot \alpha}_{i} \nabla_{\dot\alpha}^{i}  \Big|_{\rm{bosonic}} & = - \epsilon_{\dot\alpha \dot\beta} \epsilon^{\alpha \beta} \Theta^j_{\beta} \widetilde\lambda^{\dot\beta}_{j} \frac{\partial}{\partial X^{\alpha}_{\dot\alpha}}
 +  (\Theta^{\bar{i}}_{\dot\alpha} \widetilde\lambda^{\dot\alpha}_i + Y^{\bar{i}}_j \Theta^j_{\dot\alpha} \widetilde\lambda^{\dot\alpha}_i - X^{\beta}_{\dot\beta} \Theta^{\bar{i}}_{\beta} \widetilde\lambda^{\dot\beta}_i - Y^{\bar{i}}_j X^{\beta}_{\dot\beta}\Theta^j_{\beta}\widetilde\lambda^{\dot\beta}_i) \frac{\partial}{\partial Y^{\bar{i}}_{i}} \\ 
& - \frac{1}{2} (\widetilde\lambda^{\dot\alpha}_i \Theta^i_{\dot\alpha} - \widetilde\lambda^{\dot\alpha}_{i} X^{\alpha}_{\dot\alpha} \Theta^i_{\alpha})(\frac{\partial}{\partial Z} - \frac{\partial}{\partial W}) \\
& - (\widetilde\lambda^{\dot\alpha}_i \Theta^i_{\dot\beta} 
- \widetilde\lambda^{\dot\alpha}_{i} X^{\alpha}_{\dot\beta} \Theta^{i}_{\alpha} )  
\dot{M}^{\dot\beta}_{\dot\alpha} - ( \widetilde\lambda^{\dot\alpha}_j \Theta^i_{\dot\alpha}-  \widetilde\lambda^{\dot\alpha}_j X^{\alpha}_{\dot\alpha} \Theta^i_{\alpha} )\dot{M}^j_i \\
&- e^{-Z} \widetilde\lambda^{\dot\alpha}_{i} \Theta^{i}_{\alpha} (\dot{M}^{\alpha}_{\dot\alpha}+ \epsilon^{\alpha \beta} \epsilon_{\dot\alpha \dot\beta} \dot{M}^{\dot\beta}_{\beta}) \, , 
\end{aligned} 
\end{equation}
and 
\begin{equation}
\begin{aligned}
\lambda^{\alpha}_{\bar{i}} \nabla_{\alpha}^{\bar{i}}  \Big|_{\rm{bosonic}} & = 
(\widetilde\lambda^{\alpha}_{\bar{i}} \Theta^{\bar{i}}_{\dot\alpha}+Y^{\bar{i}}_j \Theta^j_{\dot\alpha} \widetilde\lambda^{\alpha}_{\bar{i}}  - X^{\beta}_{\dot\alpha} \Theta^{\bar{i}}_{\beta} \widetilde\lambda^{\alpha}_{\bar{i}}   - Y^{\bar{i}}_i X^{\beta}_{\dot\alpha} \Theta^i_{\beta} \widetilde\lambda^{\alpha}_{\bar{i}}) \frac{\partial}{\partial X^{\alpha}_{\dot\alpha}} -\epsilon_{ij} \epsilon^{\bar{i} \bar{j}} \Theta^j_{\dot\beta} \widetilde\lambda^{\dot\beta}_{\bar{j}} \frac{\partial}{\partial Y^{\bar{i}}_i} \\
& + \frac{1}{2}(\widetilde\lambda^{\alpha}_{\bar{i}} \Theta^{\bar{i}}_{\alpha}  + \widetilde\lambda^{\alpha}_{\bar{i}} Y^{\bar{i}}_{i} \Theta^i_{\alpha})(\frac{\partial}{\partial Z} - \frac{\partial}{\partial W}) 
- (\widetilde\lambda^{\alpha}_{\bar{i}} \Theta^{\bar{i}}_{\beta} + \widetilde\lambda^{\alpha}_{\bar{i}} Y^{\bar{i}}_i \Theta^i_{\beta}) \dot{M}^{\beta}_{\alpha}   
 - (\widetilde\lambda^{\alpha}_{\bar{j}} \Theta^{\bar{i}}_{\alpha} + \widetilde\lambda^{\alpha}_{\bar{j}} Y^{\bar{i}}_i \Theta^i_{\alpha}) \dot{M}^{\bar{j}}_{\bar{i}} \\
& -  e^{W} \widetilde\lambda^{\alpha}_{\bar{i}} \Theta^i_{\alpha} (\dot{M}^{\bar{i}}_i + \epsilon^{\bar{i} \bar{j}}
\epsilon_{i j} \dot{M}^{j}_{\bar{j}}) \, . 
\end{aligned} 
\end{equation}
Finally, 
\begin{equation}
\begin{aligned}
\lambda^{\dot \alpha}_{\bar{i}} \nabla_{\dot\alpha}^{\bar{i}}  \Big|_{\rm{bosonic}} & = 
   - \epsilon_{\dot\alpha \dot\beta} \epsilon^{\alpha \beta} \Theta_{\beta}^{\bar{i}} \widetilde\lambda_{\bar{i}}^{\dot\beta}  \frac{\partial}{\partial X^{\alpha}_{\dot\alpha}}
   - \epsilon_{\dot\alpha \dot\beta} \epsilon^{\alpha \beta} Y^{\bar{j}}_j \Theta^j_{\beta} \widetilde\lambda^{\dot\beta}_{\bar{j}} \frac{\partial}{\partial X^{\alpha}_{\dot\alpha}}
   - \epsilon_{i j} \epsilon^{\bar{i} \bar{j}} \Theta^j_{\dot\beta} \widetilde\lambda^{\dot\beta}_{\bar{j}}
   \frac{\partial}{\partial Y^{\bar{i}}_{i}}
   +  \epsilon_{i j} \epsilon^{\bar{i} \bar{j}} X^{\beta}_{\dot\beta} \Theta^j_{\beta} \widetilde\lambda^{\dot\beta}_{\bar{j}}\frac{\partial}{\partial Y^{\bar{i}}_{i}} \\
& - \frac{1}{2}(\widetilde\lambda^{\dot\alpha}_{\bar{i}} \Theta^{\bar{i}}_{\dot\alpha}  
- \widetilde\lambda^{\dot\alpha}_{\bar{i}} X_{\dot\alpha}^{\alpha} \Theta^{\bar{i}}_{\alpha} +\widetilde\lambda^{\dot\alpha}_{\bar{i}} Y^{\bar{i}}_i \theta^i_{\dot\alpha} 
- X^{\alpha}_{\dot\alpha} Y^{\bar{i}}_i \Theta^i_{\alpha} \widetilde\lambda^{\dot\alpha}_{\bar{i}}) 
\left(\frac{\partial}{\partial Z}+\frac{\partial}{\partial W} \right)  \\
&- ( \widetilde\lambda^{\dot\alpha}_{\bar{i}} Y^{\bar{i}}_i \Theta^i_{\dot\beta}  
- \widetilde\lambda^{\dot\alpha}_{\bar{i}} X^{\alpha}_{\dot\beta} \Theta^{\bar{i}}_{\alpha} - X^{\alpha}_{\dot\beta} Y^{\bar{i}}_i \Theta^i_{\alpha} \widetilde\lambda^{\dot\alpha}_{\bar{i}}) 
\dot{M}^{\dot\beta}_{\dot\alpha} \\
&- ( \widetilde\lambda^{\dot\alpha}_{\bar{j}} \Theta^{\bar{i}}_{\dot\alpha} - \widetilde\lambda^{\dot\alpha}_{\bar{j}} X^{\alpha}_{\dot\alpha} \Theta^{\bar{i}}_{\alpha} +
\widetilde\lambda^{\dot\alpha}_{\bar{j}} Y^{\bar{i}}_i \Theta^i_{\dot\alpha} + X^{\alpha}_{\dot\alpha} Y^{\bar{i}}_i \Theta^i_{\alpha} \widetilde\lambda^{\dot\alpha}_{\bar{j}})
\dot{M}^{\bar{j}}_{\bar{i}} \\
& - e^{-Z}( \widetilde\lambda^{\dot\alpha}_{\bar{i}} \Theta^{\bar{i}}_{\alpha} + \widetilde\lambda^{\dot\alpha}_{\bar{i}} Y^{\bar{i}}_i \Theta^i_{\alpha})(\dot{M}^{\alpha}_{\dot\alpha}+ \epsilon^{\alpha \beta} \epsilon_{\dot\alpha \dot\beta} \dot{M}^{\dot\beta}_{\beta}) \\
& -  e^{W} (\widetilde\lambda^{\dot\alpha}_{\bar{i}} \Theta^i_{\dot\alpha} - \widetilde\lambda^{\dot\alpha}_{\bar{i}} X^{\alpha}_{\dot\alpha} \Theta^i_{\alpha})(\dot{M}^{\bar{i}}_i + \epsilon^{\bar{i} \bar{j}}
\epsilon_{i j} \dot{M}^{j}_{\bar{j}}) \, . 
\end{aligned}
\end{equation}

\subsection{Change of the BRST operator under reparametrizations}
\label{BRSTreparametrizations} 

In this Subsection, we compute the transformation of the BRST operator for arbitrary reparametrizations. 
We expect these general formulas to find future applications.
The BRST transformations of the variables in $g_{\rm{new}}$, derived above by a direct calculation, 
can alternatively be obtained by the formulas here, 
combined with the results for the BRST transformations of the variables 
of $g_{\rm{old}}$ of the Appendix B of \cite{VertexOld}.   
Consider two coset representatives $g(Z)$ and $\widetilde g(\widetilde Z)$ related by the equation
  \begin{equation}\label{cosets}
  \widetilde g(\widetilde Z) = g(Z) h\ ,\quad \quad h\in USp(2,2) \times USp(4) \, . 
  \end{equation}
This defines the change of coordinates $Z = Z(\widetilde Z)$ and $h = h(\widetilde Z)$, and it gives a relation between the vielbein $E$ and spin connection $\Omega$ in both coordinate systems
  \begin{equation}\label{vielbein-spin}
  (E_{\widetilde M} + \Omega_{\widetilde M}) = \frac{\partial Z^{M}}{\partial \widetilde Z^{\widetilde M}}\ h^{-1}(E_{M} + \Omega_{M})h + (h^{-1}\partial_{\widetilde M}h)\quad\Rightarrow\quad\left\{
    \begin{aligned}
    E_{\widetilde M}^A & = \frac{\partial Z^M}{\partial \widetilde Z^{\widetilde M}}\ E_M^B\ h_{B}^A  \\
    \Omega_{\widetilde M}^I & = \frac{\partial Z^M}{\partial \widetilde Z^{\widetilde M}}\ \Omega_{M}^J\ h_J^I + (h^{-1}\partial_{\widetilde M} h)^I
    \end{aligned}
  \right.
  \end{equation}
and we used the definitions $h^{-1} t_B h =: h_B^A t_A$ and $h^{-1} t_J h =: h_J^I t_I$.

The BRST transformations of any coset representative $g$ is obtained by solving an analogous equation to the one in \eqref{BRST} for $g_{\rm{old}}$.  
They can be expressed in terms of the 
inverse vielbein and connections. Given the BRST transformations in two coordinate systems  
\begin{equation}
  g \, \lambda  = \delta_{\lambda} Z^M \partial_M g - g\ \delta \Sigma \, ,  \qquad \widetilde g\ \widetilde\lambda  = \delta_{\widetilde\lambda} \widetilde Z^{\widetilde M} \partial_{\widetilde M} \widetilde g - \widetilde g\ \delta \widetilde\Sigma \, .
  \end{equation}
These two can be related using (\ref{cosets})
  \begin{equation}
  (\delta_{\widetilde\lambda} \widetilde Z^{\widetilde M} \partial_{\widetilde M} \widetilde g - \widetilde g\ \delta \widetilde\Sigma) = g\ h\ \widetilde\lambda  = (\delta_{\lambda} Z^M \partial_M g - g\ \delta \Sigma) h \, , 
  \end{equation}
where on the right hand side of the equation $\lambda := h\widetilde\lambda h^{-1}$. Finally, using the fact that in any coordinate system $\delta\Sigma = \delta Z^M\Omega_M$, and the transformation properties of the vielbein (\ref{vielbein-spin}), we arrive at the transformation
  \begin{equation}
  \delta_{\widetilde\lambda} \widetilde Z^{\widetilde M} = \delta_{\lambda} Z^M\ \frac{\partial \widetilde Z^{\widetilde M}}{\partial Z^M} \, . 
  \end{equation}
Furthermore, the transformation of the spin connection in (\ref{vielbein-spin}) means that 
\begin{equation}
\delta\widetilde\Sigma = h^{-1}\Big(\delta\Sigma -\delta_{\lambda} Z^M (\partial_M h h^{-1}) \Big)h \, , 
\end{equation}
 which ultimately implies
  \begin{equation}
  \delta\widetilde\lambda := [\delta\widetilde\Sigma, \widetilde\lambda] = h^{-1} \Big(\delta\lambda - \left[\delta_{\lambda} Z^M (\partial_M h h^{-1}),\lambda\right] \Big) h \, . 
  \end{equation}

\section{Dilaton-like vertex operators of $g_{\rm{old}}$}
\label{DilatonOldCoset}

In the Subsection 4.2 of \cite{VertexOld}, 
it was shown that $Q$-exact contributions can be added to the
$n=0$ vertex operator constructed using $g_{\rm{old}}$, 
yielding the usual manifestly $PSU(2,2|4)$ invariant expression for the dilaton  
vertex operator -- that is, a vertex operator independent of the fermionic coordinates, see \cite{BerkovitsSimplifying}. %
This expression is given by 
(using the definitions of \eqref{lambdasso8}) 
\begin{equation}
V_{\rm{dilaton}} = (\bar{\lambda}_+ \bar{\lambda}_+) + (\lambda_+ \lambda_+),  
\end{equation}     
where 
\begin{equation}
(\bar\lambda_+ \bar\lambda_+) = \lambda^{\alpha}_{\bar{i}} \lambda^{\beta}_{\bar{j}} \epsilon_{\alpha \beta} \epsilon^{\bar{i} \bar{j}} - \lambda_{\dot{\alpha}}^{i} \lambda_{\dot{\beta}}^{j} \epsilon^{\dot{\alpha} \dot{\beta}} \epsilon_{i j} \, , \quad \quad    
(\lambda_+ \lambda_+) = \lambda^i_{\alpha} \lambda^j_{\beta} \epsilon^{\alpha \beta} \epsilon_{ij} -
\lambda^{\dot\alpha}_{\bar{i}} \lambda^{\dot\beta}_{\bar{j}} \epsilon_{\dot\alpha \dot\beta} \epsilon^{\bar{i} \bar{j}} \, . 
\end{equation}

Notice that in this notation, one of the pure spinor constraints is given by 
\begin{equation}
(\bar\lambda_+ \bar\lambda_+) + (\lambda_+ \lambda_+)-(\bar\lambda_- \bar\lambda_-)-(\lambda_- \lambda_-) =0 \, , 
\label{pureso8}
\end{equation}
where 
\begin{equation}
(\bar\lambda_- \bar\lambda_-) = \lambda^{\dot\alpha}_{i} \lambda^{\dot\beta}_{j} \epsilon_{\dot\alpha \dot\beta} \epsilon^{i j} - \lambda_{\alpha}^{\bar{i}} \lambda_{\beta}^{\bar{j}} \epsilon^{\alpha \beta} \epsilon_{\bar{i} \bar{j}} \, , \quad \quad    
(\lambda_- \lambda_-) = \lambda^{\bar{i}}_{\dot\alpha} \lambda^{\bar{j}}_{\dot\beta} \epsilon^{\dot\alpha \dot\beta} \epsilon_{\bar{i} \bar{j}} -
\lambda^{\alpha}_{i} \lambda^{\beta}_{j} \epsilon_{\alpha \beta} \epsilon^{i j} \, .  
\end{equation}

Using \eqref{pureso8},
the dilaton vertex operator can be written by contracting with the
$\sigma_6$ matrix of \eqref{sigma6} as
\begin{equation}
V_{\rm{dilaton}} = \frac{1}{2} \left( \lambda^I_R \lambda^J_S (\sigma_6)_{IJ} (\sigma_6)^{RS}  +
\lambda_I^R \lambda_J^S (\sigma_6)^{IJ} (\sigma_6)_{RS} \right) \, .
\end{equation}
It is easy to see that $Q \cdot V_{\rm{dilaton}} =0$, since the 
BRST transformations of the pure spinors \eqref{purespinortransformation} are $USp(2,2) \times USp(4)$ rotations that leaves the $\sigma_6$ matrix invariant. 

For $n \neq 0$, 
the vertices that reduce to the dilaton vertex above when
$n=0$ %
were not written down explicitly in \cite{VertexOld}. 
We present them here for completeness.
At zeroth order in
$\theta$ and $\psi$, one has 
\begin{equation}
\begin{aligned}
V_D^0 = e^{n(w+z)} V_{\rm{dilaton}} \, . 
\end{aligned}
\end{equation}
At second order, 
the vertex is given by 
\begin{equation}
\begin{aligned}
V_D^{2} = e^{n(w+z)} n & \, (2  I^2_1 +  \frac{1}{2} I^2_2 + I^3_2  + I^4_2 + I^5_2) \, , \end{aligned}
\end{equation}
where we have defined the following invariants  
\begin{equation}
I^2_1 = \widetilde\theta^i_{\alpha} \widetilde\psi^{\dot\alpha}_{\bar{i}} (\epsilon_{i j} \epsilon^{\beta \alpha} \lambda^j_{\dot\alpha} \lambda^{\bar{i}}_{\beta} + \epsilon_{\dot\beta \dot\alpha} \epsilon^{\bar{i} \bar{j}} \lambda^{\alpha}_{\bar{i}} \lambda^{\dot\beta}_i) \, , 
\end{equation}
\begin{equation}
I^2_2= \widetilde\theta^i_{\alpha} \widetilde\theta^{j}_{\gamma}  (\epsilon_{i j} \epsilon^{\beta \alpha} \lambda^k_{\beta} \lambda^{\gamma}_k + \epsilon_{j k} \epsilon^{\gamma \alpha}  \lambda^k_{\beta} \lambda^{\beta}_i) +
\widetilde\psi^{\dot\alpha}_{\bar{i}}  \widetilde\psi^{\dot\beta}_{\bar{k}} (\epsilon_{\dot\beta \dot\gamma}  \epsilon^{\bar{k} \bar{i}} \lambda^{\bar{j}}_{\dot\alpha} \lambda^{\dot\gamma}_{\bar{j}} + \epsilon_{\dot\alpha \dot\beta} \epsilon^{\bar{k} \bar{i}}  \lambda^{\bar{j}}_{\dot\gamma} \lambda^{\dot\gamma}_{\bar{k}}) \, ,
\nonumber
\end{equation}
\begin{equation}
I^3_2 =  \widetilde\theta ^{i}_{\gamma} \widetilde\theta^j_{\beta} (\epsilon_{i j} \epsilon^{\alpha \beta}  \lambda^{\bar{i}}_{\alpha} \lambda^{\gamma}_{\bar{i}} +  \epsilon_{j k} \epsilon^{\beta \gamma} \lambda^k_{\dot\alpha} \lambda^{\dot\alpha}_i) \, ,  
\nonumber
\end{equation}
\begin{equation}
I^4_2 = \widetilde\theta^i_{\alpha}  \widetilde\psi^{\dot\alpha}_{\bar{i}} (\epsilon_{\dot\alpha \dot\beta}  \epsilon^{\bar{j} \bar{i}} \lambda^{\alpha}_i \lambda^{\dot\beta}_{\bar{j}} - \epsilon_{i j} \epsilon^{\beta \alpha}  \lambda^j_{\beta} \lambda^{\bar{i}}_{\dot\alpha})\, ,  \quad 
I^5_2 =
 \widetilde\psi^{\dot\alpha}_{\bar{i}}  \widetilde\psi^{\dot\beta}_{\bar{k}} (\epsilon_{\dot\alpha \dot\beta}  \epsilon^{\bar{j} \bar{i}} \lambda^{\bar{k}}_{\gamma} \lambda^{\gamma}_{\bar{j}}-
 \epsilon_{\dot\gamma \dot\alpha} \epsilon^{\bar{j} \bar{i}} \lambda^k_{\dot\beta} \lambda^{\dot\gamma}_k) \, . 
 \nonumber
\end{equation}
The vertex at 
forth 
order 
is given by
\begin{equation}
V_D^{4} =V_D^{4,1}  +V_D^{4,2} +V_D^{4,3} \, ,
\end{equation}
where
\begin{equation}
\begin{aligned}
& V_D^{4,1} &= - \frac{1}{24} e^{n(w+z)} n ( \widetilde\theta^i_{\alpha})^4  ((-3+n) (\epsilon_{\bar{i} \bar{j}} \epsilon^{\alpha \beta} \lambda^{\bar{i}}_{\alpha} \lambda^{\bar{j}}_{\beta} + \epsilon_{\dot{\alpha} \dot{\beta}} \epsilon^{i j} \lambda^{\dot{\alpha}}_{i} \lambda^{\dot{\beta}}_{j} -\epsilon_{i j} \epsilon^{\dot\alpha \dot\beta} \lambda^{i}_{\dot\alpha} \lambda^{j}_{\dot\beta} )   \\
&  & + (-9-n) \epsilon_{\alpha \beta} \epsilon^{\bar{i} \bar{j}} \lambda^{\alpha}_{\bar{i}} \lambda^{\beta}_{\bar{j}}  + 2 \epsilon_{\alpha \beta} \epsilon^{i j} \lambda^{\alpha}_{i} \lambda^{\beta}_{j}) \, ,  \\
&V_D^{4,2} &= - \frac{1}{24} e^{n(w+z)} n ( \widetilde\psi_{\bar{i}}^{\dot\alpha})^4 ((-3+n) (\epsilon_{\bar{i} \bar{j}} \epsilon^{\alpha \beta} \lambda^{\bar{i}}_{\alpha} \lambda^{\bar{j}}_{\beta}  +\epsilon_{\dot{\alpha} \dot{\beta}} \epsilon^{i j} \lambda^{\dot{\alpha}}_{i} \lambda^{\dot{\beta}}_{j} -\epsilon_{\alpha \beta} \epsilon^{\bar{i} \bar{j}} \lambda^{\alpha}_{\bar{i}} \lambda^{\beta}_{\bar{j}})  \\
& &+(-9-n) \epsilon_{i j} \epsilon^{\dot\alpha \dot\beta} \lambda^{i}_{\dot\alpha} \lambda^{j}_{\dot\beta}  + 2 \epsilon_{\bar{i} \bar{j}} \epsilon^{\dot\alpha \dot\beta} \lambda^{\bar{i}}_{\dot\alpha} \lambda^{\bar{j}}_{\dot\beta}) \, , 
\end{aligned}
\end{equation}
and 
\begin{equation}
\begin{aligned}
V^{4,3}_D = \frac{1}{2} e^{n(w+z)} n ((3+n) (I^1_4  + \frac{4}{3} I^2_4 - \frac{2}{3}  I^3_4 - I^4_4) + \frac{2}{3} I^5_4 - \frac{2}{3} I^6_4  \\
- (-3+n) (I^{7}_4+ \frac{4}{3} I^8_4 - \frac{4}{3}  I^9_4 +I^{10}_4)) \, . 
\end{aligned} 
\end{equation}
In the expressions above, we have used the definitions 
\begin{equation}
( \widetilde\theta^i_{\alpha})^4  = 12 \,  \widetilde\theta^1_{1}  \widetilde\theta^1_{2}  \widetilde\theta^2_{1}  \widetilde\theta^2_{2} \, ,  \quad \quad
( \widetilde\psi_{\bar{i}}^{\dot\alpha})^4   = 12 \,  \widetilde\psi^{\dot{1}}_{\bar{1}}  \widetilde\psi^{\dot{1}}_{\bar{2}}  \widetilde\psi^{\dot{2}}_{\bar{1}}  \widetilde\psi^{\dot{2}}_{\bar{2}} \, .
\end{equation}
and the invariants with four 
fermionic variables 
are given by 
\begin{equation}
\begin{aligned}
& I^1_4 = \widetilde\theta^i_{\alpha} \widetilde\theta^j_{\beta} \widetilde\psi^{\dot\alpha}_{\bar{i}} \widetilde\psi^{\dot\beta}_{\bar{j}} \epsilon_{i k }  \epsilon_{j l}  \epsilon^{\bar{j} \bar{i}} \epsilon^{ \beta \alpha} \lambda^k_{\dot\alpha} \lambda^l_{\dot\beta} \, , \quad 
&I^2_4 = \widetilde\theta^i_{\alpha} \widetilde\psi^{\dot\alpha}_{\bar{i}} \widetilde\psi^{\dot\beta}_{\bar{j}} \widetilde\psi^{\dot\gamma}_{\bar{k}} \epsilon_{i l} \epsilon_{\dot\beta \dot\gamma} \epsilon^{\bar{k} \bar{i}} \epsilon^{\bar{p} \bar{j}} \lambda^l_{\dot\alpha} \lambda^{\alpha}_{\bar{p}}  \, , \quad \\
&I^3_4 = \widetilde\theta^i_{\alpha} \widetilde\theta^j_{\beta} \widetilde\theta^k_{\gamma} \widetilde\psi^{\dot\alpha}_{\bar{i}} \epsilon_{i m} \epsilon_{j k} \epsilon^{\bar{j} \bar{i}}  \epsilon^{\gamma \alpha} \lambda_{\dot\alpha}^m \lambda^{\beta}_{\bar{j}} \, , \quad 
&I^4_4 = \widetilde\theta^i_{\alpha}  \widetilde\theta^j_{\beta} \widetilde\psi^{\dot\alpha}_{\bar{i}} \widetilde\psi^{\dot\beta}_{\bar{j}} \epsilon_{i j}  \epsilon_{\dot\alpha \dot\beta}  \epsilon^{\bar{k} \bar{j}} \epsilon^{\bar{p} \bar{i}} \lambda^{\beta}_{\bar{k}} \lambda^{\alpha}_{\bar{p}}  \, , \quad \\
&I^5_4 = \widetilde\theta^i_{\alpha} \widetilde\psi^{\dot\alpha}_{\bar{i}} \widetilde\psi^{\dot\beta}_{\bar{j}} \widetilde\psi^{\dot\gamma}_{\bar{k}}  \epsilon_{\dot\beta \dot\gamma} \epsilon^{\bar{k} \bar{i}}  \lambda^{\bar{j}}_{\dot\alpha} \lambda^{\alpha}_i \, ,  
&I^6_4 = \widetilde\theta^i_{\alpha} \widetilde\theta^j_{\beta} \widetilde\theta^k_{\gamma} \widetilde\psi^{\dot\alpha}_{\bar{i}} \epsilon_{j k} \epsilon^{\gamma \alpha} \lambda^{\bar{i}}_{\dot\alpha} \lambda^{\beta}_i \, , \qquad \quad \\
& I^7_4 = 
 \widetilde\theta^i_{\alpha} \widetilde\theta^j_{\beta} \widetilde\psi^{\dot\alpha}_{\bar{i}} \widetilde\psi^{\dot\beta}_{\bar{j}} \epsilon_{\dot\delta \dot\alpha} \epsilon_{\dot\gamma \dot\beta } \epsilon^{\bar{j} \bar{i}} \epsilon^{\beta \alpha} \lambda^{\dot\delta}_i \lambda^{\dot\gamma}_j  \, , \quad 
& I^8_4 =
 \widetilde\theta^i_{\alpha}  \widetilde\psi^{\dot\alpha}_{\bar{i}} \widetilde\psi^{\dot\beta}_{\bar{j}} \widetilde\psi^{\dot\gamma}_{\bar{k}} \epsilon_{\dot\beta \dot\gamma} \epsilon_{\dot\delta \dot\alpha} \epsilon^{\bar{k} \bar{i}} \epsilon^{\delta \alpha} \lambda^{\bar{j}}_{\delta} \lambda^{\dot\delta}_i \, ,  \quad \\
 & I^9_4 = \widetilde\theta^i_{\alpha} \widetilde\theta^j_{\beta} \widetilde\theta^k_{\gamma} \widetilde\psi^{\dot\alpha}_{\bar{i}} \epsilon_{j k} \epsilon_{\dot\gamma \dot\alpha} \epsilon^{\gamma \alpha} \epsilon^{\delta \beta} \lambda^{\bar{i}}_{\delta} \lambda^{\dot\gamma}_i   \, , \quad 
 & I^{10}_4 =  \widetilde\theta^i_{\alpha} \widetilde\theta^j_{\beta} \widetilde\psi^{\dot\alpha}_{\bar{i}} \widetilde\psi^{\dot\beta}_{\bar{j}} \epsilon_{i j}  \epsilon_{\dot\alpha \dot\beta}  \epsilon^{\gamma \beta} \epsilon^{\delta \alpha} \lambda^{\bar{j}}_{\gamma} \lambda^{\bar{i}}_{\delta} \, .
 \end{aligned}
\end{equation}
At sixth order,
the vertex is given by 
\begin{equation}
V_D^6 = V_D^{6,1}+V_D^{6,2}+V_D^{6,3} \, ,
\end{equation}
where 
\begin{equation}
\begin{aligned}
V_D^{6,1} = \frac{n}{12} e^{n(w+z)} (\widetilde\psi_{\bar{i}}^{\dot\alpha})^4 ((-19-2n + n^2) \widetilde\theta^i_{\alpha} \widetilde\theta^j_{\beta}  \epsilon_{i j} \epsilon^{\gamma \beta} \lambda_{\gamma}^{\bar{i}} \lambda^{\alpha}_{\bar{i}}  - (-19+ 2 n + n^2) \widetilde\theta^i_{\alpha} \widetilde\theta^j_{\beta}   \epsilon_{j k} \epsilon^{\beta \alpha} \lambda^k_{\dot\alpha} \lambda^{\dot\alpha}_i) \, ,  \\
V_D^{6,2} = \frac{n}{12} e^{n(w+z)} (\widetilde\theta^{i}_{\alpha})^4  ((-19-2n + n^2) \widetilde\psi^{\dot\alpha}_{\bar{i}} \widetilde\psi^{\dot\beta}_{\bar{j}} \epsilon_{\dot\gamma \dot\alpha} \epsilon^{\bar{j} \bar{i}} \lambda^i_{\dot\beta} \lambda^{\dot\gamma}_i    + (-19+ 2 n + n^2) \widetilde\psi^{\dot\alpha}_{\bar{i}} \widetilde\psi^{\dot\beta}_{\bar{j}} \epsilon_{\dot\alpha \dot\beta}  \epsilon^{\bar{j} \bar{i}} \lambda_{\alpha}^{\bar{k}} \lambda^{\alpha}_{\bar{k}}  )  \, , 
\nonumber
\end{aligned}
\end{equation}
and 
\begin{equation}
V_D^{6,3} =  - \frac{2}{9} e^{n(w+z)} n(-19 + n^2) (I^6_1 + I^6_2) \, , 
\end{equation}
with 
\begin{equation*}
I^6_1 = \widetilde\theta^i_{\alpha} \widetilde\theta^j_{\beta} \widetilde\theta^k_{\gamma} \widetilde\psi^{\dot\alpha}_{\bar{i}} \widetilde\psi^{\dot\beta}_{\bar{j}} \widetilde\psi^{\dot\gamma}_{\bar{k}} \epsilon_{i l} \epsilon_{j k} \epsilon_{\dot\beta \dot\gamma} \epsilon^{\bar{k} \bar{i}} \epsilon^{\gamma \alpha} \epsilon^{\delta \beta} \lambda^l_{\dot\alpha} \lambda^{\bar{j}}_{\delta}  \, , \quad 
I^6_2 =  \widetilde\theta^i_{\alpha} \widetilde\theta^j_{\beta} \widetilde\theta^k_{\gamma} \widetilde\psi^{\dot\alpha}_{\bar{i}} \widetilde\psi^{\dot\beta}_{\bar{k}} \widetilde\psi^{\dot\gamma}_{\bar{l}} \epsilon_{j k} \epsilon_{\dot\beta \dot\gamma}  \epsilon_{\dot\delta \dot\alpha} \epsilon^{\bar{l} \bar{i}} \epsilon^{\bar{p} \bar{k}}  \epsilon^{\gamma \alpha} \lambda^{\beta}_{\bar{p}} \lambda^{\dot\delta}_i \, .   
\end{equation*}
Finally, 
one has 
\begin{equation}
V^8_D = \frac{1}{144} e^{n(w+z)}  (\widetilde\theta^i_{\alpha})^4  (\widetilde\psi_{\bar{i}}^{\dot\alpha})^4  (n^4 (\bar\lambda_- \bar\lambda_-) - 37 n^2 (\bar\lambda_+ \bar\lambda_+)) \, . 
\end{equation}

\section{The infinitesimal super-isometries of $g_{\rm{new}}$} 
\label{psu224ideal}

The super-isometries are the left translations of a coset representative $g$.
In this Appendix, we compute how these super-isometries act infinitesimally on $g_{\rm{new}}$ of \eqref{CosetIdeal}. 
We repeat here the coset representative for the reader's convenience
\begin{equation}
g_{\rm{new}} = e^{\Psi} e^{\Theta} U V  \, , \qquad 
U= e^{X^{\alpha}_{\dot\alpha} K^{\dot\alpha}_{\alpha}} e^{Z \Delta} \, , \qquad  V = 
e^{Y^{\bar{i}}_i K^i_{\bar{i}}} e^{W J} \, . 
\end{equation}
To compute the left translations, we first introduce the notation
\begin{equation}
A = A^{R}_{I}{q}^{I}_{R}\,,\qquad B = B^{I}_{R}{q}^{R}_{I}\,, \qquad 
C= C^{I}_{J}M^{J}_{I}\,,\qquad D = D^{R}_{S}M^{S}_{R}\,,
\end{equation}
and we want to solve for $\delta\Theta$, $\delta\Psi$, $\delta X$, $\delta Y$, $\delta Z$, $\delta W$ and $\Sigma$
in
\begin{equation}
\left[1+A+B+C+D\right] e^{\Psi} e^{\Theta} U V =
\end{equation}
\begin{equation*}
=e^{\Psi + \delta \Psi} e^{\Theta + \delta \Theta}e^{X^{\alpha}_{\dot\alpha} + \delta X^{\alpha}_{\dot\alpha}} e^{Y^{\bar{i}}_{i} + \delta Y^{\bar{i}}_{i}}e^{Z+\delta Z}e^{W+\delta W}\left(1+\Sigma\right)\,,    
\end{equation*}
with 
\begin{equation}
\Sigma\in USp(2,2)\times USp(4) \, ,     
\end{equation}
is a gauge compensator and $A$, $B$, $C$, $D$  are all infinitesimal transformations. 

\subsection{Transformations generated by $A^{R}_{I}{q}^{I}_{R}$}
\label{A}

We start by sending $B=C=D=0$. 
The transformations generated by $A$ are
straightforward to compute, since the
$\Psi$'s  
are %
on the left 
in $g_{\rm{new}}$. The only nonzero transformations generated by $A$ are
\begin{equation}
\delta_{A}\Psi= A \,.  
\end{equation}

\subsection{Transformations generated by $B^{I}_{R}{q}^{R}_{I}$}

The transformations generated by $B$ require some computation
\begin{equation}
\begin{aligned}
B \,   e^{\Psi} e^{\Theta} U V = e^{\Psi} e^{- \Psi} B \,   e^{\Psi} e^{\Theta} U V =
e^{\Psi} \left( B + [B,\Psi]+\frac{1}{2}[[B,\Psi],\Psi] \right) e^{\Theta} U V \, , 
\end{aligned} 
\end{equation}
where the $B$ inside the parentheses will contribute to $\delta_{B}\Theta$, while $\frac{1}{2}[[B,\Psi],\Psi]$ will contribute to $\delta_{B}\Psi$. 
We then need to compute
\begin{equation}
e^{-\Theta} [B,\Psi] e^{\Theta} = [B,\Psi]+[[B,\Psi],\Theta] \, . 
\end{equation}
The term $[B,\Psi]$ above  will contribute to the transformations of the bosonic variables, while 
$[[B,\Psi],\Theta]$ will contribute to $\delta_{B} \Theta$. 
At this point, the transformations of the fermionic variables generated by $B$ are already determined and are given by
\begin{equation}
\label{transformations B of fermions}
\delta_{B}\Psi=\frac{1}{2}[[B,\Psi],\Psi]\,,\qquad \delta_{B} \Theta=B + [[B,\Psi],\Theta]  \, .   
\end{equation}
To work out the transformations of the bosonic variables generated by $B$, we need to split $[B,\Psi]$ as follows
\begin{equation}
[B,\Psi]= [B,\Psi]^{i}_{j}\widehat  M^{j}_{i}+[B,\Psi]^{\overline i}_{\overline j}\widehat  M^{\overline j}_{\overline i}+[B,\Psi]^{\alpha}_{\beta}\widehat  M^{\beta}_{\alpha} +[B,\Psi]^{\dot\alpha}_{\dot\beta}\widehat  M^{\dot\beta}_{\dot\alpha}+   
\label{ThecontributiontoB}
\end{equation}
\begin{equation*}
+[B,\Psi]^{\alpha}_{\dot\alpha}K^{\dot\alpha}_{\alpha}+[B,\Psi]^{\overline i}_{i}K^{i}_{\overline i}+[B,\Psi]^{\Delta}\Delta+[B,\Psi]^{J}J+[B,\Psi]^{\dot\alpha}_{\alpha}P^{\alpha}_{\dot\alpha} +[B,\Psi]^{i}_{\overline i}P^{\overline i}_{i} \, , 
\end{equation*}
and the transformations are the solution of  
\begin{equation}
\left(1+[B,\Psi]\right)e^{X}e^{Z}e^{Y}e^{W}=
\label{eqBallpo} 
\end{equation}
\begin{equation*}
=e^{X+\delta_{B} X}e^{Z+\delta_{B} Z}e^{Y+\delta_{B} Y}e^{W+\delta_{B} W}\left(1+\Sigma_{B}\right)\,.
\end{equation*}

Notice that the bosonic generators organize themselves into an $\mathfrak{su}(2,2) \otimes \mathfrak{su}(4)$ subalgebra, and the computation of the variations above is identical for 
the two bosonic factors.
We give the details of the computation for the $\mathfrak{su}(2,2)$ factor. We want to solve for $\delta_{B} X$, $\delta_{B} Z$ and $\Sigma_{B}|^{{usp}(2,2)}$ in the equation below
\begin{equation}
\left(1+[B,\Psi]|^{\mathfrak{su}(2,2)}\right)e^{X}e^{Z}=e^{X+\delta_{B}X}e^{Z+\delta_{B} Z}   (1+\Sigma_{B}|^{\mathfrak{usp}(2,2)})\, ,
\label{eqforB}
\end{equation}
which is the $\mathfrak{su}(2,2)$ part of \eqref{eqBallpo}. 
Let us consider how the different terms in the decomposition of the commutator $[B,\Psi]$ of \eqref{ThecontributiontoB} contribute. 
\begin{enumerate}

\item 
Clearly, the generators of $\mathfrak{sl}(2,\mathbb{C})$ 
\begin{equation}
[B,\Psi]^{\alpha}_{\beta}\widehat  M^{\beta}_{\alpha} + [B,\Psi]^{\dot\alpha}_{\dot\beta}\widehat  M^{\dot\beta}_{\dot\alpha} \, ,  
\end{equation}
contribute to $\Sigma_{B}|^{\mathfrak{usp}(2,2)}$. They do not contribute to $\delta_{B}Z$ and
they contribute to $\delta_{B}X$ by the commutator
\begin{equation}
[\left( [B,\Psi]^{\alpha}_{\beta}\widehat  M^{\beta}_{\alpha} + [B,\Psi]^{\dot\alpha}_{\dot\beta}\widehat  M^{\dot\beta}_{\dot\alpha} \right),X]\,.    
\end{equation}

\item 

The conformal boots only contribute to $\delta_{B}X$, by adding
\begin{equation}
[B,\Psi]^{\alpha}_{\dot\alpha}K^{\dot\alpha}_{\alpha}\,.    
\end{equation}

\item 
The dilatation part of the commutator  
\begin{equation}
[B,\Psi]^{\Delta}\Delta \, , 
\end{equation}
contributes to $\delta_{B}Z$ by the term above and 
to $\delta_{B}X$ by adding 
\begin{equation}
[[B,\Psi]^{\Delta}\Delta,X]\,.    
\end{equation}

\item Finally, we are left with 
\begin{equation}
[B,\Psi]^{\dot\alpha}_{\alpha}P^{\alpha}_{\dot\alpha} \, . 
\end{equation}
Notice that, 
\begin{equation}
\left(1+[B,\Psi]^{\dot\alpha}_{\alpha}P^{\alpha}_{\dot\alpha}\right)e^{X}e^{Z}=    
\end{equation}
\begin{equation*}
=e^{X}\left(1+[B,\Psi]^{\dot\alpha}_{\alpha}P^{\alpha}_{\dot\alpha}+[[B,\Psi]^{\dot\alpha}_{\alpha}P^{\alpha}_{\dot\alpha},X]+\frac{1}{2}[[[B,\Psi]^{\dot\alpha}_{\alpha}P^{\alpha}_{\dot\alpha},X],X]\right)e^{Z}\,.    
\end{equation*}
The last term inside the parentheses contribute to $\delta_{B}X$.
The third term 
split into 
\begin{equation}
[[B,\Psi]^{\dot\alpha}_{\alpha}P^{\alpha}_{\dot\alpha},X] = \left\{[[B,\Psi]^{\dot\alpha}_{\alpha}P^{\alpha}_{\dot\alpha},X]\right\}^{\Delta}+\left\{[[B,\Psi]^{\dot\alpha}_{\alpha}P^{\alpha}_{\dot\alpha},X]\right\}^{\Sigma} \, ,     
\end{equation}
where the first term on the right hand side 
contributes to $\delta_{B}Z$ while the second
to $\Sigma_{B}|^{\mathfrak{usp}(2,2)}$. We are left with
\begin{equation}
e^{X}\left(1+[B,\Psi]^{\dot\alpha}_{\alpha}P^{\alpha}_{\dot\alpha}\right)e^{Z}=e^{X}e^{Z}\left(1+[B,\Psi]^{\dot\alpha}_{\alpha} \, e^{-Z}P^{\alpha}_{\dot\alpha}e^{Z}\right) \, , 
\end{equation}
so we compensate it by adding 
\begin{equation}
e^{-Z}[B,\Psi]^{\dot\alpha}_{\alpha}\left(P^{\alpha}_{\dot\alpha}+\varepsilon_{\dot\alpha\dot\beta}\varepsilon^{\alpha\beta}K^{\dot\beta}_{\beta}\right)
\end{equation}
to $\Sigma_{B}|^{\mathfrak{usp}(2,2)}$. We are left with
\begin{equation}
e^{X}e^{Z}\left(1 - e^{-Z}\varepsilon^{\dot\alpha\dot\beta}\varepsilon_{\alpha\beta}[B,\Psi]^{\dot\alpha}_{\alpha}K^{\dot\beta}_{\beta}\right)=e^{X}\left(1- e^{-2Z}\varepsilon_{\dot\alpha\dot\beta}\varepsilon^{\alpha\beta}[B,\Psi]^{\dot\alpha}_{\alpha}K^{\dot\beta}_{\beta}\right)e^{Z} \, ,     
\end{equation}
and the second term inside the parentheses on the right-hand side 
contributes to $\delta_{B}X$. 
In this way, we have computed $\delta_{B}X$, $\delta_{B}Z$, and $\Sigma_{B}|^{\mathfrak{usp}(2,2)}$. 
The remaining quantities
$\delta_{B}Y$, $\delta_{B}W$ and $\Sigma_{B}|^{\mathfrak{usp}(4)}$ %
can be obtained by performing the same manipulations.

\end{enumerate}

\noindent {\bf{Transformations generated by B: results}}

\vspace{2mm}

The transformations of the fermionic variables $\delta_{B}\Psi$ and $\delta_{B}\Theta$ are
\begin{equation}
\delta_{B}\Psi=\frac{1}{2}[[B,\Psi],\Psi]\,,\qquad \delta_{B} \Theta=B + [[B,\Psi],\Theta]\,.    
\end{equation}
In what follows the compact notation 
\begin{equation}
[B,\Psi]^{\alpha}_{\beta}\widehat  M^{\beta}_{\alpha} = B_{\Psi}^{\widehat{M}_{\alpha}^{\beta}} \, , 
\end{equation}
will be used and similarly for the other generators. The transformations of the variables $\delta_{B}X$, $\delta_{B}Z$, and $\Sigma_{B}|^{\mathfrak{usp}(2,2)}$ are
\begin{equation}
\begin{aligned}
(\delta_{{B}} X^{\alpha}_{\dot\alpha}) K^{\dot\alpha}_{\alpha} & = 
- e^{2 Z} (B_{\Psi})^{\dot\alpha}_{\alpha} \epsilon^{\alpha \beta} \epsilon_{\dot\alpha \dot\beta} K^{\dot\beta}_{\beta} + B_{\Psi}^{K^{\dot\alpha}_{\alpha}} + [B_{\Psi}^{\widehat{M}^{\alpha}_{\alpha}}+B_{\Psi}^{\widehat{M}^{\dot\alpha}_{\dot\alpha}} + B_{\Psi}^{\Delta},X^{\beta}_{\dot\beta} K^{\dot\beta}_{\beta}] \\ 
& + \frac{1}{2} [ [B_{\Psi}^{K^{\dot\alpha}_{\alpha}},X^{\beta}_{\dot\beta} K^{\dot\beta}_{\beta}],X^{\gamma}_{\dot\gamma} K^{\dot\gamma}_{\gamma}] \, ,    \\
\delta_B Z  & =   B_{\Psi}^{\Delta} + [B_{\Psi}^{P^{\alpha}_{\dot\alpha}},X^{\alpha}_{\dot\alpha} K^{\dot\alpha}_{\alpha}]^{\Delta}\,.
\end{aligned} 
\end{equation}
and 
\begin{equation}
\Sigma_{B}|^{\mathfrak{usp}(2,2)}  = B_{\Psi}^{\widehat{M}^{\alpha}_{\alpha}}+B_{\Psi}^{\widehat{M}^{\dot\alpha}_{\dot\alpha}} + [B_{\Psi}^{P^{\alpha}_{\dot\alpha}}, X^{\beta}_{\dot\beta} K^{\dot\beta}_{\beta}] +
e^{-Z}(B_{\Psi})^{\dot\alpha}_{\alpha}\left(P^{\alpha}_{\dot\alpha}+\varepsilon_{\dot\alpha\dot\beta}\varepsilon^{\alpha\beta}K^{\dot\beta}_{\beta}\right) \, .
\end{equation}

The transformations of the variables $\delta_{B}Y$, $\delta_{B}W$ and $\Sigma_{B}|^{\mathfrak{usp}(4)}$ are
\begin{equation}
\begin{aligned}
(\delta_{{B}\,} Y^{\bar{i}}_{i}) K^{i}_{\bar{i}} & = - e^{- 2 W} (B_{\Psi})^{i}_{\bar{i}} \epsilon^{\bar{i} \bar{j}} \epsilon_{i j} M^{j}_{\bar{i}} + B_{\Psi}^{K^{i}_{\bar{i}}} - [B_{\Psi}^{\widehat{M}^{i}_{i}}-B_{\Psi}^{\widehat{M}^{\bar{i}}_{\bar{i}}} - B_{\Psi}^{J},Y^{\bar{i}}_{i} K^{i}_{\bar{i}}] \\
& + \frac{1}{2} [ [B_{\Psi}^{K^{i}_{\bar{i}}},Y^{\bar{j}}_{j} K^{j}_{\bar{j}}],Y^{\bar{k}}_{k} K^{k}_{\bar{k}}] \, . \\ 
\delta_{B} W &=   B_{\Psi}^{J} + [B_{\Psi}^{M^{\bar{i}}_i},y]^{J}\,,
\end{aligned} 
\end{equation}
and 
\begin{equation}
\begin{aligned}
\Sigma_{B}|^{\mathfrak{usp}(4)} & =   B_{\Psi}^{\widehat{M}^i_i} + B_{\Psi}^{\widehat{M}^{\bar{i}}_{\bar{i}}} + [B_{\Psi}^{P^{\bar{i}}_i}, X^{\bar{i}}_i K^{i}_{\bar{i}}] 
+e^{W}(B_{\Psi})^{i}_{\overline i}\left(P^{\overline i}_{i}+\varepsilon_{ij}\varepsilon^{\overline i\overline j}K^{j}_{\overline j}\right) \,.   
\end{aligned} 
\end{equation}

\subsection{Transformations generated by $C^{I}_{J}M^{J}_{I}$}

The transformations of the fermionic variables generated by $C$ 
are straightforward and are given by
\begin{equation}
\delta_{C}\Psi=[C,\Psi]\,,\qquad \delta_{C}\Theta=[C,\Theta]\,,
\end{equation}
while 
\begin{equation}
\delta_{C}Y=0\,,\qquad \delta_{C}W=0\,,\qquad \Sigma_{C}|^{\mathfrak{usp}(4)}=0\,.    
\end{equation}
To obtain $\delta_{C}X$, $\delta_{C}Z$ and $\Sigma_{C}|^{\mathfrak{usp}(2,2)}$, we need to solve
\begin{equation}
(1+C)e^{X}e^{Z}=e^{X+\delta_{C}X}e^{Z+\delta_{C}Z}\left(1+\Sigma_{C}|^{\mathfrak{usp}(2,2)}\right)\,.   
\end{equation}

Notice that the equation above is similar to \eqref{eqforB}, and we can borrow the results of the previous subsection by simply replacing 
$[B,\Psi] \rightarrow  C$.

\subsection{Transformations generated by $D^{R}_{S}M^{S}_{R}$}

The computation of the transformations generated by $D$ is identical to that of the $C$ 
transformations above. However, the variations $\delta_{D}Y, \delta_{D}W,$ and $\Sigma_{D}|^{\mathfrak{usp}(4)}$ are nonzero now. 
The corresponding formulas are similar to those for the $B$ transformations, with the replacement $[B,\Psi] \rightarrow  D$.

\section{Vertex operators of $g^{\prime}_{\text{new}}$} 
\label{vertexgprime}

To write the remaining contributions to the vertex operators using $g^{\prime}_{\text{new}}$ at maximal powers of $n$, 
we will use the following definitions
\begin{equation}
\widehat\Theta^4  = 12 \widehat\Theta^1_{1} \widehat\Theta^1_{2} \widehat\Theta^2_{1} \widehat\Theta^2_{2} \, ,  \quad \quad
\widehat\Theta^{\prime 4}  = 12 \widehat\Theta_{\dot{1}}^{\bar{1}} \widehat\Theta_{\dot{1}}^{\bar{2}} \widehat\Theta_{\dot{2}}^{\bar{1}} \widehat\Theta_{\dot{2}}^{\bar{2}} \, , \quad \quad 
\widehat\Psi^{4}  = 12 \widehat\Psi^{\dot{1}}_{\bar{1}} \widehat\Psi^{\dot{1}}_{\bar{2}} \widehat\Psi^{\dot{2}}_{\bar{1}} \widehat\Psi^{\dot{2}}_{\bar{2}} \, . 
\end{equation}

Preceding with the transformations of the $g_{\rm{old}}$ vertex operators, 
one obtains the following results at fourth order in the $\Psi$'s and $\Theta$'s and order $n^2$, 
\begingroup
\allowdisplaybreaks
\begin{align}
V^4_{\text{new}^{\prime},n^2} & = -\frac{1}{24} e^{n (W+Z)} n^2(
\widehat\Theta^4 + \widehat\Psi^4) 
(\lambda^{\bar {i}} _ {\alpha}\epsilon^{\alpha\beta}\epsilon_ {\bar \
{i}\bar {j}}\lambda^{\bar {j}} _ {\beta}+\lambda_ {i}^{\dot\alpha}\epsilon_ {\dot\alpha\dot\beta}\epsilon^{i \
j}\lambda_ {j}^{\dot\beta}-\lambda^{i} _ {\alpha}\epsilon^{\alpha\beta}\epsilon_ {i \
j}\lambda^{j} _ {\beta}-
\lambda_ {\bar {i}}^{\dot\alpha}\epsilon_{\dot\alpha\dot\beta}\epsilon^{\bar {i}\bar {j}}\lambda_ {\bar {j}}^{\dot\beta}) \nonumber \\
 & +  
\frac{1}{3} e^{n (W+Z)} n^2 \epsilon ^{\bar{j} \bar{k}} \epsilon _{\dot{\beta } \dot{\gamma }} \widehat{\Theta }_{\dot{\delta }}^{\bar{l}} \widehat{\Psi }_{\bar{j}}^{\dot{\gamma }} \widehat{\Psi }_{\bar{k}}^{\dot{\delta }} \widehat{\Psi }_{\bar{l}}^{\dot{\beta }} (\lambda^{\bar {i}} _ {\alpha}\lambda^{\alpha} _ {\bar {i}} +
\lambda^i_{\dot\alpha} \lambda^{\dot\alpha}_i)
\\
& 
-\frac{1}{4} e^{n (W+Z)} n^2 \epsilon ^{\bar{l} \bar{m}} \epsilon _{\bar{j} \bar{k}} \widehat{\Theta }_{\dot{\beta }}^{\bar{j}} \widehat{\Theta }_{\dot{\gamma }}^{\bar{k}} \widehat{\Psi }_{\bar{l}}^{\dot{\beta }} \widehat{\Psi }_{\bar{m}}^{\dot{\gamma }}(\lambda_ {i}^{\dot\alpha}\epsilon_ {\dot\alpha\dot\beta}\epsilon^{i \
j}\lambda_ {j}^{\dot\beta}+ \lambda^{i} _ {\alpha}\epsilon^{\alpha\beta}\epsilon_ {i j}\lambda^{j}_{\beta} 
-  \lambda_ {\bar {i}}^{\dot\alpha}\epsilon_{\dot\alpha\dot\beta}\epsilon^{\bar {i}\bar {j}}\lambda_ {\bar {j}}^{\dot\beta}
-  \lambda^{\bar {i}} _ {\alpha}\epsilon^{\alpha\beta}\epsilon_ {\bar{i}\bar {j}}\lambda^{\bar {j}} _ {\beta})  \nonumber  \\
& +
\frac{1}{4} e^{n (W+Z)} n^2 \epsilon ^{\dot{\delta } \dot{\rho }} \epsilon _{\dot{\beta } \dot{\gamma }} \widehat{\Theta }_{\dot{\delta }}^{\bar{k}} \widehat{\Theta }_{\dot{\rho }}^{\bar{j}} \widehat{\Psi }_{\bar{j}}^{\dot{\gamma }} \widehat{\Psi }_{\bar{k}}^{\dot{\beta }} (\lambda_ {i}^{\dot\alpha}\epsilon_ {\dot\alpha\dot\beta}\epsilon^{i \
j}\lambda_ {j}^{\dot\beta}+ \lambda^{i} _ {\alpha}\epsilon^{\alpha\beta}\epsilon_ {i j}\lambda^{j}_{\beta} 
-  \lambda_ {\bar {i}}^{\dot\alpha}\epsilon_{\dot\alpha\dot\beta}\epsilon^{\bar {i}\bar {j}}\lambda_ {\bar {j}}^{\dot\beta}
-  \lambda^{\bar {i}} _ {\alpha}\epsilon^{\alpha\beta}\epsilon_ {\bar{i}\bar {j}}\lambda^{\bar {j}} _ {\beta})  \nonumber  \\
& +\frac{1}{2} e^{n (W+Z)} n^2 \epsilon _{i j} (\epsilon ^{\beta  \gamma} \epsilon^{\bar{j} \bar{k}} \epsilon _{l s} \lambda _{\dot{\alpha }}^i \lambda _{\dot{\eta }}^s \widehat{\Theta }_{\beta }^l \widehat{\Theta }_{\gamma }^j \widehat{\Psi }_{\bar{j}}^{\dot{\eta }} \widehat{\Psi }_{\bar{k}}^{\dot{\alpha }}
+ \epsilon ^{\alpha  \gamma} \epsilon^{\bar{l} \bar{m}} \epsilon _{\bar{i} \bar{k}} \lambda _{\alpha }^i \lambda _{\dot{\alpha }}^{\bar{i}} \widehat{\Theta }_{\gamma }^j \widehat{\Theta }_{\dot{\beta }}^{\bar{k}} \widehat{\Psi }_{\bar{l}}^{\dot{\alpha }} \widehat{\Psi }_{\bar{m}}^{\dot{\beta }} )  \nonumber  \\
&-e^{n (W+Z)} n^2 \epsilon ^{\alpha  \gamma}  \epsilon _{i j} ( \epsilon^{\bar{l} \bar{m}} \epsilon _{\bar{i} \bar{k}} \lambda _{\alpha }^{\bar{i}} \lambda _{\dot{\alpha }}^i \widehat{\Theta }_{\gamma }^j \widehat{\Theta }_{\dot{\beta }}^{\bar{k}} \widehat{\Psi }_{\bar{l}}^{\dot{\alpha }} \widehat{\Psi }_{\bar{m}}^{\dot{\beta }} -
  \epsilon^{\dot{\alpha } \dot{\delta }} \epsilon _{\dot{\beta } \dot{\gamma }} \lambda _{\alpha }^{\bar{i}} \lambda _{\dot{\alpha }}^i \widehat{\Theta }_{\gamma }^j \widehat{\Theta }_{\dot{\delta }}^{\bar{j}} \widehat{\Psi }_{\bar{i}}^{\dot{\beta }} \widehat{\Psi }_{\bar{j}}^{\dot{\gamma }} )  \nonumber  \\
& -\frac{1}{2} e^{n (W+Z)} n^2 \epsilon _{\dot{\beta } \dot{\gamma }} (\epsilon ^{\alpha  \gamma} \epsilon^{\dot{\alpha } \dot{\delta }} \epsilon _{i j}  \lambda _{\alpha }^i \lambda _{\dot{\alpha }}^{\bar{i}} \widehat{\Theta }_{\gamma }^j \widehat{\Theta }_{\dot{\delta }}^{\bar{j}} \widehat{\Psi }_{\bar{i}}^{\dot{\beta }} \widehat{\Psi }_{\bar{j}}^{\dot{\gamma }}
-  \epsilon ^{\gamma  \eta} \epsilon^{\alpha  \rho } \epsilon _{j k}  \lambda _{\alpha }^{\bar{i}} \lambda _{\eta }^{\bar{s}} \widehat{\Theta }_{\gamma }^k \widehat{\Theta }_{\rho }^j \widehat{\Psi }_{\bar{i}}^{\dot{\beta }} \widehat{\Psi }_{\bar{s}}^{\dot{\gamma }} )  \nonumber  \\
& + 
\frac{2}{3} e^{n (W+Z)} n^2 \epsilon _{\dot{\beta } \dot{\gamma }} (\epsilon ^{\bar{j} \bar{k}} \epsilon^{\bar{i} \bar{m}} \epsilon _{i j} \lambda _{\bar{i}}^{\alpha } \lambda _{\dot{\alpha }}^i \widehat{\Theta }_{\alpha }^j \widehat{\Psi }_{\bar{j}}^{\dot{\gamma }} \widehat{\Psi }_{\bar{k}}^{\dot{\alpha }} \widehat{\Psi }_{\bar{m}}^{\dot{\beta }}
- \epsilon ^{\bar{l} \bar{m}} \epsilon^{\bar{o} \bar{s}} \epsilon _{\bar{i} \bar{k}}  \lambda _{\alpha }^{\bar{i}} \lambda _{\bar{s}}^{\alpha } \widehat{\Theta }_{\dot{\delta }}^{\bar{k}} \widehat{\Psi }_{\bar{l}}^{\dot{\gamma }} \widehat{\Psi }_{\bar{m}}^{\dot{\delta }} \widehat{\Psi }_{\bar{o}}^{\dot{\beta }} )  \nonumber  \\
& + \frac{1}{2} e^{n (W+Z)} n^2 \lambda _{\alpha }^i (\epsilon ^{\beta  \gamma } \epsilon _{i j} \lambda _s^{\alpha }  \widehat{\Theta }_{\beta }^j \widehat{\Theta }_{\gamma }^s \widehat{\Theta }_{\dot{\beta }}^{\bar{j}} \widehat{\Psi }_{\bar{j}}^{\dot{\beta }}
+ \epsilon ^{\alpha  \gamma } \epsilon _{j k} \lambda _i^{\eta } \widehat{\Theta }_{\gamma }^j \widehat{\Theta }_{\eta }^k \widehat{\Theta }_{\dot{\beta }}^{\bar{j}} \widehat{\Psi }_{\bar{j}}^{\dot{\beta }} )  \nonumber  \\
& -\frac{2}{3} e^{n (W+Z)} n^2 \epsilon ^{\beta  \gamma}  (\epsilon^{\bar{i} \bar{k}} \epsilon _{i j} \epsilon _{l m} \lambda _{\bar{i}}^{\alpha } \lambda _{\dot{\alpha }}^i \widehat{\Theta }_{\alpha }^l \widehat{\Theta }_{\beta }^m \widehat{\Theta }_{\gamma }^j \widehat{\Psi }_{\bar{k}}^{\dot{\alpha }}
- \epsilon ^{\alpha  \rho } \epsilon _{j k} \epsilon _{\dot{\alpha } \dot{\gamma }} \lambda _i^{\dot{\alpha }} \lambda _{\alpha }^{\bar{i}} \widehat{\Theta }_{\beta }^k \widehat{\Theta }_{\gamma }^i \widehat{\Theta }_{\rho }^j \widehat{\Psi }_{\bar{i}}^{\dot{\gamma }} )  \nonumber  \\
& -e^{n (W+Z)} n^2 (\epsilon ^{\alpha  \gamma } \epsilon _{j k} \lambda _{\alpha }^{\bar{i}} \lambda _{\bar{i}}^{\eta } \widehat{\Theta }_{\gamma }^k \widehat{\Theta }_{\eta }^j \widehat{\Theta }_{\dot{\beta }}^{\bar{j}} \widehat{\Psi }_{\bar{j}}^{\dot{\beta }} 
+ \epsilon ^{\bar{j} \bar{k}} \epsilon _{\dot{\alpha } \dot{\gamma }} \lambda _i^{\dot{\alpha }} \lambda _{\bar{i}}^{\alpha } \widehat{\Theta }_{\alpha }^i \widehat{\Theta }_{\dot{\delta }}^{\bar{i}} \widehat{\Psi }_{\bar{j}}^{\dot{\gamma }} \widehat{\Psi }_{\bar{k}}^{\dot{\delta }})  \nonumber   \\
& -\frac{1}{2} e^{n (W+Z)} n^2 ( \epsilon ^{\bar{i} \bar{m}} \epsilon^{\bar{k} \bar{s}} \epsilon _{j k}  \epsilon _{\dot{\beta } \dot{\gamma }} \lambda _{\bar{i}}^{\alpha } \lambda _{\bar{s}}^{\eta } \widehat{\Theta }_{\alpha }^j \widehat{\Theta }_{\eta }^k \widehat{\Psi }_{\bar{k}}^{\dot{\gamma }} \widehat{\Psi }_{\bar{m}}^{\dot{\beta }}
- \epsilon ^{\bar{j} \bar{k}} \epsilon _{\dot{\alpha } \dot{\beta }} \lambda _i^{\alpha } \lambda _{\bar{i}}^{\dot{\alpha }} \widehat{\Theta }_{\alpha }^i \widehat{\Theta }_{\dot{\delta }}^{\bar{i}} \widehat{\Psi }_{\bar{j}}^{\dot{\beta }} \widehat{\Psi }_{\bar{k}}^{\dot{\delta }} )  \nonumber  \\
& +\frac{1}{3} e^{n (W+Z)} n^2 (\epsilon ^{\bar{j} \bar{k}} \epsilon^{\dot{\alpha } \dot{\sigma }} \epsilon _{\dot{\beta } \dot{\eta }} \epsilon _{\dot{\delta } \dot{\rho }} \lambda _{\bar{i}}^{\dot{\eta }} \lambda _{\dot{\alpha }}^{\bar{i}} \widehat{\Theta }_{\dot{\sigma }}^{\bar{l}} \widehat{\Psi }_{\bar{j}}^{\dot{\rho }} \widehat{\Psi }_{\bar{k}}^{\dot{\beta }} \widehat{\Psi }_{\bar{l}}^{\dot{\delta }} 
-\epsilon ^{\bar{l} \bar{m}} \epsilon^{\bar{o} \bar{s}} \epsilon _{\bar{i} \bar{k}} \epsilon _{\dot{\beta } \dot{\gamma }} \lambda _{\bar{s}}^{\dot{\alpha }} \lambda _{\dot{\alpha }}^{\bar{i}} \widehat{\Theta }_{\dot{\delta }}^{\bar{k}} \widehat{\Psi }_{\bar{l}}^{\dot{\gamma }} \widehat{\Psi }_{\bar{m}}^{\dot{\delta }} \widehat{\Psi }_{\bar{o}}^{\dot{\beta }})  \nonumber \\
& -\frac{2}{3} e^{n (W+Z)} n^2 (\epsilon ^{\bar{j} \bar{k}} \epsilon _{\dot{\beta } \dot{\gamma }} \lambda _i^{\dot{\eta }} \lambda _{\dot{\alpha }}^i \widehat{\Theta }_{\dot{\eta }}^{\bar{l}} \widehat{\Psi }_{\bar{j}}^{\dot{\gamma }} \widehat{\Psi }_{\bar{k}}^{\dot{\alpha }} \widehat{\Psi }_{\bar{l}}^{\dot{\beta }}
+ \epsilon ^{\alpha  \gamma} \epsilon^{\bar{j} \bar{k}} \epsilon _{\dot{\beta } \dot{\gamma }} \epsilon _{\dot{\alpha } \dot{\rho }} \lambda _i^{\dot{\alpha }} \lambda _{\alpha }^{\bar{i}} \widehat{\Theta }_{\gamma }^i \widehat{\Psi }_{\bar{i}}^{\dot{\beta }} \widehat{\Psi }_{\bar{j}}^{\dot{\gamma }} \widehat{\Psi }_{\bar{k}}^{\dot{\rho }})  \nonumber  \\
& 
-\frac{1}{2} e^{n (W+Z)} n^2 (\epsilon ^{\beta  \gamma} \epsilon^{\bar{j} \bar{k}} \epsilon _{\dot{\alpha } \dot{\gamma }} \epsilon _{\dot{\eta } \dot{\rho }} \lambda _i^{\dot{\alpha }} \lambda _s^{\dot{\eta }} \widehat{\Theta }_{\beta }^s \widehat{\Theta }_{\gamma }^i \widehat{\Psi }_{\bar{j}}^{\dot{\rho }} \widehat{\Psi }_{\bar{k}}^{\dot{\gamma }} 
+\epsilon ^{\bar{i} \bar{k}} \epsilon^{\dot{\delta } \dot{\sigma }} \epsilon _{\dot{\beta } \dot{\gamma }} \epsilon _{\dot{\alpha } \dot{\delta }} \lambda _i^{\alpha } \lambda _{\bar{i}}^{\dot{\alpha }} \widehat{\Theta }_{\alpha }^i \widehat{\Theta }_{\dot{\sigma }}^{\bar{l}} \widehat{\Psi }_{\bar{k}}^{\dot{\beta }} \widehat{\Psi }_{\bar{l}}^{\dot{\gamma }} )  \nonumber  \\
& + e^{n (W+Z)} n^2 (\epsilon ^{\beta  \gamma } \epsilon _{i j} \lambda _s^{\dot{\alpha }} \lambda _{\dot{\alpha }}^i \widehat{\Theta }_{\beta }^j \widehat{\Theta }_{\gamma }^s \widehat{\Theta }_{\dot{\beta }}^{\bar{j}} \widehat{\Psi }_{\bar{j}}^{\dot{\beta }} 
- \epsilon ^{\bar{i} \bar{k}}  \epsilon _{\dot{\beta } \dot{\gamma }} \lambda _i^{\dot{\alpha }} \lambda _{\bar{i}}^{\alpha } \widehat{\Theta }_{\alpha }^i \widehat{\Theta }_{\dot{\alpha }}^{\bar{l}} \widehat{\Psi }_{\bar{k}}^{\dot{\beta }} \widehat{\Psi }_{\bar{l}}^{\dot{\gamma }}) \, . \nonumber 
\end{align} 
Note that we have written down only the 
$n^2$ contribution to $V^4$. It also has 
a term proportional to $n$.  At sixth order in the ferminic variables and order $n^3$, the result is 
\begingroup
\allowdisplaybreaks
\begin{align}
\allowdisplaybreaks
V^{6}_{\text{new}^{\prime},n^3} & = \frac{1}{24} e^{n (W+Z)} n^3 
 \widehat\Theta^4
\widehat{\Theta }_{\dot{\beta }}^{\bar{j}} \widehat{\Psi }_{\bar{j}}^{\dot{\beta }}
(
\lambda^{\bar{i}}_{\alpha} \epsilon^{\alpha \beta} \epsilon_{\bar{i} \bar{j}} \lambda^{\bar{j}}_{\beta}
+\lambda_{i}^{\dot\alpha} \epsilon_{\dot\alpha \dot\beta} \epsilon^{i j} \lambda_{j}^{\dot\beta}
- \lambda^{i}_{\alpha} \epsilon^{\alpha \beta} \epsilon_{i j} \lambda^{j}_{\beta}
-\lambda_{\bar{i}}^{\dot\alpha} \epsilon_{\dot\alpha \dot\beta} \epsilon^{\bar{i} \bar{j}} \lambda_{\bar{j}}^{\dot\beta})  \\
& + \frac{1}{9} e^{n (W+Z)} n^3  \epsilon^{\bar{l} \bar{m}} \epsilon^{\dot{\delta } \dot{\rho }} \epsilon _{\bar{j} \bar{k}} \epsilon _{\dot{\beta } \dot{\gamma }} \widehat{\Theta }_{\dot{\delta }}^{\bar{k}} \widehat{\Theta }_{\dot{\rho }}^{\bar{n}} \widehat{\Theta }_{\dot{\sigma }}^{\bar{j}} \widehat{\Psi }_{\bar{l}}^{\dot{\gamma }} \widehat{\Psi }_{\bar{m}}^{\dot{\sigma }} \widehat{\Psi }_{\bar{n}}^{\dot{\beta }} \nonumber \\
& \quad \quad \times (
\lambda^{\bar{i}}_{\alpha} \epsilon^{\alpha \beta} \epsilon_{\bar{i} \bar{j}} \lambda^{\bar{j}}_{\beta}
+\lambda_{i}^{\dot\alpha} \epsilon_{\dot\alpha \dot\beta} \epsilon^{i j} \lambda_{j}^{\dot\beta}
- \lambda^{i}_{\alpha} \epsilon^{\alpha \beta} \epsilon_{i j} \lambda^{j}_{\beta}
-\lambda_{\bar{i}}^{\dot\alpha} \epsilon_{\dot\alpha \dot\beta} \epsilon^{\bar{i} \bar{j}} \lambda_{\bar{j}}^{\dot\beta}) \nonumber \\
& -\frac{1}{6} e^{n (W+Z)} n^3 \widehat\Psi^4 (\epsilon ^{\dot{\alpha } \dot{\beta }} \epsilon _{i j} \lambda _{\bar{i}}^{\alpha } \lambda _{\dot{\alpha }}^i \widehat{\Theta }_{\alpha }^j \widehat{\Theta }_{\dot{\beta }}^{\bar{i}}
-  \epsilon ^{\alpha  \gamma } \epsilon _{\bar{i} \bar{k}} \lambda _i^{\dot{\alpha }} \lambda _{\alpha }^{\bar{i}} \widehat{\Theta }_{\gamma }^i \widehat{\Theta }_{\dot{\alpha }}^{\bar{k}} ) \nonumber \\
&-\frac{1}{12} e^{n (W+Z)} n^3  \widehat\Theta^4 (\epsilon ^{\bar{k} \bar{s}} \epsilon _{\dot{\beta } \dot{\gamma }} \lambda _{\alpha }^{\bar{i}} \lambda _{\bar{s}}^{\alpha } \widehat{\Psi }_{\bar{i}}^{\dot{\gamma }} \widehat{\Psi }_{\bar{k}}^{\dot{\beta }}
- \epsilon ^{\bar{j} \bar{k}} \epsilon _{\dot{\gamma } \dot{\eta }} \lambda _i^{\dot{\eta }} \lambda _{\dot{\alpha }}^i \widehat{\Psi }_{\bar{j}}^{\dot{\alpha }} \widehat{\Psi }_{\bar{k}}^{\dot{\gamma }} ) \nonumber \\
& + \frac{1}{24} e^{n (W+Z)} n^3 \widehat\Psi^4 (\epsilon ^{\dot{\beta } \dot{\gamma }} \epsilon _{\bar{i} \bar{k}} \lambda _{\bar{s}}^{\dot{\alpha }} \lambda _{\dot{\alpha }}^{\bar{i}} \widehat{\Theta }_{\dot{\beta }}^{\bar{s}} \widehat{\Theta }_{\dot{\gamma }}^{\bar{k}} +
 \epsilon ^{\dot{\alpha } \dot{\beta }} \epsilon _{\bar{j} \bar{k}} \lambda _{\bar{i}}^{\dot{\eta }} \lambda _{\dot{\alpha }}^{\bar{i}} \widehat{\Theta }_{\dot{\beta }}^{\bar{k}} \widehat{\Theta }_{\dot{\eta }}^{\bar{j}} ) \nonumber \\
& + \frac{1}{12} e^{n (W+Z)} n^3 \widehat\Theta^{\prime 4} (\epsilon ^{\dot{\beta } \dot{\gamma }} \epsilon _{\bar{i} \bar{k}} \lambda _{\alpha }^{\bar{i}} \lambda _{\bar{s}}^{\alpha } \widehat{\Theta }_{\dot{\beta }}^{\bar{s}} \widehat{\Theta }_{\dot{\gamma }}^{\bar{k}} 
+ \epsilon ^{\alpha  \gamma } \epsilon _{j k} \lambda _{\alpha }^{\bar{i}} \lambda _{\bar{i}}^{\eta } \widehat{\Theta }_{\gamma }^k \widehat{\Theta }_{\eta }^j ) \nonumber \\
& + \frac{1}{12} e^{n (W+Z)} n^3 \widehat\Theta^{\prime 4}(
\epsilon ^{\beta  \gamma } \epsilon _{i j} \lambda _s^{\dot{\alpha }} \lambda _{\dot{\alpha }}^i \widehat{\Theta }_{\beta }^j \widehat{\Theta }_{\gamma }^s 
- \epsilon ^{\dot{\alpha } \dot{\beta }} \epsilon _{\bar{j} \bar{k}} \lambda _i^{\dot{\eta }} \lambda _{\dot{\alpha }}^i \widehat{\Theta }_{\dot{\beta }}^{\bar{j}} \widehat{\Theta }_{\dot{\eta }}^{\bar{k}}) \nonumber \\
& 
+ \frac{1}{4} e^{n (W+Z)} n^3 \lambda _{\alpha }^{\bar{i}} \lambda _{\bar{i}}^{\eta } \epsilon ^{\alpha  \gamma} \epsilon _{j k} (\epsilon^{\dot{\delta } \dot{\rho }}  \epsilon _{\dot{\beta } \dot{\gamma }} \widehat{\Theta }_{\gamma }^k \widehat{\Theta }_{\eta }^j \widehat{\Theta }_{\dot{\delta }}^{\bar{k}} \widehat{\Theta }_{\dot{\rho }}^{\bar{j}} \widehat{\Psi }_{\bar{j}}^{\dot{\gamma }} \widehat{\Psi }_{\bar{k}}^{\dot{\beta }}
- \epsilon^{\bar{l} \bar{m}} \epsilon _{\bar{j} \bar{k}} \widehat{\Theta }_{\gamma }^k \widehat{\Theta }_{\eta }^j \widehat{\Theta }_{\dot{\beta }}^{\bar{j}} \widehat{\Theta }_{\dot{\gamma }}^{\bar{k}} \widehat{\Psi }_{\bar{l}}^{\dot{\beta }} \widehat{\Psi }_{\bar{m}}^{\dot{\gamma }}) \nonumber \\
& + \frac{1}{4} e^{n (W+Z)} n^3 \lambda _s^{\dot{\alpha }} \lambda _{\dot{\alpha }}^i \epsilon ^{\beta  \gamma} \epsilon _{i j} ( \epsilon^{\bar{l} \bar{m}} \epsilon _{\bar{j} \bar{k}}  \widehat{\Theta }_{\beta }^j \widehat{\Theta }_{\gamma }^s \widehat{\Theta }_{\dot{\beta }}^{\bar{j}} \widehat{\Theta }_{\dot{\gamma }}^{\bar{k}} \widehat{\Psi }_{\bar{l}}^{\dot{\beta }} \widehat{\Psi }_{\bar{m}}^{\dot{\gamma }} 
- \epsilon ^{\dot{\delta } \dot{\rho }} \epsilon _{\dot{\beta } \dot{\gamma }} \widehat{\Theta }_{\beta }^j \widehat{\Theta }_{\gamma }^s \widehat{\Theta }_{\dot{\delta }}^{\bar{k}} \widehat{\Theta }_{\dot{\rho }}^{\bar{j}} \widehat{\Psi }_{\bar{j}}^{\dot{\gamma }} \widehat{\Psi }_{\bar{k}}^{\dot{\beta }} ) \nonumber \\
& +\frac{1}{8} e^{n (W+Z)} n^3 \lambda _s^{\alpha } \lambda _{\alpha }^i \epsilon ^{\beta  \gamma} \epsilon _{i j} ( \epsilon^{\bar{l} \bar{m}}  \epsilon _{\bar{j} \bar{k}}  \widehat{\Theta }_{\beta }^j \widehat{\Theta }_{\gamma }^s \widehat{\Theta }_{\dot{\beta }}^{\bar{j}} \widehat{\Theta }_{\dot{\gamma }}^{\bar{k}} \widehat{\Psi }_{\bar{l}}^{\dot{\beta }} \widehat{\Psi }_{\bar{m}}^{\dot{\gamma }}
- \epsilon ^{\dot{\delta } \dot{\rho }} \epsilon _{\dot{\beta } \dot{\gamma }} \widehat{\Theta }_{\beta }^j \widehat{\Theta }_{\gamma }^s \widehat{\Theta }_{\dot{\delta }}^{\bar{k}} \widehat{\Theta }_{\dot{\rho }}^{\bar{j}} \widehat{\Psi }_{\bar{j}}^{\dot{\gamma }} \widehat{\Psi }_{\bar{k}}^{\dot{\beta }}) \nonumber \\
& +
\frac{1}{8} e^{n (W+Z)} n^3 \lambda _i^{\eta } \lambda _{\alpha }^i \epsilon ^{\alpha  \gamma} \epsilon _{j k} ( \epsilon^{\bar{l} \bar{m}} \epsilon _{\bar{j} \bar{k}}  \widehat{\Theta }_{\gamma }^j \widehat{\Theta }_{\eta }^k \widehat{\Theta }_{\dot{\beta }}^{\bar{j}} \widehat{\Theta }_{\dot{\gamma }}^{\bar{k}} \widehat{\Psi }_{\bar{l}}^{\dot{\beta }} \widehat{\Psi }_{\bar{m}}^{\dot{\gamma }}
-\epsilon ^{\dot{\delta } \dot{\rho }}  \epsilon _{\dot{\beta } \dot{\gamma }} \widehat{\Theta }_{\gamma }^j \widehat{\Theta }_{\eta }^k \widehat{\Theta }_{\dot{\delta }}^{\bar{k}} \widehat{\Theta }_{\dot{\rho }}^{\bar{j}} \widehat{\Psi }_{\bar{j}}^{\dot{\gamma }} \widehat{\Psi }_{\bar{k}}^{\dot{\beta }} ) \nonumber \\
&+\frac{1}{3} e^{n (W+Z)} n^3 \lambda _{\bar{i}}^{\alpha } \lambda _{\dot{\alpha }}^i \epsilon ^{\beta  \gamma} \epsilon _{i j} (\epsilon^{\bar{i} \bar{k}} \epsilon^{\dot{\alpha } \dot{\delta }}  \epsilon _{l m} \epsilon _{\dot{\beta } \dot{\gamma }}  \widehat{\Theta }_{\alpha }^l \widehat{\Theta }_{\beta }^m \widehat{\Theta }_{\gamma }^j \widehat{\Theta }_{\dot{\delta }}^{\bar{l}} \widehat{\Psi }_{\bar{k}}^{\dot{\beta }} \widehat{\Psi }_{\bar{l}}^{\dot{\gamma }}
- \epsilon ^{\bar{j} \bar{k}}  \epsilon _{l m} \widehat{\Theta }_{\alpha }^l \widehat{\Theta }_{\beta }^m \widehat{\Theta }_{\gamma }^j \widehat{\Theta }_{\dot{\beta }}^{\bar{i}} \widehat{\Psi }_{\bar{j}}^{\dot{\alpha }} \widehat{\Psi }_{\bar{k}}^{\dot{\beta }} ) \nonumber \\
&-\frac{1}{3} e^{n (W+Z)} n^3 \lambda _i^{\dot{\alpha }} \lambda _{\alpha }^{\bar{i}} \epsilon ^{\beta  \gamma}\epsilon^{\alpha  \rho}  \epsilon _{j k} (\epsilon^{\bar{l} \bar{m}}  \epsilon _{\bar{i} \bar{k}} \epsilon _{\dot{\alpha } \dot{\gamma }}  \widehat{\Theta }_{\beta }^k \widehat{\Theta }_{\gamma }^i \widehat{\Theta }_{\rho }^j \widehat{\Theta }_{\dot{\delta }}^{\bar{k}} \widehat{\Psi }_{\bar{l}}^{\dot{\gamma }} \widehat{\Psi }_{\bar{m}}^{\dot{\delta }} +
 \epsilon ^{\dot{\rho } \dot{\sigma }}  \epsilon _{\dot{\beta } \dot{\gamma }} \epsilon _{\dot{\alpha } \dot{\rho }}  \widehat{\Theta }_{\beta }^k \widehat{\Theta }_{\gamma }^i \widehat{\Theta }_{\rho }^j \widehat{\Theta }_{\dot{\sigma }}^{\bar{j}} \widehat{\Psi }_{\bar{i}}^{\dot{\beta }} \widehat{\Psi }_{\bar{j}}^{\dot{\gamma }}) \nonumber \\
 &-\frac{1}{6} e^{n (W+Z)} n^3 \lambda _{\alpha }^i \lambda _{\dot{\alpha }}^{\bar{i}} \epsilon ^{\alpha  \gamma +\bar{l} \bar{m}}  \epsilon _{i j} \epsilon _{\dot{\beta } \dot{\gamma }}  ( \epsilon^{\dot{\delta } \dot{\rho }} \epsilon _{\bar{i} \bar{k}}  \widehat{\Theta }_{\gamma }^j \widehat{\Theta }_{\dot{\delta }}^{\bar{k}} \widehat{\Theta }_{\dot{\rho }}^{\bar{n}} \widehat{\Psi }_{\bar{l}}^{\dot{\gamma }} \widehat{\Psi }_{\bar{m}}^{\dot{\alpha }} \widehat{\Psi }_{\bar{n}}^{\dot{\beta }}
- \epsilon ^{\dot{\alpha } \dot{\delta }}  \epsilon _{\bar{j} \bar{k}}  \widehat{\Theta }_{\gamma }^j \widehat{\Theta }_{\dot{\delta }}^{\bar{j}} \widehat{\Theta }_{\dot{\sigma }}^{\bar{k}} \widehat{\Psi }_{\bar{i}}^{\dot{\beta }} \widehat{\Psi }_{\bar{l}}^{\dot{\gamma }} \widehat{\Psi }_{\bar{m}}^{\dot{\sigma }}) \nonumber \\
& -\frac{1}{6} e^{n (W+Z)} n^3 \lambda _i^{\alpha } \lambda _{\bar{i}}^{\dot{\alpha }} (\epsilon ^{\bar{j} \bar{k}} \epsilon^{\dot{\sigma } \dot{\tau }} \epsilon _{\dot{\alpha } \dot{\beta }} \epsilon _{\dot{\delta } \dot{\rho }}  \widehat{\Theta }_{\alpha }^i \widehat{\Theta }_{\dot{\sigma }}^{\bar{i}} \widehat{\Theta }_{\dot{\tau }}^{\bar{l}} \widehat{\Psi }_{\bar{j}}^{\dot{\rho }} \widehat{\Psi }_{\bar{k}}^{\dot{\beta }} \widehat{\Psi }_{\bar{l}}^{\dot{\delta }}
-  \epsilon ^{\bar{l} \bar{m}} \epsilon^{\bar{i} \bar{o}} \epsilon^{\dot{\delta } \dot{\sigma }} \epsilon _{\bar{j} \bar{k}} \epsilon _{\dot{\beta } \dot{\gamma }} \epsilon _{\dot{\alpha } \dot{\delta }} \widehat{\Theta }_{\alpha }^i \widehat{\Theta }_{\dot{\sigma }}^{\bar{j}} \widehat{\Theta }_{\dot{\tau }}^{\bar{k}} \widehat{\Psi }_{\bar{l}}^{\dot{\gamma }} \widehat{\Psi }_{\bar{m}}^{\dot{\tau }} \widehat{\Psi }_{\bar{o}}^{\dot{\beta }}) \nonumber \\
& + \frac{1}{3} e^{n (W+Z)} n^3 \lambda _{\dot{\alpha }}^i \epsilon _{i j} \epsilon _{\dot{\beta } \dot{\gamma }} (\epsilon ^{\beta  \gamma} \epsilon^{\bar{j} \bar{k}} \epsilon^{\dot{\alpha } \dot{\delta }}  \epsilon _{l s}   \lambda _{\dot{\eta }}^s \widehat{\Theta }_{\beta }^l \widehat{\Theta }_{\gamma }^j \widehat{\Theta }_{\dot{\delta }}^{\bar{l}} \widehat{\Psi }_{\bar{j}}^{\dot{\gamma }} \widehat{\Psi }_{\bar{k}}^{\dot{\eta }} \widehat{\Psi }_{\bar{l}}^{\dot{\beta }} +
\epsilon ^{\alpha  \gamma} \epsilon^{\bar{l} \bar{m}} \epsilon^{\dot{\delta } \dot{\rho }} \epsilon _{\bar{i} \bar{k}} \lambda _{\alpha }^{\bar{i}}  \widehat{\Theta }_{\gamma }^j \widehat{\Theta }_{\dot{\delta }}^{\bar{k}} \widehat{\Theta }_{\dot{\rho }}^{\bar{n}} \widehat{\Psi }_{\bar{l}}^{\dot{\gamma }} \widehat{\Psi }_{\bar{m}}^{\dot{\alpha }} \widehat{\Psi }_{\bar{n}}^{\dot{\beta }} ) \nonumber \\
& -\frac{1}{3} e^{n (W+Z)} n^3 \epsilon ^{\alpha  \gamma}  \epsilon _{\dot{\beta } \dot{\gamma }} \epsilon^{\bar{l} \bar{m}} \lambda _{\alpha }^{\bar{i}} ( \epsilon^{\dot{\alpha } \dot{\delta }} \epsilon _{i j} \epsilon _{\bar{j} \bar{k}} \lambda _{\dot{\alpha }}^i \widehat{\Theta }_{\gamma }^j \widehat{\Theta }_{\dot{\delta }}^{\bar{j}} \widehat{\Theta }_{\dot{\sigma }}^{\bar{k}} \widehat{\Psi }_{\bar{i}}^{\dot{\beta }} \widehat{\Psi }_{\bar{l}}^{\dot{\gamma }} \widehat{\Psi }_{\bar{m}}^{\dot{\sigma }}
+ \epsilon ^{\eta  \rho} \epsilon _{j k} \epsilon _{\bar{i} \bar{k}}   \lambda _{\eta }^{\bar{s}} \widehat{\Theta }_{\gamma }^k \widehat{\Theta }_{\rho }^j \widehat{\Theta }_{\dot{\delta }}^{\bar{k}} \widehat{\Psi }_{\bar{l}}^{\dot{\gamma }} \widehat{\Psi }_{\bar{m}}^{\dot{\delta }} \widehat{\Psi }_{\bar{s}}^{\dot{\beta }}) \nonumber \\
& -\frac{1}{3} e^{n (W+Z)} n^3 \epsilon _{\dot{\beta } \dot{\gamma }} \epsilon^{\bar{j} \bar{k}} \lambda _{\bar{i}}^{\alpha } (\epsilon ^{\beta  \delta}  \epsilon^{\bar{i} \bar{m}} \epsilon _{j k} \epsilon _{\alpha  \beta }  \lambda _{\bar{s}}^{\eta } \widehat{\Theta }_{\delta }^j \widehat{\Theta }_{\eta }^k \widehat{\Theta }_{\dot{\delta }}^{\bar{s}} \widehat{\Psi }_{\bar{j}}^{\dot{\gamma }} \widehat{\Psi }_{\bar{k}}^{\dot{\delta }} \widehat{\Psi }_{\bar{m}}^{\dot{\beta }}
- 
\epsilon ^{\dot{\sigma } \dot{\tau }}  \epsilon _{\dot{\alpha } \dot{\rho }} \lambda _i^{\dot{\alpha }}  \widehat{\Theta }_{\alpha }^i \widehat{\Theta }_{\dot{\sigma }}^{\bar{i}} \widehat{\Theta }_{\dot{\tau }}^{\bar{l}} \widehat{\Psi }_{\bar{j}}^{\dot{\gamma }} \widehat{\Psi }_{\bar{k}}^{\dot{\rho }} \widehat{\Psi }_{\bar{l}}^{\dot{\beta }} ) \nonumber \\
& -\frac{1}{3} e^{n (W+Z)} n^3 \epsilon _{\dot{\beta } \dot{\gamma }} \lambda _i^{\dot{\alpha }}  ( \epsilon ^{\bar{l} \bar{m}} \epsilon^{\bar{i} \bar{o}}  \epsilon _{\bar{j} \bar{k}} \lambda _{\bar{i}}^{\alpha } \widehat{\Theta }_{\alpha }^i \widehat{\Theta }_{\dot{\alpha }}^{\bar{j}} \widehat{\Theta }_{\dot{\rho }}^{\bar{k}} \widehat{\Psi }_{\bar{l}}^{\dot{\gamma }} \widehat{\Psi }_{\bar{m}}^{\dot{\rho }} \widehat{\Psi }_{\bar{o}}^{\dot{\beta }}
+ \epsilon ^{\beta  \gamma} \epsilon^{\bar{j} \bar{k}} \epsilon _{\dot{\alpha } \dot{\rho }} \lambda _s^{\dot{\eta }} \widehat{\Theta }_{\beta }^i \widehat{\Theta }_{\gamma }^s \widehat{\Theta }_{\dot{\eta }}^{\bar{l}} \widehat{\Psi }_{\bar{j}}^{\dot{\gamma }} \widehat{\Psi }_{\bar{k}}^{\dot{\rho }} \widehat{\Psi }_{\bar{l}}^{\dot{\beta }} ) \nonumber \\
& + \frac{2}{9} e^{n (W+Z)} n^3 \epsilon ^{\beta  \gamma} \epsilon^{\alpha  \rho} \epsilon^{\bar{j} \bar{k}} \epsilon _{i j} \epsilon _{l m} \epsilon _{\dot{\beta } \dot{\gamma }} \lambda _{\alpha }^{\bar{i}} \lambda _{\dot{\alpha }}^i \widehat{\Theta }_{\beta }^m \widehat{\Theta }_{\gamma }^j \widehat{\Theta }_{\rho }^l \widehat{\Psi }_{\bar{i}}^{\dot{\beta }} \widehat{\Psi }_{\bar{j}}^{\dot{\gamma }} \widehat{\Psi }_{\bar{k}}^{\dot{\alpha }} \nonumber \\
& -\frac{2}{9} e^{n (W+Z)}  n^3 \epsilon ^{\beta  \gamma} \epsilon^{\bar{j} \bar{k}} \epsilon^{\bar{i} \bar{m}} \epsilon _{j k} \epsilon _{\dot{\beta } \dot{\gamma }} \epsilon _{\dot{\alpha } \dot{\rho }} \lambda _i^{\dot{\alpha }} \lambda _{\bar{i}}^{\alpha } \widehat{\Theta }_{\alpha }^j \widehat{\Theta }_{\beta }^k \widehat{\Theta }_{\gamma }^i \widehat{\Psi }_{\bar{j}}^{\dot{\gamma }} \widehat{\Psi }_{\bar{k}}^{\dot{\rho }} \widehat{\Psi }_{\bar{m}}^{\dot{\beta }} \nonumber \, .
\end{align}
\endgroup
At order eighth and order $n^4$, one has  
\begingroup
\allowdisplaybreaks
\begin{align}
V^{8}_{\text{new}^{\prime},n^4} & =   \frac{1}{144} e^{n (W+Z)} n^4 
\widehat\Theta^4 \widehat\Psi^4 
(\lambda^{\dot\alpha}_i \epsilon^{i j} \epsilon_{\dot\alpha \dot\beta}
\lambda^{\dot\beta}_j  - 
\lambda_{\alpha}^{\bar{i}} \epsilon_{\bar{i} \bar{j}} \epsilon^{\alpha \beta}
\lambda_{\beta}^{\bar{j}} ) \\
& -\frac{1}{144} e^{n (W+Z)} n^4 
\widehat\Theta^{\prime 4} \widehat\Psi^4  
(\lambda^{\dot\alpha}_i \epsilon^{i j} \epsilon_{\dot\alpha \dot\beta}
\lambda^{\dot\beta}_j  - 
\lambda_{\alpha}^{\bar{i}} \epsilon_{\bar{i} \bar{j}} \epsilon^{\alpha \beta}
\lambda_{\beta}^{\bar{j}} 
+ \lambda_{\alpha}^i \epsilon_{i j} \epsilon^{\alpha \beta}
\lambda_{\beta}^j - \lambda^{\dot\alpha}_{\bar{i}} \epsilon^{\bar{i} 
\bar{j}} \epsilon_{\dot\alpha \dot\beta}
\lambda^{\dot\beta}_{\bar{j}}) \nonumber \\
& + \frac{1}{36} e^{n (W+Z)} n^4 
\widehat\Theta^4  \epsilon ^{\bar{j} \bar{k}} \epsilon _{\dot{\beta } \dot{\gamma }} \widehat{\Theta }_{\dot{\delta }}^{\bar{l}} \widehat{\Psi }_{\bar{j}}^{\dot{\gamma }} \widehat{\Psi }_{\bar{k}}^{\dot{\delta }} \widehat{\Psi }_{\bar{l}}^{\dot{\beta }} 
(\lambda^{\bar{i}}_{\alpha}\lambda^{\alpha}_{\bar{i}} - 
\lambda^{i}_{\dot\alpha}\lambda^{\dot\alpha}_{i})
\nonumber \\
& -\frac{1}{18} e^{n (W+Z)} n^4 
\widehat\Psi^4
\lambda _{\alpha }^{\bar{i}} \lambda _{\dot{\alpha }}^i  \epsilon^{\dot{\alpha } \dot{\beta }} \epsilon _{i j} (\epsilon ^{\beta  \gamma} \epsilon^{\alpha  \rho}  \epsilon _{l m} \epsilon _{\bar{i} \bar{k}} \widehat{\Theta }_{\beta }^m \widehat{\Theta }_{\gamma }^j \widehat{\Theta }_{\rho }^l \widehat{\Theta }_{\dot{\beta }}^{\bar{k}} 
+ \epsilon ^{\alpha  \gamma} \epsilon^{\dot{\delta } \dot{\rho }}  \epsilon _{\bar{j} \bar{k}} \epsilon _{\bar{i} \bar{m}}  \widehat{\Theta }_{\gamma }^j \widehat{\Theta }_{\dot{\beta }}^{\bar{j}} \widehat{\Theta }_{\dot{\delta }}^{\bar{k}} \widehat{\Theta }_{\dot{\rho }}^{\bar{m}} ) \nonumber \\
&+\frac{1}{18} e^{n (W+Z)} n^4 \widehat\Theta^{\prime 4}  \lambda _i^{\dot{\alpha }} \lambda _{\bar{i}}^{\alpha } \epsilon ^{\beta  \delta} \epsilon _{\alpha  \beta } ( \epsilon^{\rho  \sigma } \epsilon _{j k}  \widehat{\Theta }_{\delta }^j \widehat{\Theta }_{\rho }^k \widehat{\Theta }_{\sigma }^i \widehat{\Theta }_{\dot{\alpha }}^{\bar{i}}
+ \epsilon^{\dot{\beta } \dot{\gamma }}  \epsilon _{\bar{j} \bar{k}} \widehat{\Theta }_{\delta }^i \widehat{\Theta }_{\dot{\alpha }}^{\bar{j}} \widehat{\Theta }_{\dot{\beta }}^{\bar{k}} \widehat{\Theta }_{\dot{\gamma }}^{\bar{i}} ) \nonumber \\
& -\frac{1}{18} e^{n (W+Z)} n^4\widehat\Theta^4 (\epsilon ^{\bar{l} \bar{m}} \epsilon^{\bar{o} \bar{s}} \epsilon _{\bar{i} \bar{k}} \epsilon _{\dot{\beta } \dot{\gamma }} \lambda _{\alpha }^{\bar{i}} \lambda _{\bar{s}}^{\alpha } \widehat{\Theta }_{\dot{\delta }}^{\bar{k}} \widehat{\Psi }_{\bar{l}}^{\dot{\gamma }} \widehat{\Psi }_{\bar{m}}^{\dot{\delta }} \widehat{\Psi }_{\bar{o}}^{\dot{\beta }}
-\epsilon ^{\bar{j} \bar{k}} \epsilon _{\dot{\beta } \dot{\gamma }} \lambda _i^{\dot{\eta }} \lambda _{\dot{\alpha }}^i \widehat{\Theta }_{\dot{\eta }}^{\bar{l}} \widehat{\Psi }_{\bar{j}}^{\dot{\gamma }} \widehat{\Psi }_{\bar{k}}^{\dot{\alpha }} \widehat{\Psi }_{\bar{l}}^{\dot{\beta }}) \nonumber \\
& +\frac{1}{36} e^{n (W+Z)} n^4 \widehat\Theta^{\prime 4} (\epsilon ^{\alpha  \gamma} \epsilon^{\dot{\alpha } \dot{\beta }} \epsilon^{\dot{\delta } \dot{\rho }} \epsilon _{i j} \epsilon _{\bar{j} \bar{k}} \epsilon _{\bar{i} \bar{m}} \lambda _{\alpha }^i \lambda _{\dot{\alpha }}^{\bar{i}} \widehat{\Theta }_{\gamma }^j \widehat{\Theta }_{\dot{\beta }}^{\bar{j}} \widehat{\Theta }_{\dot{\delta }}^{\bar{k}} \widehat{\Theta }_{\dot{\rho }}^{\bar{m}}
+  \epsilon ^{\dot{\beta } \dot{\delta }} \epsilon^{\dot{\rho } \dot{\sigma }} \epsilon _{\bar{j} \bar{k}} \epsilon _{\dot{\alpha } \dot{\beta }} \lambda _i^{\alpha } \lambda _{\bar{i}}^{\dot{\alpha }} \widehat{\Theta }_{\alpha }^i \widehat{\Theta }_{\dot{\delta }}^{\bar{j}} \widehat{\Theta }_{\dot{\rho }}^{\bar{k}} \widehat{\Theta }_{\dot{\sigma }}^{\bar{i}} ) \nonumber \\
& + \frac{1}{18} e^{n (W+Z)} n^4 \epsilon ^{\beta  \gamma} \epsilon^{\bar{l} \bar{m}} \epsilon^{\dot{\delta } \dot{\rho }} \epsilon _{i j} \epsilon _{\bar{j} \bar{k}} \epsilon _{\dot{\beta } \dot{\gamma }} \lambda _s^{\alpha } \lambda _{\alpha }^i \widehat{\Theta }_{\beta }^j \widehat{\Theta }_{\gamma }^s \widehat{\Theta }_{\dot{\delta }}^{\bar{k}} \widehat{\Theta }_{\dot{\rho }}^{\bar{n}} \widehat{\Theta }_{\dot{\sigma }}^{\bar{j}} \widehat{\Psi }_{\bar{l}}^{\dot{\gamma }} \widehat{\Psi }_{\bar{m}}^{\dot{\sigma }} \widehat{\Psi }_{\bar{n}}^{\dot{\beta }} \nonumber  \\
& + 
\frac{1}{18} e^{n (W+Z)} n^4 
\epsilon ^{\alpha  \gamma} \epsilon^{\bar{l} \bar{m}} \epsilon^{\dot{\delta } \dot{\rho }} \epsilon _{j k} \epsilon _{\bar{j} \bar{k}} \epsilon _{\dot{\beta } \dot{\gamma }} \lambda _i^{\eta } \lambda _{\alpha }^i \widehat{\Theta }_{\gamma }^j \widehat{\Theta }_{\eta }^k \widehat{\Theta }_{\dot{\delta }}^{\bar{k}} \widehat{\Theta }_{\dot{\rho }}^{\bar{n}} \widehat{\Theta }_{\dot{\sigma }}^{\bar{j}} \widehat{\Psi }_{\bar{l}}^{\dot{\gamma }} \widehat{\Psi }_{\bar{m}}^{\dot{\sigma }} \widehat{\Psi }_{\bar{n}}^{\dot{\beta }} \nonumber \\
& + \frac{1}{9} e^{n (W+Z)} n^4 \epsilon ^{\beta  \gamma} \epsilon^{\bar{l} \bar{m}} \epsilon^{\dot{\delta } \dot{\rho }} \epsilon _{i j} \epsilon _{\bar{j} \bar{k}} \epsilon _{\dot{\beta } \dot{\gamma }} \lambda _s^{\dot{\alpha }} \lambda _{\dot{\alpha }}^i \widehat{\Theta }_{\beta }^j \widehat{\Theta }_{\gamma }^s \widehat{\Theta }_{\dot{\delta }}^{\bar{k}} \widehat{\Theta }_{\dot{\rho }}^{\bar{n}} \widehat{\Theta }_{\dot{\sigma }}^{\bar{j}} \widehat{\Psi }_{\bar{l}}^{\dot{\gamma }} \widehat{\Psi }_{\bar{m}}^{\dot{\sigma }} \widehat{\Psi }_{\bar{n}}^{\dot{\beta }} \nonumber \\
& -\frac{1}{9} e^{n (W+Z)} n^4 \epsilon ^{\alpha  \gamma} \epsilon^{\bar{l} \bar{m}} \epsilon^{\dot{\delta } \dot{\rho }} \epsilon _{j k} \epsilon _{\bar{j} \bar{k}} \epsilon _{\dot{\beta } \dot{\gamma }} \lambda _{\alpha }^{\bar{i}} \lambda _{\bar{i}}^{\eta } \widehat{\Theta }_{\gamma }^k \widehat{\Theta }_{\eta }^j \widehat{\Theta }_{\dot{\delta }}^{\bar{k}} \widehat{\Theta }_{\dot{\rho }}^{\bar{n}} \widehat{\Theta }_{\dot{\sigma }}^{\bar{j}} \widehat{\Psi }_{\bar{l}}^{\dot{\gamma }} \widehat{\Psi }_{\bar{m}}^{\dot{\sigma }} \widehat{\Psi }_{\bar{n}}^{\dot{\beta }} \nonumber \, . 
\end{align}
\endgroup
Finally, at order tenth and order $n^5$, we obtain the following results
\begingroup
\allowdisplaybreaks
\begin{align}
V^{10}_{\text{new}^{\prime},n^5} = & \frac{1}{216} e^{n (W+Z)} n^5 \widehat\Theta^4 
\epsilon ^{\bar{l} \bar{m}} \epsilon^{\dot{\delta } \dot{\rho }} \epsilon _{\bar{j} \bar{k}} \epsilon _{\dot{\beta } \dot{\gamma }} \widehat{\Theta }_{\dot{\delta }}^{\bar{k}} \widehat{\Theta }_{\dot{\rho }}^{\bar{n}} \widehat{\Theta }_{\dot{\sigma }}^{\bar{j}} \widehat{\Psi }_{\bar{l}}^{\dot{\gamma }} \widehat{\Psi }_{\bar{m}}^{\dot{\sigma }} \widehat{\Psi }_{\bar{n}}^{\dot{\beta }} \\
& \times (\lambda^{\bar{i}}_{\alpha} \epsilon_{\bar{i} \bar{j}} \epsilon^{\alpha \beta} \lambda^{\bar{j}}_{\beta} + 
\lambda^{\dot{\alpha}}_{i} \epsilon^{ i j} \epsilon_{\dot\alpha \dot\beta} \lambda_{j}^{\dot\beta}
-\lambda_{\bar{i}}^{\alpha} \epsilon^{\bar{i} \bar{j}} \epsilon_{\alpha \beta} \lambda_{\bar{j}}^{\beta} - \lambda_{i}^{\dot\alpha} \epsilon^{i j} \epsilon_{\dot\alpha \dot\beta} \lambda_{j}^{\dot\beta}) \nonumber \\
& + \frac{1}{54} e^{n (W+Z)} n^5 \widehat\Psi^4( \epsilon ^{\beta  \gamma} \epsilon^{\alpha  \rho} \epsilon^ {\dot{\beta } \dot{\gamma }} \epsilon _{j k} \epsilon _{\bar{j} \bar{k}} \epsilon _{\bar{i} \bar{m}} \lambda _i^{\dot{\alpha }} \lambda _{\alpha }^{\bar{i}} \widehat{\Theta }_{\beta }^k \widehat{\Theta }_{\gamma }^i \widehat{\Theta }_{\rho }^j \widehat{\Theta }_{\dot{\alpha }}^{\bar{j}} \widehat{\Theta }_{\dot{\beta }}^{\bar{k}} \widehat{\Theta }_{\dot{\gamma }}^{\bar{m}} \nonumber \\
& \quad \quad \quad \quad  - \epsilon ^{\beta  \gamma} \epsilon^{\dot{\alpha } \dot{\beta }} \epsilon^{\dot{\delta } \dot{\rho }} \epsilon _{i j} \epsilon _{l m} \epsilon _{\bar{j} \bar{k}} \lambda _{\bar{i}}^{\alpha } \lambda _{\dot{\alpha }}^i \widehat{\Theta }_{\alpha }^l \widehat{\Theta }_{\beta }^m \widehat{\Theta }_{\gamma }^j \widehat{\Theta }_{\dot{\beta }}^{\bar{j}} \widehat{\Theta }_{\dot{\delta }}^{\bar{k}} \widehat{\Theta }_{\dot{\rho }}^{\bar{i}}) \nonumber \\
&+ \frac{1}{144} e^{n (W+Z)} n^5 \widehat\Theta^4   \widehat\Psi^4 (\epsilon ^{\dot{\beta } \dot{\gamma }} \epsilon _{\bar{i} \bar{k}} \lambda _{\alpha }^{\bar{i}} \lambda _{\bar{s}}^{\alpha } \widehat{\Theta }_{\dot{\beta }}^{\bar{s}} \widehat{\Theta }_{\dot{\gamma }}^{\bar{k}}+
\epsilon ^{\dot{\alpha } \dot{\beta }} \epsilon _{\bar{j} \bar{k}} \lambda _i^{\dot{\eta }} \lambda _{\dot{\alpha }}^i \widehat{\Theta }_{\dot{\beta }}^{\bar{j}} \widehat{\Theta }_{\dot{\eta }}^{\bar{k}}) \nonumber \\
& -\frac{1}{144} e^{n (W+Z)} n^5 
\widehat\Psi^4  \widehat\Theta^{\prime 4} ( \epsilon ^{\alpha  \gamma } \epsilon _{j k} \lambda _{\alpha }^{\bar{i}} \lambda _{\bar{i}}^{\eta } \widehat{\Theta }_{\gamma }^k \widehat{\Theta }_{\eta }^j
-  \epsilon ^{\beta  \gamma } \epsilon _{i j} \lambda _s^{\dot{\alpha }} \lambda _{\dot{\alpha }}^i \widehat{\Theta }_{\beta }^j \widehat{\Theta }_{\gamma }^s) \nonumber \\
& +\frac{1}{288} e^{n (W+Z)} n^5 \widehat\Psi^4  \widehat\Theta^{\prime 4}( \epsilon ^{\beta  \gamma } \epsilon _{i j} \lambda _s^{\alpha } \lambda _{\alpha }^i \widehat{\Theta }_{\beta }^j \widehat{\Theta }_{\gamma }^s
+ \epsilon ^{\alpha  \gamma } \epsilon _{j k} \lambda _i^{\eta } \lambda _{\alpha }^i \widehat{\Theta }_{\gamma }^j \widehat{\Theta }_{\eta }^k) \, ,  \nonumber
\end{align} 
\endgroup
and at order twelfth and order $n^6$,
\begin{equation}
V^{12}_{\text{new}^{\prime},n^6} = \frac{1}{3456} n^6 e^{n(Z+W)} \widehat\Psi^4\widehat\Theta^4\widehat\Theta^{\prime 4} (\lambda^{\bar{i}}_{\alpha} \epsilon_{\bar{i} \bar{j}} \epsilon^{\alpha \beta} \lambda^{\bar{j}}_{\beta} + 
\lambda^{\dot{\alpha}}_{i} \epsilon^{ i j} \epsilon_{\dot\alpha \dot\beta} \lambda_{j}^{\dot\beta}
-\lambda_{\bar{i}}^{\alpha} \epsilon^{\bar{i} \bar{j}} \epsilon_{\alpha \beta} \lambda_{\bar{j}}^{\beta} - \lambda_{i}^{\dot\alpha} \epsilon^{i j} \epsilon_{\dot\alpha \dot\beta} \lambda_{j}^{\dot\beta}) \, .    
\end{equation}

\end{document}